\documentclass{article}

\usepackage{arxiv}

\usepackage[utf8]{inputenc} % allow utf-8 input
\usepackage[T1]{fontenc}    % use 8-bit T1 fonts
\usepackage{hyperref}       % hyperlinks
\usepackage{url}            % simple URL typesetting
\usepackage{amsfonts,amsmath,amsthm}       % blackboard math symbols
\usepackage{nicefrac}       % compact symbols for 1/2, etc.
\usepackage{graphicx}
\usepackage{subfigure}
\usepackage{color}
\usepackage[english]{babel}
\usepackage{soul}
\newtheorem{definition}{Definition}[section]
\newtheorem{theorem}{Theorem}[section]
\newtheorem{proposition}{Proposition}[section]

\DeclareMathOperator{\Tr}{Tr} 
\DeclareMathOperator{\proj}{proj}
\def\R{\mathbb{R}}

\newcommand{\der}[2]{\frac{\mathrm{d}#1}{\mathrm{d}#2}}
\newcommand{\dder}[2]{\frac{\mathrm{d}^2#1}{{\mathrm{d}#2}^2}}
\newcommand{\pd}[2]{\frac{\partial#1}{\partial#2}}

\newcommand{\wrt}{\, \mathrm{d}}
\newcommand*{\NNet}{\mathrm{NNet}}
\newcommand*{\VPNet}{\mathrm{VPNNs}}
\newcommand*{\SNet}{\mathrm{SympNets}}
\newcommand*{\GNet}{\mathrm{G-SympNet}}
\newcommand*{\LANet}{\mathrm{LA-SympNet}}
\newcommand*{\LSNet}{\mathrm{LocSympNets}}
\newcommand*{\SLSNet}{\mathrm{SymLocSympNets}}
\newcommand*{\NICE}{\mathrm{NICE}}

\title{Locally-symplectic neural networks for learning volume-preserving dynamics \thanks{Article was accepted for publication in the Journal of Computational Physics, \protect\href{https://doi.org/10.1016/j.jcp.2023.111911}{DOI: https://doi.org/10.1016/j.jcp.2023.111911} $\newline$ 
$\copyright$ 2023.~This manuscript version is made available under the CC-BY-NC-ND 4.0 license \protect\url{https://creativecommons.org/licenses/by-nc-nd/4.0/}}}

\author{
 J\={a}nis Baj\={a}rs \\
 Faculty of Physics, Mathematics and Optometry \\
 University of Latvia \\
 Jelgavas Street 3\\
 Riga, LV-1004, Latvia \\
 \texttt{janis.bajars@lu.lv}  
}

\begin{document}
\maketitle

\begin{abstract}
We propose locally-symplectic neural networks $\LSNet$ for learning the flow of phase volume-preserving dynamics. The construction of $\LSNet$ stems from the theorem of the local Hamiltonian description of the divergence-free vector field and the splitting methods based on symplectic integrators. Symplectic gradient modules of the recently proposed symplecticity-preserving neural networks $\SNet$ are used to construct invertible locally-symplectic modules, which compositions result in volume-preserving neural networks $\LSNet$. To further preserve properties of the flow of a dynamical system $\LSNet$ are extended to symmetric locally-symplectic neural networks $\SLSNet$, such that the inverse of $\SLSNet$ is equal to the feed-forward propagation of $\SLSNet$ with the negative time step, which is a general property of the flow of a dynamical system. $\LSNet$ and $\SLSNet$ are studied numerically considering learning linear and nonlinear volume-preserving dynamics. In particular, we demonstrate learning of linear traveling wave solutions to the semi-discretized advection equation, periodic trajectories of the Euler equations of the motion of a free rigid body, and quasi-periodic solutions of the charged particle motion in an electromagnetic field. $\LSNet$ and $\SLSNet$ can learn linear and nonlinear dynamics to a high degree of accuracy even when random noise is added to the training data. In all numerical experiments, $\SLSNet$ have produced smaller errors in long-time predictions compared to the $\LSNet$. When learning a single trajectory of the rigid body dynamics locally-symplectic neural networks can learn both quadratic invariants of the system with absolute relative errors below $1\%$. In addition, $\SLSNet$ produce qualitatively good long-time predictions, when the learning of the whole system from randomly sampled data is considered. $\LSNet$ and $\SLSNet$ can produce accurate short-time predictions of quasi-periodic solutions, which is illustrated in the example of the charged particle motion in an electromagnetic field.
\end{abstract}


\section{Introduction}\label{sec:Intro}
The study of dynamical systems is one of the most fundamental disciplines in mathematics and applied sciences. It has been well recognized that numerical methods for differential equations, which incorporate structural knowledge of the original problem, lead to qualitatively better (long-time) numerical results and have been one of the most inspiring research disciples in the past decades \cite{Hairer}. With emerging the fourth paradigm of scientific discovery \cite{4ParaD}, learning algorithms, such as neural networks, have gained high attention and interest from the development and application point of view \cite{DDbook,Montans19}. Data-driven methods allow, essentially, to learn and explore dynamical systems from given data alone \cite{Toth20,Bondesan19,Yang20}. Incorporation of the prior knowledge of the dynamical system into the data-driven method results in the use of fewer training data samples. In addition to that, structure-preserving neural networks have been shown to generalize better than regular neural networks and produce qualitatively better predictions \cite{Sam19,SympNets,PoisNets,Zhong20,Bertalan19,Xiong21}.  

The dynamical system view of deep learning has recently also gained significant recognition \cite{Weinan17,Chen18,celledoni21}. Structure-preserving, in particular, Hamiltonian dynamics-inspired neural network architectures have been put forward to improve the stability of feed-forward propagation \cite{Haber17,Chang18} and to address the problem of exploding and vanishing gradients in deep learning \cite{Galimberti21,MacDonald21}, where non-vanishing gradients are achieved by design while exploding gradients can be controlled through regularization or avoided for particular neural network architectures. Neural network architectures for learning dynamical systems have also been derived from structure-preserving numerical integrators \cite{Xiong21,Chen20,Zhu20,Tong20}. Deep learning-based integrators may provide competitive alternatives to conventional numerical integrators, e.g., by allowing the use of larger time steps \cite{Kadupitiya20}. In addition to \cite{MacDonald21}, volume preservation by neural networks has also been considered in \cite{NICE}, where $\NICE$ may be considered one of the most known efficient frameworks for learning continuous bijective nonlinear transformations.      

In this work, the emphasis is on learning the flow of phase volume-preserving dynamics with locally-symplectic neural networks $\LSNet$. This work has been inspired by the recent work on symplecticity-preserving neural networks for learning Hamiltonian dynamics: $\SNet$ \cite{SympNets}, and volume-preserving numerical methods based on the local Hamiltonian structure of the divergence-free vector fields and symplectic splitting methods \cite{Feng95}. The construction of the phase volume-preserving numerical methods relies on the theorem by Feng \& Shang \cite{Feng95}, which demonstrates that the right-hand side vector field of a volume-preserving dynamical system, i.e., divergence-free vector field, can be split into the sum of local Hamiltonian vector fields. Thus, compositions of symplectic numerical integrator flow maps applied to each subsystem of the Hamiltonian dynamics lead to phase volume-preserving numerical methods. In general, such numerical methods are implicit, while $\SNet$ neural networks are explicit maps with separable Hamiltonian structures. $\SNet$ are universal approximators and also can learn nonseparable Hamiltonian dynamics, as was demonstrated in \cite{SympNets} for the double-pendulum problem together with the proof of the universal approximation theorem. 

The phase volume-preserving neural networks $\LSNet$ are constructed using symplectic gradient modules of $\SNet$, adapted for modeling the flow of local Hamiltonian systems of the divergence-free vector field of the dynamical system. Derived locally-symplectic modules are phase volume-preserving and, importantly, invertible, with exactly and easily computable inverse maps (modules). There is great flexibility for composing such locally-symplectic modules. For the construction of $\LSNet$, we propose module compositions resembling the volume-preserving numerical splitting methods \cite{Feng95}.

So far, limited attention has been placed on the development of symmetric neural networks mimicking geometric properties of the exact flow of a dynamical system, such that the inverse of the flow is equal to the flow with the reversed direction of time. For example, in numerical integration symmetric methods are obtained from the Lie-Trotter or Strang splitting of exact or numerical flow maps \cite{Hairer}. Since locally-symplectic modules have efficiently computable inverse maps, with the same amount of work as the module computation itself, in this work, we also propose symmetric locally-symplectic neural networks $\SLSNet$. $\SLSNet$ are directly constructed from $\LSNet$, where we form a composition of $\LSNet$ with its adjoint map, which shares the same weight and bias values.

Performance of the proposed neural networks $\LSNet$ and $\SLSNet$ is demonstrated and compared with the volume-preserving neural networks $\VPNet$, which consist of invertible modules adopted from $\NICE$ \cite{NICE} and was proposed in \cite{PoisNets}, considering examples of phase volume-preserving linear and nonlinear dynamics, in particular, the semi-discretized advection equation, the Euler equations of the motion of a free rigid body and the motion of the charged particle in an electromagnetic potential. In this work, a large emphasis is placed on learning the dynamics of the motion of a free rigid body, which is constrained to the unit sphere. A such (quadratic) constraint is not explicitly built into the neural networks $\LSNet$, $\SLSNet$, and $\VPNet$ but learned during the training from provided data.

For the rigid body dynamics, we consider three numerical experiments with the following learning objectives, i.e., to learn a single periodic trajectory, recover stable predictions in learning with noisy data, and demonstrate learning of the whole rigid body dynamics from randomly sampled training data points on the unit sphere. Obtained results demonstrate that the best long-time predictions are obtained with the symmetric phase volume-preserving neural networks $\SLSNet$, even when trained with noisy data. The example of learning linear systems may be viewed in its entirety as an academic example providing an alternative view for the multivariate linear regression in linear algorithms, where the time step is an additional input data. The concluding numerical experiment, concerning learning the motion of the charged particle in an electromagnetic potential, addresses the questions of learning quasi-periodic solutions in higher dimensions with the proposed locally-symplectic neural networks $\LSNet$ and $\SLSNet$.

The manuscript is organized as follows. In Section \ref{sec:VPodes} we describe phase volume-preserving dynamics and discuss the main properties of the Hamiltonian systems. The local Hamiltonian description of the phase volume-preserving dynamics is shown, and the definition of a symmetric map is also provided. Locally-symplectic neural networks $\LSNet$ and $\SLSNet$ are presented in Section \ref{sec:VolNets} with $\VPNet$ being described at the end of this section. The training set-up of neural networks is explained at the beginning of Section \ref{sec:NumResults}. The remaining of Section \ref{sec:NumResults} contains numerical results of learning linear and nonlinear phase volume-preserving dynamics. Discussion and conclusions are provided in Section \ref{sec:Conclusions}.

\section{Volume-preserving dynamics}\label{sec:VPodes}
In this work, we are concerned with learning dynamical systems
\begin{equation}\label{eq:ODE}
\der{y}{t} = f(y), 
\end{equation}
where $y: [0,+\infty) \to \R^n$ is the time-dependent function in $n$-dimensional phase space $\Omega \subset \R^{n}$. With $\der{y}{t}$ we denote ordinary derivative with respect to time variable $t \geq 0$. The dynamical system \eqref{eq:ODE} is said to be phase volume-preserving if
\begin{equation}\label{eq:div}
\nabla \cdot f(y) = 0,
\end{equation}
where $\nabla$ denotes the gradient operator, while $\nabla \cdot$ denotes the divergence of the vector field $f(y)$. Examples of volume-preserving dynamical systems include all linear autonomous systems of differential equations with $\Tr(A)=0$, where $A$ is the system's matrix, see \eqref{eq:VolODE}, canonical Hamiltonian dynamics \eqref{eq:HamQP}, the Euler equations for the motion of a free rigid body \eqref{eq:RBody} \cite{Hairer,Arnold}, and the motion of the charged particle in an electromagnetic potential \eqref{eq:ChargeEq} \cite{PoisNets,Zhu22}.

A special case of \eqref{eq:ODE}, when the system is of even dimension $n=2d$, is the canonical Hamiltonian dynamics
\begin{align}\label{eq:HamQP}
\begin{split}
\der{q}{t} &= \nabla_p H(q,p), \\
\der{p}{t} &= -\nabla_q H(q,p),
\end{split}
\end{align}
where $q,p: [0,+\infty) \to \R^d $ are time-dependent functions, commonly describing position and momentum, respectively, in $2d$-dimensional phase space $\Omega_H \subset \R^{2d}$. The Hamiltonian dynamics \eqref{eq:HamQP} can also be written in the general form:
\begin{equation} \label{eq:HamJ}
\der{z}{t} = J \nabla_z H(z), \quad J = - J^T,
\end{equation}
where $z=(q,p)^T\in\R^{2d}$ and $J$ is a skew-symmetric matrix. In the case of the canonical Hamiltonian dynamics \eqref{eq:HamQP} the matrix $J$ is nonsingular with inverse $J^{-1}=-J$ and is in the following form:
\begin{equation} \label{eq:J}
J = \begin{pmatrix}
0 & I_d \\
-I_d & 0
\end{pmatrix}, 
\end{equation}
where $I_d\in \R^{d \times d}$ is the $d$-dimensional identity matrix. The Hamiltonian $H$ is the first integral of the systems \eqref{eq:HamQP} and \eqref{eq:HamJ}, or simply a conserved quantity, i.e., $\der{H}{t} = 0$. A special case of $H$ is a separable Hamiltonian 
\begin{equation}\label{eq:sepH}
H(q,p) = K(p) + U(q),
\end{equation}
where $K$ and $U$ are commonly the kinetic and potential energies, respectively. It is easy to verify that the right-hand side vector fields of \eqref{eq:HamQP} and \eqref{eq:HamJ} are divergence-free \eqref{eq:div}.  

\subsection{Symplecticity of Hamiltonian dynamics}\label{sec:HamDyn}
The Hamiltonian dynamics \eqref{eq:HamQP}, i.e., \eqref{eq:HamJ} with \eqref{eq:J}, is also known to be symplectic \cite{Arnold}. To define symplecticity we introduce the flow $\phi_t:\Omega_H\to\R^{2d}$ of the Hamiltonian system, which advances the solution at time $t$, i.e., for any given initial condition $(q(0),p(0))^T=(q_0,p_0)^T\in\Omega_H$ $\phi_t(q_0,p_0)=(q(t),p(t))^T$, where $(q(t),p(t))^T$ is the solution of the Hamiltonian system \eqref{eq:HamQP}. The flow $\phi_t$ exists as long as the solution exists to the system of differential equations. It is easy to check that the flow is continuous function of $t$ and $\phi_0$ is the identity map, i.e., $\phi_0(q_0,p_0)=(q_0,p_0)^T$ for all $(q_0,p_0)^T\in\Omega_H$. Since the Hamiltonian system \eqref{eq:HamQP} is invariant to a translation of time, then $\phi_{t} \circ \phi_{s} = \phi_{t+s}$ for all $s,t,t+s\geq 0$ as long the solution exists, where $\circ$ denotes function composition. In addition, if the solution exists for all $t\in\R$, then $\phi_{t} \circ \phi_{-t} = \phi_0$. Hence, the inverse of the flow $\phi_t$ is 
\begin{equation}\label{eq:FlowProp}
\phi_t^{-1}=\phi_{-t}.
\end{equation}
Clearly, all the discussion above applies to the flow $\varphi_t:\Omega\to\R^{n}$ of the dynamical system \eqref{eq:ODE}. 

The introduction of the flows $\phi_t$ and $\varphi_t$ is not only useful for studying dynamical systems but also to construct numerical methods \cite{Hairer} as well as structure-preserving neural networks \cite{SympNets,PoisNets}, which we discuss in more detail in Section \ref{sec:VolNets}. Note that the property \eqref{eq:FlowProp} is not commonly shared by numerical flow maps or regular neural networks approximating the flow $\phi_{t}$. Thus, in learning dynamics, we will address this with the construction of symmetric neural networks in Section \ref{sec:SymLocSympNets}.

We proceed by deriving the variational equation for the Hamiltonian dynamics \eqref{eq:HamQP} in the form \eqref{eq:HamJ} by differentiating the Jacobian of the flow $Y(t)=\frac{\partial \phi_t(z_0)}{\partial z_0 }$ with respect to time, i.e.,
\begin{equation} \label{eq:var}
\der{Y(t)}{t} = J \nabla_{z_0z_0} H(z_0) Y(t),
\end{equation}
for all $z_0=(q_0,p_0)^T\in\Omega_H$, where $J$ is given in \eqref{eq:J} and $\nabla_{z_0z_0} H(z_0)$ is the symmetric Hessian matrix of the Hamiltonian $H(z_0)$. The variational equation \eqref{eq:var} describes the propagation of variations of the initial condition $z_0$ along the dynamics in time. Recently, the variational equation \eqref{eq:var} has attracted great attention in deep learning community and has inspired the definition and construction of neural networks with stable feed-forward propagation \cite{Haber17,Chang18}. 

Assuming that the Hamiltonian $H$ is twice continuously differentiable with respect to $q$ and $p$ on $\Omega_H$, then by the Poincar\'{e} theorem \cite{Hairer} the variational equation \eqref{eq:var} will imply symplecticy of the Hamiltonian dynamics defined as follows.

\begin{definition}
The flow $\phi_t$ of the canonical Hamiltonian system \eqref{eq:HamQP} is symplectic if
\begin{equation} \label{eq:symp}
\frac{\partial \phi_t(z_0)}{\partial z_0 }^T J^{-1} \frac{\partial \phi_t(z_0)}{\partial z_0 } = J^{-1}
\end{equation}
holds for any value of $t$ and $z_0=(q_0,p_0)^T\in\Omega_H$ for which the flow $\phi_t$ is defined.
\end{definition}

\subsection{Phase volume preservation}\label{sec:PhaseVol}
From the symplecticity of the Hamiltonian dynamics \eqref{eq:HamQP} it is easy to see that by computing the determinants of both sides of the equation \eqref{eq:symp} and from the flow property $\phi_0 = \mathrm{id}$, the symplecticity \eqref{eq:symp} implies phase volume preservation in the phase space $\Omega_H$, i.e.,
\begin{equation}\label{eq:VolDet}
\det \left( \frac{\partial \phi_t(z_0)}{\partial z_0 } \right) = 1, \quad \forall \, t, \, z_0. 
\end{equation}

Under the phase volume preservation by the flow $\phi_t$ (or $\varphi_t$) we understand that for any bounded subset $U\subset \Omega_H$ for which $\phi_t(U)$ exists, volumes and orientations of $U$ and $\phi_t(U)$ are the same, i.e.,
\[
\int_U \wrt z_0 = \int_{\phi_t(U)} \wrt z.
\]
From the change of variables rule under the integral sign, for the transformation to be phase volume-preserving, the determinant identity \eqref{eq:VolDet} must hold. Then differentiating \eqref{eq:VolDet} with respect to $t$ and applying Abel-Liouville-Jacobi-Ostrogradskii identity \cite{Hairer} we can show that \eqref{eq:VolDet} holds if the vector field of a dynamical system is divergence-free \eqref{eq:div}.

The contrary is not always the case, i.e., phase volume-preserving dynamics may not be symplectic, especially, when the dynamical system \eqref{eq:ODE} is of an odd dimension. In that case, we can consider the local Hamiltonian description of the phase volume-preserving dynamics \eqref{eq:ODE}, which we describe in the following section.

\subsection{Local Hamiltonian description}\label{sec:LocHam}
The phase volume-preserving dynamics \eqref{eq:ODE} can be described by the local Hamiltonian functions as stated in the following theorem by Feng \& Shang \cite{Feng95}.
\begin{theorem}\label{TheoreFeng}
Every divergence-free vector field $f:\R^n\to\R^n$ can be written as the sum of $n-1$ vector fields
\begin{equation}
f = f_{1,2} + f_{2,3} + \dots + f_{n-1,n},
\end{equation}
where each of $f_{k,k+1}$ is Hamiltonian in the variables $(y_k,y_{k+1})$, i.e., there exist functions $H_{k,k+1}:\R^n\to\R$ such that
\begin{equation}
f_{k,k+1} = \left(0, \dots, 0, \pd{H_{k,k+1}}{y_{k+1}}, -\pd{H_{k,k+1}}{y_{k}}, 0, \dots, 0 \right)^T.
\end{equation}
\end{theorem}

In geometric numerical integration \cite{Hairer} the result of the Theorem \ref{TheoreFeng} has been used to construct phase volume-preserving numerical methods by splitting the dynamical system \eqref{eq:ODE} into $n-1$ subsystems:
\begin{equation}\label{eq:ykODE}
\der{\bar{y}}{t} = f_{k,k+1}(\bar{y}),
\end{equation}
where each subsystem \eqref{eq:ykODE} is solved numerically by a symplectic numerical method \cite{Hairer,Feng95,Xue14}. We say that the flow map $\psi_\tau$, where $\tau>0$ is the time step, of a numerical method is symplectic if it satisfies the symplecticity condition \eqref{eq:symp} for all time step $\tau$ values. 

For example, \eqref{eq:ykODE} could be solved numerically with the symplectic Euler method \cite{Hairer}:
\begin{align}\label{eq:SE}
\begin{split}
\bar{y}_k^{n+1} &= \bar{y}^n_k + \tau \pd{H_{k,k+1}}{\bar{y}_{k+1}}(\bar{y}^{n}_{1}, \dots,\bar{y}^{n}_{k-1}, \bar{y}^{n+1}_{k}, \bar{y}^n_{k+1},\bar{y}^{n}_{k+2}, \dots,\bar{y}^{n}_{n}), \\
\bar{y}_{k+1}^{n+1} &= \bar{y}^n_{k+1} - \tau \pd{H_{k,k+1}}{\bar{y}_{k}}(\bar{y}^{n}_{1}, \dots,\bar{y}^{n}_{k-1}, \bar{y}^{n+1}_{k}, \bar{y}^n_{k+1},\bar{y}^{n}_{k+2}, \dots,\bar{y}^{n}_{n}),\\
\bar{y}_{i}^{n+1} &= \bar{y}^n_{i}, \quad \forall \, i\neq k,k+1, 
\end{split}
\end{align}
which becomes an explicit method if the Hamiltonian function $H_{k,k+1}$ is separable \eqref{eq:sepH} in $(y_k,y_{k+1})$ variables. 

Since the numerical flow map $\psi^{k,k+1}_\tau$ of \eqref{eq:SE} is symplectic in $(y_k,y_{k+1})$ variables, the method is phase volume-preserving, which follows from the calculation: 
\[
\det \left( \frac{\partial \psi^{k,k+1}_\tau(y_0)}{\partial y_0 } \right) =
\det \left( \frac{\partial \psi^{k,k+1}_\tau({y_0}_k,{y_0}_{k+1})}{\partial ({y_0}_k,{y_0}_{k+1}) } \right) = 1, \quad \forall \, \tau, \, y_0. 
\]

Then different compositions of the symplectic numerical flow maps $\psi^{k,k+1}_\tau$, e.g.,
\begin{equation}\label{eq:compos}
\varphi_\tau \approx \psi_\tau = \psi^{1,2}_\tau \circ \psi^{2,3}_\tau \circ \dots \circ \psi^{k,k+1}_\tau \circ \dots \circ \psi^{n-1,n}_\tau,
\end{equation}
are phase volume-preserving maps and approximate the analytic flow $\varphi_t$ with accuracy depending on the method's \eqref{eq:compos} approximation order and the choice of the time step $\tau$. Such an approach has inspired to construct locally-symplectic neural networks $\LSNet$ for learning phase volume-preserving dynamical systems \eqref{eq:ODE} by applying and combining symplecticity-preserving neural network modules for approximating the flow of each subsystem \eqref{eq:ykODE}, see Section \ref{sec:VolNets}.

\subsection{Symmetric maps}\label{sec:SymMap}
It is well known that the property \eqref{eq:FlowProp} of the analytic flow $\phi_t$ is not generally shared by discrete (numerical) invertible flow maps $\psi_\tau$ approximating $\phi_t$. The flow map $\psi_\tau$ is called symmetric or time-reversible \cite{Hairer} if 
\[
\psi_\tau^{-1} = \psi_{-\tau}, \quad \forall \, \tau.
\] 
The flow map $\psi_\tau^{*} := \psi^{-1}_{-\tau}$ is commonly referred to as the adjoint map of the flow map $\psi_\tau$. Thus, the composition of any flow map $\psi_{\tau/2}$ with its adjoint map $\psi_{\tau/2}^{*}$ leads to a symmetric map
\[
\Psi_\tau = \psi_{\tau/2}^* \circ \psi_{\tau/2}
\]  
such that
\[
\Psi_\tau^{-1} = \Psi_{-\tau}, \quad \forall \, \tau.
\]

Note that the composition \eqref{eq:compos} of symmetric flow maps $\psi^{k,k+1}_\tau={\psi^{k,k+1}_\tau}^*$ does not lead to the symmetric flow map $\psi_\tau$ since we obtain that
\begin{align*}
\psi_\tau^* &= \left(\psi^{1,2}_\tau \circ \psi^{2,3}_\tau \circ \dots \circ \psi^{k,k+1}_\tau \circ \dots \circ \psi^{n-1,n}_\tau \right)^* \\
& = {\psi^{n-1,n}_\tau}^* \circ \dots \circ {\psi^{k,k+1}_\tau}^* \circ \dots \circ {\psi^{2,3}_\tau}^* \circ {\psi^{1,2}_\tau}^* \\
& = {\psi^{n-1,n}_\tau} \circ \dots \circ {\psi^{k,k+1}_\tau} \circ \dots \circ {\psi^{2,3}_\tau} \circ {\psi^{1,2}_\tau \neq \psi_\tau}.
\end{align*}
Thus, to obtain a symmetric flow map $\psi_\tau$ we require to compose it with its adjoint map $\psi_\tau^*$. Importantly, the proposed locally-symplectic neural networks $\LSNet$ of the following section are constructed from easily invertible modules and provide efficient means to construct symmetric locally-symplectic phase volume-preserving neural networks $\SLSNet$, see Section \ref{sec:SymLocSympNets}.

\section{Volume-preserving neural networks}\label{sec:VolNets}
In this section, we describe phase volume-preserving neural networks considered in this work. We proceed by recalling and summarizing symplecticity-preserving and Hamiltonian dynamics-inspired neural networks. In this work, we are only concerned with structure-preserving neural networks modeling the flow of a dynamical system.

\subsection{Symplectic neural networks}\label{sec:SympNets}
Recently, in the article \cite{SympNets}, several symplecticity preserving neural network architectures $\SNet$ were put forward illustrating their high abilities for learning Hamiltonian dynamics. Such neural network architectures are built relying on the fact that the composition of symplectic maps is also a symplectic map. Thus, the neural networks are compositions of symplecticity-preserving modules, e.g., gradient modules:
\begin{align}\label{eq:Up}
\begin{split}
Q &= q + h W^T \mbox{diag}(w) \underline{\sigma}(Wp+b), \\
P &= p, 
\end{split}
\end{align}
and
\begin{align}\label{eq:Low}
\begin{split}
Q&= q, \\
P &= p - h W^T \mbox{diag}(w) \underline{\sigma}(Wq+b), 
\end{split}
\end{align}
which are referred to as {\it Up} and {\it Low} modules in \cite{SympNets}, respectively. In the modules \eqref{eq:Up}--\eqref{eq:Low} $(q,p)^T$ and $(Q,P)^T$ of the dimension $2d$ are input and output vectors, respectively, $h$ is a free input parameter and can be taken to be equal to the time step $\tau$. $W\in\R^{m\times d}$ and $w\in\R^m$ are weight matrix and vector, respectively, where $m$ specifies the width of the neural network, and $b\in\R^m$ is a bias vector. With $\mbox{diag}(w)\in\R^{m\times m}$ we define a diagonal matrix containing the weight vector $w$ components on the diagonal. $\sigma$ is a differentiable activation function defined in the vector form, i.e., 
\[
\underline{\sigma}(x)=(\sigma(x_1),\sigma(x_2),...,\sigma(x_m))^T,
\]
where $x\in\R^m$. 

Identifying modules \eqref{eq:Up}--\eqref{eq:Low} as maps $\mathcal{M}_{Up}^h$ and $\mathcal{M}_{Low}^h$, i.e.,
\begin{equation}\label{eq:SympMaps}
(Q,P)^T=\mathcal{M}_{Up}^h(q,p), \quad (Q,P)^T=\mathcal{M}_{Low}^h(q,p), 
\end{equation}
it is easy to verify that both maps \eqref{eq:SympMaps} satisfy the equation \eqref{eq:symp}, i.e.,
\begin{equation*}
\frac{\partial \mathcal{M}_{Up,Low}^h(q,p)}{\partial (q,p) }^T J^{-1} \frac{\partial \mathcal{M}_{Up,Low}^h(q,p)}{\partial (q,p) } = J^{-1},
\end{equation*}
for all $h$, $\sigma$, $W$, $w$ and $b$, with $J$ matrix \eqref{eq:J}.

An alternating composition of modules \eqref{eq:Up}--\eqref{eq:Low} leads to the symplecticity-preserving neural networks $\GNet$, see \cite{SympNets}, and satisfies the universal approximation theorem if the activation function $\sigma$ is sigmoidal, as proven in \cite{SympNets}. Importantly, such modules \eqref{eq:Up}--\eqref{eq:Low} simultaneously approximate a function and its derivative.

A particular composition of the gradient modules \eqref{eq:Up}--\eqref{eq:Low}, which bears the resemblance to the symplectic Euler method \eqref{eq:SE}, leads to the residual type neural network with layer equations in the following form:
\begin{align}\label{eq:SympNets}
\begin{split}
q_{j+1} & = q_j + h W_{p,j}^T \mbox{diag}(w_{p,j}) \underline{\sigma_p}\left( W_{p,j} p_j + b_{p,j}\right),\\
p_{j+1} & = p_j - h W_{q,j}^T \mbox{diag}(w_{q,j}) \underline{\sigma_q}\left( W_{q,j} q_{j+1} + b_{q,j}\right),
\end{split} 
\end{align}
where the index $j=0,1,...,L-1$ refers to the feature values in the $j$'s layer of the network. Thus, in total, we have a neural network composed of $2L$ number of modules with $j=0$ indicating the input data $(q_0,p_0)^T$, while with the index $L$ indicating the output data $(q_L,p_L)^T$. For learning Hamiltonian dynamics \eqref{eq:HamQP} with the neural network \eqref{eq:SympNets}, where for a given state $(q_0,p_0)^T$ we predict the state $(q_L,p_L)^T$ after a time interval $\tau$, we can set $h=\tau$ or $h=\tau/L$. This approach allows us to use $\tau$ as an additional input to the neural network and train with different values of $\tau$, i.e., with irregularly in time sampled data.     
 
Hamiltonian system-inspired neural networks, such as \eqref{eq:SympNets}, in general, with applications to classification problems, have been put forward by several authors \cite{Haber17,Chang18,Galimberti21} to address the questions regarding stable deep learning, exploding and non-vanishing gradient problems. In \cite{SympNets} authors also proposed activation modules with $W_{q,p}=I_d$ and $b_{q,p}=0$. In combination with symplectic linear modules, authors constructed neural networks $\LANet$, which showed superior performance over the gradient networks \eqref{eq:SympNets} in their chosen examples. 

Arguing for stable deep neural networks, i.e., stable feed-forward propagation, authors in \cite{Haber17,Chang18} considered Hamiltonian modules with $\mbox{diag}(w)=I_m$ and the leapfrog network, e.g., with $W_{p}=I_d$, $b_{p}=0$ and the linear activation function $\sigma_p$. In our experiments with $\mbox{diag}(w)=I_m$ for such a simple example as the mathematical pendulum we observed poor performance compared to the method \eqref{eq:SympNets}. This may be explained by looking at the eigenvalues of the system matrix $J \nabla_{z_0z_0} H(z_0)$ in the variational equation \eqref{eq:var}. Imposing $\mbox{diag}(w)=I_m$ leads to matrix $J \nabla_{z_0z_0} H(z_0)$ having purely imaginary eigenvalues \cite{Chang18}, which may guarantee the stability in the feed-forward propagation, but may be in contradiction with an actual Hamiltonian dynamics, where the eigenvalues of $J \nabla_{z_0z_0} H(z_0)$ are not always purely imaginary for all $z_0$ and $H(z_0)$.  

Recently, alternative Hamiltonian neural network architectures inspired by the time-varying Hamiltonian systems were put forward in \cite{Galimberti21}, which guarantee non-vanishing gradients by design, while exploding gradients can be either controlled by introducing regularization or avoided completely. In contrast to learning the flow of the Hamiltonian dynamics, alternatively, the Hamiltonian itself can be learned, see \cite{Sam19}, or the framework of Neural ODEs \cite{Chen18} can be adapted for large-scale nonseparable Hamiltonian systems \cite{Xiong21}.  

\subsection{Locally-symplectic neural networks}\label{sec:LocSympNets}
Based on the symplectic gradient modules \eqref{eq:Up}--\eqref{eq:Low}, in this work, we propose locally-symplectic neural network modules to approximate the flow of the subsystems \eqref{eq:ykODE} for learning phase volume-preserving dynamics \eqref{eq:ODE}. We proceed by defining a projection operator for the state variable $y$, i.e.,
\begin{equation}\label{eq:ProjY}
\hat{y}^k = \proj_k y = (y_1, \dots, y_{k-1},y_{k+1},\dots,y_n)^T \in \R^{n-1}, \quad k=1,\dots,n,
\end{equation}
and a projection operator for the weight matrix $W\in\R^{m \times n-1}$, i.e.,
\begin{equation}\label{eq:ProjW}
\hat{W}^k = \proj_k W = (W_{1k}, \dots, W_{mk})^T \in \R^{m},
\quad k=1,\dots,n-1.
\end{equation}
Then the locally-symplectic modules are defined as follows:
\begin{align}\label{eq:LocUp}
\begin{split}
Y_k &= y_k + h {\hat{W}^k}\,^T \mbox{diag}(w) \underline{\sigma}(W \hat{y}^k + b), \\
Y_i &= y_i, \quad \forall \, i\neq k,
\end{split}
\end{align}
and
\begin{align}\label{eq:LocLow}
\begin{split}
Y_i &= y_i, \quad \forall \, i\neq k+1,\\
Y_{k+1} &= y_{k+1} - h {\hat{W}^k}\,^T \mbox{diag}(w) \underline{\sigma}(W \hat{y}^{k+1} + b), 
\end{split}
\end{align}
where $k=1,\dots,n-1$, $W\in\R^{m\times n-1}$, $w\in\R^m$ and $b\in\R^m$. Each module \eqref{eq:LocUp} or \eqref{eq:LocLow} contains $m(n+1)$ parameter values to be identified. Locally-symplectic modules \eqref{eq:LocUp}--\eqref{eq:LocLow} may also be viewed as the auxiliary measure-preserving modules discussed in \cite{Zhu22}. Notice that in both modules \eqref{eq:LocUp}--\eqref{eq:LocLow} projections \eqref{eq:ProjW} of the weight matrix $W$ are with the same index $k$ and in the special case when $n=2$ we recover symplectic gradient modules \eqref{eq:Up}--\eqref{eq:Low}. In addition, not excluding other components of $y$ in $\hat{y}^k$ \eqref{eq:ProjY} we ensure the necessary coupling between different components of the system \eqref{eq:ODE}. 

Similarly to the symplectic gradient modules \eqref{eq:Up}--\eqref{eq:Low}, we will refer to both modules \eqref{eq:LocUp}--\eqref{eq:LocLow} as $Up$ and $Low$ modules for variable pair $(y_k,y_{k+1})$. Thus, we identify both modules \eqref{eq:LocUp}--\eqref{eq:LocLow} as maps $\mathcal{V}_{k,Up}^h$ and $\mathcal{V}_{k,Low}^h$, i.e.,
\begin{equation}\label{eq:LocMaps}
Y=\mathcal{V}_{k,Up}^h(y), \quad Y=\mathcal{V}_{k,Low}^h(y). 
\end{equation}

We can prove the following proposition.
\begin{proposition}
Maps $\mathcal{V}_{k,Up}^h$ and $\mathcal{V}_{k,Low}^h$ \eqref{eq:LocMaps} are symplectic with respect to variable pair $(y_k,y_{k+1})$ and phase volume-preserving. 
\end{proposition}
\begin{proof}
The Jacobian of the map $\mathcal{V}_{k,Up}^h$ with respect to variable pair $(y_k,y_{k+1})$ is
\[
\pd{\mathcal{V}_{k,Up}^h(y)}{(y_k,y_{k+1})} = 
\begin{pmatrix}
1 & h {\hat{W}^k}\,^T \mbox{diag}(w) \mbox{diag}(\underline{\sigma'}(W \hat{y}^k + b))\hat{W}^k \\
0 & 1
\end{pmatrix}
=: 
\begin{pmatrix}
1 & A \\
0 & 1
\end{pmatrix},
\]
where $\displaystyle A = h \sum_{i=1}^{m} w_i \sigma'(W \hat{y}^k + b)_i \left(\hat{W}_i^k\right)^2 \in\R$ is scalar and the symplecticity condition \eqref{eq:symp}:
\[
\begin{pmatrix}
1 & A \\
0 & 1
\end{pmatrix}^T
\begin{pmatrix}
0 & -1 \\
1 & 0
\end{pmatrix}
\begin{pmatrix}
1 & A \\
0 & 1
\end{pmatrix} = 
\begin{pmatrix}
0 & -1 \\
1 & 0
\end{pmatrix},
\]
is automatically satisfied. Volume preservation follows from the determinant properties and the fact that all diagonal elements of the Jacobian matrix $\pd{\mathcal{V}_{k,Up}^h(y)}{y}$ are equal to one, which completes the proof since the proof for the map $\mathcal{V}_{k,Low}^h$ is identical.
\end{proof}
Notice the importance of the weight vector $w$, which allows for the scalar $A$ to take positive and negative values, if the monotone activation function $\sigma$ is used.

As already stated above, when $n=2$ the locally-symplectic modules \eqref{eq:LocUp}--\eqref{eq:LocLow} coincide with the symplectic gradient modules \eqref{eq:Up}--\eqref{eq:Low}. Unfortunately, when $n>2$ and the objective is to learn Hamiltonian dynamics, then modules \eqref{eq:LocUp}--\eqref{eq:LocLow} may not be suited since they are not symplectic with respect to variables $q$ and $p$, i.e., the Jacobian matrices $\pd{\mathcal{V}_{k,Up}^h(y)}{y}$ and $\pd{\mathcal{V}_{k,Low}^h(y)}{y}$ do not satisfy symplecticity condition \eqref{eq:symp} when the system's dimension $n$ is even. In that case, symplectic neural networks, e.g., as $\SNet$ \cite{SympNets}, are better suited and should be used. 

\begin{figure}[t]
\centering 
\includegraphics[trim=0cm 0cm 0cm 0cm,clip=true,width=0.6\textwidth]{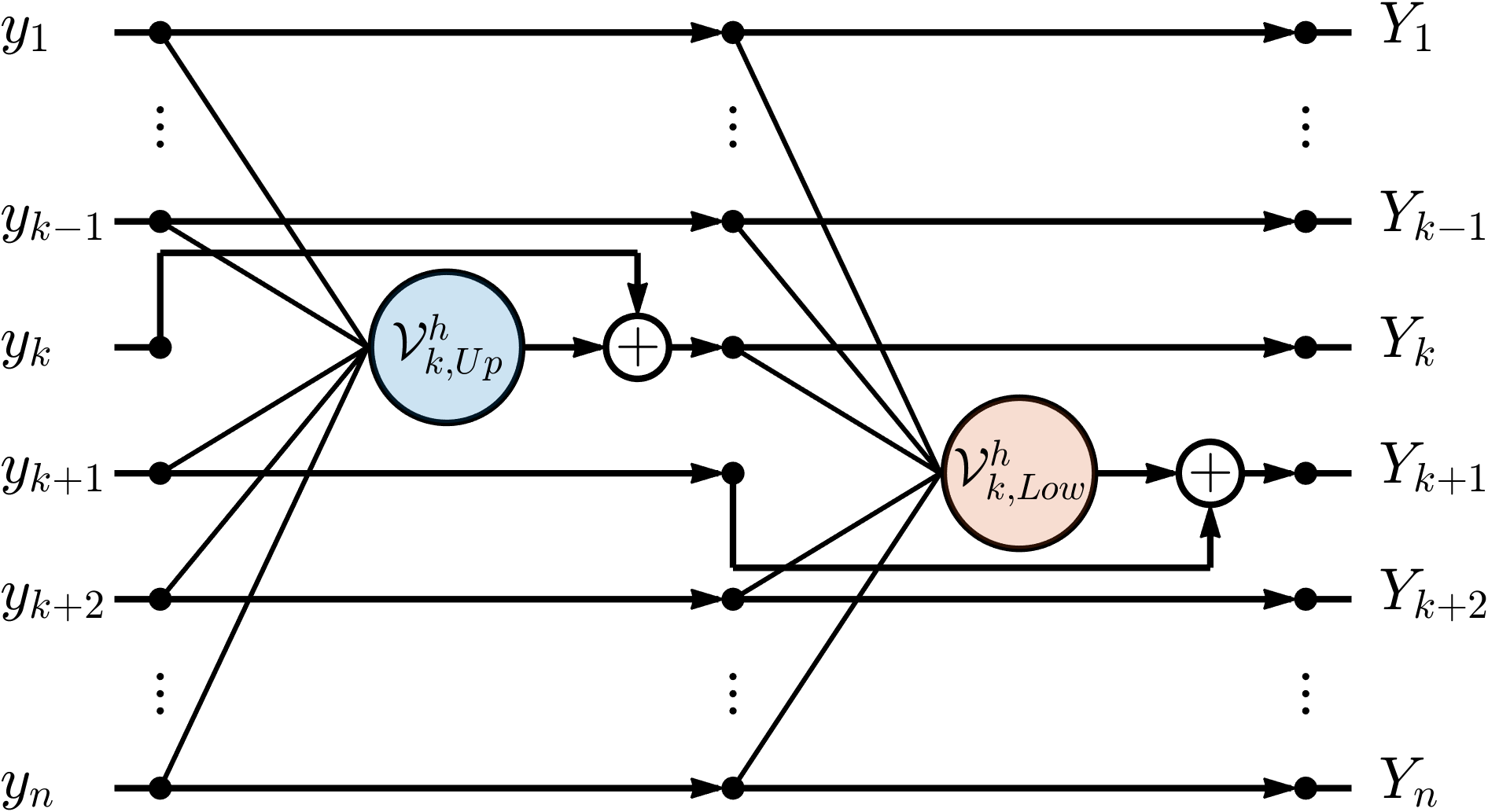}
\caption{Schematic diagram of the combined map $\mathcal{V}_{k}^h$ \eqref{eq:CombUpLow} for index $k$. Arrows indicate the flow of data through locally-symplectic modules $\mathcal{V}_{k,Up}^h$ and $\mathcal{V}_{k,Low}^h$, where $(Y_1,\dots,Y_n)^T$ is the output vector to an input $(y_1,\dots,y_n)^T$.}\label{fig:Diagram}
\end{figure}

Similarly to the symplectic gradient neural networks \eqref{eq:SympNets}, we consider the composition of modules \eqref{eq:LocUp}--\eqref{eq:LocLow}: 
\begin{align*} %\label{eq:LocSympNets}
\begin{split}
Y_k &= y_k + h {\hat{W_1}^k}\,^T \mbox{diag}(w_1) \underline{\sigma}_1(W_1 \hat{y}^k + b_1), \\
Y_{k+1} &= y_{k+1} - h {\hat{W_2}^k}\,^T \mbox{diag}(w_2) \underline{\sigma}_2(W_2 \hat{Y}^{k+1} + b_2), \\
Y_i &= y_i, \quad \forall \, i\neq k, k+1,
\end{split} 
\end{align*}
with the combined map 
\begin{equation}\label{eq:CombUpLow}
\mathcal{V}_{k}^h = \mathcal{V}_{k,Low}^h \circ \mathcal{V}_{k,Up}^h.
\end{equation}
A schematic diagram of the combined map \eqref{eq:CombUpLow} is illustrated in Figure \ref{fig:Diagram}.

Then locally-symplectic phase volume-preserving neural networks $\LSNet$ are constructed of module \eqref{eq:CombUpLow} compositions through index $k$, i.e.,
\begin{equation}\label{eq:LSNet}
\LSNet_h = \left(\mathcal{V}_{1}^h \circ \mathcal{V}_{2}^h \circ\dots \circ \mathcal{V}_{n-1}^h \right) \circ \left(\mathcal{V}_{1}^h \circ \mathcal{V}_{2}^h \circ\dots \circ \mathcal{V}_{n-1}^h \right) \circ {\dots} \circ
\left(\mathcal{V}_{1}^h \circ \mathcal{V}_{2}^h \circ\dots \circ \mathcal{V}_{n-1}^h \right),
\end{equation}
where we take $h=\tau$. If we repeat the composition $\mathcal{V}_{1}^h \circ \mathcal{V}_{2}^h \circ\dots \circ \mathcal{V}_{n-1}^h$ $K$-times, then the total number of neural network layers is $L = 2 (n-1)K$. $\LSNet$ can be exceptionally deep, but recall that in each module \eqref{eq:LocUp}--\eqref{eq:LocLow} we only update one component of $y$. Alternative compositions of \eqref{eq:LSNet} can also be explored, e.g., symmetric locally-symplectic neural networks $\SLSNet$ of the following section. 

\subsection{Symmetric volume-preserving neural networks}\label{sec:SymLocSympNets}
As we already stated in Section \ref{sec:SymMap}, symmetric maps preserve property \eqref{eq:FlowProp} of the analytic flow. We argue and demonstrate in Section \ref{sec:NumResults} that it is also a desirable property to be preserved by the phase volume-preserving neural networks $\LSNet$. This can be efficiently achieved by symmetric phase volume-preserving neural networks $\SLSNet$ defined as
\begin{equation}\label{eq:SymLSNet}
\SLSNet_h = \LSNet_{h/2}^* \circ \LSNet_{h/2},
\end{equation}
where 
\[
\LSNet_{h/2}^* = \left(\mathcal{V}_{n-1}^{{h/2}^*} \circ \mathcal{V}_{n-2}^{{h/2}^*} \circ\dots \circ \mathcal{V}_{1}^{{h/2}^*} \right) \circ {\dots} \circ
\left(\mathcal{V}_{n-1}^{{h/2}^*} \circ \mathcal{V}_{n-2}^{{h/2}^*} \circ\dots \circ \mathcal{V}_{1}^{{h/2}^*} \right)
\]
and
\[
\mathcal{V}_{k}^{{h/2}^*} = \left(\mathcal{V}_{k,Low}^{h/2} \circ \mathcal{V}_{k,Up}^{h/2} \right)^* = 
\mathcal{V}_{k,Up}^{{h/2}^*} \circ \mathcal{V}_{k,Low}^{{h/2}^*} = 
\mathcal{V}_{k,Up}^{{h/2}} \circ \mathcal{V}_{k,Low}^{{h/2}},
\]
for all $k=1,\dots,n-1$, since both locally-symplectic modules \eqref{eq:LocUp}--\eqref{eq:LocLow} are symmetric. The same is true for the symplecticity-preserving gradient modules \eqref{eq:Up}--\eqref{eq:Low}.   

Importantly, in the definition of $\SLSNet_h$ \eqref{eq:SymLSNet} the composition is of the same neural network $\LSNet_{h/2}$, i.e., with the same weight and bias values. Thus, $\SLSNet$ have the same number of parameter values as the neural networks $\LSNet$ for the specified network $K$ and $m$ values, despite having twice the number of layers $L$. Calculation of $\LSNet_{h/2}^*$ only requires one forward propagation of $\LSNet_{h/2}$ performed backward.
   
\subsection{Volume-preserving modules of $\NICE$ and $\VPNet$}\label{sec:ModNice}
To enrich the presentation of the $\LSNet$ and $\SLSNet$ performance, in Section \ref{sec:RigidBody} we compare both neural networks with adopted $\NICE$ \cite{NICE} volume-preserving neural networks $\VPNet$ composed of $\mathrm{L}$ alternating {\it Up} and {\it Low} invertible modules \cite{PoisNets}:
\begin{equation}\label{eq:NICEmodules}
\mathcal{U}_{Up}\begin{pmatrix} y^1 \\ y^2\end{pmatrix} = 
\begin{pmatrix}y^1 + \NNet_1(y^2) \\ y^2 \end{pmatrix}, \quad
\mathcal{U}_{Low}\begin{pmatrix} y^1 \\ y^2\end{pmatrix} = 
\begin{pmatrix}y^1 \\ y^2 + \NNet_2(y^1) \end{pmatrix},
\end{equation}
respectively, where $y^1\in\R^{s}$ and $y^2\in\R^{n-s}$ are partitions of the state vector $y\in\R^n$ with the partition dimension $s$, i.e., $y = (y^1,y^2)^T$, and $\NNet_1: \R^{n-s} \to \R^{s}$ and $\NNet_2: \R^{s} \to \R^{n-s}$ are fully-connected neural networks with $l$ number of hidden layers of feature dimension $m$. Similarly to the $\LSNet$ and $\SLSNet$, we consider the sigmoid activation function in $\NNet_{1,2}$. The module $\mathcal{U}_{Up}$ contains $[n+1+(m+1)(l-1)]m+s$ parameter values to be determined, while the module $\mathcal{U}_{Low}$ contains $[n+1+(m+1)(l-1)]m+n-s$ parameter values.

In the following section, we consider numerical experiments, where we train $\LSNet$ \eqref{eq:LSNet}, $\SLSNet$ \eqref{eq:SymLSNet}, and $\VPNet$ to learn phase volume-preserving dynamics \eqref{eq:ODE}. 

\section{Numerical results} \label{sec:NumResults}
In this section, we provide numerical examples for learning phase volume-preserving linear and nonlinear dynamics with the neural networks $\LSNet$, $\SLSNet$, and $\VPNet$. In particular, we consider learning solutions of the semi-discretized linear advection equation, the Euler equations for the motion of a free rigid-body dynamics and the motion of the charged particle in an electromagnetic potential. All calculations are performed in PyTorch\footnote{\url{https://pytorch.org/}}, which provides a flexible and efficient platform for designing and testing neural networks. 

In all examples mean squared error (MSE) loss function is considered and minimized:
\begin{equation}\label{eq:loss}
\mathcal{L} = \frac{1}{N} \sum_{j=1}^{N} \|y(t_j) - Y_j \|_2^2,
\end{equation}
where $y(t_j)=\phi_{\tau_j}(y(t_{j-1}))$ is the ground truth or analytic solution of the dynamical system \eqref{eq:ODE} at time $t_j$, where $\tau_j=t_j-t_{j-1}$, while $Y_j$ is the predicted value by the neural network for the given input $y(t_{j-1})$, $N$ is the number of training data samples, and $\|\cdot\|_2^2$ is the Euclidean distance squared. Data samples $y(t_{j-1})$, $j=1,\dots,N$, are obtained from a single solution trajectory, or are randomly sampled in the phase space, e.g., see numerical results of Section \ref{sec:RB_Whole}. In all examples, without loss of generality, we consider constant time step $\tau$ values.

In addition to the loss function \eqref{eq:loss}, to evaluate the learning abilities of the neural networks on unseen data during the training we form a validation data set and compute the MSE accuracy function:
\begin{equation}\label{eq:acc}
\mathcal{A} = \frac{1}{M} \sum_{i=1}^{M} \|y(t_i) - Y_i \|_2^2,
\end{equation}    
where $y(t_i)=\phi_{\tau_i}(y(t_{i-1}))$. The validation data samples $y(t_{i-1})$, $i=1,\dots,M$, are sampled from the continuation of the trajectory $y(t_{j})$ above, i.e., $y(t_{i})=y(t_{N})$ when $i=0$, or are randomly sampled in the phase space. Accordingly, $M$ is the number of validation data samples. In what follows, we evaluate and visualize both mean squared errors \eqref{eq:loss} and \eqref{eq:acc} for each epoch. Similarly, the ground truth values for further testing, i.e., for comparison to the neural network predictions, are obtained. In numerical experiments, all training, validation, and testing data were obtained by solving differential equations numerically to high precision with adaptive step size $5(4)$ order Runge-Kutta method, provided by the SciPy\footnote{\url{https://www.scipy.org/}} library and {\it solve$\_$ivp} solver. 

In training neural networks, the loss function \eqref{eq:loss} is minimized with the batch Adam optimization method \cite{Adam}. We use batch Adam since our training data sets and problem sizes are relatively small. We consider the Adam algorithm with the standard parameter values. For the linear problem of Section \ref{sec:LinProblem} we consider constant learning rate $\eta=10^{-3}$. For the nonlinear problems, we employ exponential scheduling for the learning rate $\eta$, where the decay rate parameter $\gamma=e^{\log(\eta_2/\eta_1)/N_{e}}$ is obtained from specifying the initial and final learning rate values $\eta_1$ and $\eta_2$, respectively, for the fixed number of epochs $N_{e}$. In our experiments, we set $\eta_1=10^{-2}$ and $\eta_2=10^{-6}$ and vary $N_e$. With the exponential decay scheduling, we observed improved convergence of the loss \eqref{eq:loss} and accuracy \eqref{eq:acc} functions, but the initial manual tuning of the decay rate $\gamma$, performing test runs with different $\eta_1$, $\eta_2$, and $N_e$ values, is generally required. For training $\LSNet$ and $\SLSNet$, we initialize the weight values $w$ and $W$ from the standard normal distribution with mean zero and variance $0.01$ while the initial bias values are chosen to be zero. For training the volume-preserving neural networks $\VPNet$, we also initialize all bias values to be zero, but the weights of the linear layers are initialized by Xavier uniform initialization \cite{Glorot10}.

For linear problems, we consider linear activation function with bias $b$ set to zero. For nonlinear problems, we consider the sigmoid activation function $\sigma(x)=1/(1+e^{-x})$. Comparable results (not shown) were also obtained with the Swish activation function \cite{Swish}. For nonlinear problems, the number of epochs $N_e$ was significantly larger compared to the linear problem considering the increased complexity of the dynamics. For all problems, neural networks were trained and tested with the different number of network layers, hidden layers (only for $\VPNet$), and width parameter values. In general, objectively good results for $\LSNet$ were only observed when $K>1$, while $K=1$ was sufficient in most cases for $\SLSNet$. 

\subsection{Linear problems}\label{sec:LinProblem}
Linear dynamical system
\begin{equation}\label{eq:VolODE}
\der{y}{t} = Ay,
\end{equation}
where $y\in\R^n$ and $A\in\R^{n \times n}$, is phase volume-preserving if $\Tr(A)=0$, and has the solution in the following form:
\[
y(t) = e^{At}y_0,
\]
for any initial condition $y_0\in\R^n$. Thus, the flow $\varphi_t = e^{At}$. Since $\Tr(A)=0$, it is easy to see that
\[
\det\left( \frac{\partial \varphi_t(y_0)}{\partial y_0} \right) = 
\det\left( e^{At} \right) = 
 e^{\Tr(A)t} = 1. 
\] 

For a fixed time step $\tau$, the matrix exponential $e^{A \tau}$ can be found (learned) from the given dynamics data considering multivariate linear regression, i.e.,
\begin{equation}\label{eq:LinReg}
e^{A \tau} = \left(\sum_{i=1}^{N} Y_i X_i^T \right)
\left(\sum_{i=1}^{N} X_i X_i^T \right)^{-1},
\end{equation}
where $X_i\in\R^n$ and $Y_i\in\R^n$ are i-$th$ input and output vectors, respectively, in the training data set of $N$ samples, such that
\[
Y_i = e^{A \tau} X_i. 
\]
As long as the matrix on the right-hand side in \eqref{eq:LinReg} is invertible, the linear regression problem has a unique solution. Thus, it may seem even not necessary to consider locally-symplectic neural networks \eqref{eq:LSNet} and \eqref{eq:SymLSNet} for learning phase volume-preserving linear systems \eqref{eq:VolODE}. Despite that, neural networks \eqref{eq:LSNet} and \eqref{eq:SymLSNet} provide a different interpretation for the matrix $e^{A \tau}$ compared to the linear regression problem \eqref{eq:LinReg}, i.e., linear neural networks \eqref{eq:LSNet} and \eqref{eq:SymLSNet} are learning matrices $B_1,B_2,\dots$ in the Taylor series expansion with respect to $\tau$ of the matrix exponential
\[
e^{A \tau} = I + \tau B_1 + \frac{\tau^2}{2!} B_2 + \frac{\tau^3}{3!} B_3 + \dots, 
\]   
which follows from the composition of the phase volume-preserving (linear) modules \eqref{eq:LocMaps} with $h=\tau$. 

Without the proof, we state that $\Tr(B_1)=0$, which follows from the constructions \eqref{eq:LSNet} and \eqref{eq:SymLSNet}, and the neural networks will aim to learn $B_1=A, \, B_2=A^2, \dots$ if the training data are provided with different time step $\tau$ values. It is important to state that by construction we do not obtain that $B_2=B_1^2, \, B_3=B_1^3, \dots$ A closer investigation of these properties is required and left for future work.  

\begin{figure}[t]
\centering 
\subfigure[]{\label{fig:AdvSol}
\includegraphics[trim=0cm 0cm 0cm 0cm,clip=true,width=0.49\textwidth]{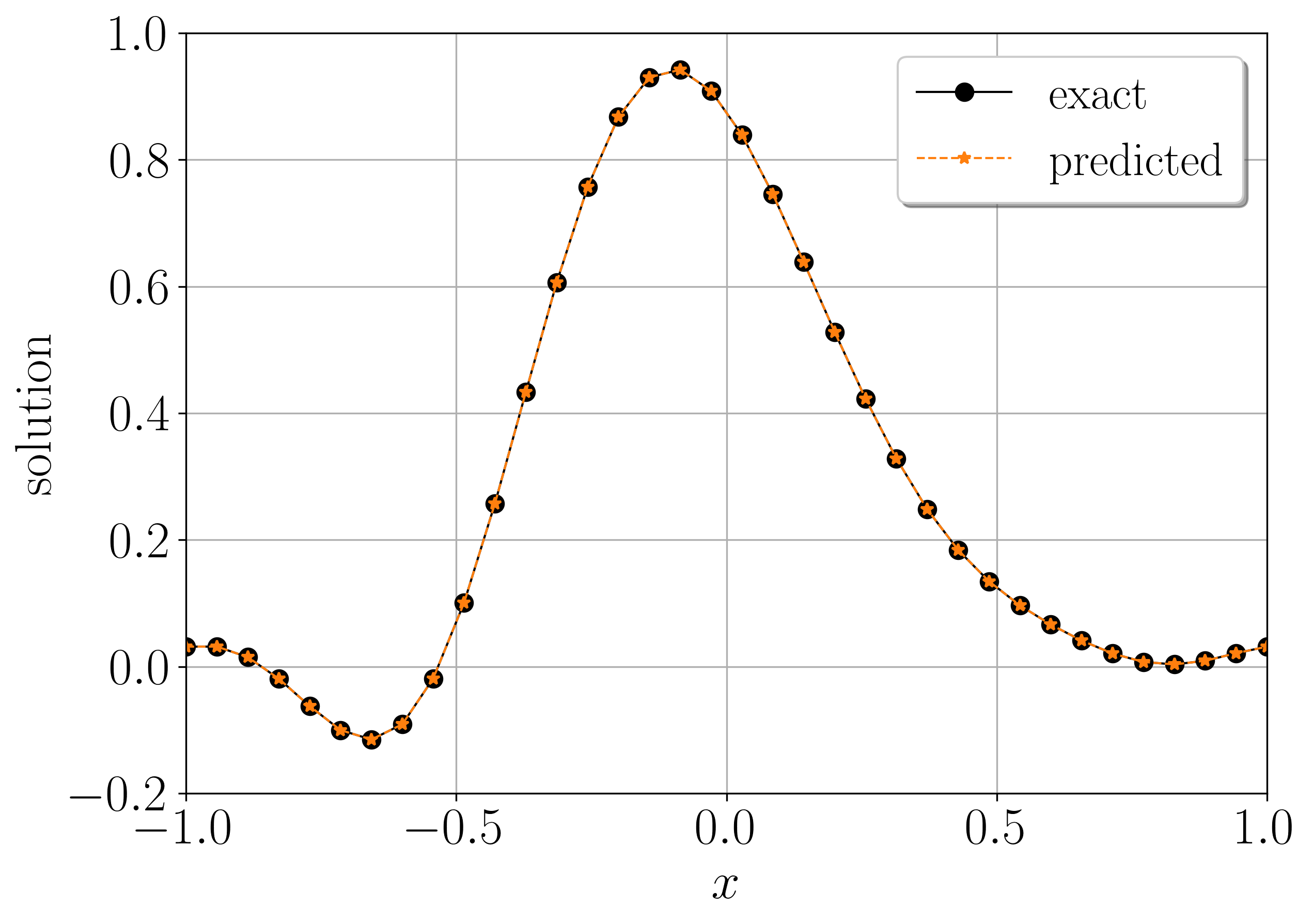}}
\subfigure[]{\label{fig:AdvLoss}
\includegraphics[trim=0cm 0cm 0cm 0cm,clip=true,width=0.49\textwidth]{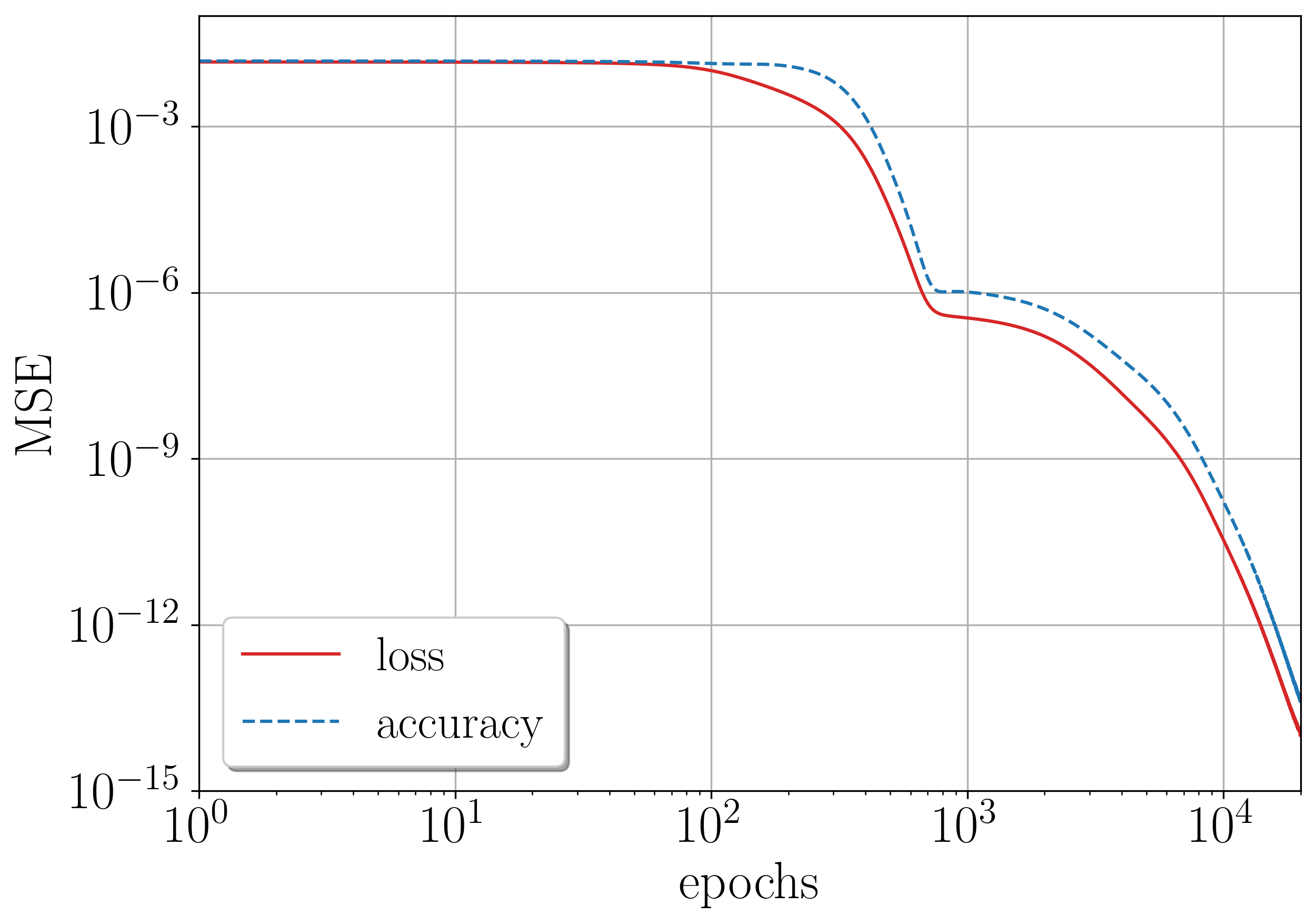}}
\subfigure[]{\label{fig:AdvSol_Sym}
\includegraphics[trim=0cm 0cm 0cm 0cm,clip=true,width=0.49\textwidth]{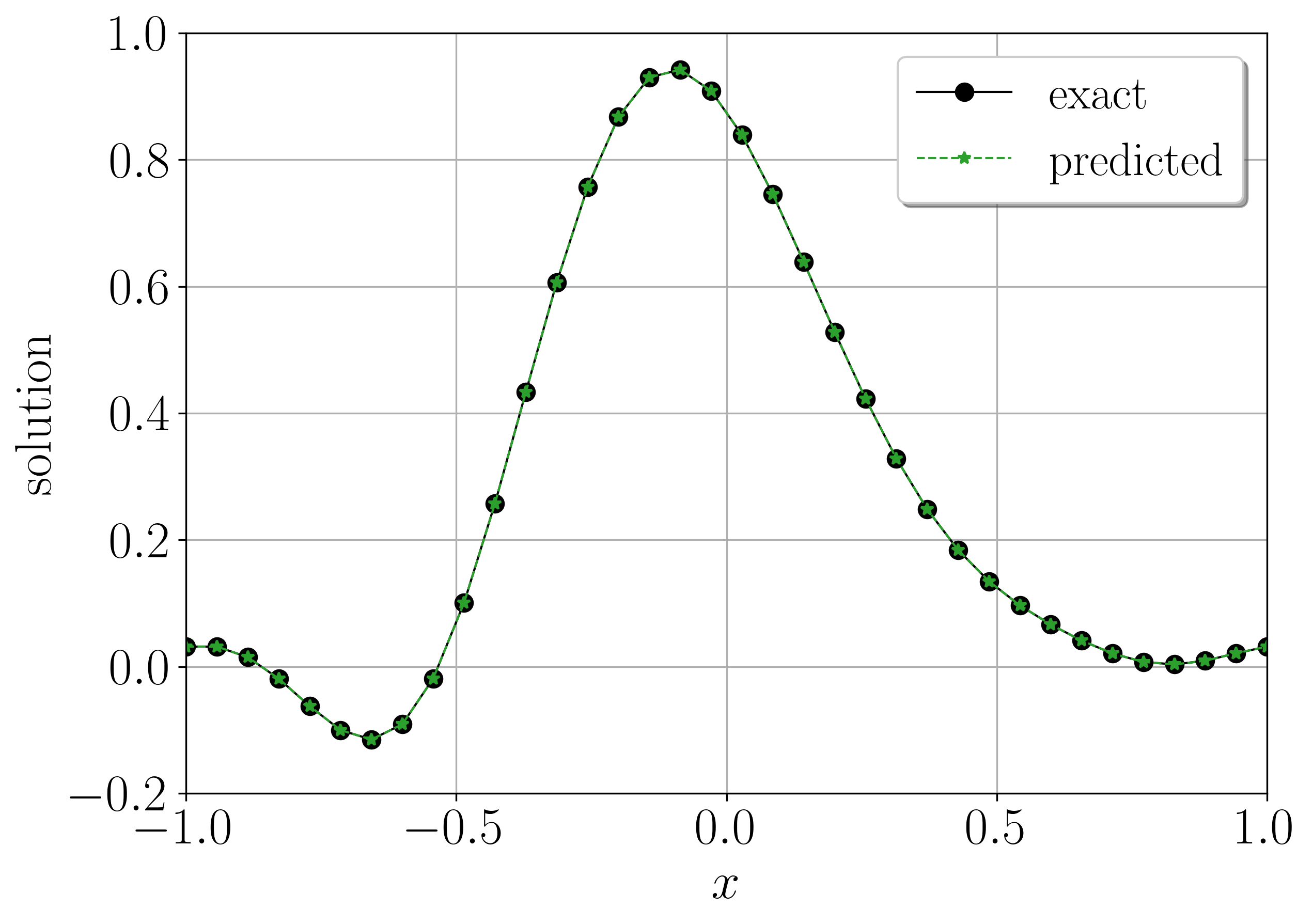}}
\subfigure[]{\label{fig:AdvLoss_Sym}
\includegraphics[trim=0cm 0cm 0cm 0cm,clip=true,width=0.49\textwidth]{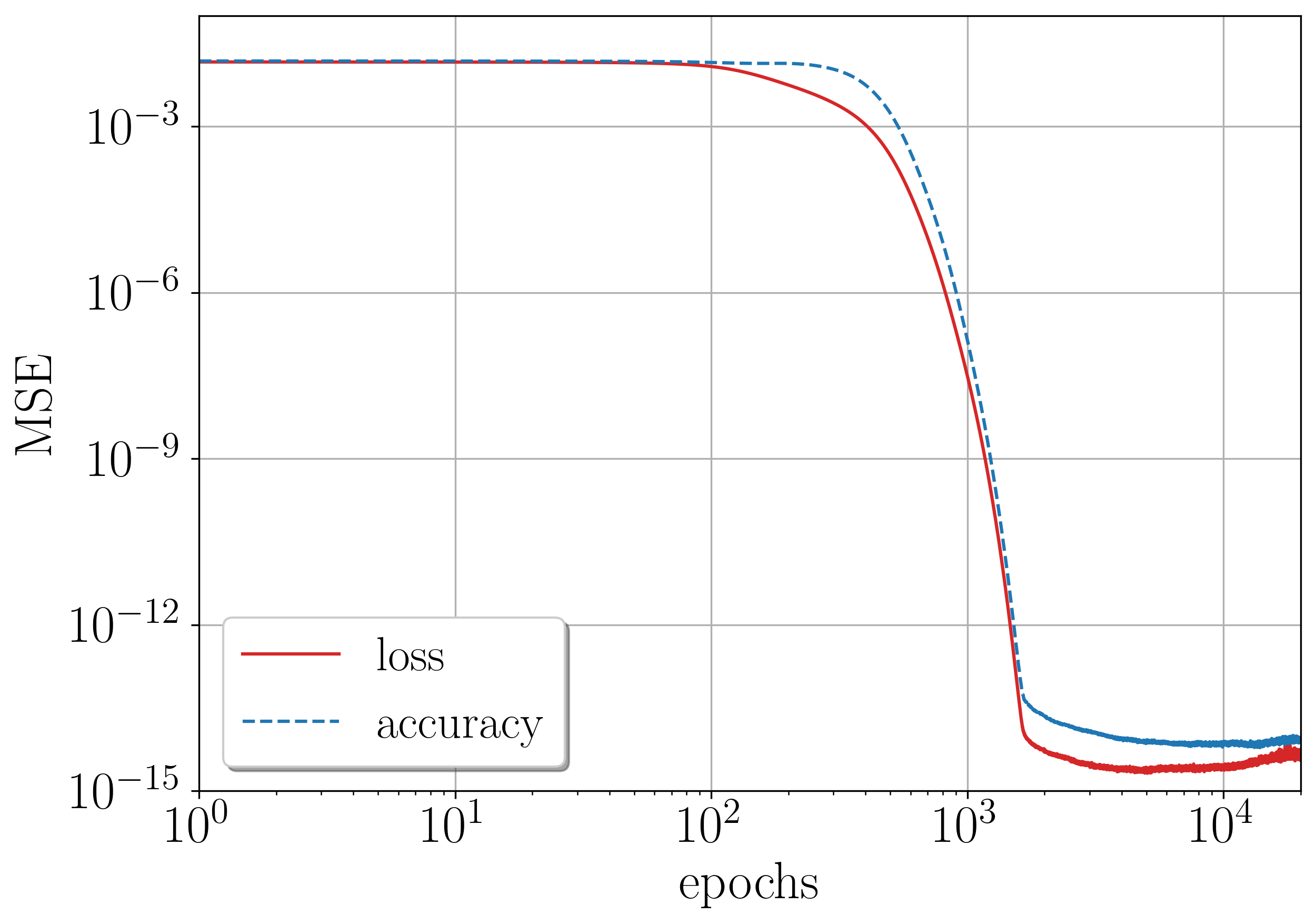}}
\caption{(a)\&(c) predicted and exact solutions of the semi-discretized advection equations \eqref{eq:SemiAdv} at $t=4$ with $n=35$, $c=1$ and the initial condition \eqref{eq:IC}. (b)\&(d) loss \eqref{eq:loss} and accuracy \eqref{eq:acc} function values at each epoch. (a)-(b) $\LSNet$, $K=2$ and $m=n$. (c)-(d) $\SLSNet$, $K=1$ and $m=n$.}\label{fig:Advection}
\end{figure}

To demonstrate the neural network \eqref{eq:LSNet} and \eqref{eq:SymLSNet} capabilities of learning linear phase volume-preserving dynamics \eqref{eq:VolODE} we consider the semi-discretized advection equation
\begin{equation}\label{eq:SemiAdv}
\der{u_i}{t} = -c \frac{u_{i+1}-u_{i-1}}{2 \Delta_x}, \quad i=0,\dots,n-1,
\end{equation}
where $c\in\R$ is the constant wave speed, $u_i(t)$ is the time-dependent grid function, $n$ is the number of grid points (or the dimension of the system) and $\Delta_x=\frac{2}{n}$ is the grid size in the domain $[-1,1]$. We solve \eqref{eq:SemiAdv} with the periodic boundary conditions, i.e., $u_{-1}=u_{n-1}$ and $u_{n}=u_0$. In the limit, when $\Delta_x \to 0$, we recover the advection equation $u_t + c u_x = 0$ with traveling wave solutions $u(t,x)=v(x-ct)$, where $u(0,x)=v(x)$. Note that the finite difference approximation \eqref{eq:SemiAdv} is conservative but dispersive \cite{LeVeque}, as can be seen in Figures \ref{fig:AdvSol} and \ref{fig:AdvSol_Sym}. If the objective is to learn the actual traveling wave solutions of the advection equation from the training data provided by the semi-discretized equations \eqref{eq:SemiAdv}, then a larger value of $n$ should be considered. In our experiments $n=35$ and $c=1$.  

For training data, we consider $N=60$ randomly generated initial conditions from the normal distribution with mean zero and variance one, i.e., $u_i(0)\sim \mathcal{N}(0,1)$, $i=0,\dots,n-1$. Then the semi-discretized equations \eqref{eq:SemiAdv} are solved to collect output data after the time instance $\tau=0.01$. In addition to the training data, we also collected $M=20$ random initial conditions with their respective outputs to validate the learning capabilities of the neural network during the training by evaluating the prediction accuracy function \eqref{eq:acc} at each epoch. Then the testing is performed by predicting a (dispersive) traveling wave solution of the semi-discretized advection equations \eqref{eq:SemiAdv} with the Gaussian initial condition
\begin{equation}\label{eq:IC}
u_i(0) = e^{-10(-1+i\Delta_x)^2}, \quad i=0,\dots,n-1.
\end{equation}

In Figures \ref{fig:AdvSol} and \ref{fig:AdvSol_Sym} we plot predicted outputs by the linear phase volume-preserving neural networks $\LSNet$ ($K=2$ and $m=n$) and $\SLSNet$ ($K=1$ and $m=n$) at time $t=4$, respectively, and compare to the exact solution of \eqref{eq:SemiAdv} after training for $N_e=2\times 10^4$ epochs, see also Figures \ref{fig:AdvLoss} and \ref{fig:AdvLoss_Sym}. In Figures \ref{fig:AdvLoss} and \ref{fig:AdvLoss_Sym} we demonstrate both MSEs \eqref{eq:loss} and \eqref{eq:loss} at each epoch. Not only the loss MSE tends to zero but accuracy MSE as well, which illustrates good generalization of the neural networks.

To obtain satisfactory results with $\LSNet$ we required $K>1$, while it was sufficient to consider $K=1$ for the $\SLSNet$. Recall that both neural networks $\LSNet$ with $K=2$ and $\SLSNet$ with $K=1$ have the same number of layers $L$, while $\SLSNet$ have twice fewer parameter values, i.e., weight and bias values. 

Predicted solutions in Figures \ref{fig:AdvSol} and \ref{fig:AdvSol_Sym} are indistinguishable from the exact solution. To measure the difference between exact and predicted solutions, we consider $L_2$ grid norm:
\begin{equation}\label{eq:GridNorm}
\| u \|_2 = \sqrt{\Delta_x \sum_{i=0}^{n-1} |u_i|^2},
\end{equation}
which is equal to $4.1862\times 10^{-6}$ for $\LSNet$ and $2.4687\times 10^{-6}$ for $\SLSNet$ in the illustrated experiments of Figure \ref{fig:Advection} when $t=4$. Considering 10 different random initializations of weight values, we calculated predictions by 10 (different) trained neural networks and arrived at mean $L_2$ grid error values $4.9778\times 10^{-6}$ and $2.6168\times 10^{-6}$ for the $\LSNet$ and $\SLSNet$, respectively. Notice that $\SLSNet$ have not only produced smaller solution errors compared to $\LSNet$ but also require ten times fewer training epochs to achieve the desired accuracy in the loss \eqref{eq:loss} and accuracy \eqref{eq:acc} functions, compare Figures \ref{fig:AdvLoss} and \ref{fig:AdvLoss_Sym}.

Overall, Figure \ref{fig:Advection} demonstrates that the linear phase volume-preserving neural networks $\LSNet$ and $\SLSNet$ are capable of learning linear dynamics \eqref{eq:VolODE}, and $\SLSNet$ outperform $\LSNet$. We will arrive at similar conclusions considering learning of nonlinear dynamics, which we present in the following sections.
 
\subsection{Rigid body dynamics}\label{sec:RigidBody}
As the first example for learning nonlinear dynamics, we consider the Euler equations of the motion of a free rigid body \cite{Hairer,Arnold}:
\begin{align}\label{eq:RBody}
\begin{split}
\der{y_1}{t} &= a_1 y_2 y_3, \quad a_1 = \frac{I_2-I_3}{I_2 I_3}, \\
\der{y_2}{t} &= a_2 y_3 y_1, \quad a_2 = \frac{I_3-I_1}{I_3 I_1}, \\
\der{y_3}{t} &= a_3 y_1 y_2, \quad a_3 = \frac{I_1-I_2}{I_1 I_2},
\end{split}
\end{align}
where the state vector $y=(y_1,y_2,y_3)^T\in\R^3$, i.e., $n=3$, describes the angular momentum in the body frame. $I_{1,2,3}>0$ are the principal components of inertia. It is easy to see that the rigid body dynamics \eqref{eq:RBody} is phase volume-preserving \eqref{eq:div} as well as has two quadratic conserved quantities, i.e., the kinetic energy  
\begin{equation}\label{eq:KinEn}
H(y_1,y_2,y_3) = \frac{1}{2} \left( 
\frac{y_1^2}{I_1} +
\frac{y_2^2}{I_2} +
\frac{y_3^2}{I_3}
\right)
\end{equation}
and the invariant
\begin{equation}\label{eq:Iinv}
I(y_1,y_2,y_3) = y_1^2 + y_2^2 + y_3^2.
\end{equation}
Thus, the solution of \eqref{eq:RBody} lies on the intersection of the sphere \eqref{eq:Iinv} with the ellipsoid given by \eqref{eq:KinEn}. In what follows, without loss of generality, we set $I_1 =2$, $I_2=1$ and $I_3=\frac{2}{3}$.

We further split this section in four parts. In the first part, we train the phase volume-preserving neural networks $\LSNet$, $\SLSNet$, and $\VPNet$ to learn a single periodic solution trajectory of \eqref{eq:RBody}, while in the second part we consider the same example but induce random noise into the training data. In the third part of this section we demonstrate learning of the whole dynamics of a free rigid body \eqref{eq:RBody}. We conclude this section with the numerical results of learning rigid body dynamics with different time steps $\tau$ and number of training data samples $N$.  

\subsubsection{Learning a single periodic trajectory}\label{sec:RB_Single}
To learn a single periodic solution trajectory of \eqref{eq:RBody}, we consider an initial condition $y_0:=y(0)=(\cos(1.1),0,\sin(1.1))^T$ and time step $\tau=0.1$. We collect $N=120$ training data points, i.e., from the time interval $[0,12]$, followed by $M=40$ solution values to form the validation data set on the time interval $[12,16]$ of the same trajectory for the computation of the MSE accuracy function \eqref{eq:acc}. The testing of neural networks is performed by predicting the solution for $t > 12$ with the initial condition $y_{12}:=y(12)=\varphi_{12}(y_0)$.

\begin{figure}[t]
\centering 
\subfigure[]{\label{fig:RB_sol_A}
\includegraphics[trim=0cm 0cm 0cm 0cm,clip=true,width=0.49\textwidth]{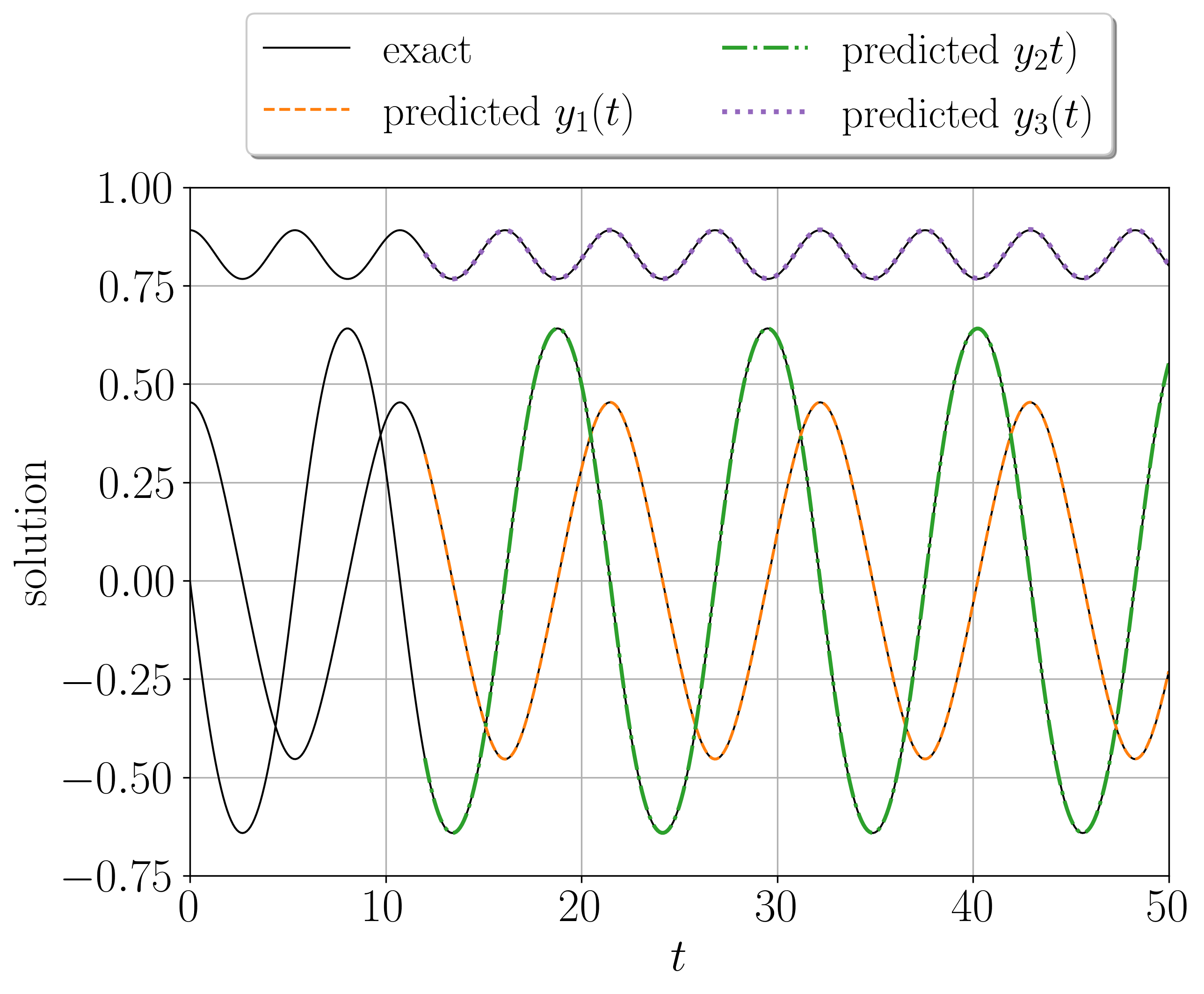}}
\subfigure[]{\label{fig:RB_sol_B}
\includegraphics[trim=0cm 0cm 0cm 0cm,clip=true,width=0.49\textwidth]{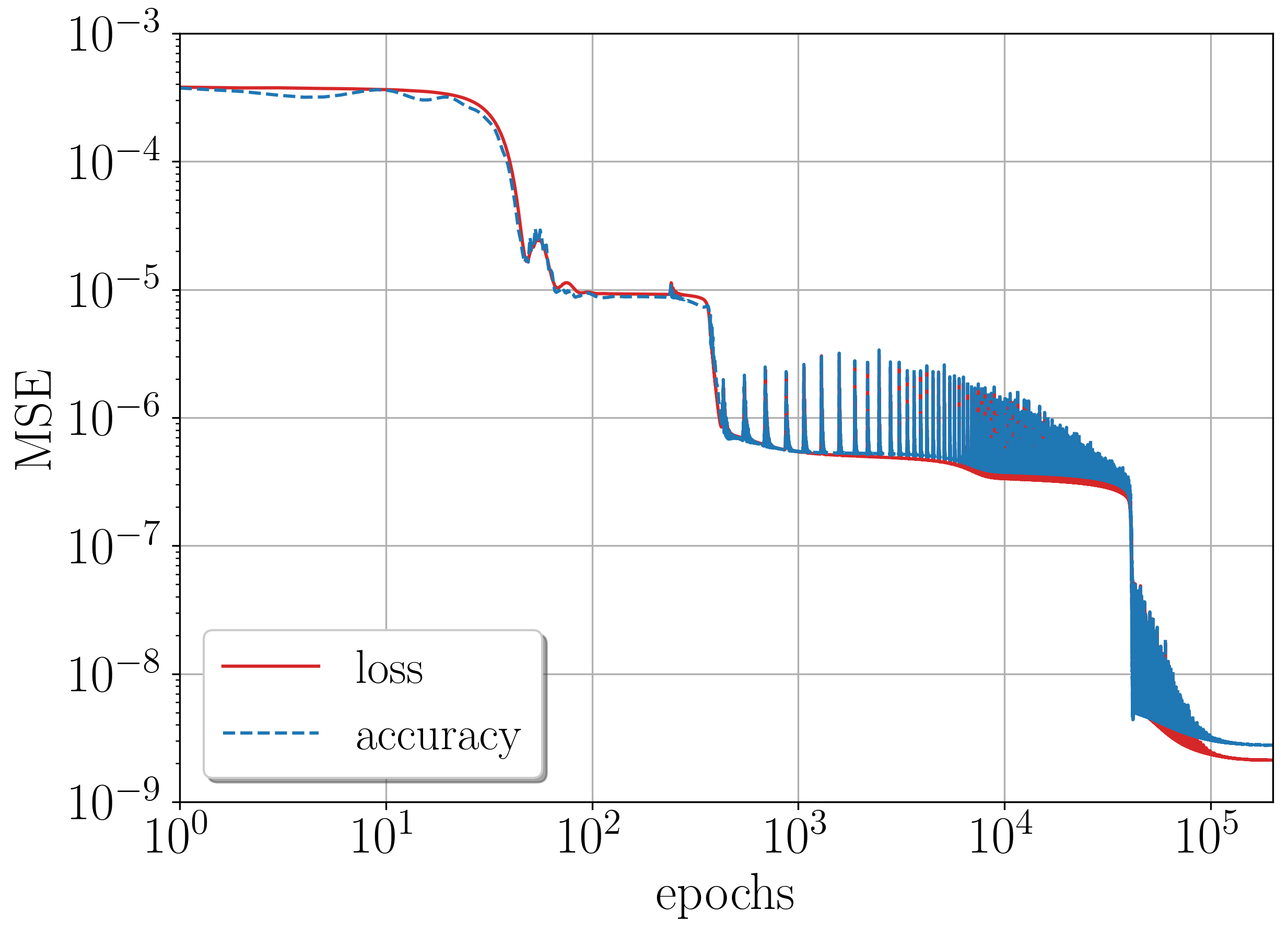}}
\caption{Learning a single periodic solution of the rigid body dynamics \eqref{eq:RBody} with $\LSNet$ \eqref{eq:LSNet}, $K=2$ and $m=16$. (a) exact solution with the initial condition $y_0$ and predicted solution by $\LSNet$ for $t>12$. (b) mean squared error loss \eqref{eq:loss} and accuracy \eqref{eq:acc} function values at each epoch.}\label{fig:RB_sol}
\end{figure} 

In Figure \ref{fig:RB_sol_A} we plot exact periodic solution of \eqref{eq:RBody} on the time interval $[0,50]$ together with the predicted solution for $t>12$ by the volume-preserving neural network $\LSNet$ \eqref{eq:LSNet} with $K=2$ and $m=16$. Both solutions appear indistinguishable. To obtain such accuracy $N_e=2\times 10^5$ epochs for training was required, see Figure \ref{fig:RB_sol_B}. Figure \ref{fig:RB_sol_B} illustrates the loss function \eqref{eq:loss} values together with network's prediction accuracy function \eqref{eq:acc} evaluated on the validation data set demonstrating good generalization of the neural network. Tests with more epochs did not give a significant improvement in the predictions and we were not able to obtain satisfactory results by $\LSNet$ with $K=1$.

\begin{figure}[t]
\centering 
\subfigure[]{\label{fig:RB_LongSol}
\includegraphics[trim=0cm 0cm 0cm 0cm,clip=true,width=0.4\textwidth]{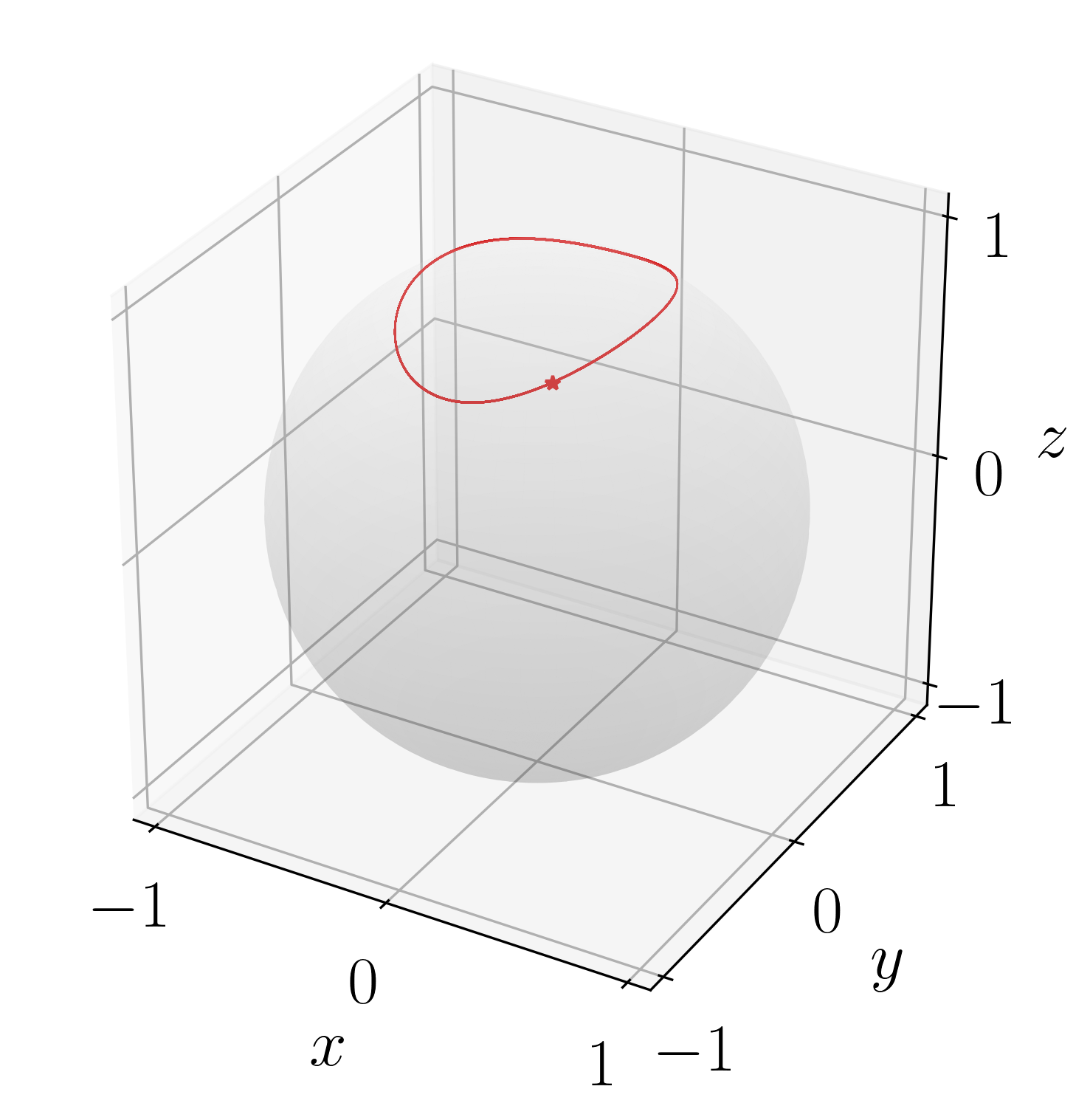}}\quad 
\subfigure[]{\label{fig:RB_LongGE}
\includegraphics[trim=0cm 0cm 0cm 0cm,clip=true,width=0.54\textwidth]{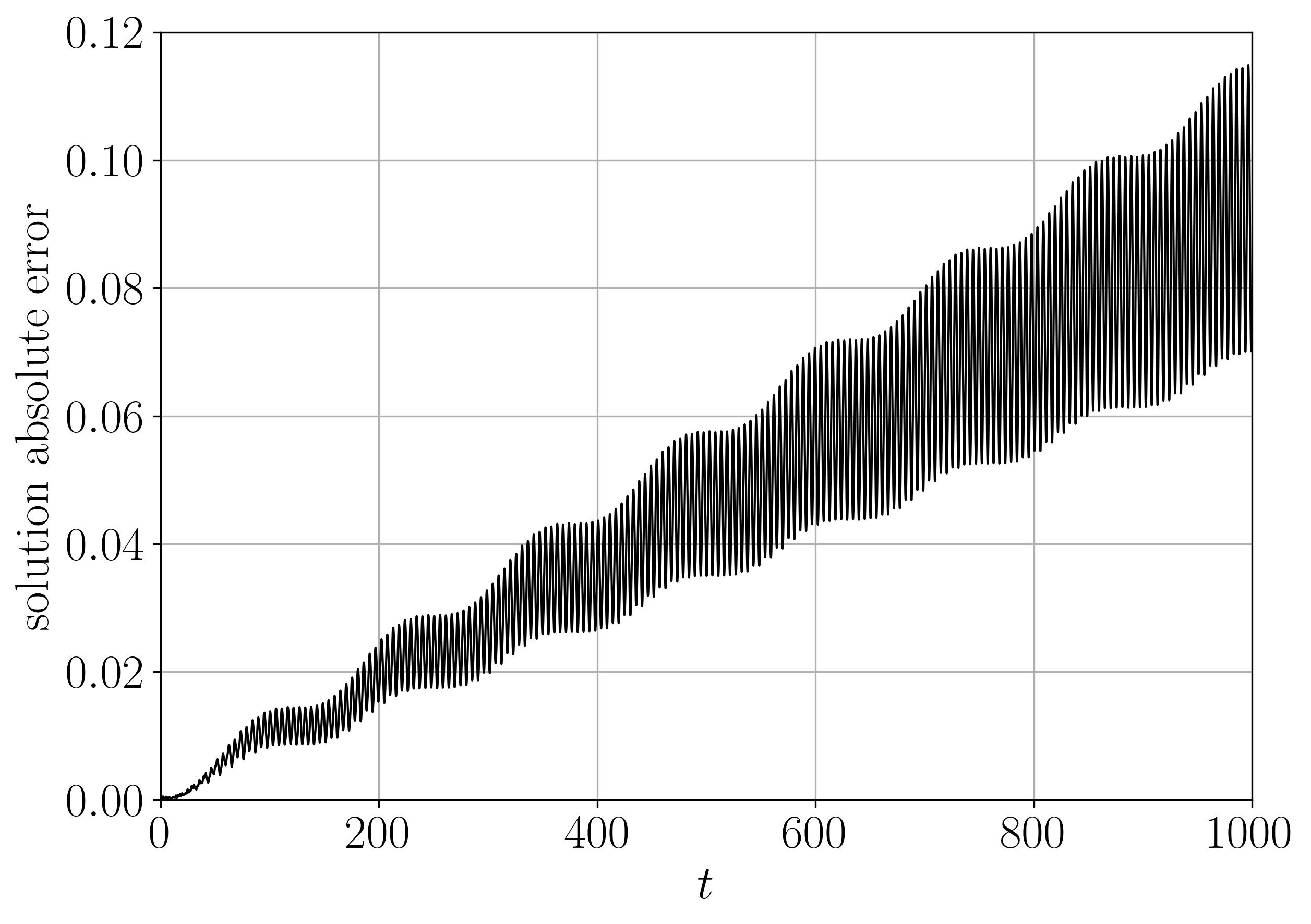}}
\subfigure[]{\label{fig:RB_LongHE}
\includegraphics[trim=0cm 0cm 0cm 0cm,clip=true,width=0.49\textwidth]{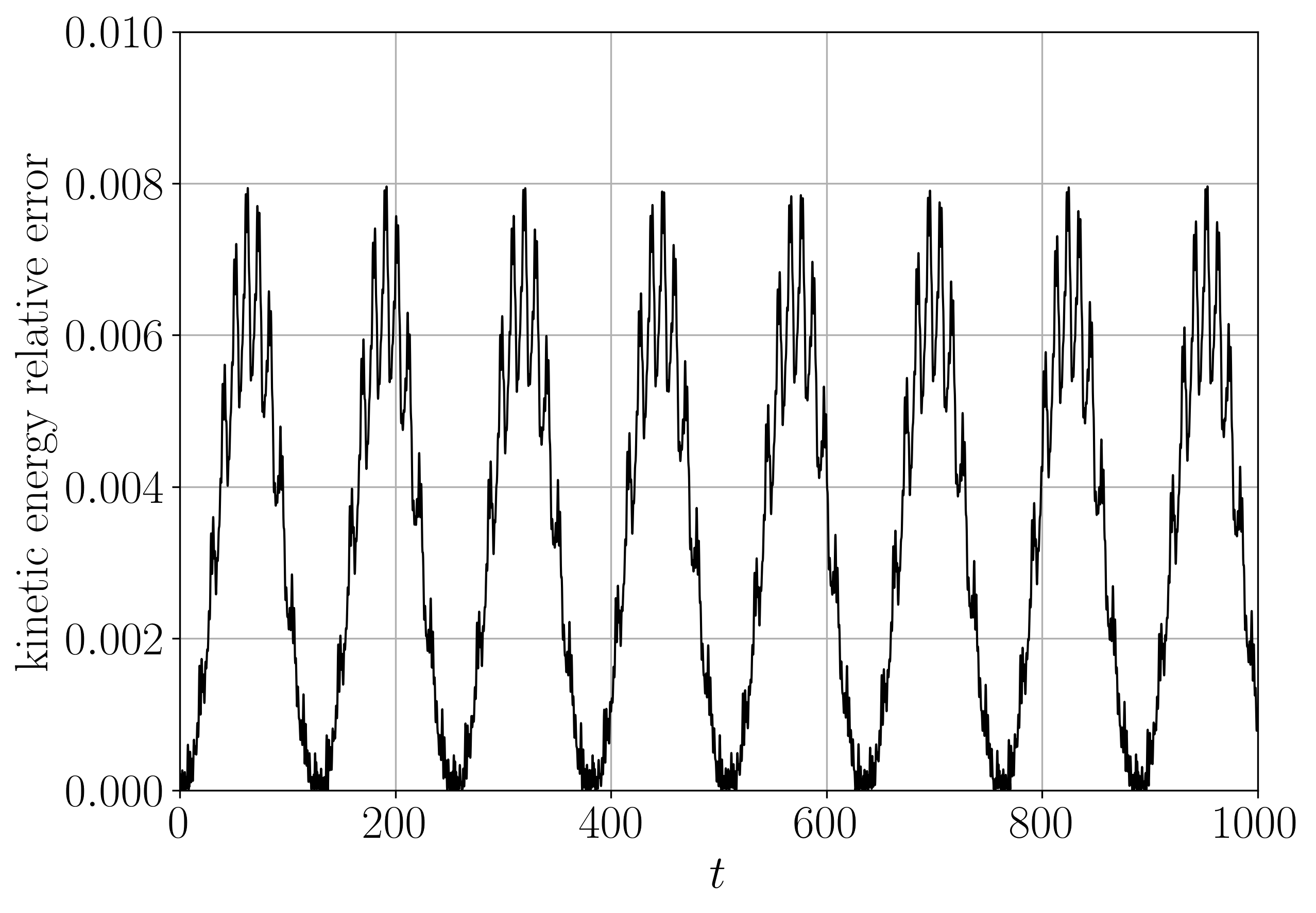}}
\subfigure[]{\label{fig:RB_LongIE}
\includegraphics[trim=0cm 0cm 0cm 0cm,clip=true,width=0.49\textwidth]{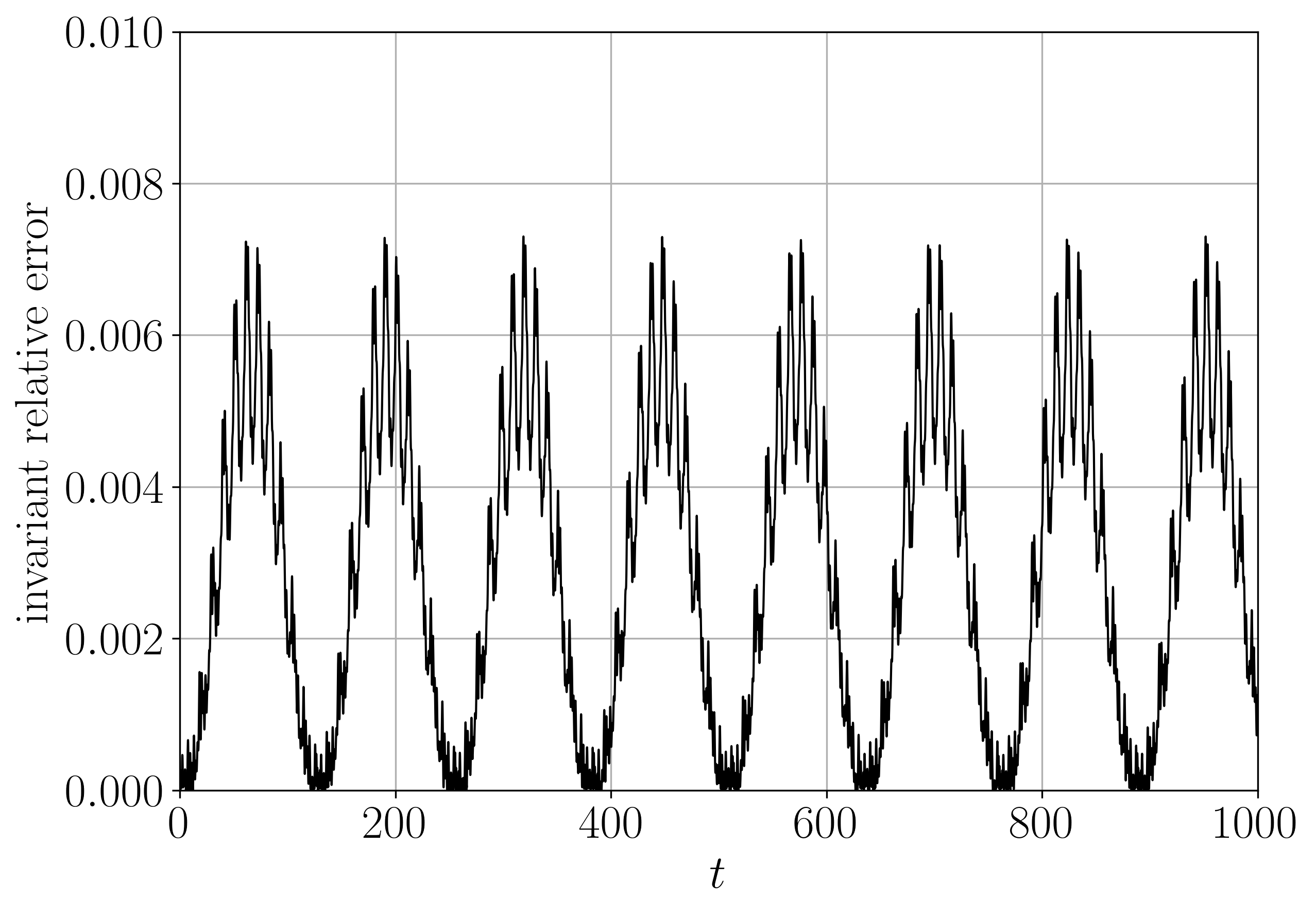}}
\caption{Long-time predictions with the phase volume-preserving neural network $\LSNet$ \eqref{eq:LSNet}, $K=2$ and $m=16$. (a) predicted trajectory by $\LSNet$, which is indistinguishable from the exact solution. The dot indicates the initial condition $y_{12}$. (b) absolute global error as a function of time. (c) absolute value of the relative error of the kinetic energy \eqref{eq:KinEn} in time. (d) absolute value of the relative error of the invariant \eqref{eq:Iinv} in time.}\label{fig:RB_LongSim}
\end{figure}

To further demonstrate network's generalization capabilities, we consider long-time predictions by the trained network of Figure \ref{fig:RB_sol} with the initial condition $y_{12}$ on the time interval $[0,1000]$, see Figure \ref{fig:RB_LongSim}. Figure \ref{fig:RB_LongSol} illustrates predicted solution by $\LSNet$ plotted on the unit sphere, where the dot indicates the initial condition $y_{12}$. Figure \ref{fig:RB_LongGE} demonstrates linear growth in absolute global error, i.e., the Euclidean distance between the exact and predicted solution at each time $t$. Linear growth in global error can be anticipated, as it is common in structure-preserving numerical methods \cite{Hairer} and has been proven for learning with exactly-symplectic maps that in long-time predictions the global error grows at most linearly \cite{Chen21}. Importantly, Figures \ref{fig:RB_LongHE}--\ref{fig:RB_LongIE} show that the learned neural network was able to learn and preserve to high accuracy both invariants \eqref{eq:KinEn}--\eqref{eq:Iinv}. Note that the phase volume-preserving neural networks $\LSNet$ \eqref{eq:LSNet} do not preserve quadratic constraints by construction. Notice that absolute values of the relative errors of both invariants \eqref{eq:KinEn}--\eqref{eq:Iinv} on the whole time interval $[0,1000]$ are below $1\%$, see Figures \ref{fig:RB_LongHE}--\ref{fig:RB_LongIE}.

Figures \ref{fig:RB_sol} and \ref{fig:RB_LongSim} demonstrate that the phase volume-preserving neural networks $\LSNet$ can be trained to learn single periodic solutions of a free rigid body dynamics \eqref{eq:RBody}. So far, we have only illustrated results for one particular trained neural network $\LSNet$ with $K=2$ and $m=16$. In the following, we explore the benefits of considering symmetric locally-symplectic neural networks $\SLSNet$ \eqref{eq:SymLSNet}, and compare obtained results with $\VPNet$ of Section \ref{sec:ModNice}. To compare all three neural networks $\LSNet$, $\SLSNet$ and $\VPNet$, we trained them $100$ times with different randomly generated initial weight values for different $K$, $m$, and $l$ network parameter values. Recall that bias values at the start of the training are always set to zero.   

\begin{figure}[t]
\centering 
\subfigure[]{\label{fig:RB_LossLoc}
\includegraphics[trim=0cm 0cm 0cm 0cm,clip=true,width=0.32\textwidth]{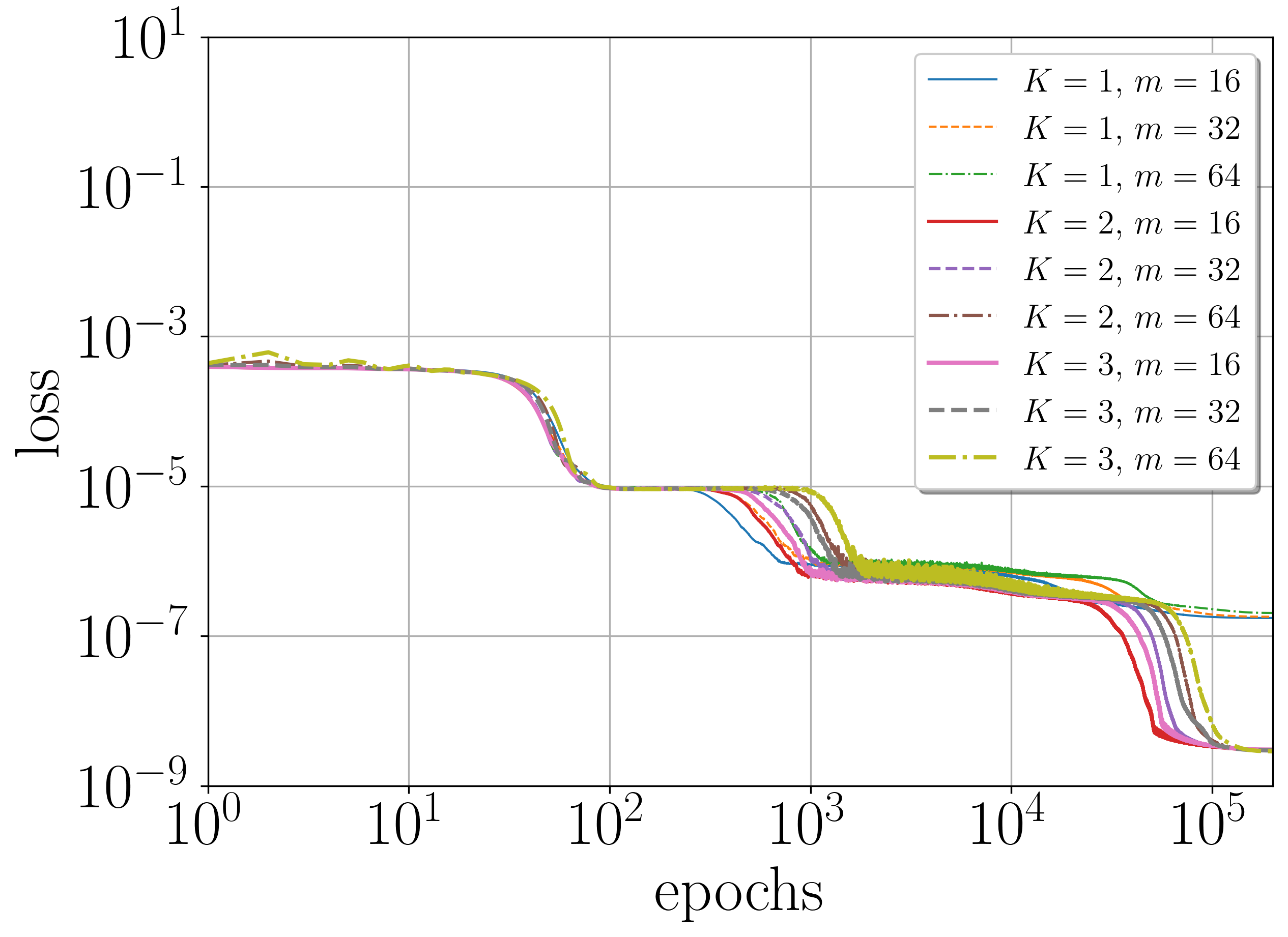}}
\subfigure[]{\label{fig:RB_LossSym}
\includegraphics[trim=0cm 0cm 0cm 0cm,clip=true,width=0.32\textwidth]{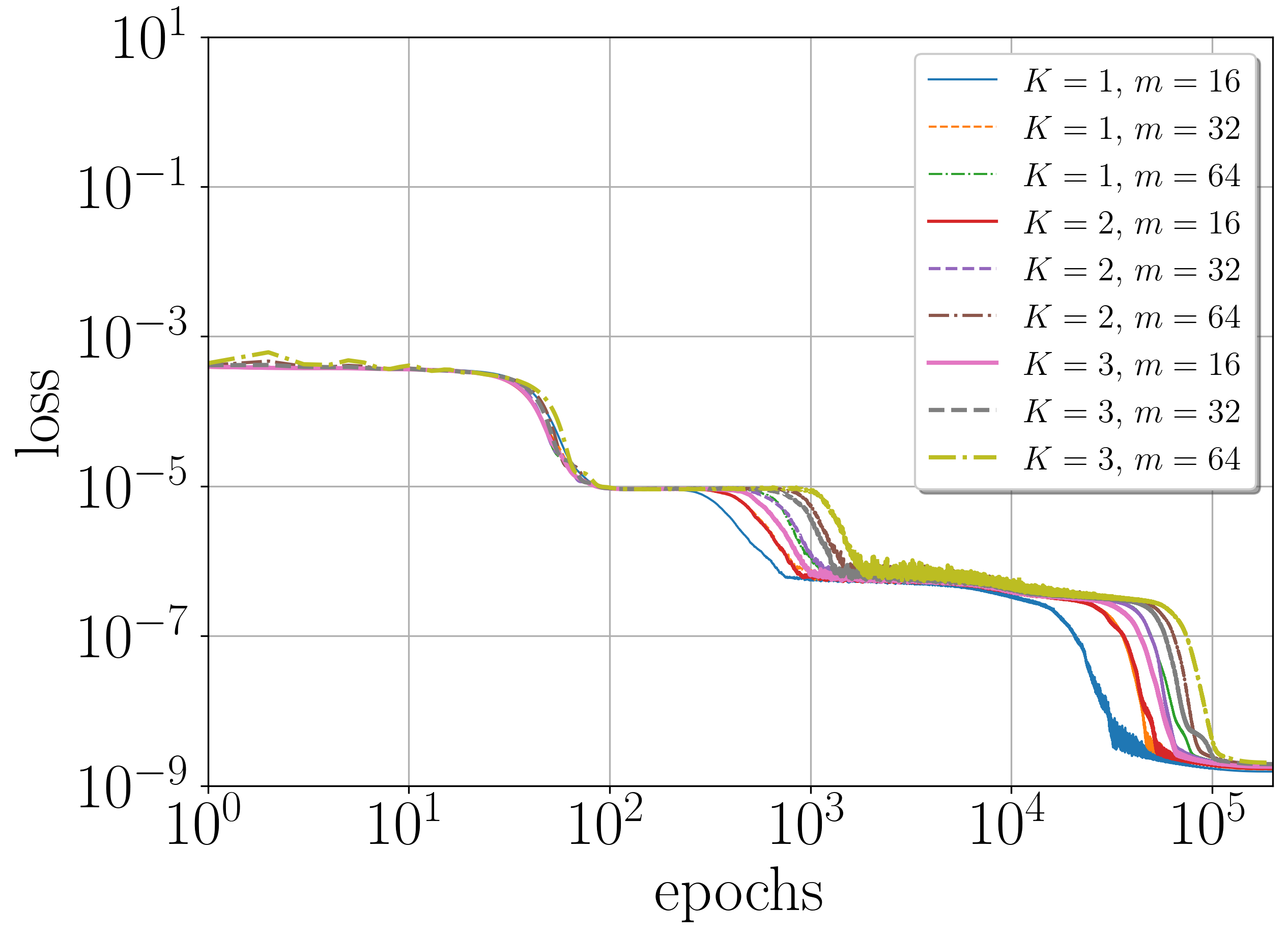}}
\subfigure[]{\label{fig:RB_LossNice}
\includegraphics[trim=0cm 0cm 0cm 0cm,clip=true,width=0.32\textwidth]{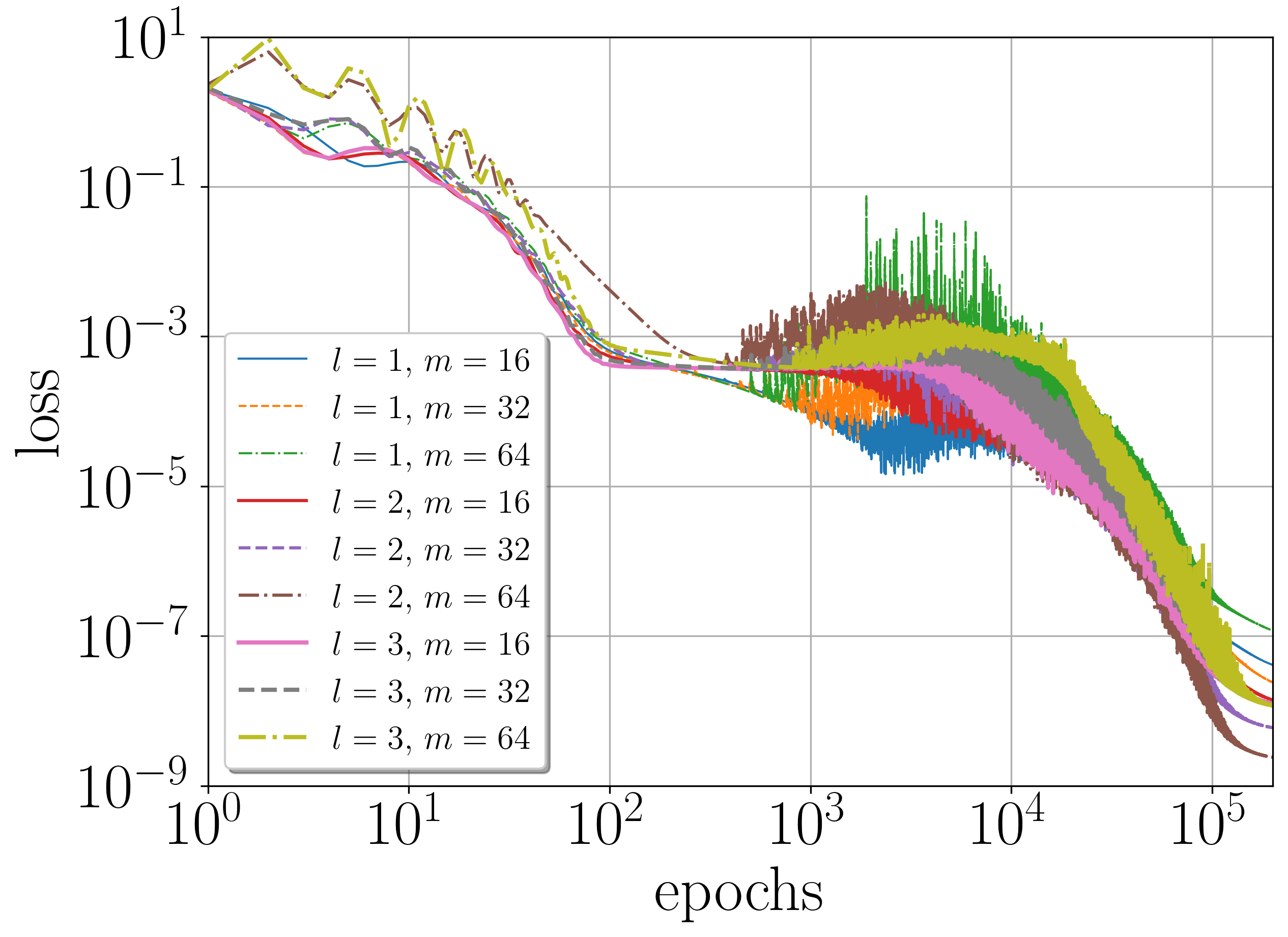}}
\subfigure[]{\label{fig:RB_AccLoc}
\includegraphics[trim=0cm 0cm 0cm 0cm,clip=true,width=0.32\textwidth]{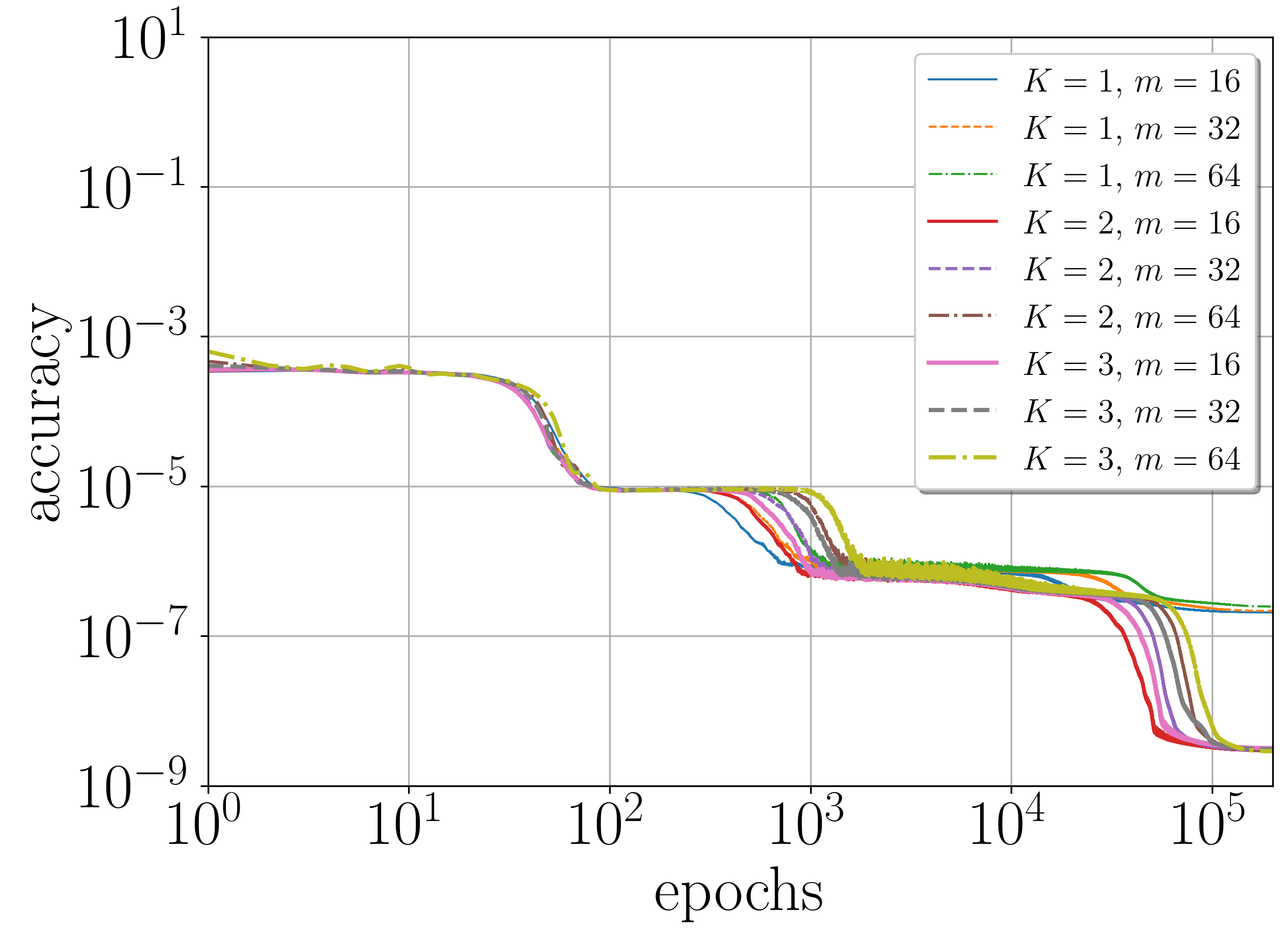}}
\subfigure[]{\label{fig:RB_AccSym}
\includegraphics[trim=0cm 0cm 0cm 0cm,clip=true,width=0.32\textwidth]{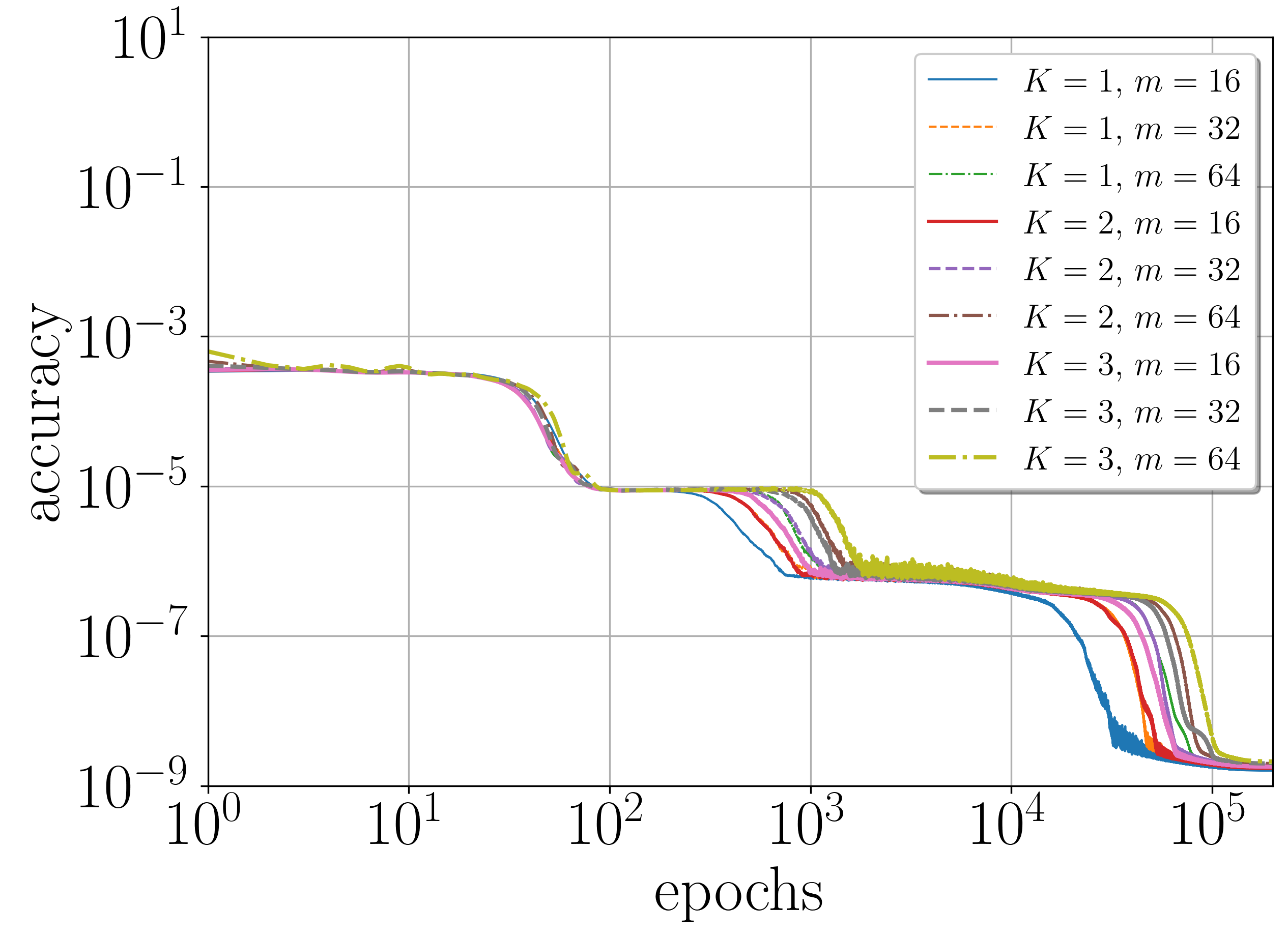}}
\subfigure[]{\label{fig:RB_AccNice}
\includegraphics[trim=0cm 0cm 0cm 0cm,clip=true,width=0.32\textwidth]{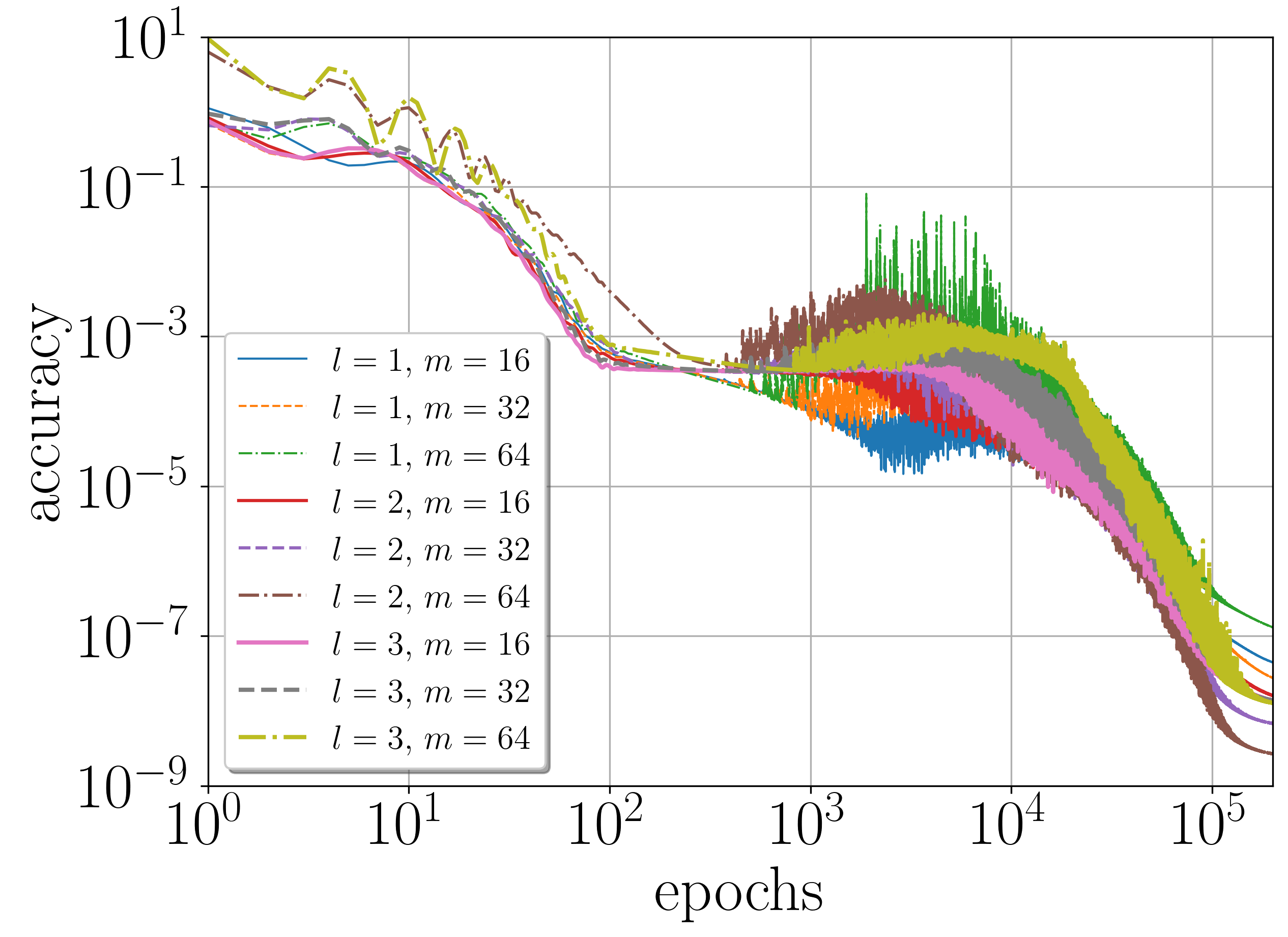}}
\caption{Averaged mean squared error loss \eqref{eq:loss} and accuracy \eqref{eq:acc} function values of the phase volume-preserving neural networks $\LSNet$, $\SLSNet$, and $\VPNet$. (a)-(c) loss values of $\LSNet$, $\SLSNet$, and $\VPNet$, respectively. (d)-(f) accuracy values of $\LSNet$, $\SLSNet$, and $\VPNet$, respectively.}\label{fig:RB_LossAcc}
\end{figure}

\begin{figure}[t]
\centering 
\subfigure[]{\label{fig:RB_RecLocS}
\includegraphics[trim=0cm 0cm 0cm 0cm,clip=true,width=0.32\textwidth]{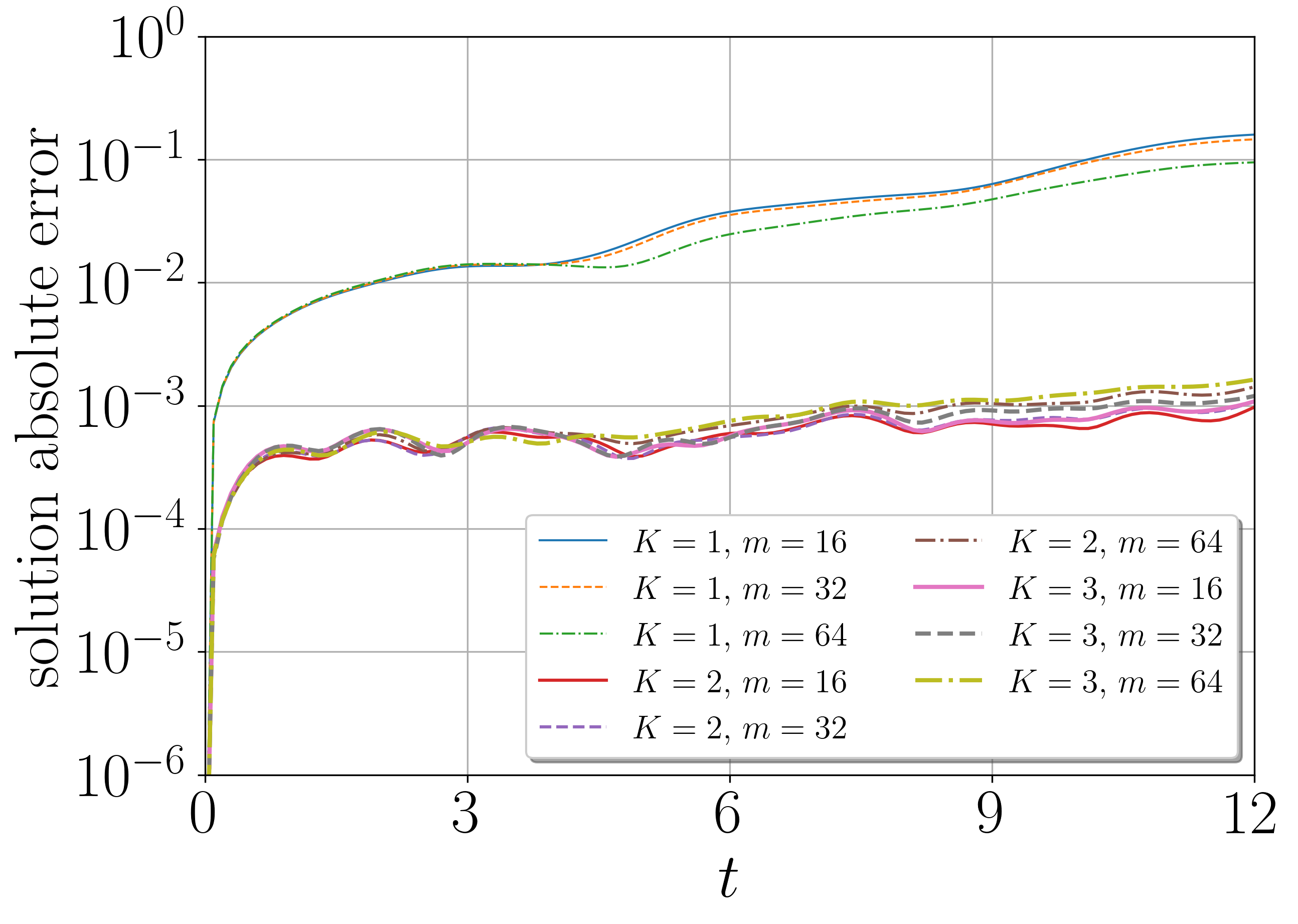}}
\subfigure[]{\label{fig:RB_RecSymS}
\includegraphics[trim=0cm 0cm 0cm 0cm,clip=true,width=0.32\textwidth]{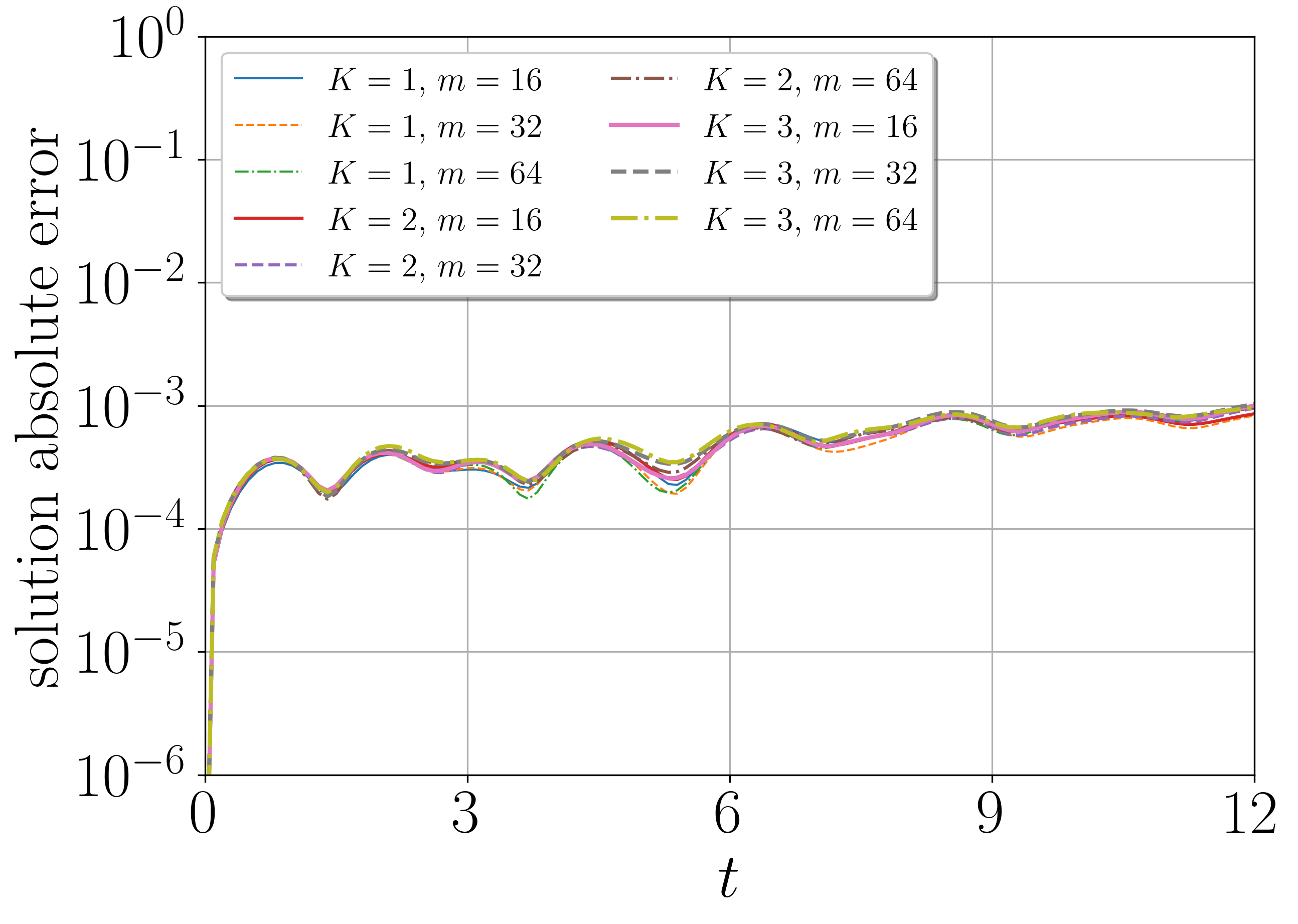}}
\subfigure[]{\label{fig:RB_RecNiceS}
\includegraphics[trim=0cm 0cm 0cm 0cm,clip=true,width=0.32\textwidth]{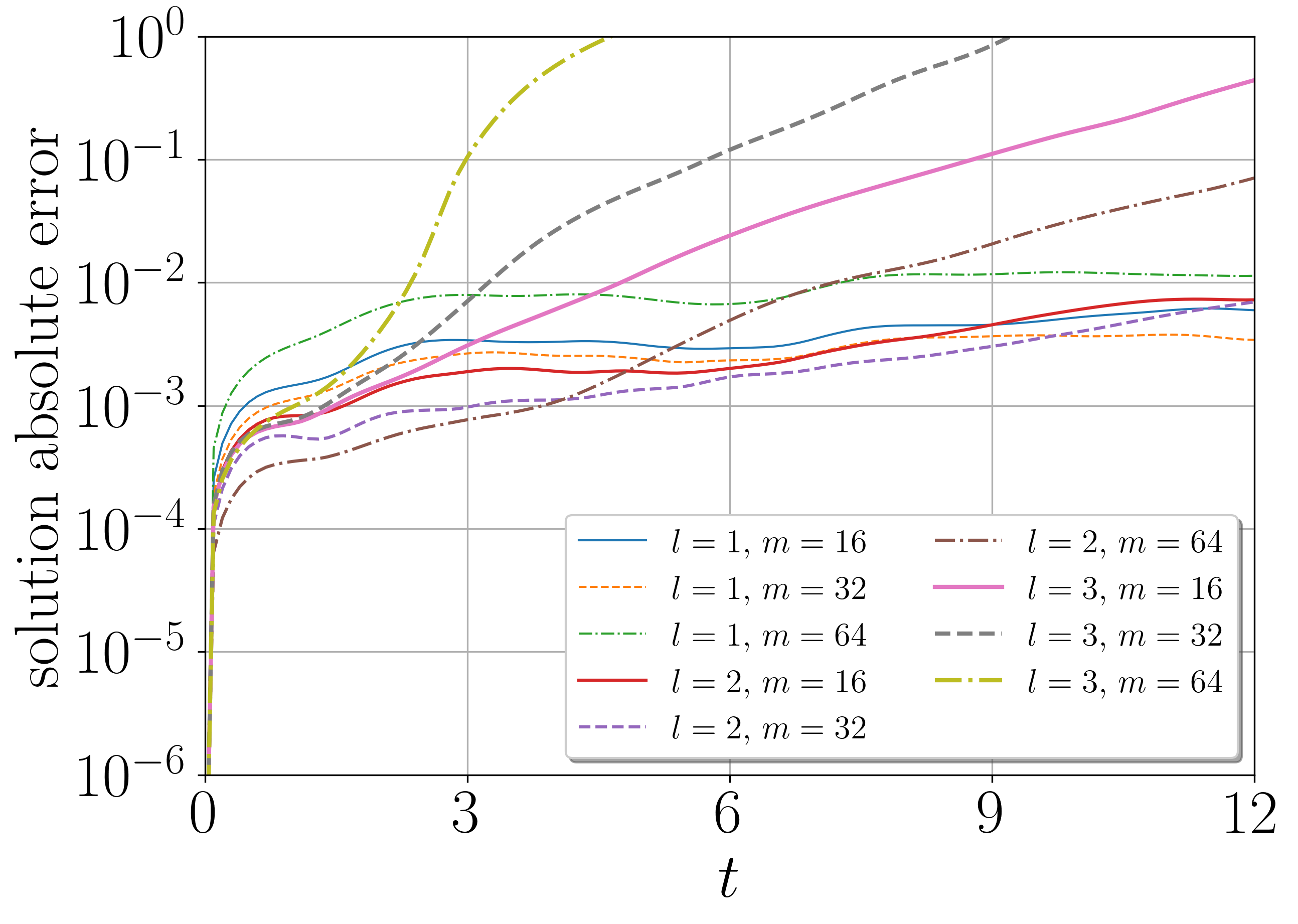}}
\subfigure[]{\label{fig:RB_RecLocH}
\includegraphics[trim=0cm 0cm 0cm 0cm,clip=true,width=0.32\textwidth]{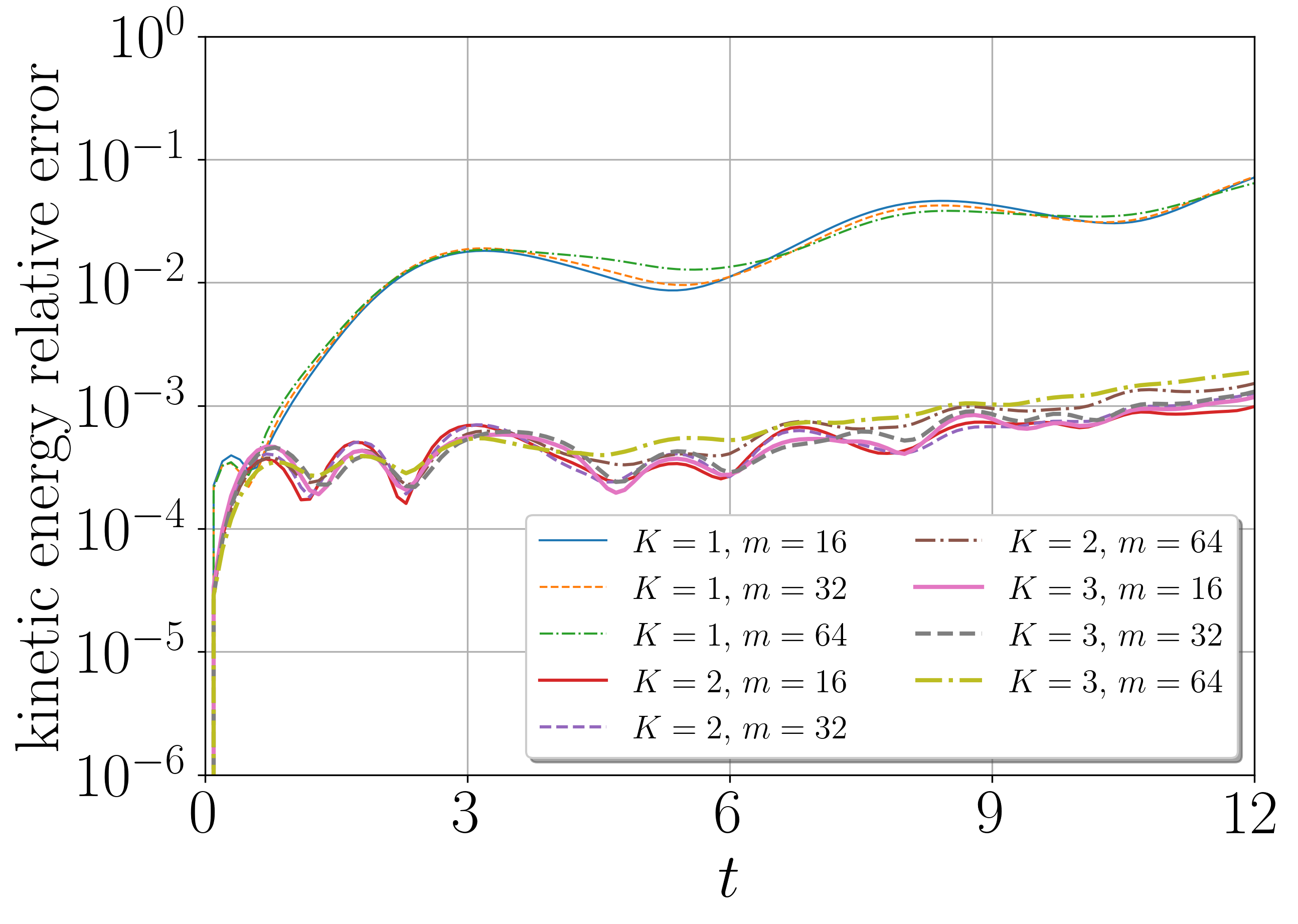}}
\subfigure[]{\label{fig:RB_RecSymH}
\includegraphics[trim=0cm 0cm 0cm 0cm,clip=true,width=0.32\textwidth]{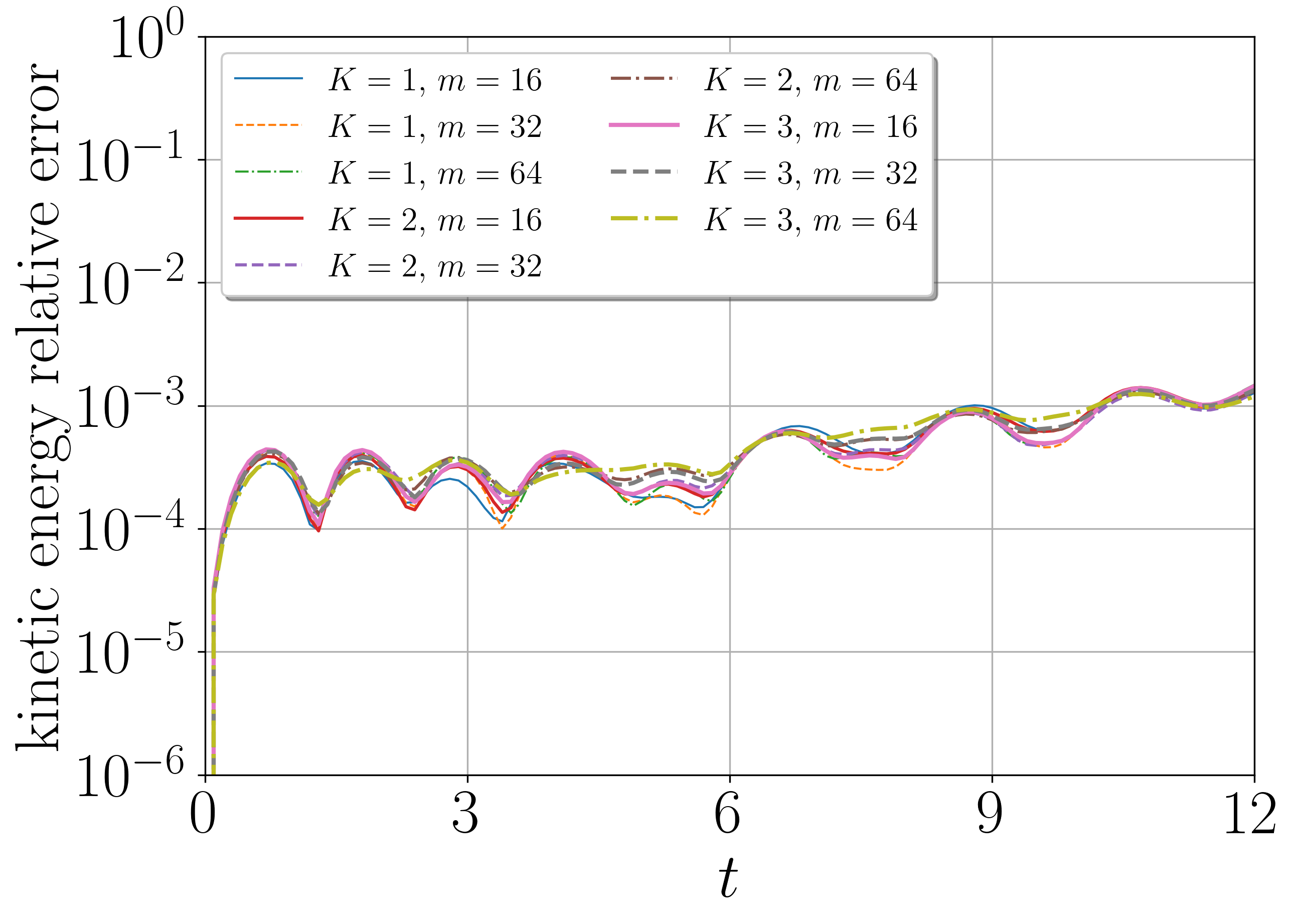}}
\subfigure[]{\label{fig:RB_RecNiceH}
\includegraphics[trim=0cm 0cm 0cm 0cm,clip=true,width=0.32\textwidth]{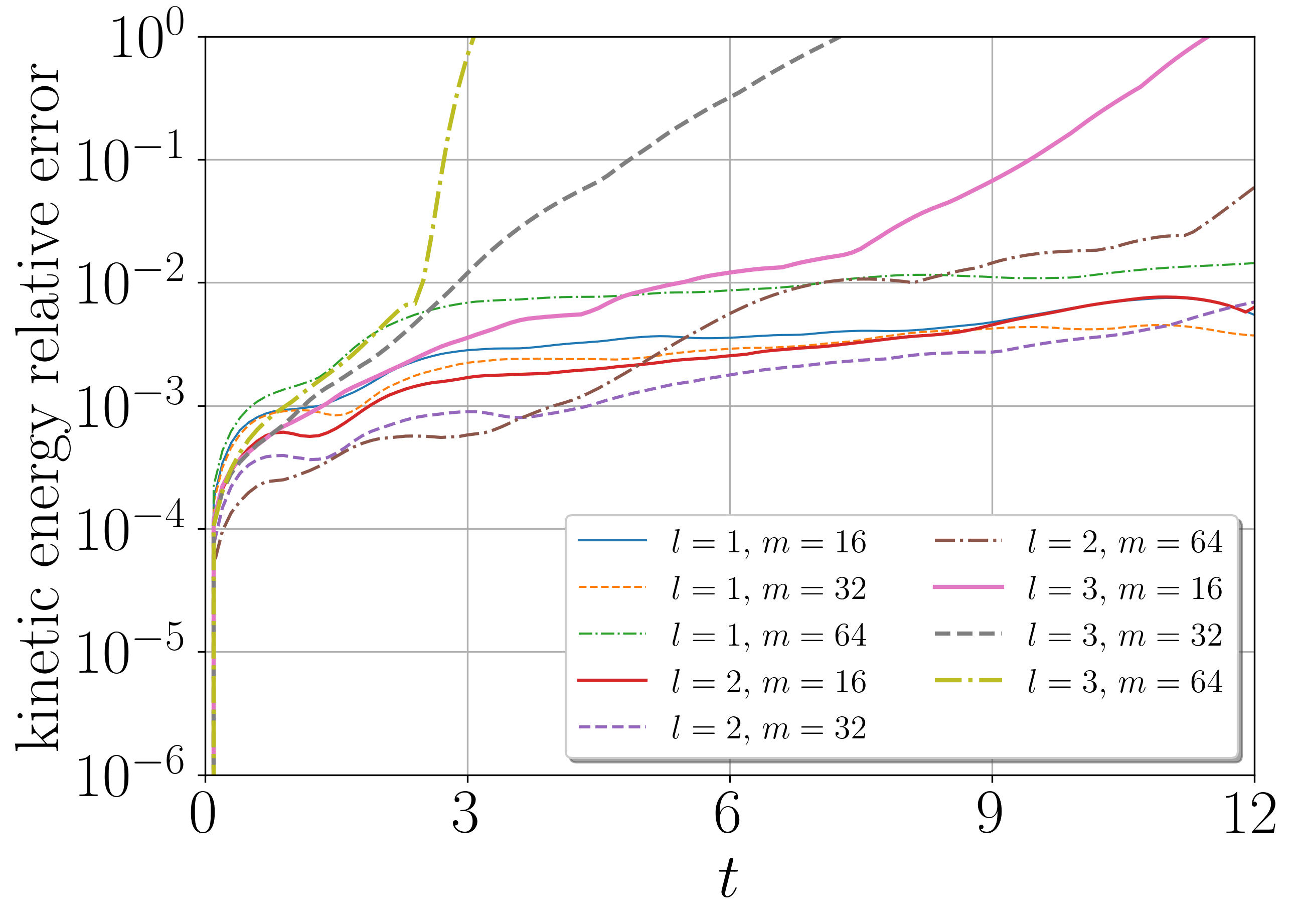}}
\caption{Averaged solution and kinetic energy \eqref{eq:KinEn} errors on the solution reconstruction time interval $[0,12]$ with the initial condition $y_0$ of the phase volume-preserving neural networks $\LSNet$, $\SLSNet$, and $\VPNet$. (a)-(c) solution absolute errors of $\LSNet$, $\SLSNet$, and $\VPNet$, respectively. (d)-(f) absolute values of the kinetic energy relative errors of $\LSNet$, $\SLSNet$, and $\VPNet$, respectively.}\label{fig:RB_Rec}
\end{figure}

\begin{figure}[t]
\centering 
\subfigure[]{\label{fig:RB_PredLocS}
\includegraphics[trim=0cm 0cm 0cm 0cm,clip=true,width=0.32\textwidth]{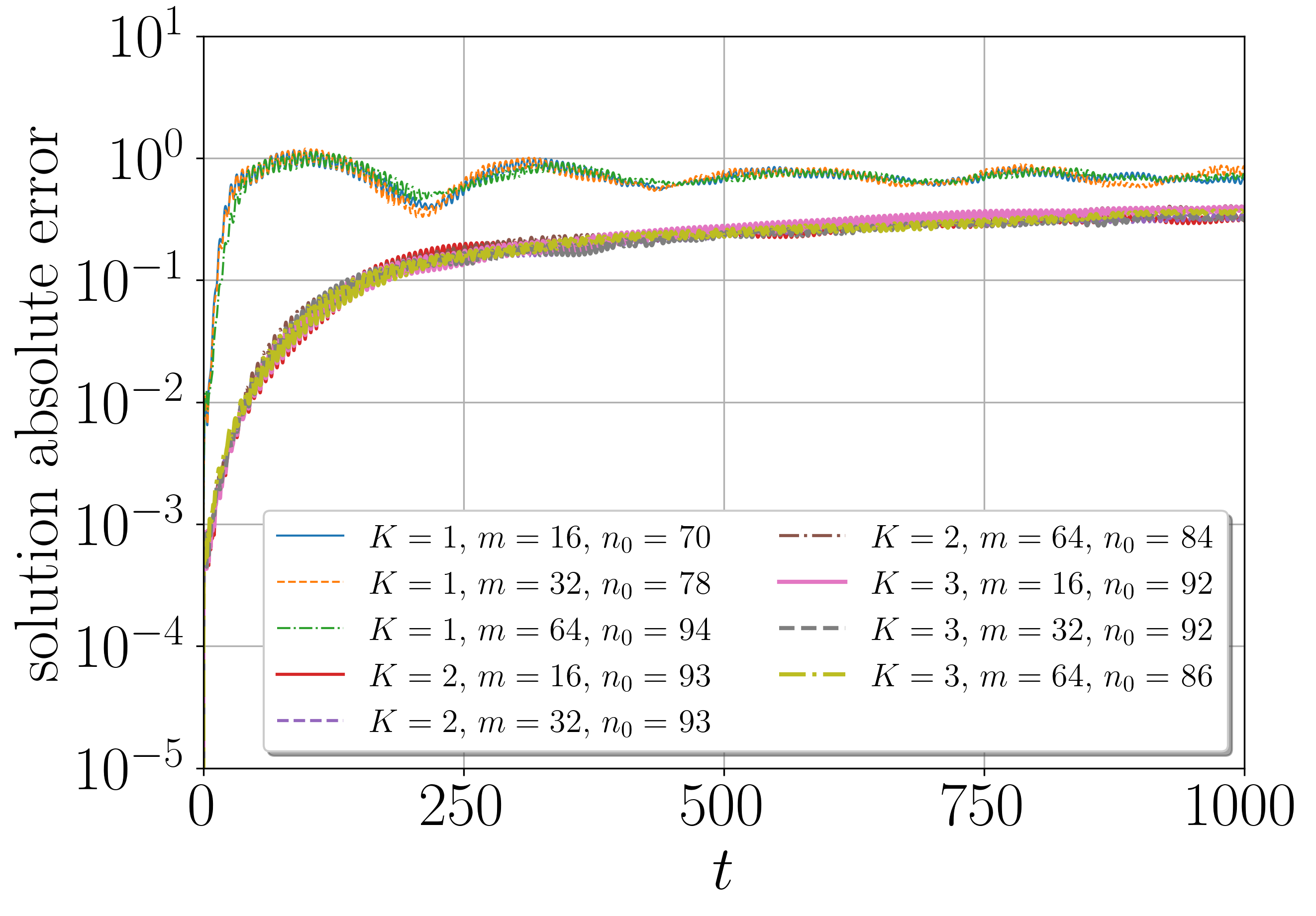}}
\subfigure[]{\label{fig:RB_PredSymS}
\includegraphics[trim=0cm 0cm 0cm 0cm,clip=true,width=0.32\textwidth]{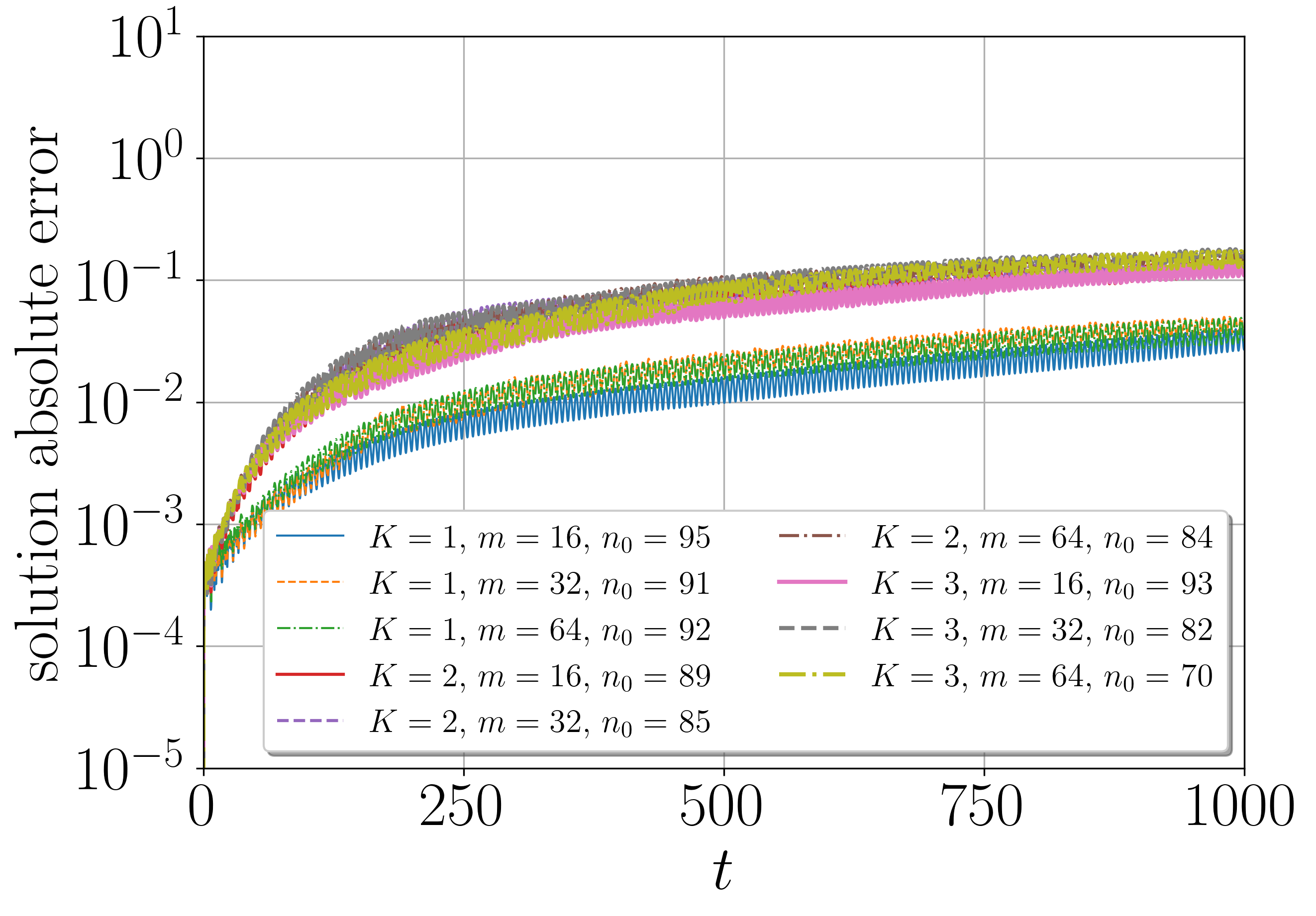}}
\subfigure[]{\label{fig:RB_PredNiceS}
\includegraphics[trim=0cm 0cm 0cm 0cm,clip=true,width=0.32\textwidth]{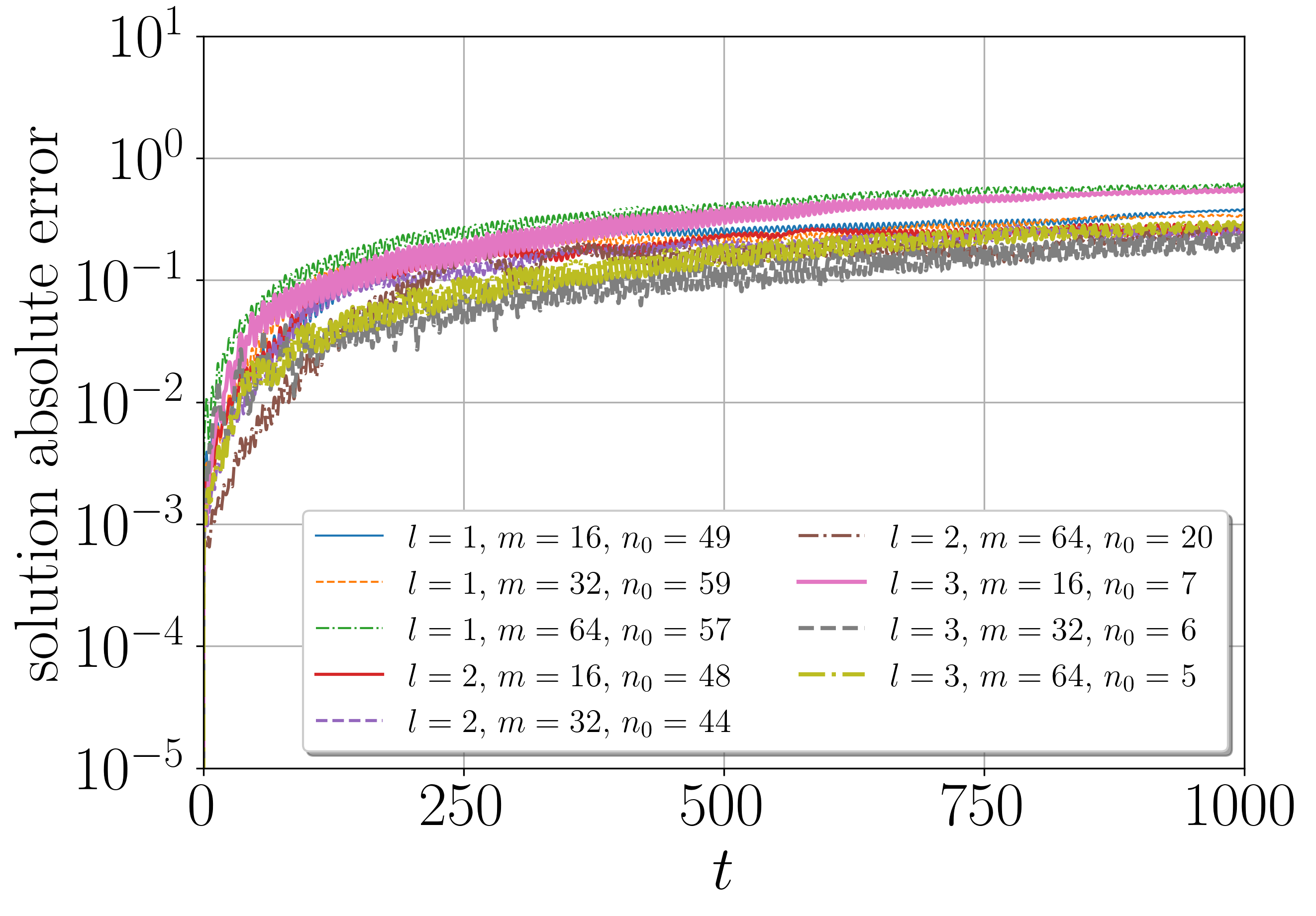}}
\subfigure[]{\label{fig:RB_PredLocH}
\includegraphics[trim=0cm 0cm 0cm 0cm,clip=true,width=0.32\textwidth]{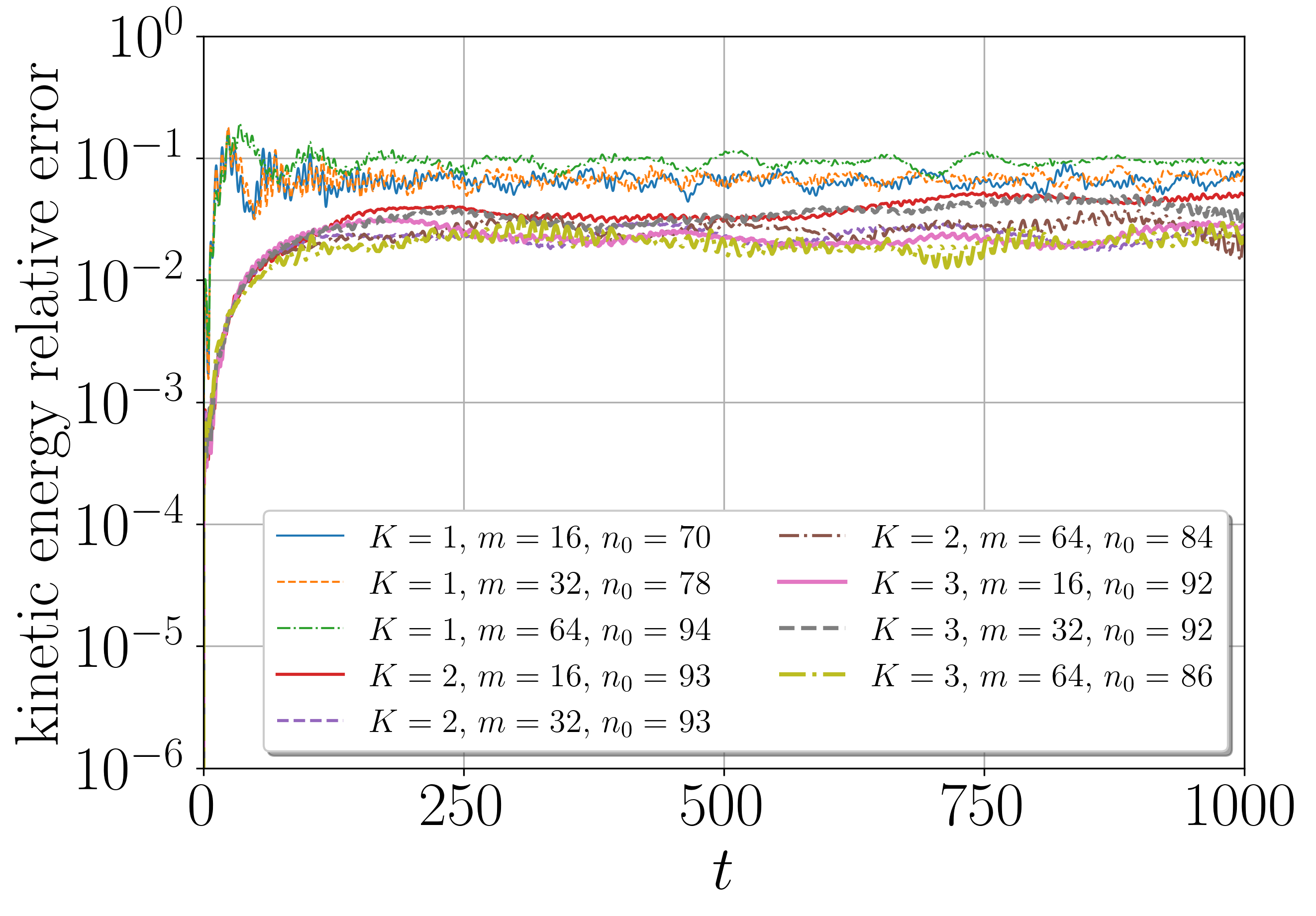}}
\subfigure[]{\label{fig:RB_PredSymH}
\includegraphics[trim=0cm 0cm 0cm 0cm,clip=true,width=0.32\textwidth]{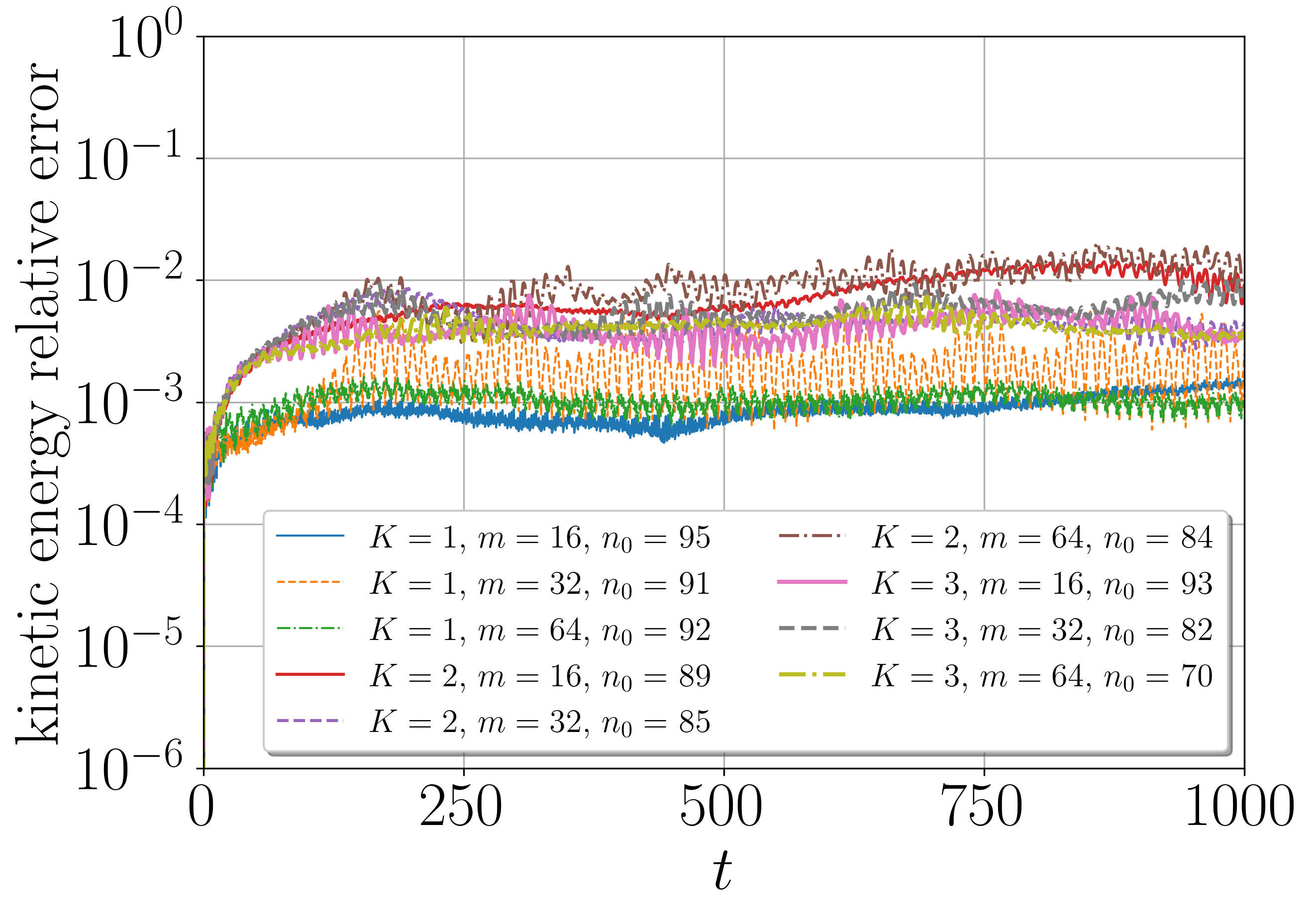}}
\subfigure[]{\label{fig:RB_PredNiceH}
\includegraphics[trim=0cm 0cm 0cm 0cm,clip=true,width=0.32\textwidth]{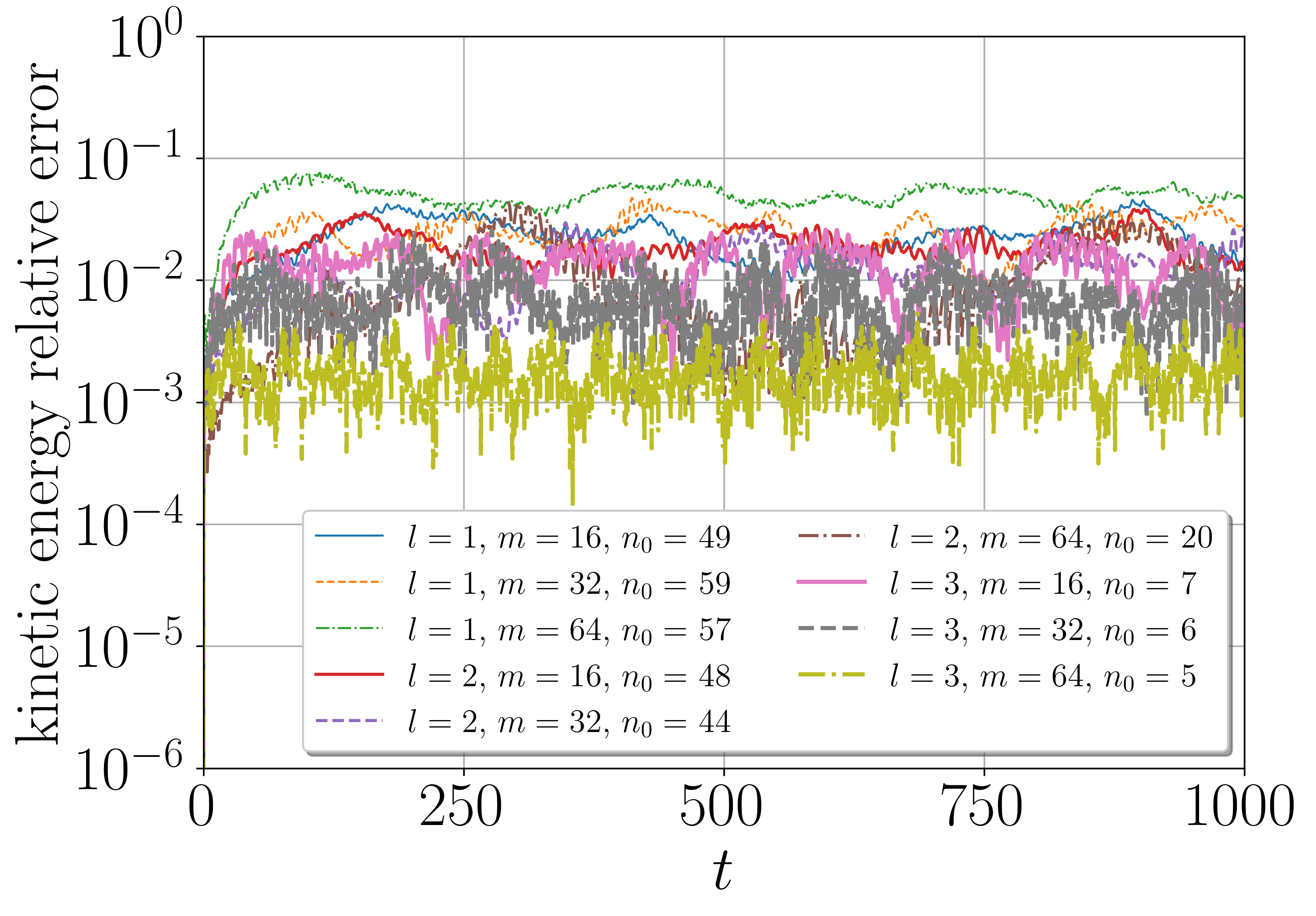}}
\caption{Averaged solution and kinetic energy \eqref{eq:KinEn} errors on the solution prediction time interval $[0,1000]$ with the initial condition $y_{12}$ of the phase volume-preserving neural networks $\LSNet$, $\SLSNet$, and $\VPNet$. From $100$ trained neural networks $n_0$ indicates the number of produced stable long-time predictions. (a)-(c) solution absolute errors of $\LSNet$, $\SLSNet$, and $\VPNet$, respectively. (d)-(f) absolute values of the kinetic energy relative errors of $\LSNet$, $\SLSNet$, and $\VPNet$, respectively.}\label{fig:RB_Pred}
\end{figure}

As in the example of Figure \ref{fig:RB_sol}, we train all three neural networks $\LSNet$, $\SLSNet$, and $\VPNet$ for $N_e=2\times 10^5$ epochs using exponential scheduling with $\eta_1=10^{-2}$ and $\eta_2=10^{-6}$. We required at least six alternating phase volume-preserving modules \eqref{eq:NICEmodules} in $\VPNet$ for the neural network to be able to learn a single periodic trajectory of \eqref{eq:RBody}. In Figures \ref{fig:RB_LossAcc}-\ref{fig:RB_Pred} we illustrate results of $\VPNet$ with composition of $\mathrm{L}=8$ number of alternating modules \eqref{eq:NICEmodules}. We obtained qualitatively similar, but not better, results with $\mathrm{L}=6$ and $\mathrm{L}=7$, suggesting that no potential significant improvements can be obtained in this numerical experiment with $\VPNet$ considering a larger number of {\it Up} and {\it Low} modules \eqref{eq:NICEmodules}.      

All three neural networks $\LSNet$, $\SLSNet$, and $\VPNet$ were trained with three different network width values $m=16$, $32$, $64$. For $\VPNet$ we considered fully-connected networks $\NNet_{1,2}$, see \eqref{eq:NICEmodules}, with one, two and three hidden layers, i.e., $l=1$, $2$, $3$. We trained locally-symplectic neural networks $\LSNet$ and $\SLSNet$ with $K=1$, $2$, $3$. In Figures \ref{fig:RB_LossAcc}-\ref{fig:RB_Rec} we show results averaged over all results produced by $100$ trained neural networks with the same network parameters $K$, $m$ and $l$. Similarly, we also illustrate averaged results in Figure \ref{fig:RB_Pred}, where the averaging is done only over the stable predicted long-time solutions by the neural networks. The number of predicted stable solutions is noted by $n_0$ and indicated in the figures' legends. The criteria for a predicted solution to be stable was imposed with the requirement that the maximal absolute relative errors of the kinetic energy \eqref{eq:KinEn} and invariant \eqref{eq:Iinv} over the whole prediction time interval are smaller than one. Considering sufficient long-time predictions, over the time interval $[0,1000]$ in our example, we were able to quantify a rough percentage of trained neural networks which were able to learn to some approximation the single periodic solution of the rigid body problem \eqref{eq:RBody}. Thus, the number of long-time stable predictions $n_0$ provides another means for the quantitative comparison of the neural networks $\LSNet$, $\SLSNet$, and $\VPNet$.       

In Figure \ref{fig:RB_LossAcc} we show averaged MSE loss \eqref{eq:loss} and accuracy \eqref{eq:acc} for all three neural networks $\LSNet$, $\SLSNet$, and $\VPNet$ trained with different network parameter values. Notice settle differences for the loss and accuracy values between all three neural networks. At initial epochs $\VPNet$ have significantly larger error values compared to both locally-symplectic neural networks $\LSNet$ and $\SLSNet$ and slightly larger error values at the final epoch, compare Figures \ref{fig:RB_LossNice} and \ref{fig:RB_AccNice} with Figures \ref{fig:RB_LossLoc}--\ref{fig:RB_LossSym} and \ref{fig:RB_AccLoc}--\ref{fig:RB_AccSym}. It appears that $\SLSNet$ have the smallest loss and accuracy values at the final epoch, see Figures \ref{fig:RB_LossSym} and \ref{fig:RB_AccSym}. In Figures \ref{fig:RB_LossNice} and \ref{fig:RB_AccNice} noticeable large fluctuations during the training in the loss and accuracy values are visible compared to the function values in Figures \ref{fig:RB_LossLoc}--\ref{fig:RB_LossSym} and \ref{fig:RB_AccLoc}--\ref{fig:RB_AccSym}, where the averaged functions appear more smooth in nature. Such fluctuations in the loss and accuracy functions can be reduced with the use of a smaller learning rate, but that may result in the need for training with more epochs.  

Notice that the loss and accuracy values for $\LSNet$ in Figures \ref{fig:RB_LossLoc} and \ref{fig:RB_AccLoc}, respectively, with $K=1$ are much greater compared to the values with $K=2$ and $K=3$, which explains why we were not able to obtain very good predictions by $\LSNet$ with $K=1$. On the contrary, $\SLSNet$ loss and accuracy errors, Figures \ref{fig:RB_LossSym} and \ref{fig:RB_AccSym}, for the case with $K=1$ are much smaller and indistinguishable from the cases $K=2$ and $K=3$ at the final epoch. Since we applied exponential scheduling with the final learning rate value $\eta_2=10^{-6}$, approaching the final epoch the learning has essentially slowed down and undesirable fluctuations have disappeared. 

Importantly, not only do the loss function values tend to zero but also the accuracy function values, Figures \ref{fig:RB_AccLoc}--\ref{fig:RB_AccNice}, which are computed using validation data set samples not seen by the neural networks during training. This indicates that all three neural networks $\LSNet$, $\SLSNet$, and $\VPNet$ will be able to predict part of the trajectory formed by the validation data samples, i.e., will be able to produce at least short-time good predictions. Initially seamed settle differences in the loss and accuracy values at the final epoch will essentially reflect on the accuracy of the reconstructed trajectory and long-time predictions, which we illustrate in Figures \ref{fig:RB_Rec}--\ref{fig:RB_Pred}. To keep the presentation more concise in both Figures \ref{fig:RB_Rec}--\ref{fig:RB_Pred} we have excluded error results for the conservation of the invariant \eqref{eq:Iinv} by the neural networks. Error results for the invariant \eqref{eq:Iinv} are equivalent to the error plots for the kinetic energy \eqref{eq:KinEn}, which can be attributed to the fact that both conserved quantities \eqref{eq:KinEn}--\eqref{eq:Iinv} are quadratic functions.    

In Figure \ref{fig:RB_Rec} we demonstrate averaged solution absolute global errors, Figures \ref{fig:RB_RecLocS}--\ref{fig:RB_RecNiceS}, and the absolute values of the relative errors of the kinetic energy \eqref{eq:KinEn}, Figures \ref{fig:RB_RecLocH}--\ref{fig:RB_RecNiceH}, for all three phase volume-preserving neural networks $\LSNet$, $\SLSNet$ and $\VPNet$ on the solution reconstruction time interval $[0,12]$. Reconstructions of the periodic solution are obtained iteratively by the neural networks with the initial condition $y_0$. Noticeably, not all reconstructed solutions by $\VPNet$ are bounded solutions, especially, for $\VPNet$ with three hidden layers ($l=3$). As a consequence, not all averaged kinetic energy \eqref{eq:KinEn} relative errors stay bounded over the whole reconstruction time interval $[0,12]$, see Figure \ref{fig:RB_RecNiceH}. As expected from the loss and accuracy values in Figures \ref{fig:RB_LossLoc} and \ref{fig:RB_AccLoc}, respectively, for the $\LSNet$ with $K=1$ solution reconstruction and kinetic energy conservation errors in Figures \ref{fig:RB_RecLocS} and \ref{fig:RB_RecLocH} are significantly worse compared to the results of $\LSNet$ with $K=2$, $3$. Evidently, see Figures \ref{fig:RB_RecSymS} and \ref{fig:RB_RecSymH}, symmetric locally-symplectic neural networks $\SLSNet$ outperform $\LSNet$ with $K=1$ and have comparable results with $\LSNet$ when $K=2$ or $K=3$.

Averaged stable solution global errors in Figures \ref{fig:RB_RecLocS}--\ref{fig:RB_RecNiceS} still grow in a linear fashion as in Figure \ref{fig:RB_LongGE}. Note that errors in Figure \ref{fig:RB_Rec} (as well as in Figure \ref{fig:RB_Pred}) are visualized on a logarithmic scale. Comparing reconstruction error results in Figure \ref{fig:RB_Rec} for all three neural networks we can observe that all $\LSNet$ (when $K>1$) and $\SLSNet$ have produced stable reconstructions with smaller errors compared to $\VPNet$. In addition, interestingly, we do not observe significant differences between averaged reconstruction errors depending on $K$ and width $m$ values, $K=1$ for $\LSNet$ being the exception. Despite $\SLSNet$ preserving the flow property \eqref{eq:FlowProp} there is no evident reduction in the averaged reconstruction errors compared to the non-symmetric $\LSNet$.    

Results of Figure \ref{fig:RB_Rec} significantly differ when long-time predictions by $\LSNet$, $\SLSNet$, and $\VPNet$ are considered, see Figure \ref{fig:RB_Pred}. In Figure \ref{fig:RB_Pred} we illustrate averaged solution absolute global errors, Figures \ref{fig:RB_PredLocS}--\ref{fig:RB_PredNiceS}, and the absolute values of the relative errors of the kinetic energy \eqref{eq:KinEn}, Figures \ref{fig:RB_PredLocH}--\ref{fig:RB_PredNiceH}, for all three neural networks on the solution prediction time interval $[0,1000]$. Predictions of the periodic solution are obtained iteratively with the initial condition $y_{12}$, where the initial condition $y_{12}$ is the first ground truth value (sample) in the validation data set. As already stated above, in contrast to the results of Figures \ref{fig:RB_LossAcc}--\ref{fig:RB_Rec}, averaging in Figure \ref{fig:RB_Pred} is performed only over the predicted stable solutions. The number of obtained stable solutions by the neural networks with the same network parameter values is indicated by $n_0$ in the figures' legends. 

Observing numbers $n_0$ of predicted stable solutions over the whole time interval $[0,1000]$ in Figure \ref{fig:RB_Pred} it is evident that not all trained neural networks were able to produce stable long-time predictions. Notice that $\VPNet$ in Figures \ref{fig:RB_PredNiceS} and \ref{fig:RB_PredNiceH} were able to produce the least amount of stable solutions, where the number $n_0$ decreases with increased number of hidden layers $l$. Interestingly, the most accurate predicted stable solutions by $\VPNet$, judged by the relative errors of the kinetic energy \eqref{eq:KinEn}, see Figure \ref{fig:RB_PredNiceH}, are obtained with fully-connected neural networks $\NNet_{1,2}$ having three hidden layers while at the same time these $\VPNet$ have produced the least amount of stable long-time predictions. Evidently, Figures \ref{fig:RB_PredLocS}--\ref{fig:RB_PredSymS} and \ref{fig:RB_PredLocH}--\ref{fig:RB_PredSymH} show that $\LSNet$ and $\SLSNet$ were able to produce much greater equivalent number of stable long-time predicted solutions compared to $\VPNet$. A striking difference is observed when the solution and kinetic energy conservation errors are compared between $\LSNet$ and $\SLSNet$, where the errors for the symmetric locally-symplectic neural networks have errors of significant magnitude smaller, e.g., compare Figures \ref{fig:RB_PredLocH}--\ref{fig:RB_PredSymH}. In particular, $\SLSNet$ with $K=1$ produced the smallest errors.

In conclusion, $\SLSNet$ have produced an equivalent number of long-time stable solutions compared to the $\LSNet$ but also have the smallest averaged solution and kinetic energy conservation relative errors compared to $\LSNet$ and $\VPNet$. We already observed $\SLSNet$ superior performance over the $\LSNet$ when learning linear dynamics in Section \ref{sec:LinProblem}, see also Figure \ref{fig:Advection}. Thus, we advocate that symmetric neural networks, i.e., the flow property \eqref{eq:FlowProp}-preserving neural networks, may provide significant performance gains in long-time predictions.

\subsubsection{Learning a single periodic trajectory with noisy data}\label{sec:RB_Noisy} 
In the previous section, we trained neural networks $\LSNet$, $\SLSNet$, and $\VPNet$ for learning a single periodic trajectory of the rigid body dynamics using training data obtained from the high precision numerical integration of the equations \eqref{eq:RBody}. In practice, training data may be subject to noise or, simply, to round-off numerical errors. In this section, we repeat the numerical experiment of Section \ref{sec:RB_Single} but with induced random noise into the training data. We consider the same training data set of $N=120$ samples obtained from the initial condition $y_0$ with time step $\tau=0.1$. Then we add a random perturbation to each ground truth input and output data sample drawn from the uniform distribution $\mathcal{U}(-\delta,\delta)$, where $\delta\geq 0$ indicates the magnitude of noise. We do not add any random perturbation to the validation data set samples, i.e., the MSE accuracy function \eqref{eq:acc} is evaluated using the actual ground truth solution values. 

\begin{figure}[t]
\centering 
\subfigure[]{\label{fig:Noisy_LossLoc}
\includegraphics[trim=0cm 0cm 0cm 0cm,clip=true,width=0.32\textwidth]{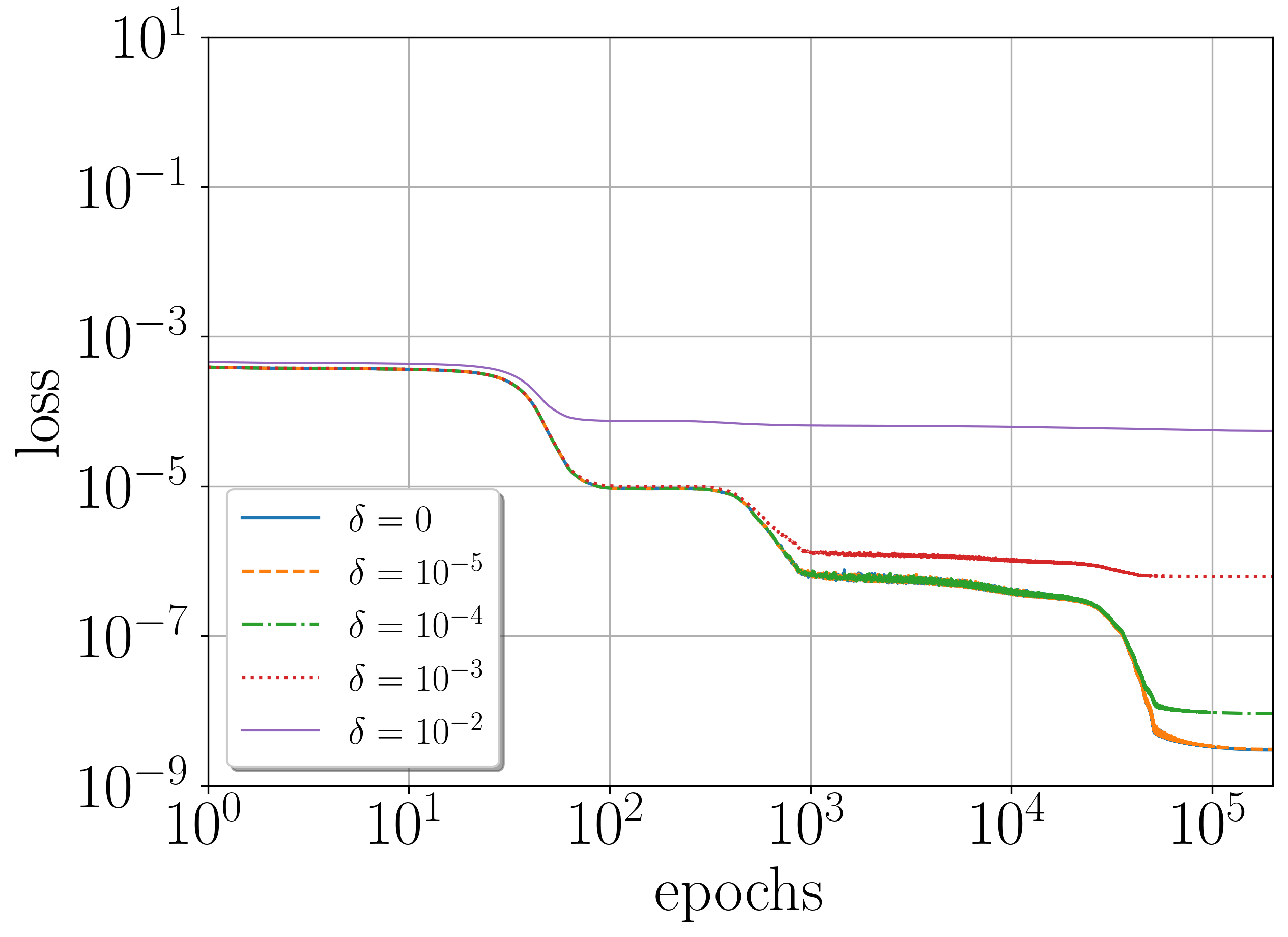}}
\subfigure[]{\label{fig:Noisy_LossSym}
\includegraphics[trim=0cm 0cm 0cm 0cm,clip=true,width=0.32\textwidth]{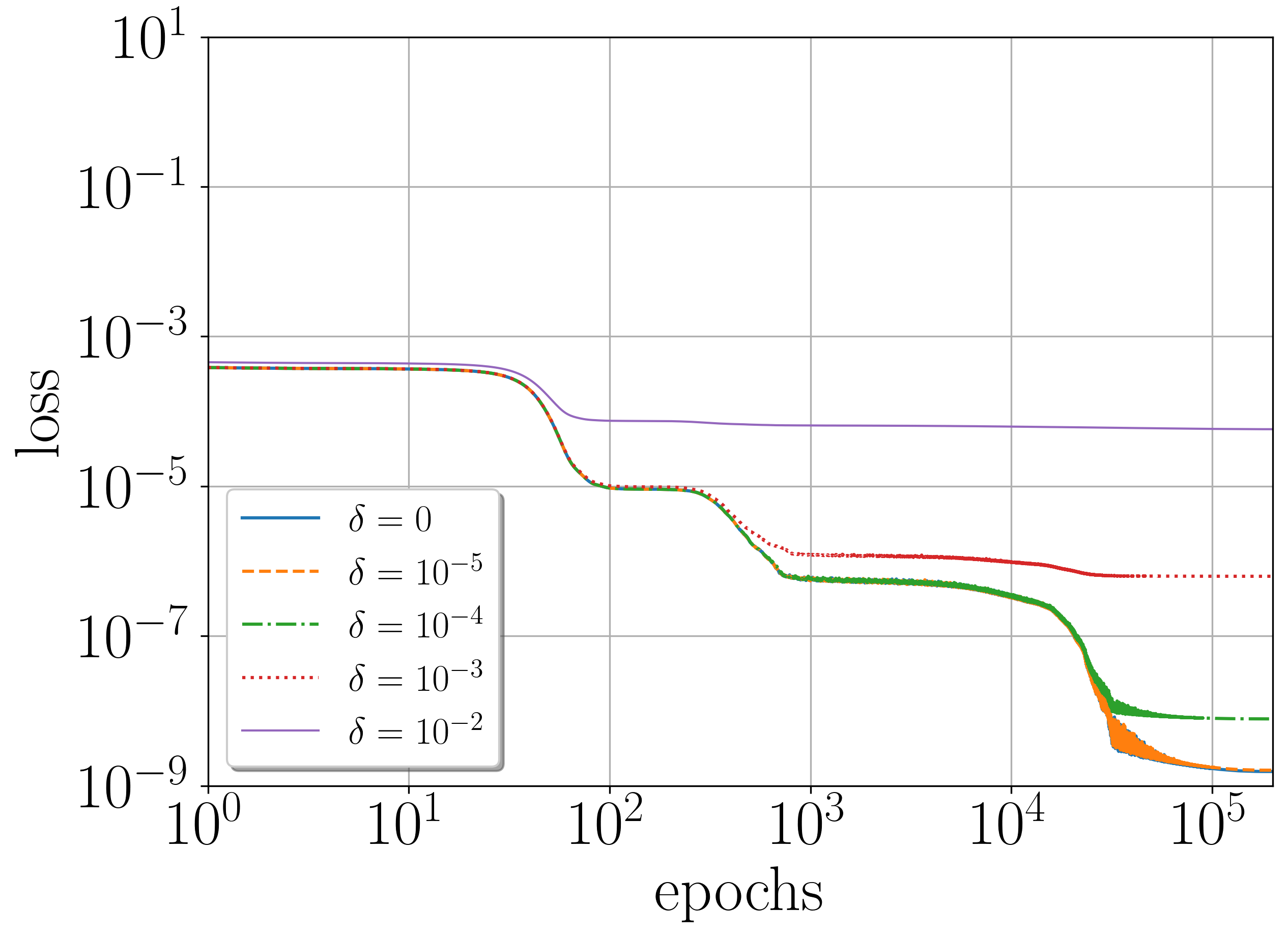}}
\subfigure[]{\label{fig:Noisy_LossNice}
\includegraphics[trim=0cm 0cm 0cm 0cm,clip=true,width=0.32\textwidth]{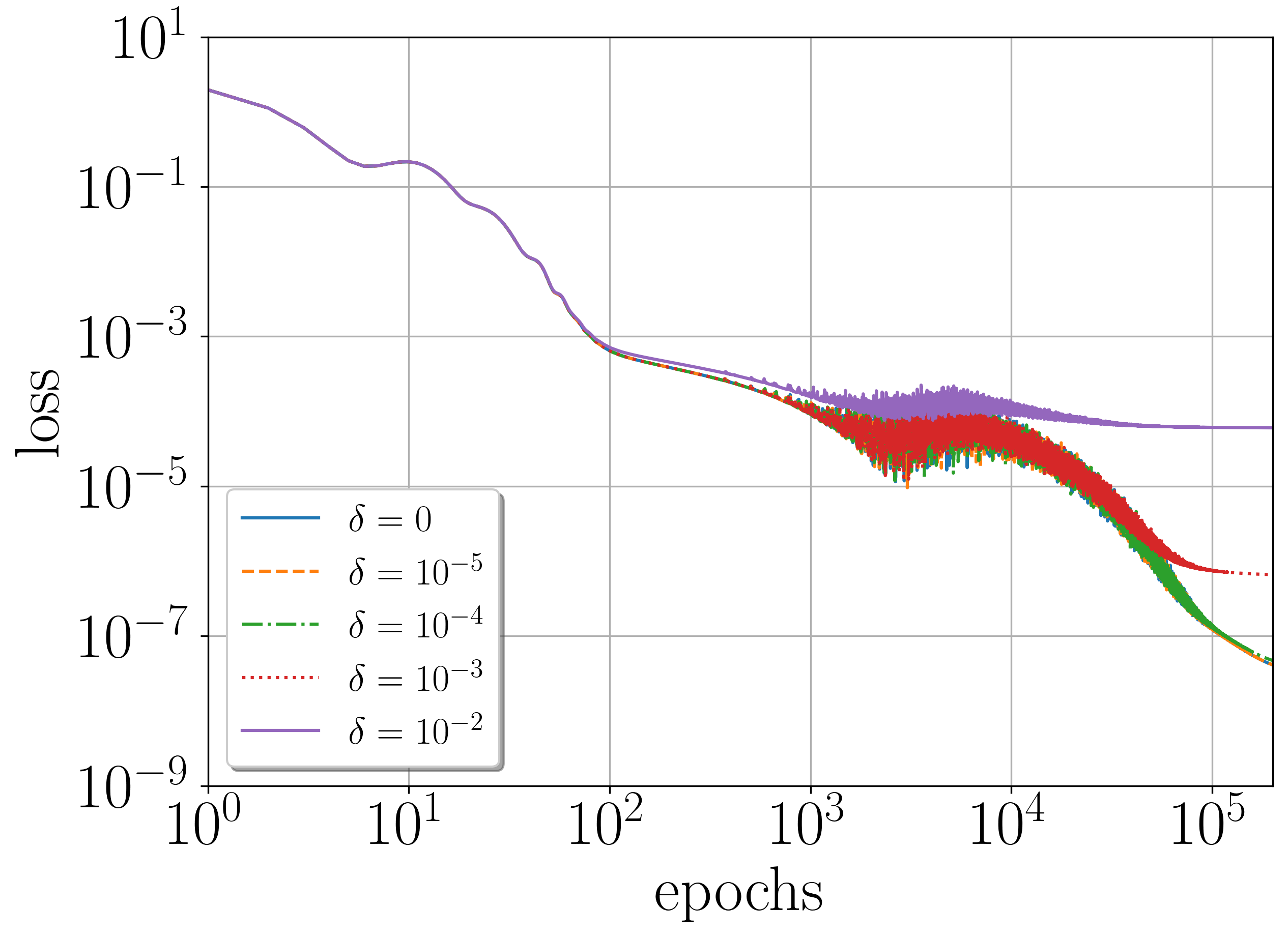}}
\subfigure[]{\label{fig:Noisy_AccLoc}
\includegraphics[trim=0cm 0cm 0cm 0cm,clip=true,width=0.32\textwidth]{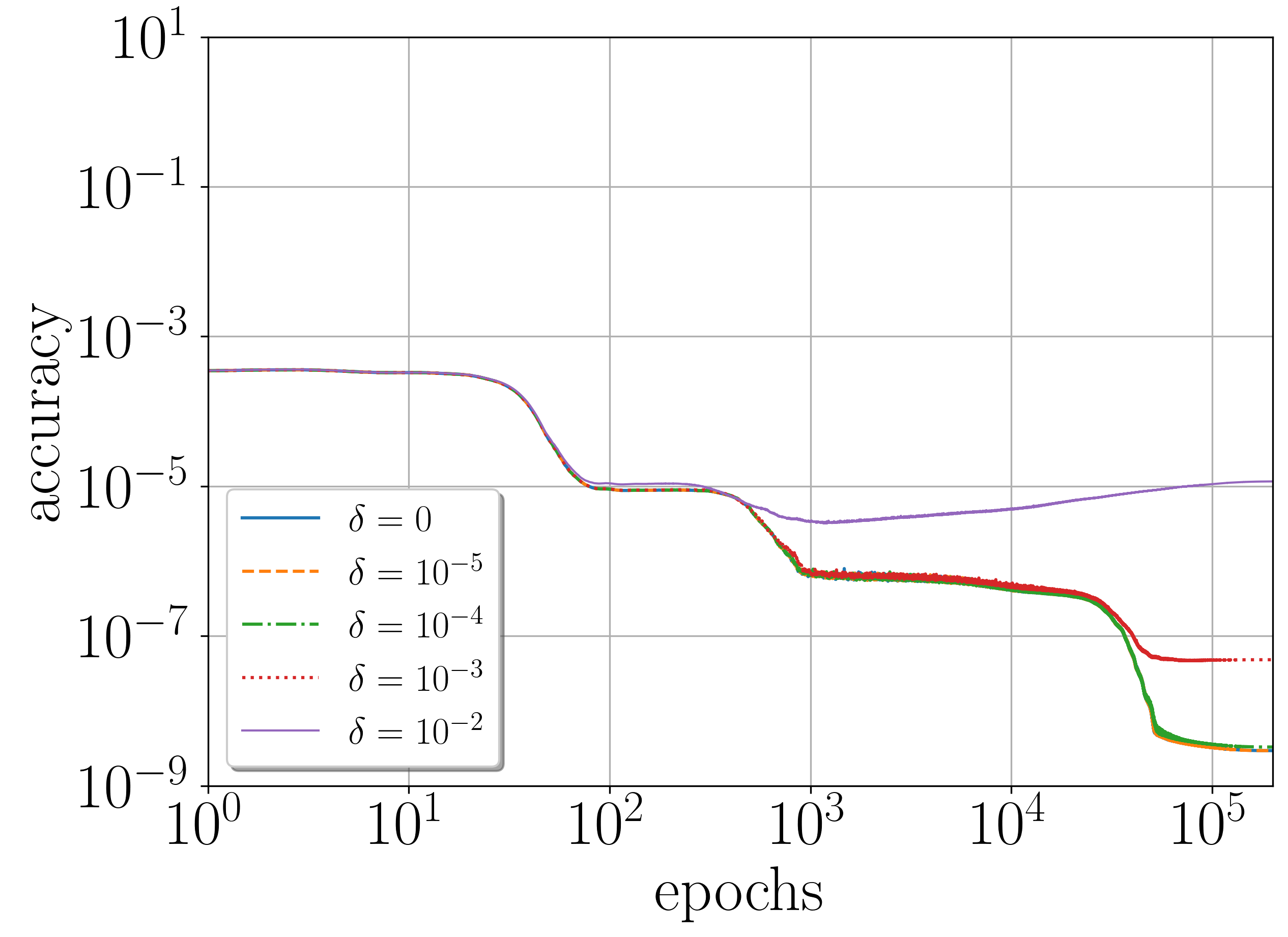}}
\subfigure[]{\label{fig:Noisy_AccSym}
\includegraphics[trim=0cm 0cm 0cm 0cm,clip=true,width=0.32\textwidth]{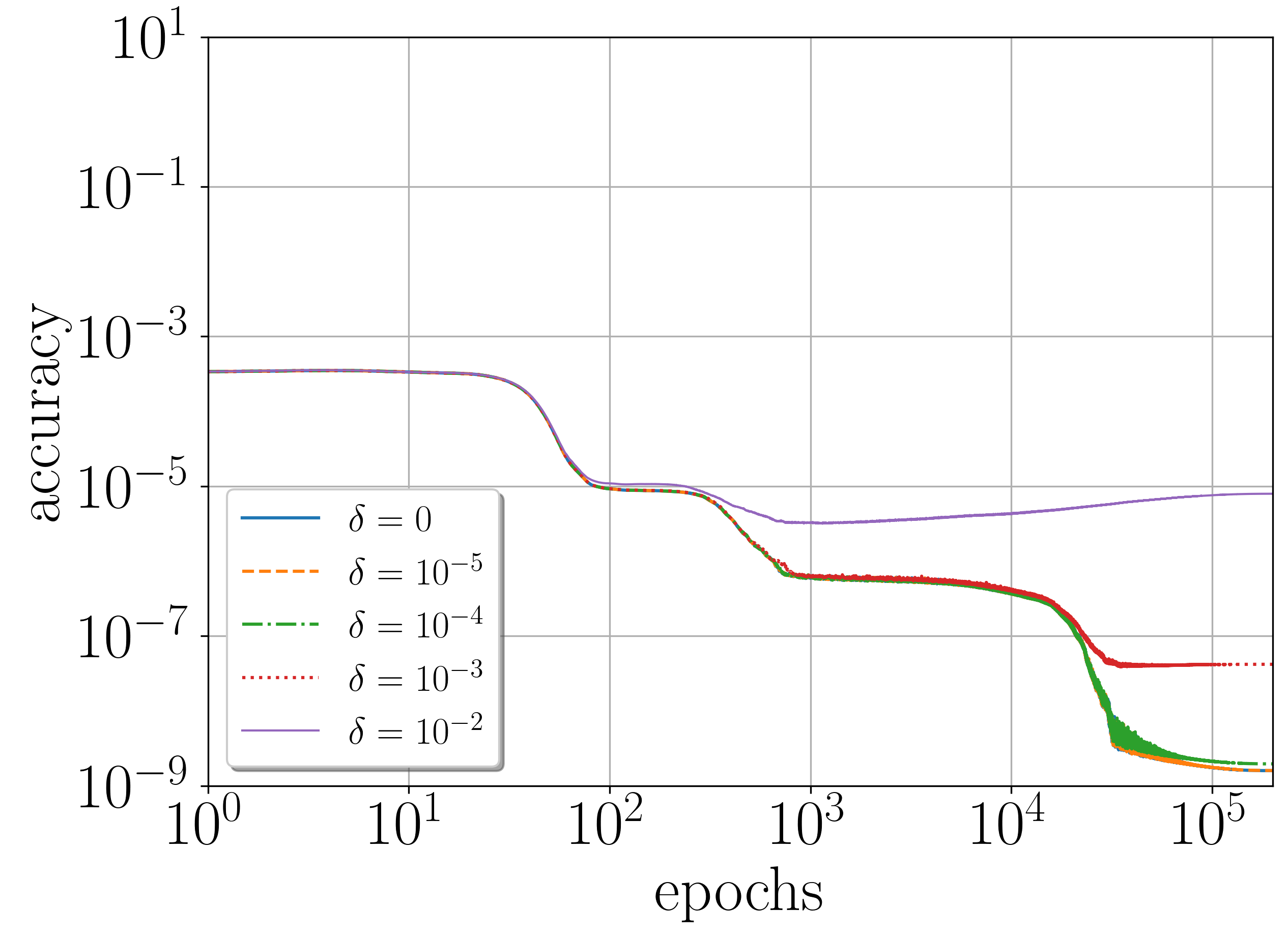}}
\subfigure[]{\label{fig:Noisy_AccNice}
\includegraphics[trim=0cm 0cm 0cm 0cm,clip=true,width=0.32\textwidth]{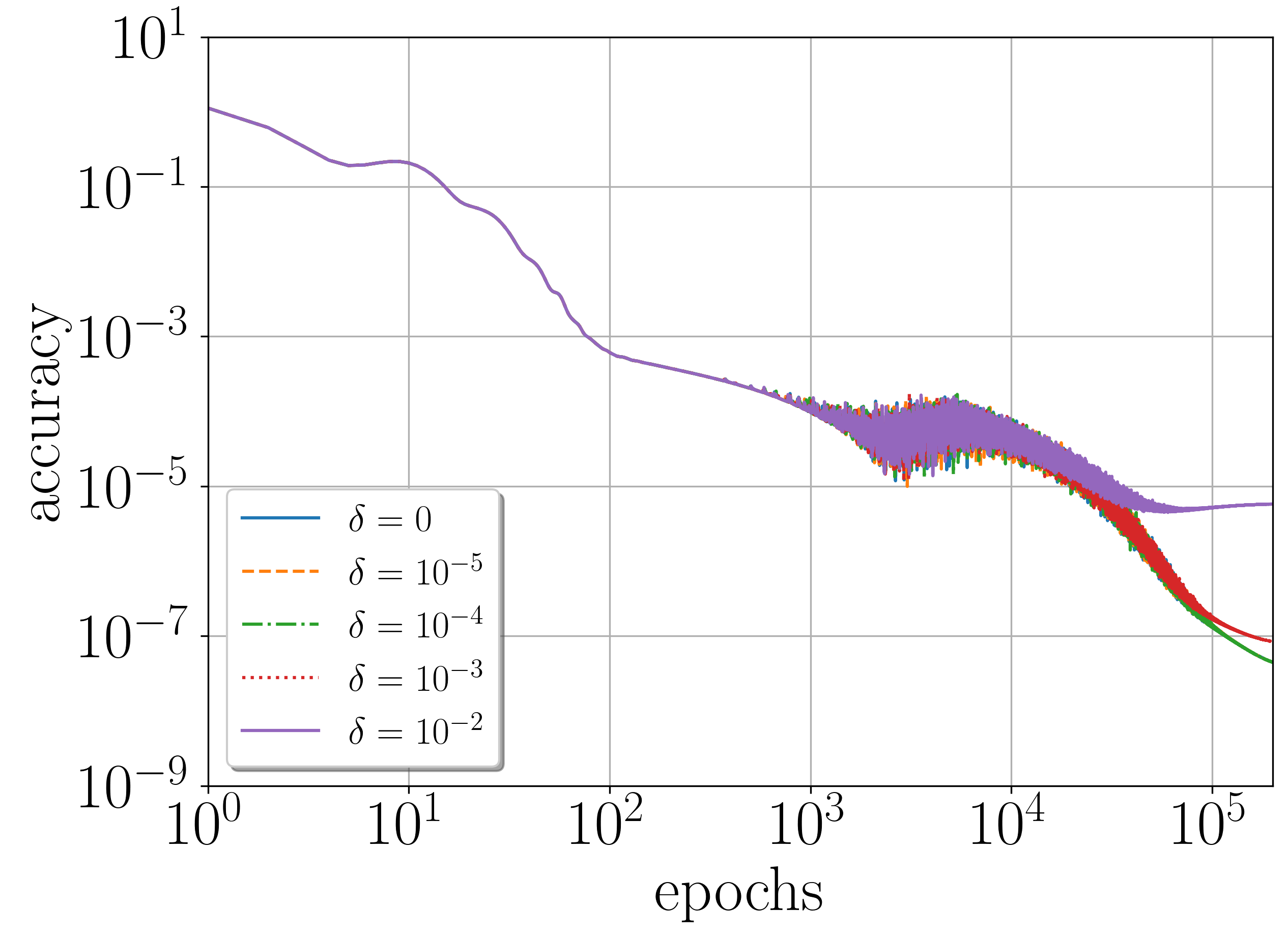}}
\caption{Averaged mean squared error loss \eqref{eq:loss} and accuracy \eqref{eq:acc} function values of the phase volume-preserving neural networks $\LSNet$ ($K=2$ and $m=16$), $\SLSNet$ ($K=1$ and $m=16$), and $\VPNet$ ($\mathrm{L}=8$, $l=1$ and $m=16$) trained with noisy data, where the parameter $\delta$ characterizes the added amount of noise. (a)-(c) loss values of $\LSNet$, $\SLSNet$, and $\VPNet$, respectively. (d)-(f) accuracy values of $\LSNet$, $\SLSNet$, and $\VPNet$, respectively.
}\label{fig:Noisy_LossAcc}
\end{figure}

\begin{figure}[t]
\centering 
\subfigure[]{\label{fig:Noisy_RecLocS}
\includegraphics[trim=0cm 0cm 0cm 0cm,clip=true,width=0.32\textwidth]{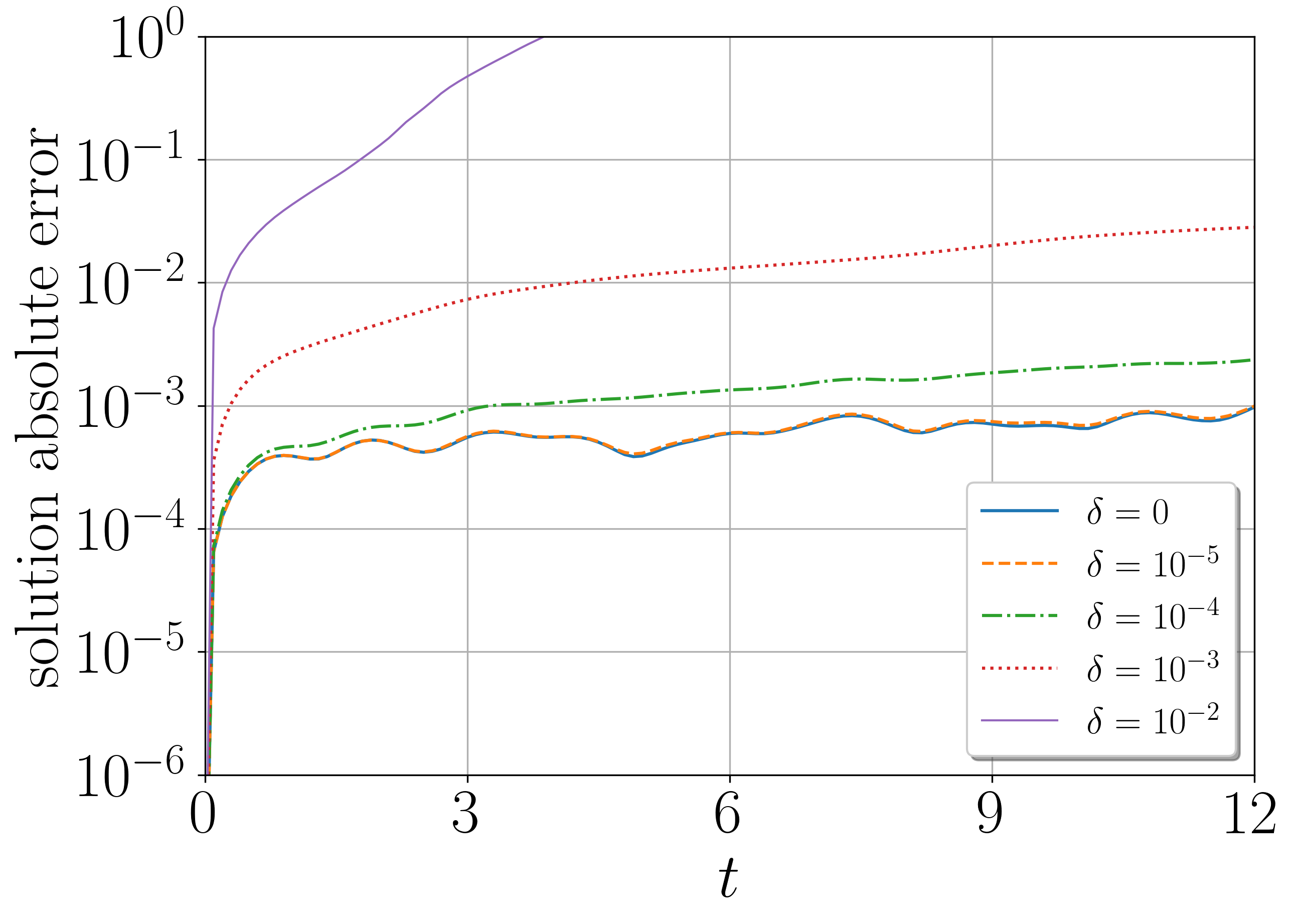}}
\subfigure[]{\label{fig:Noisy_RecSymS}
\includegraphics[trim=0cm 0cm 0cm 0cm,clip=true,width=0.32\textwidth]{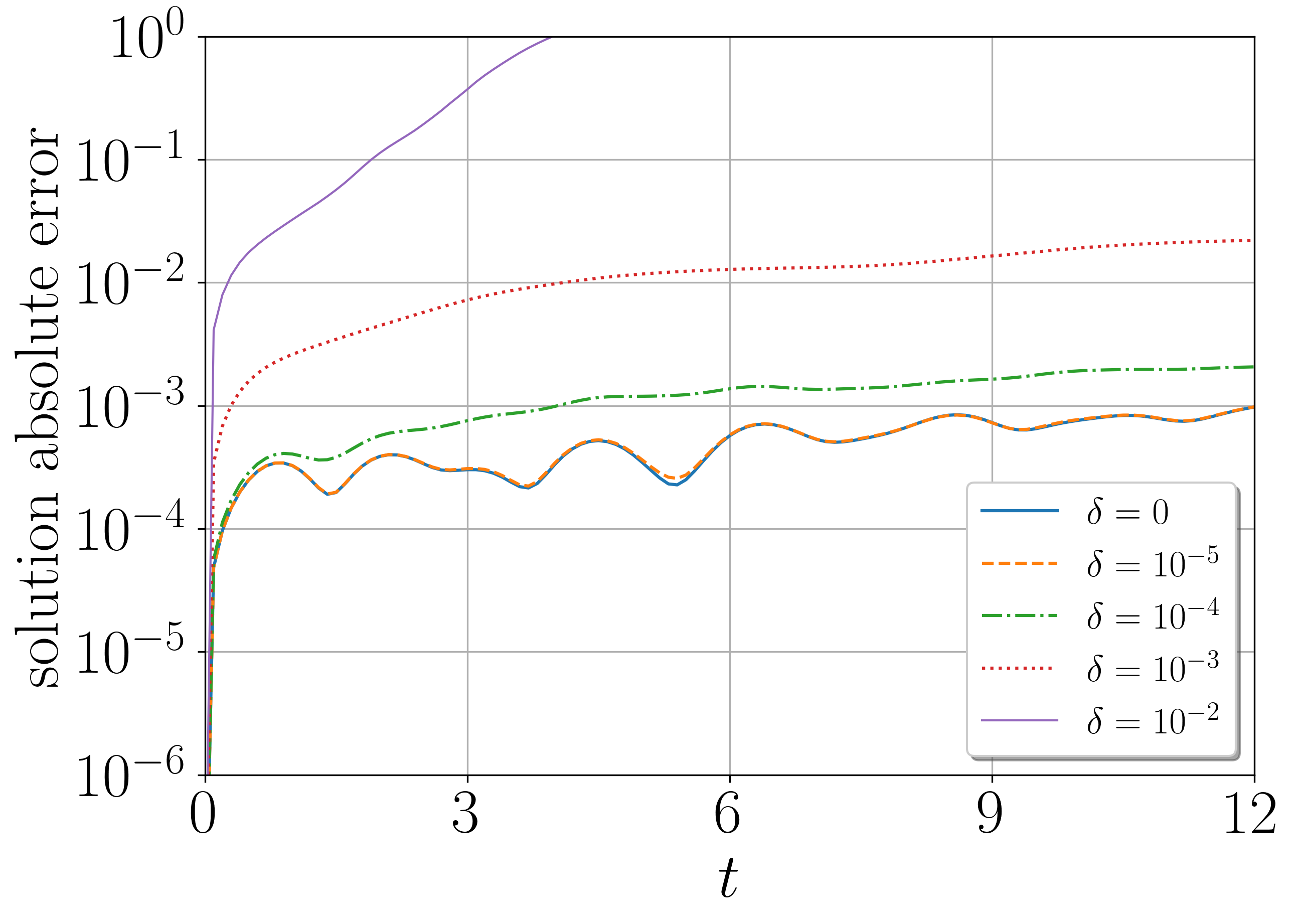}}
\subfigure[]{\label{fig:Noisy_RecNiceS}
\includegraphics[trim=0cm 0cm 0cm 0cm,clip=true,width=0.32\textwidth]{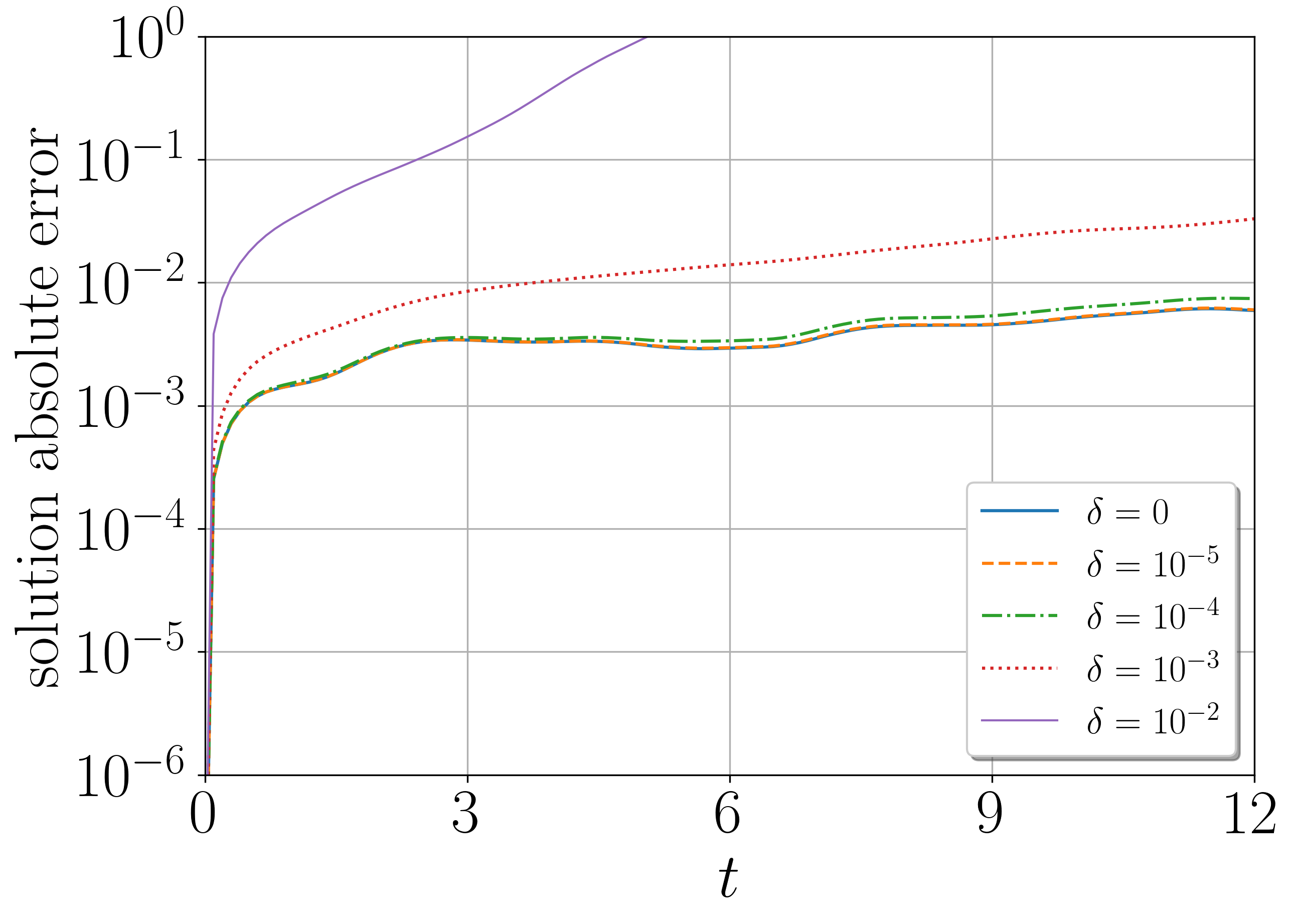}}
\subfigure[]{\label{fig:Noisy_RecLocH}
\includegraphics[trim=0cm 0cm 0cm 0cm,clip=true,width=0.32\textwidth]{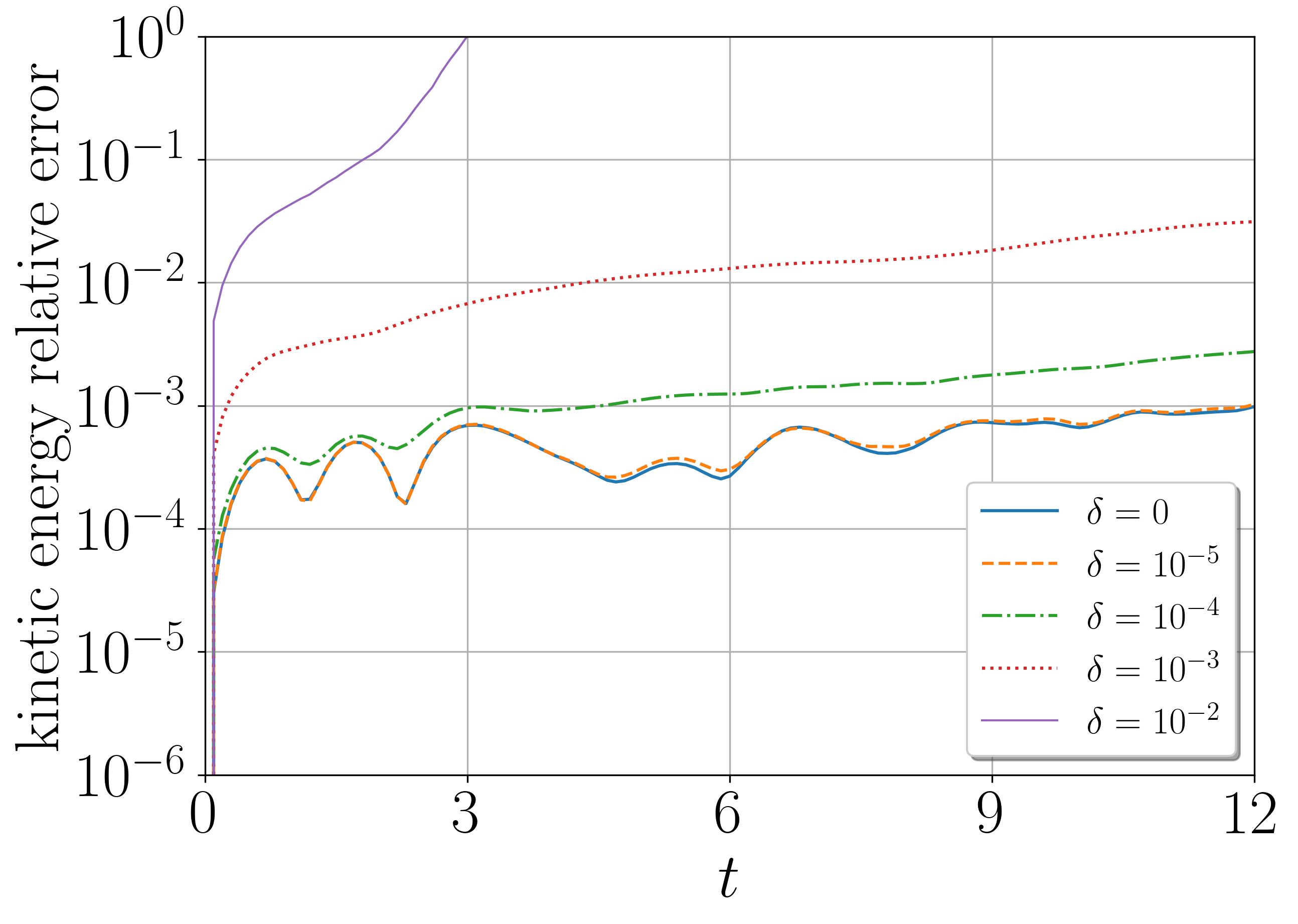}}
\subfigure[]{\label{fig:Noisy_RecSymH}
\includegraphics[trim=0cm 0cm 0cm 0cm,clip=true,width=0.32\textwidth]{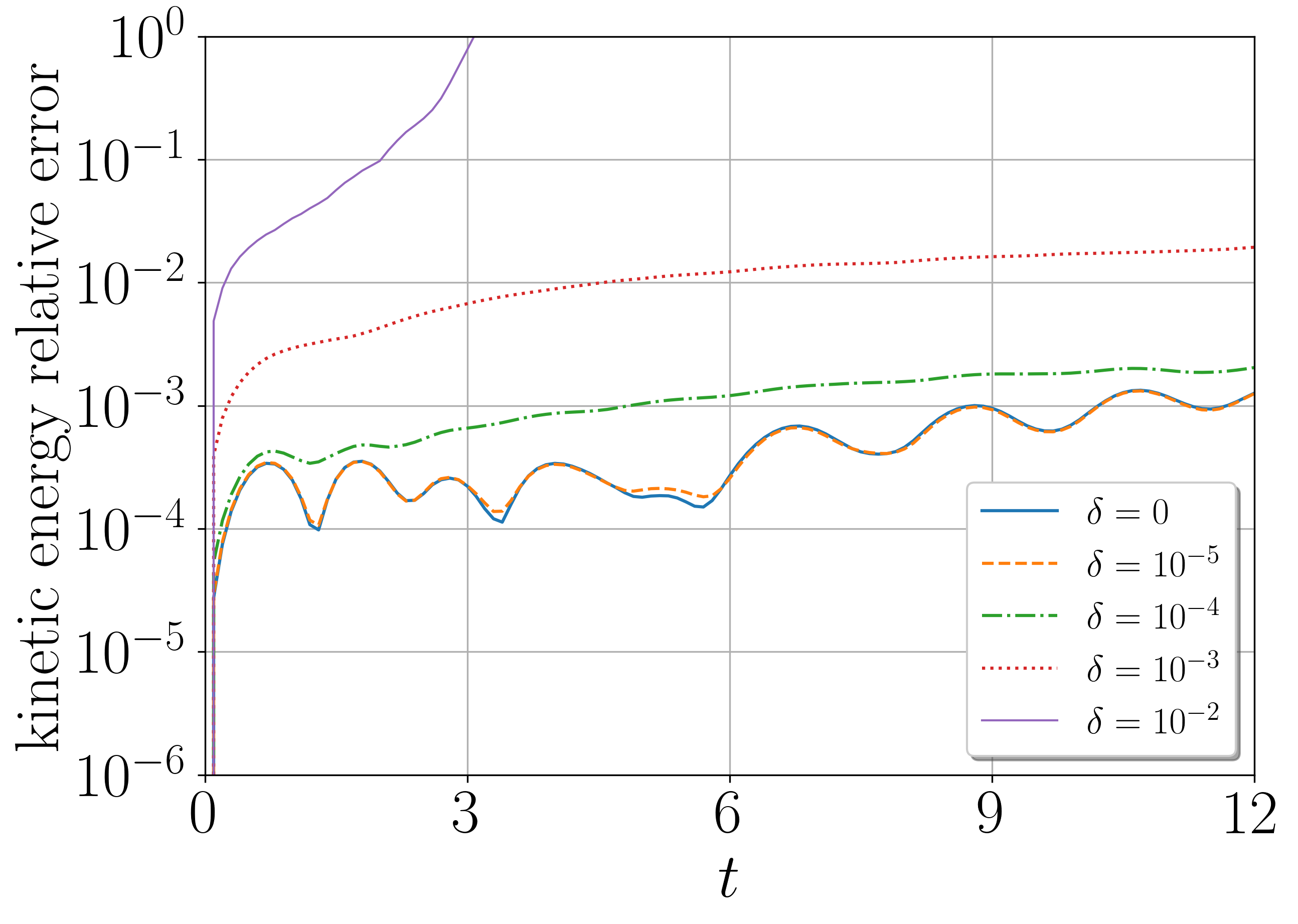}}
\subfigure[]{\label{fig:Noisy_RecNiceH}
\includegraphics[trim=0cm 0cm 0cm 0cm,clip=true,width=0.32\textwidth]{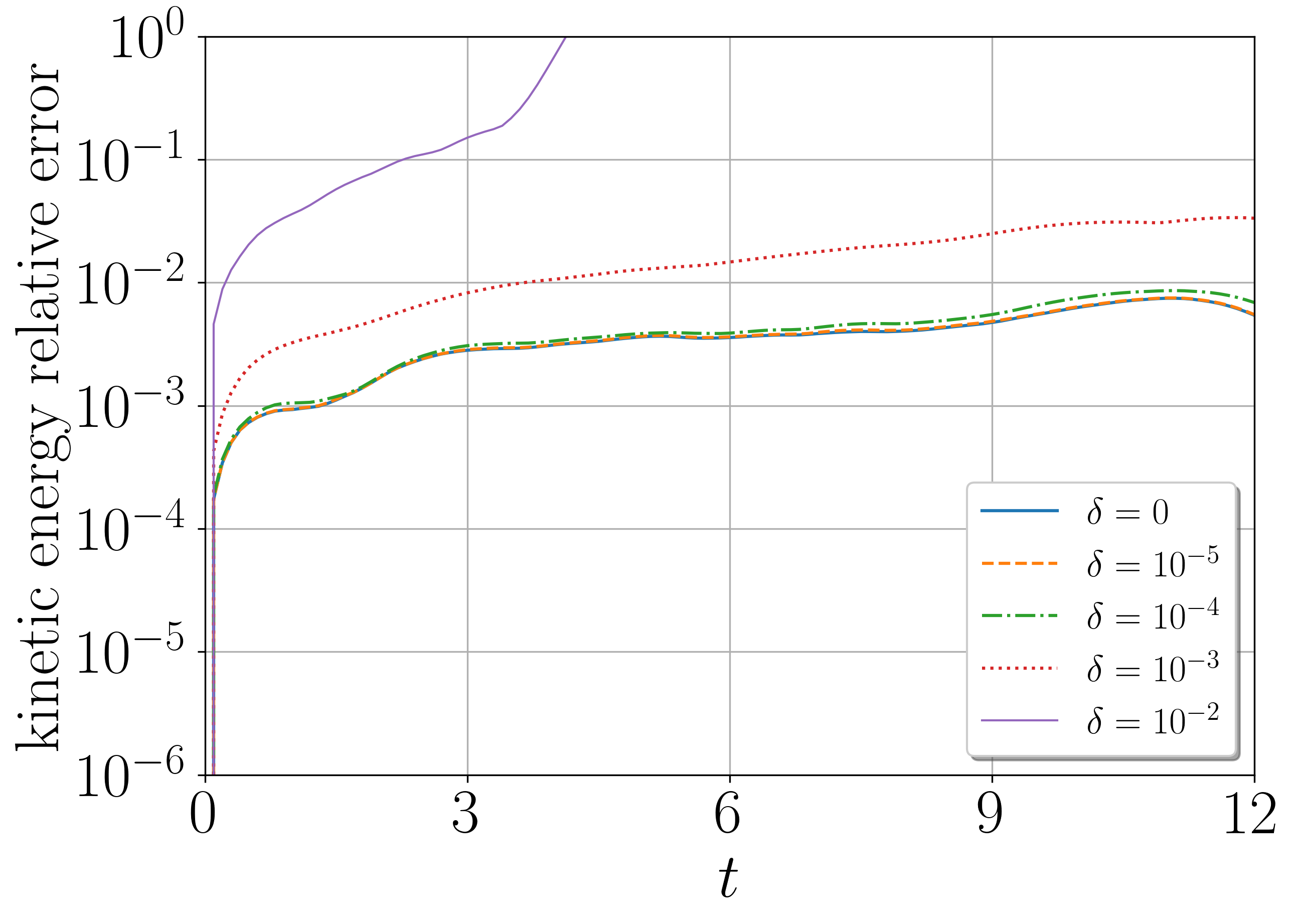}}
\caption{Averaged solution and kinetic energy \eqref{eq:KinEn} errors on the solution reconstruction time interval $[0,12]$ with the initial condition $y_0$ of the phase volume-preserving neural networks $\LSNet$ ($K=2$ and $m=16$), $\SLSNet$ ($K=1$ and $m=16$), and $\VPNet$ ($\mathrm{L}=8$, $l=1$ and $m=16$) trained with noisy data, where the parameter $\delta$ characterizes the added amount of noise. (a)-(c) solution absolute errors of $\LSNet$, $\SLSNet$, and $\VPNet$, respectively. (d)-(f) absolute values of the kinetic energy relative errors of $\LSNet$, $\SLSNet$, and $\VPNet$, respectively.
}\label{fig:Noisy_Rec}
\end{figure}

\begin{figure}[t]
\centering 
\subfigure[]{\label{fig:Noisy_PredLocS}
\includegraphics[trim=0cm 0cm 0cm 0cm,clip=true,width=0.32\textwidth]{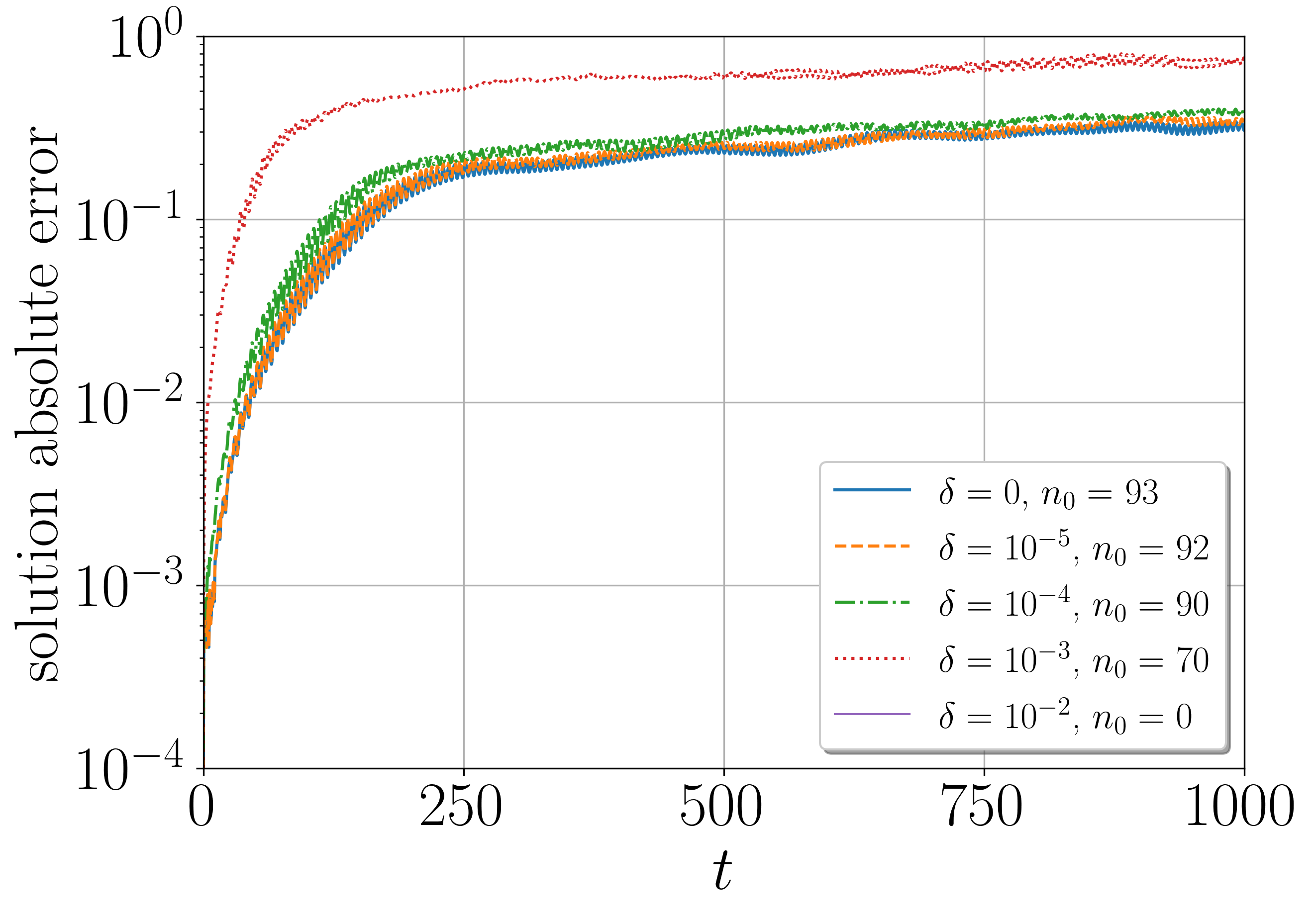}}
\subfigure[]{\label{fig:Noisy_PredSymS}
\includegraphics[trim=0cm 0cm 0cm 0cm,clip=true,width=0.32\textwidth]{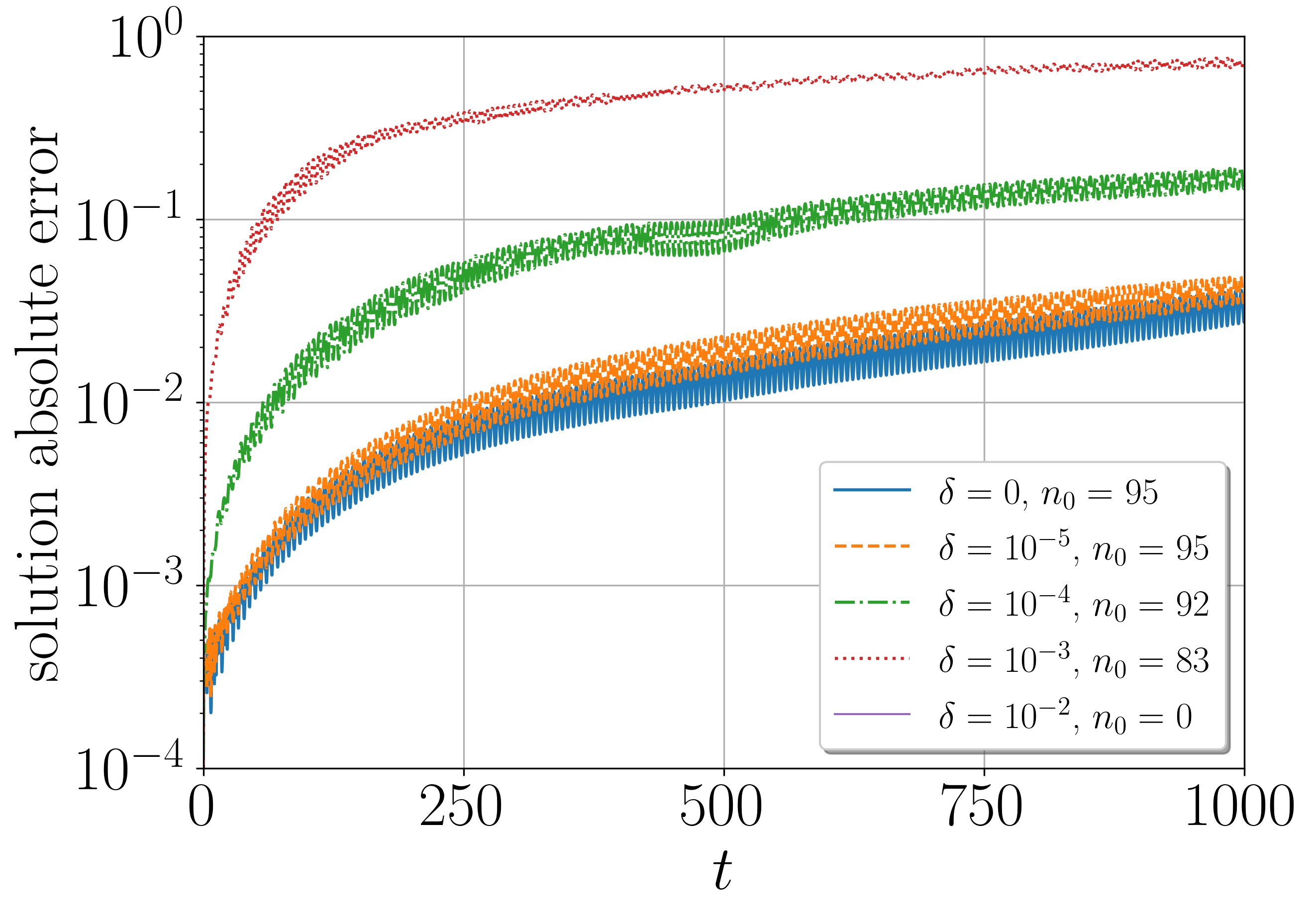}}
\subfigure[]{\label{fig:Noisy_PredNiceS}
\includegraphics[trim=0cm 0cm 0cm 0cm,clip=true,width=0.32\textwidth]{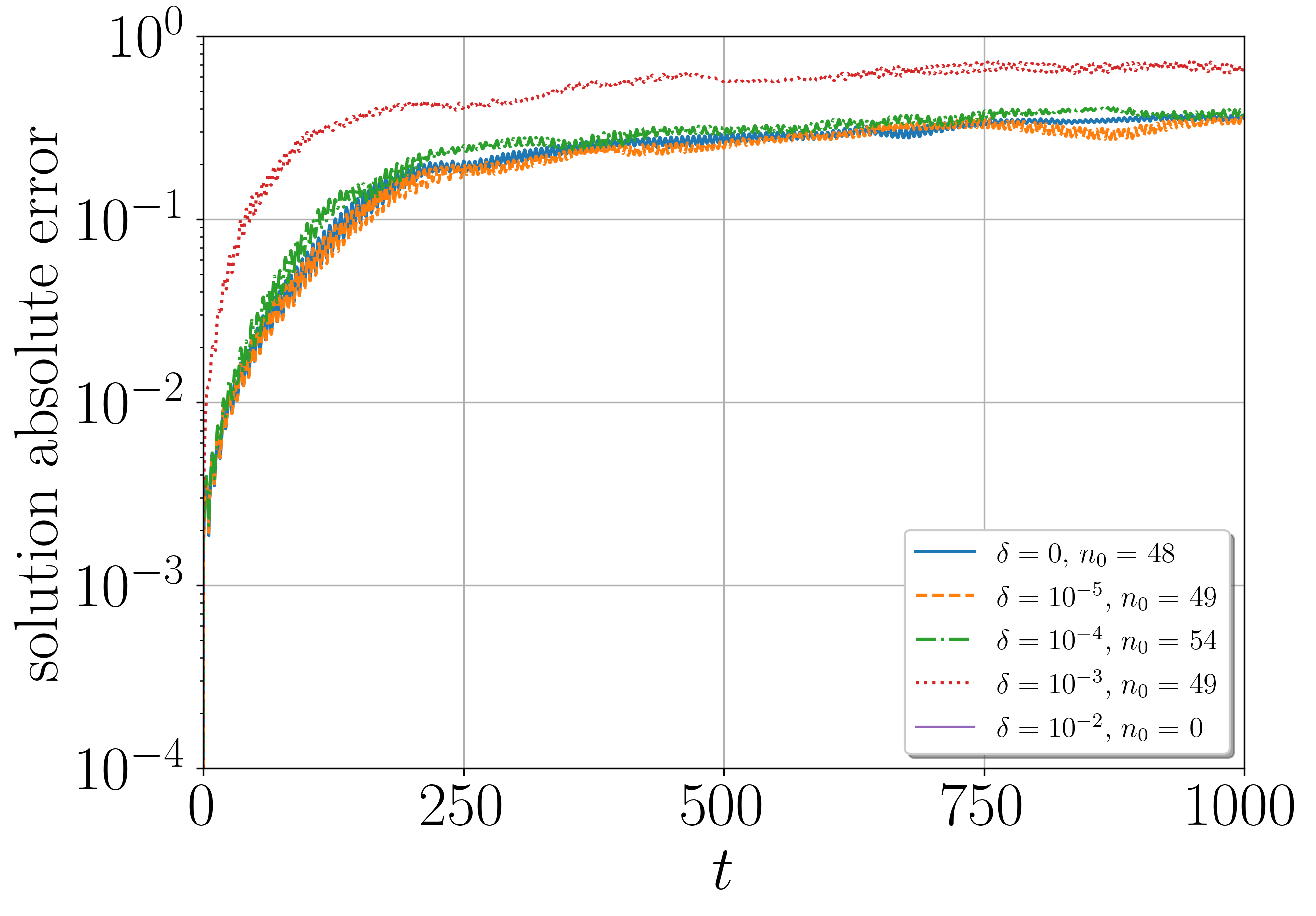}}
\subfigure[]{\label{fig:Noisy_PredLocH}
\includegraphics[trim=0cm 0cm 0cm 0cm,clip=true,width=0.32\textwidth]{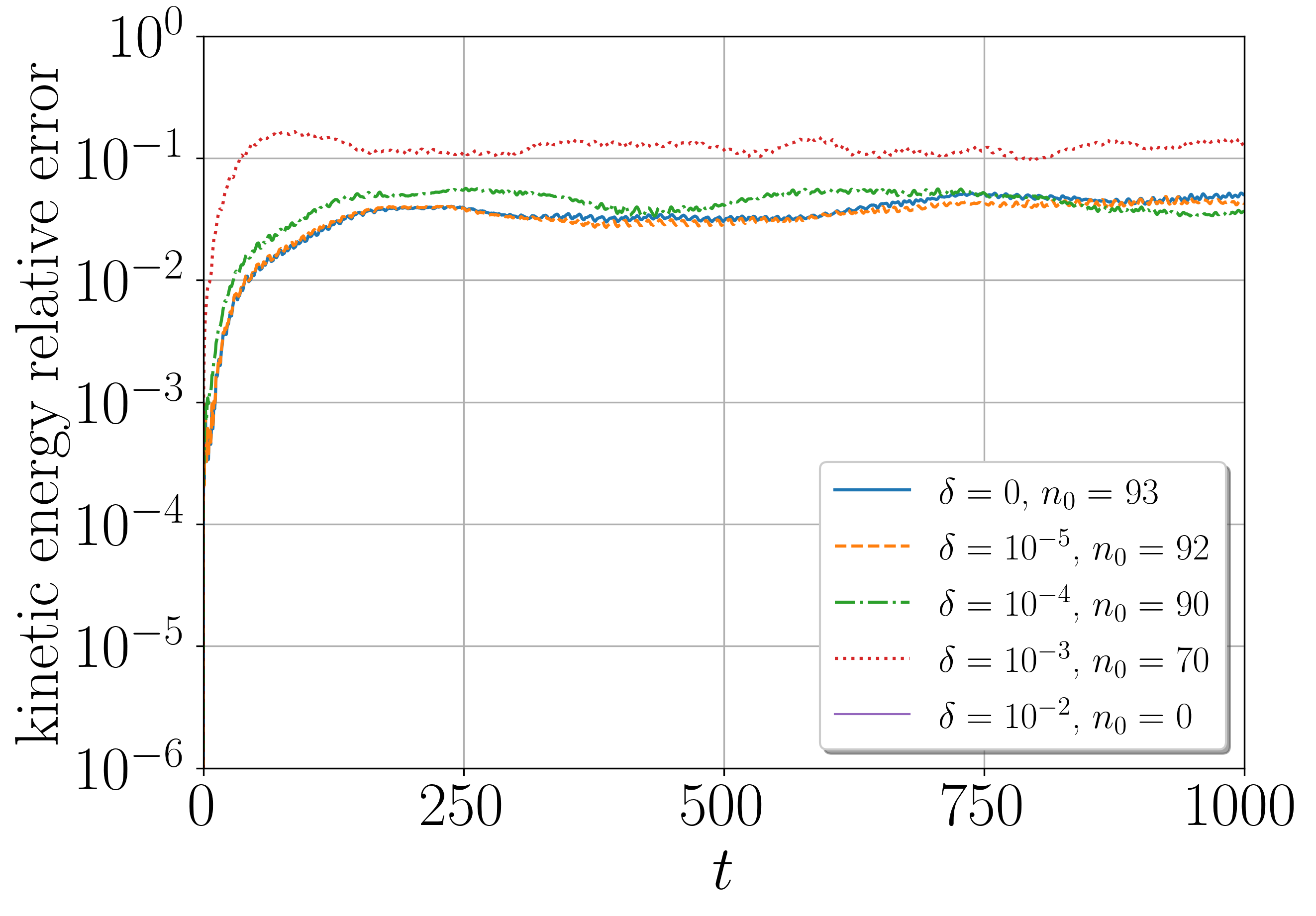}}
\subfigure[]{\label{fig:Noisy_PredSymH}
\includegraphics[trim=0cm 0cm 0cm 0cm,clip=true,width=0.32\textwidth]{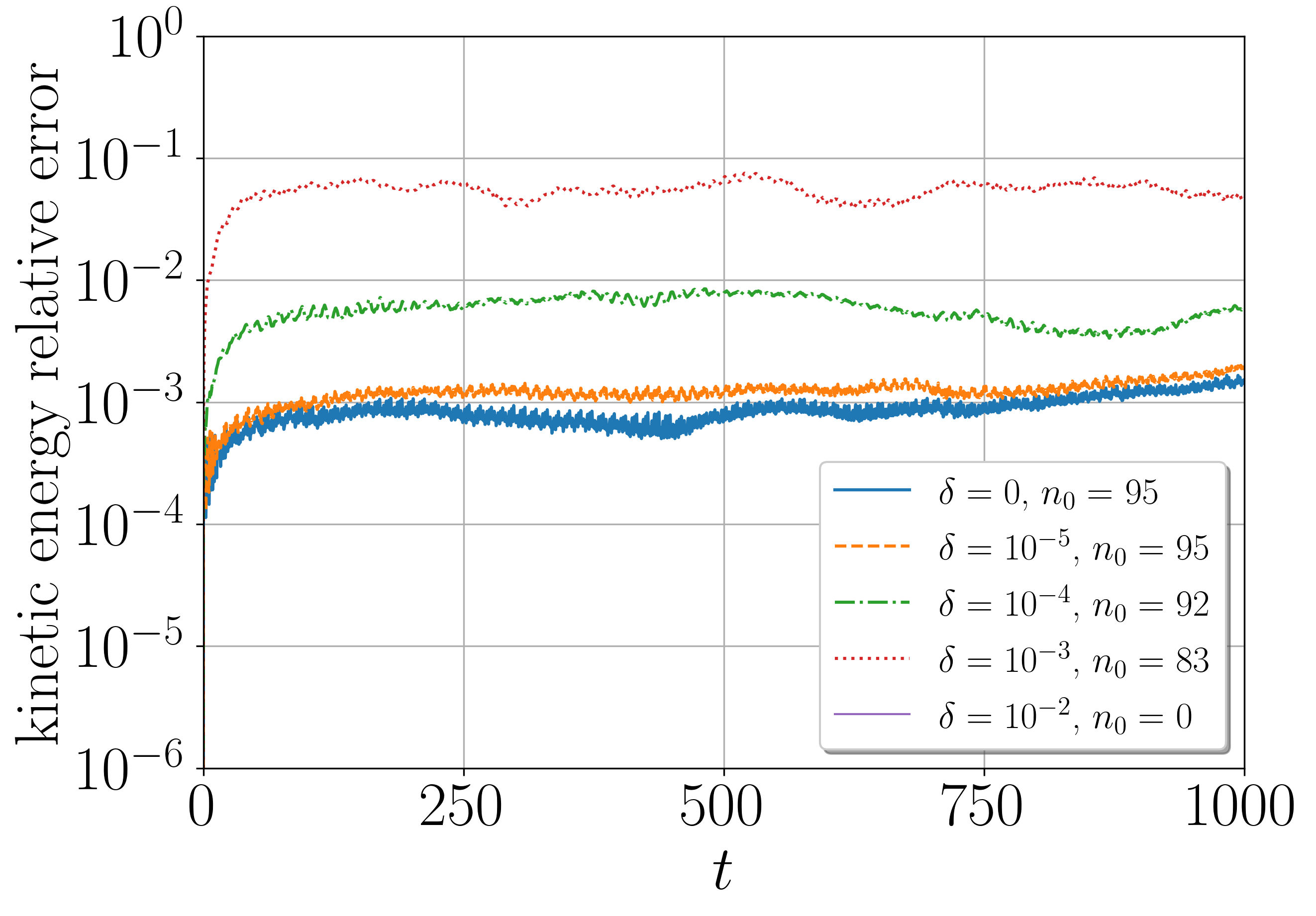}}
\subfigure[]{\label{fig:Noisy_PredNiceH}
\includegraphics[trim=0cm 0cm 0cm 0cm,clip=true,width=0.32\textwidth]{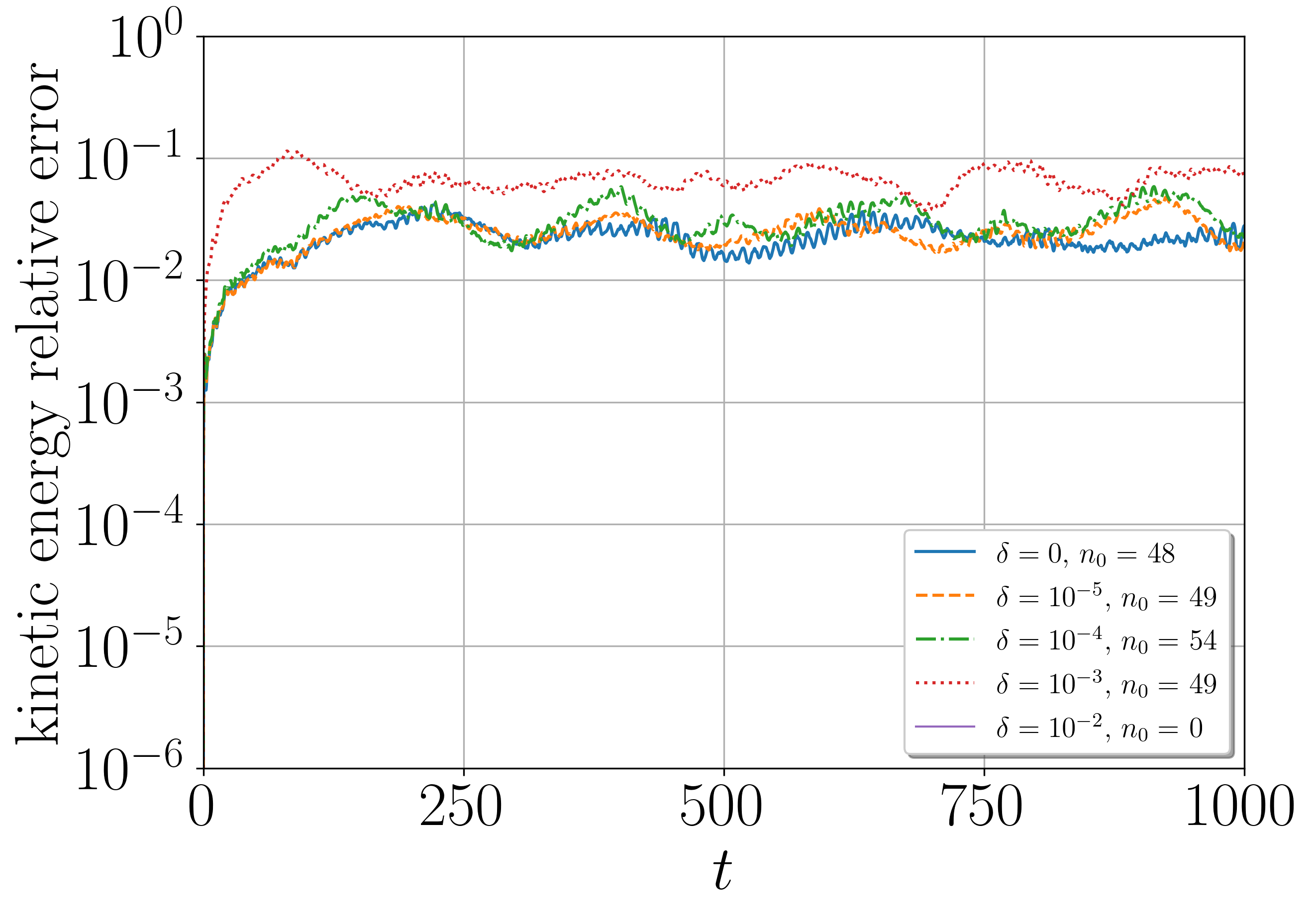}}
\caption{Averaged solution and kinetic energy \eqref{eq:KinEn} errors on the solution prediction time interval $[0,1000]$ with the initial condition $y_{12}$ of the phase volume-preserving neural networks $\LSNet$ ($K=2$ and $m=16$), $\SLSNet$ ($K=1$ and $m=16$), and $\VPNet$ ($\mathrm{L}=8$, $l=1$ and $m=16$) trained with noisy data, where the parameter $\delta$ characterizes the added amount of noise. From $100$ trained neural networks $n_0$ indicates the number of produced stable long-time predictions. (a)-(c) solution absolute errors of $\LSNet$, $\SLSNet$, and $\VPNet$, respectively. (d)-(f) absolute values of the kinetic energy relative errors of $\LSNet$, $\SLSNet$, and $\VPNet$, respectively.}\label{fig:Noisy_Pred}
\end{figure}

In Figures \ref{fig:Noisy_LossNice}, \ref{fig:Noisy_AccNice}, \ref{fig:Noisy_RecNiceS}, \ref{fig:Noisy_RecNiceH}, \ref{fig:Noisy_PredNiceS} and \ref{fig:Noisy_PredNiceH} we visualize results for the phase volume-preserving neural network $\VPNet$ with $\mathrm{L}=8$ alternating modules \eqref{eq:NICEmodules}, where we have used fully-connected neural networks $\NNet_{1,2}$ with one ($l=1$) hidden layer and width $m=16$. For $\LSNet$ we have considered the neural network with $K=2$ and $m=16$, see Figures \ref{fig:Noisy_LossLoc}, \ref{fig:Noisy_AccLoc}, \ref{fig:Noisy_RecLocS}, \ref{fig:Noisy_RecLocH}, \ref{fig:Noisy_PredLocS} and \ref{fig:Noisy_PredLocH}, while we have chosen the neural network $\SLSNet$ with $K=1$ and $m=16$. Results of 
$\SLSNet$ are shown in Figures \ref{fig:Noisy_LossSym}, \ref{fig:Noisy_AccSym}, \ref{fig:Noisy_RecSymS}, \ref{fig:Noisy_RecSymH}, \ref{fig:Noisy_PredSymS} and \ref{fig:Noisy_PredSymH}.

As in the previous Section \ref{sec:RB_Single}, for each above specified three neural networks we have trained $100$ neural networks with different random initial weight values and random perturbations of the training data set samples. In Figures \ref{fig:Noisy_LossAcc}--\ref{fig:Noisy_Pred} we illustrate averaged results for five different $\delta$ values, i.e., $\delta=0$, $10^{-5}$, $10^{-4}$, $10^{-3}$, $10^{-2}$, where $\delta=0$ indicates the case without noise in the training data. Averaged mean square error loss \eqref{eq:loss} and accuracy \eqref{eq:acc} values are shown in Figure \ref{fig:Noisy_LossAcc}, recall also Figure \ref{fig:RB_LossAcc} for comparison. It is evident that as $\delta$ value increases, i.e., the amount of noise in the training data, the loss and accuracy values increase, showing the decrease of learning abilities by the neural networks. Notice that the averaged loss values for $\delta=0$ and $\delta=10^{-5}$ are indistinguishable and slightly increased for $\delta=10^{-4}$, Figures \ref{fig:Noisy_LossLoc}--\ref{fig:Noisy_LossNice}, while the averaged accuracy values appear indistinguishable even for three $\delta$ values, i.e., $\delta=0$, $10^{-5}$, $10^{-4}$, see Figures \ref{fig:Noisy_AccLoc}--\ref{fig:Noisy_AccNice}. Recall that the MSE accuracy function \eqref{eq:acc} in this numerical experiment is evaluated using unperturbed ground truth validation data set samples. Thus, all three neural networks trained with noisy data when $\delta=10^{-5}$ and $\delta=10^{-4}$ are still expected to produce qualitatively good predictions, at least for short-time, which we demonstrate in Figure \ref{fig:Noisy_Rec}.        

In Figure \ref{fig:Noisy_Rec} we demonstrate averaged solution absolute errors, Figures \ref{fig:Noisy_RecLocS}--\ref{fig:Noisy_RecNiceS}, and the absolute values of the relative errors of the kinetic energy \eqref{eq:KinEn}, Figures \ref{fig:RB_RecLocH}--\ref{fig:RB_RecNiceH}. Periodic solution reconstruction on the time interval $[0,12]$ with the initial condition $y_0$ is illustrated for all three phase volume-preserving neural networks $\LSNet$, $\SLSNet$, and $\VPNet$ trained with noisy data. Compare Figure \ref{fig:Noisy_Rec} with Figure \ref{fig:RB_Rec}. From Figures \ref{fig:Noisy_RecLocS}--\ref{fig:Noisy_RecNiceS} it is easy to see that the solution global errors grow linearly, where the rate increases with an increase of $\delta$ value, i.e., with the increased amount of noise in the training data. Similar observations can also be seen in Figures \ref{fig:Noisy_PredLocS}--\ref{fig:Noisy_PredNiceS}.

Despite the presence of the noise in the training data all $100$ trained neural networks of $\LSNet$, $\SLSNet$, and $\VPNet$ for $\delta=10^{-5}$, $10^{-4}$, $10^{-3}$ can reconstruct an approximate solution to the rigid body dynamics. That can be seen by investigating the kinetic energy \eqref{eq:KinEn} conservation relative errors in Figures \ref{fig:Noisy_RecLocH}--\ref{fig:Noisy_RecNiceH}. We can observe that the kinetic energy errors for $\LSNet$ and $\SLSNet$ are comparable and slightly smaller compared to $\VPNet$ for $\delta=10^{-5}$, $10^{-4}$. As expected, the errors increase as the $\delta$ value increases since the noisy training data samples do not exactly satisfy both constraints \eqref{eq:KinEn} and \eqref{eq:Iinv}. It is worth mentioning that the invariant \eqref{eq:Iinv} averaged relative errors are equivalent to the kinetic energy averaged relative errors in Figures \ref{fig:Noisy_RecLocH}--\ref{fig:Noisy_RecNiceH}. The fact that reconstruction errors in Figures \ref{fig:Noisy_RecLocH}--\ref{fig:Noisy_RecNiceH} agree very well for $\delta=0$ and $\delta=10^{-5}$ may suggest that the amount of noise into the training data is equivalent to the reconstruction errors induced by the iterations of the neural networks. 

In Figure \ref{fig:Noisy_Pred} we demonstrate long-time predictions by the neural networks $\LSNet$, $\SLSNet$, and $\VPNet$ trained with noisy data. Figures \ref{fig:Noisy_PredLocS}--\ref{fig:Noisy_PredNiceS} illustrate averaged solution absolute errors, while Figures \ref{fig:Noisy_PredLocH}--\ref{fig:Noisy_PredNiceH} illustrate averaged absolute values of the relative errors of the kinetic energy \eqref{eq:KinEn} on the solution prediction time interval $[0,1000]$. As in Figure \ref{fig:RB_Pred}, predictions of the periodic solution are obtained iteratively with the (unperturbed) initial condition $y_{12}$ and errors are only averaged over the number $n_0$ of predicted stable long-time solutions using the same criteria described in Section \ref{sec:RB_Single}. Evidently, in Figures \ref{fig:Noisy_PredNiceS} and \ref{fig:Noisy_PredNiceH}, as can also be seen in Figures \ref{fig:RB_PredNiceS} and \ref{fig:RB_PredNiceH}, $\VPNet$ have produced significantly less stable solutions compared to the locally-symplectic neural networks $\LSNet$ and $\SLSNet$, which results are shown in Figures \ref{fig:Noisy_PredLocS}--\ref{fig:Noisy_PredSymS} and \ref{fig:Noisy_PredLocH}--\ref{fig:Noisy_PredSymH}. Investigating Figures \ref{fig:Noisy_PredLocH}--\ref{fig:Noisy_PredNiceH} we find that averaged kinetic energy relative errors in this experiment are comparable for the neural networks $\VPNet$ and $\LSNet$, while the $\SLSNet$ have significantly smaller errors even when trained with noisy data. Recall that we already observed this in Figures \ref{fig:RB_PredLocH}--\ref{fig:RB_PredNiceH} comparing the performance of the neural networks when trained with highly accurate ground truth training data.

To summarize the results of this section, from the results presented in Figures \ref{fig:Noisy_LossAcc}--\ref{fig:Noisy_Pred} we can conclude that all three neural networks $\LSNet$, $\SLSNet$, and $\VPNet$ are quite robust to the presence of a small amount of noise in the training data. Even with noisy training data, the phase volume-preserving neural networks are still able to learn rigid body dynamics and produce stable and approximately quadratic invariant conserving solutions in long-time predictions. In addition, we have again observed, as in Section \ref{sec:RB_Single}, indicative importance for the neural networks to preserve the flow property \eqref{eq:FlowProp}.   

\subsubsection{Learning the whole rigid body dynamics}\label{sec:RB_Whole}
In the previous Section \ref{sec:RB_Single}, we considered learning of a single periodic trajectory of the rigid body equations \eqref{eq:RBody}. In this section, we investigate and demonstrate the locally-symplectic neural networks' $\LSNet$ \eqref{eq:LSNet} and $\SLSNet$ \eqref{eq:SymLSNet} capabilities of learning the whole rigid body dynamics. To keep the presentation concise we have excluded results of $\VPNet$ since we were not able to obtain better results compared to the obtained results by $\LSNet$ and $\SLSNet$.

\begin{figure}[t]
\centering 
\includegraphics[trim=0cm 0cm 0cm 0cm,clip=true,width=0.5\textwidth]{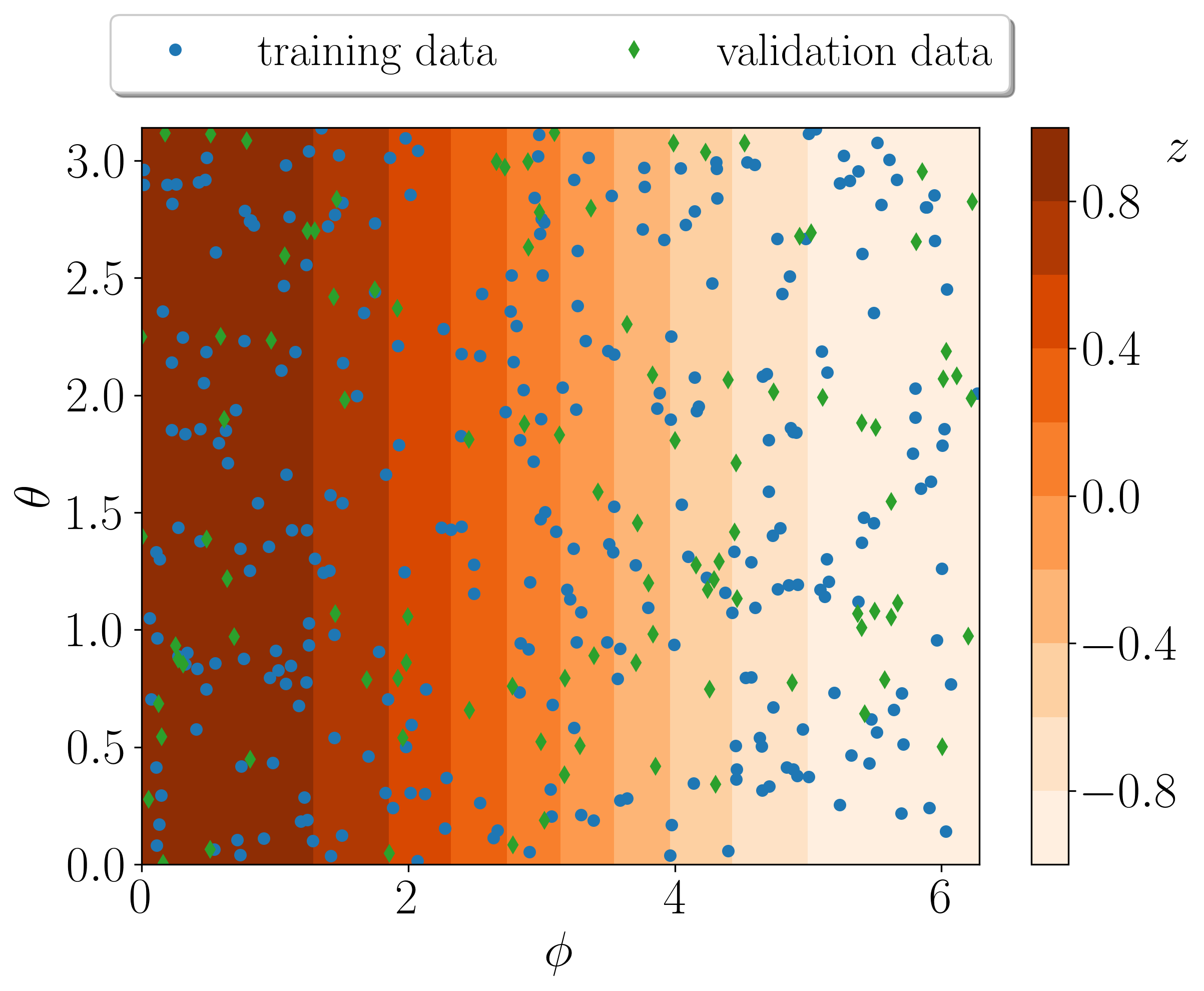}
\caption{Randomly chosen initial conditions on the unit sphere to form the training data (dots) and validation data (diamonds) sets for learning the whole rigid body dynamics \eqref{eq:RBody}. The pair of angles $(\phi,\theta)$ indicates the spherical polar coordinates, while the contour lines indicate $z$ values in the Cartesian $(x,y,z)$ coordinates.}\label{fig:RB_RandIC}
\end{figure}

To learn the whole dynamics of \eqref{eq:RBody} we can consider training data set samples from multiple different solution trajectories. Alternatively, we can consider irregularly sampled one-time step data in the phase space. In addition, we can consider irregularly sampled data with constant or randomly chosen time steps. We have chosen to sample training data from $N=300$ randomly chosen initial conditions on the unit sphere, see Figure \ref{fig:RB_RandIC}, with constant time step $\tau=0.1$. Similarly, $M=100$ random initial conditions are chosen to form the validation data set. Figure \ref{fig:RB_RandIC} illustrates randomly chosen initial conditions in the spherical polar coordinates $(\phi,\theta)$, where the dots and diamonds indicate the initial conditions for the training and validation data sets, respectively, while the contour lines indicate $z$ values in the Cartesian $(x,y,z)$ coordinates. Testing of the neural networks is performed as follows. To perform long-time predictions we consider $\mathrm{J}=12$ initial conditions on the unit sphere defined by the spherical polar angles:
\begin{equation}\label{eq:ICond}
\phi_i = \frac{1+2j}{2}\Delta, \quad
\theta_i = -\frac{1+2j}{4}\Delta, \quad 
\Delta=\frac{\pi}{\mathrm{J}}, \quad j=0,\dots,\mathrm{J}-1.
\end{equation}

For learning the whole dynamics we may require more complex networks, more training data, and more epochs to acquire the necessary accuracy in predictions. In this experiment, we considered $N_e=10^6$ number of epochs. We trained both phase volume-preserving neural networks $\LSNet$ and $\SLSNet$ with different width values $m=16$, $32$, $64$, $128$. We considered $\LSNet$ with $K=2$, $3$, $4$, where it was already shown in Section \ref{sec:RB_Single} that the case $K=1$ gives poor results. For $\SLSNet$ we considered networks with $K=1$, $2$, $3$. Each neural network was trained $10$ times with different initial randomly generated weight values. Then we perform long-time predictions over the time interval $[0,1000]$ by all trained neural networks with $12$ different initial conditions \eqref{eq:ICond}. 

\begin{figure}[t]
\centering 
\subfigure[]{\label{fig:RB_Rand_LocLoss}
\includegraphics[trim=0cm 0cm 0cm 0cm,clip=true,width=0.32\textwidth]{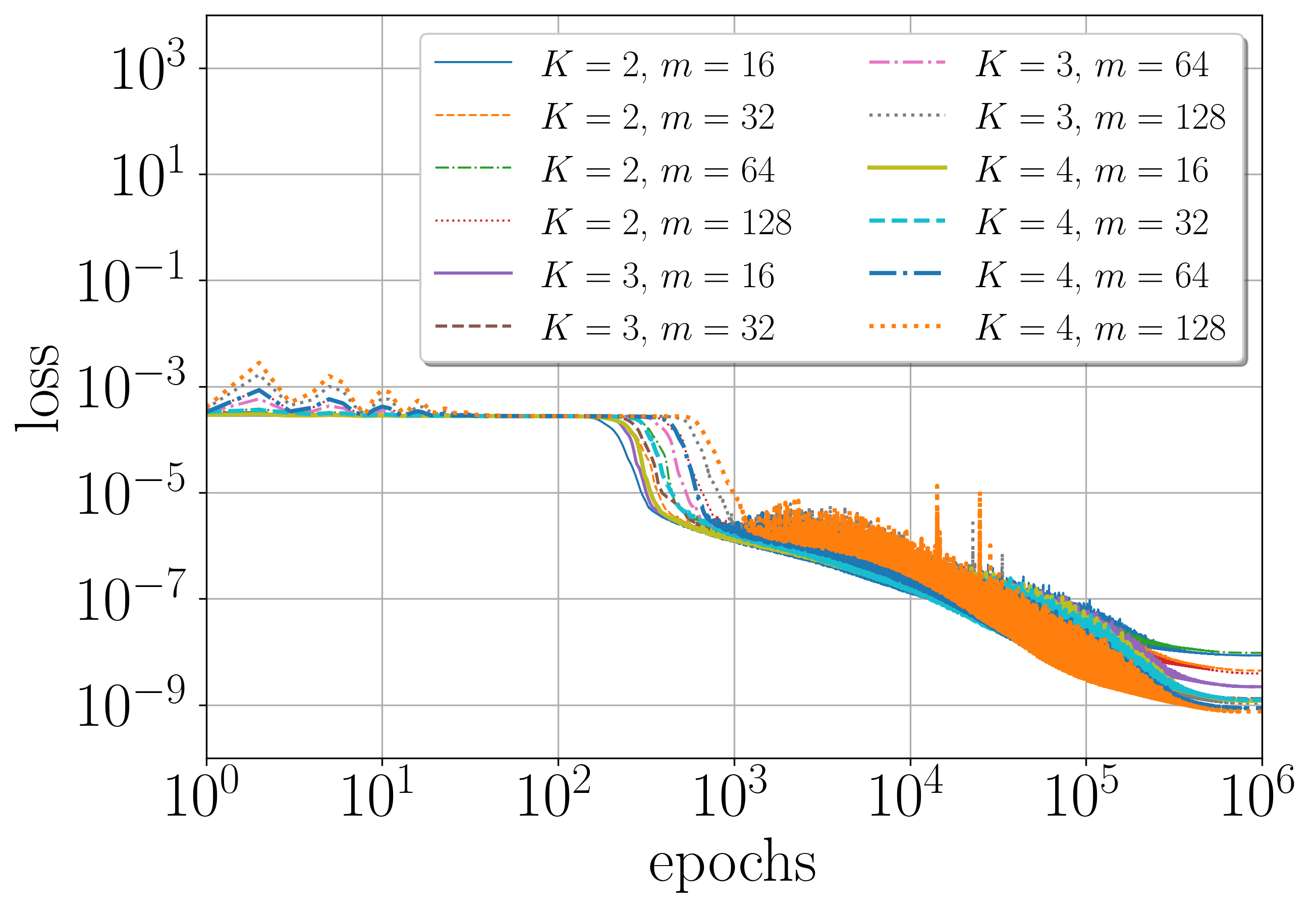}}
\subfigure[]{\label{fig:RB_Rand_LocS}
\includegraphics[trim=0cm 0cm 0cm 0cm,clip=true,width=0.32\textwidth]{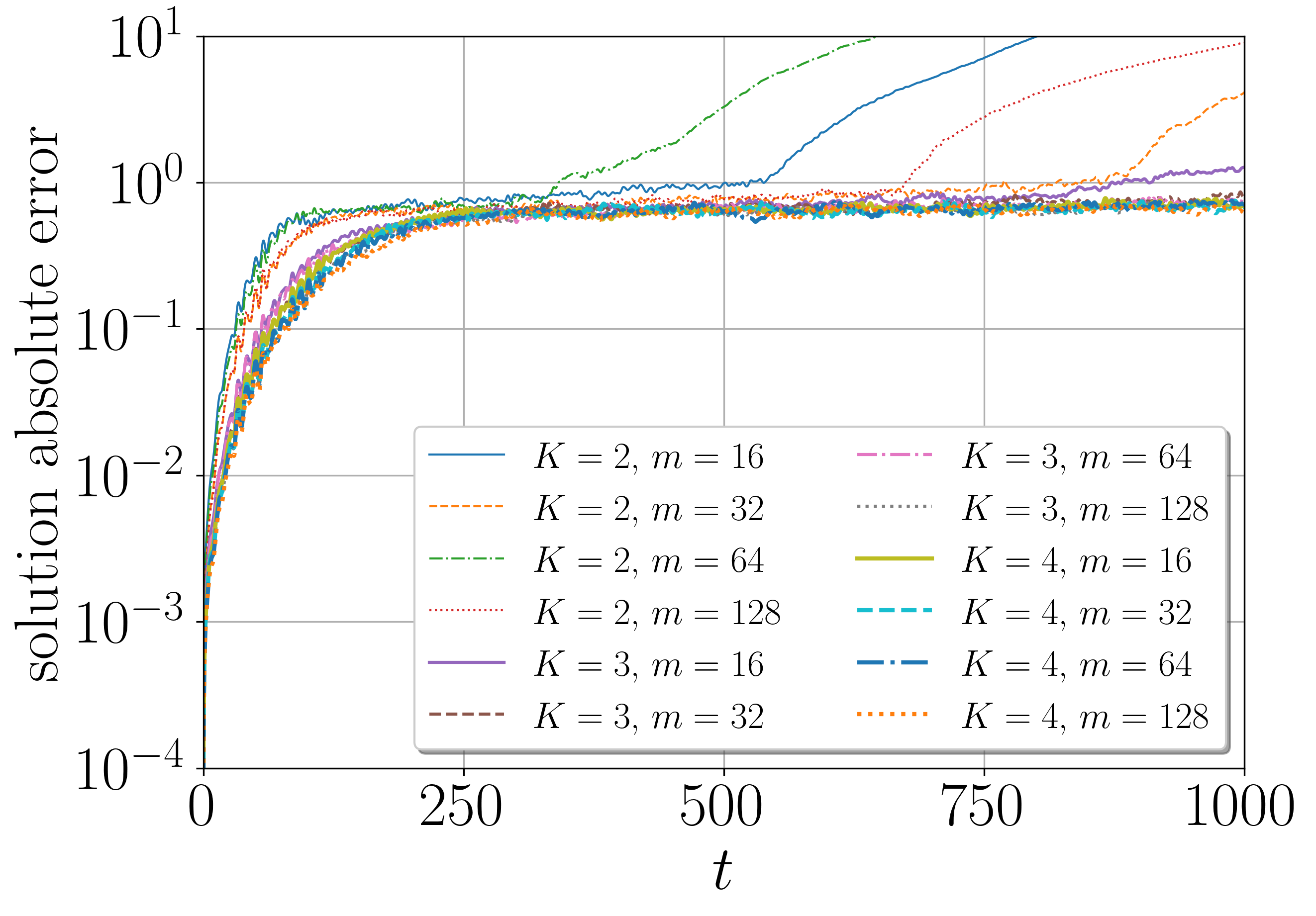}}
\subfigure[]{\label{fig:RB_Rand_LocH}
\includegraphics[trim=0cm 0cm 0cm 0cm,clip=true,width=0.32\textwidth]{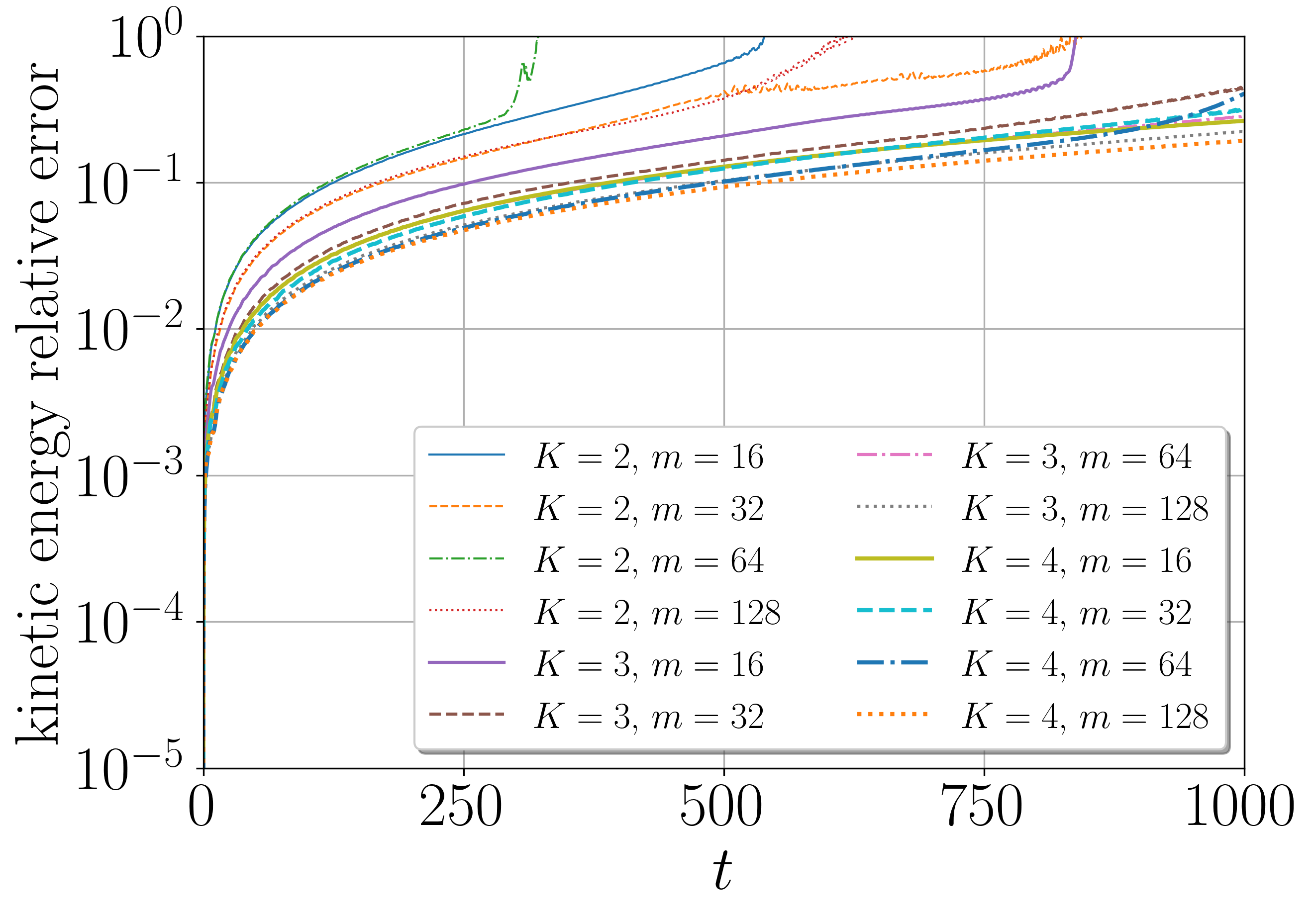}}
\subfigure[]{\label{fig:RB_Rand_SymLoss}
\includegraphics[trim=0cm 0cm 0cm 0cm,clip=true,width=0.32\textwidth]{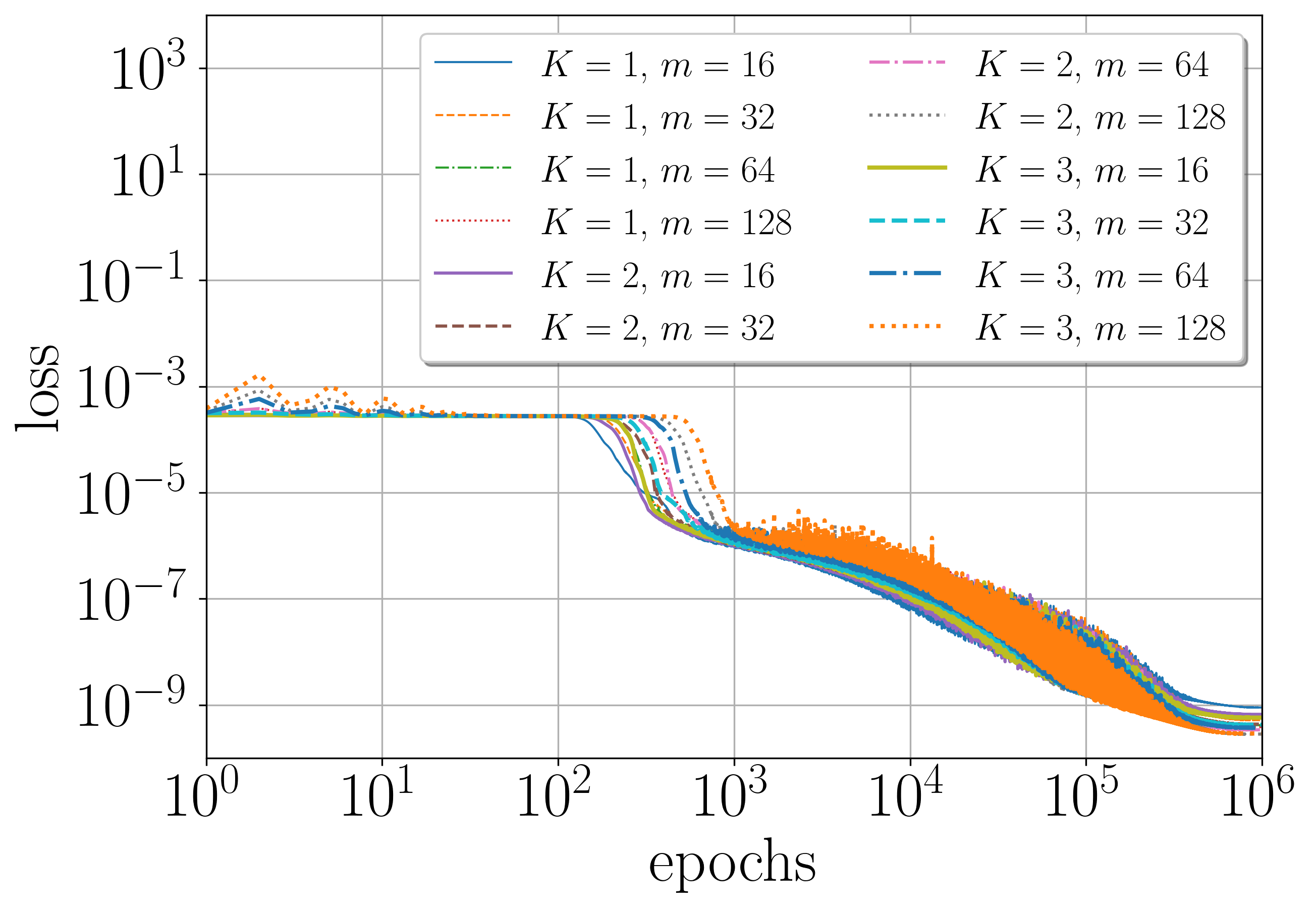}}
\subfigure[]{\label{fig:RB_Rand_SymS}
\includegraphics[trim=0cm 0cm 0cm 0cm,clip=true,width=0.32\textwidth]{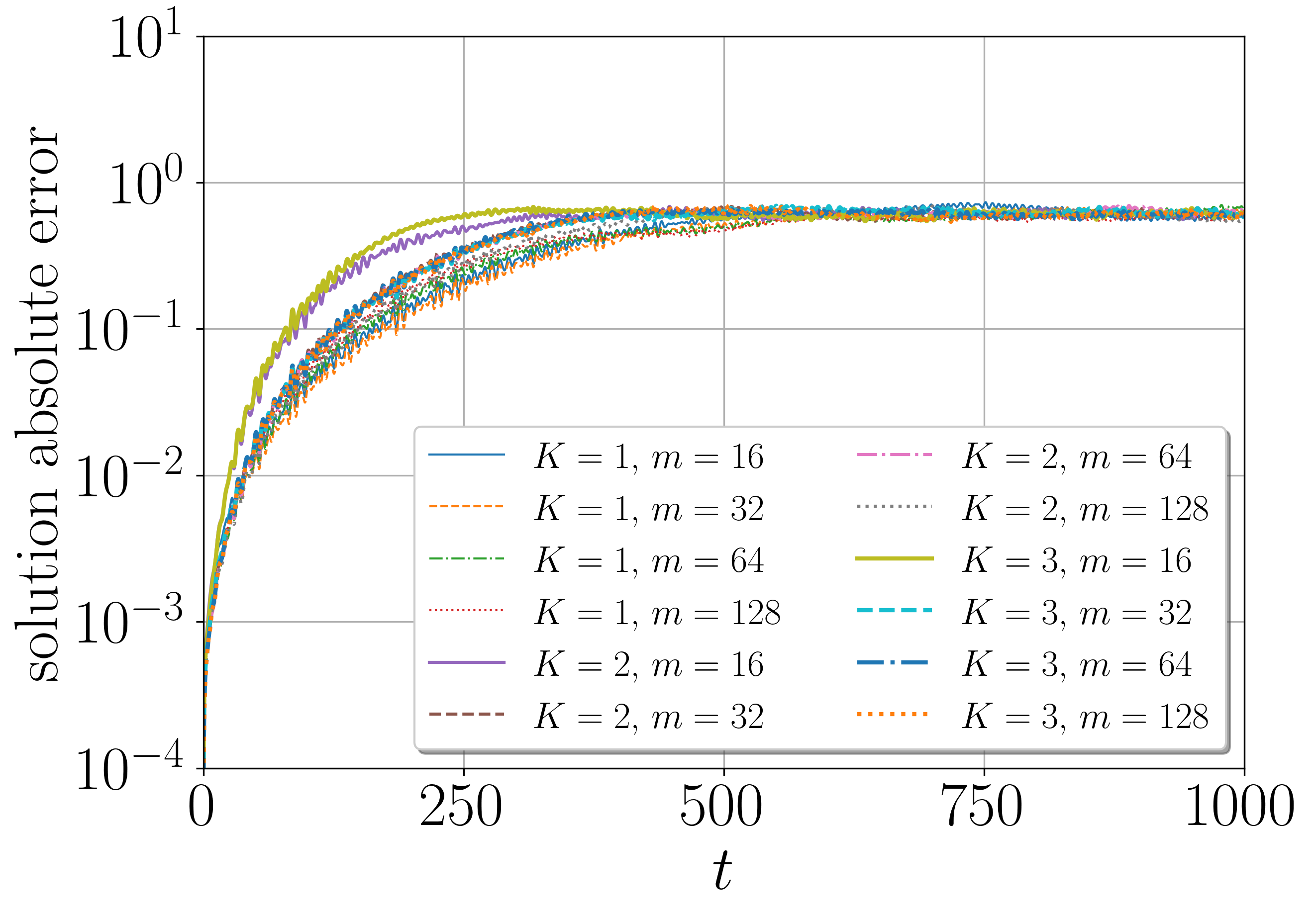}}
\subfigure[]{\label{fig:RB_Rand_SymH}
\includegraphics[trim=0cm 0cm 0cm 0cm,clip=true,width=0.32\textwidth]{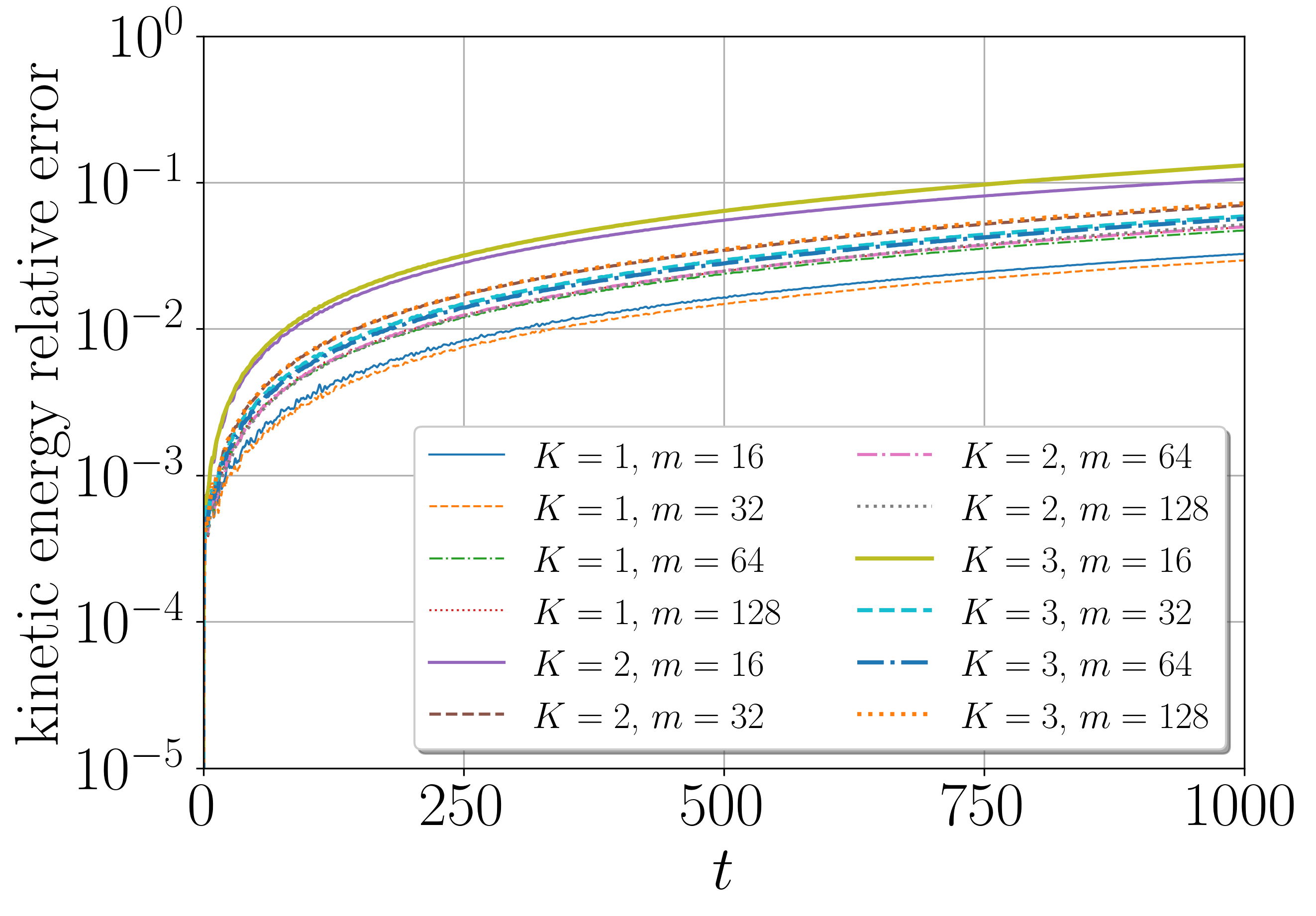}}
\caption{(a)\&(d) averaged MSE loss \eqref{eq:loss} function values of $\LSNet$ and $\SLSNet$, respectively. (b)\&(e) averaged solution absolute errors of $\LSNet$ and $\SLSNet$, respectively. (c)\&(f) averaged kinetic energy \eqref{eq:KinEn} relative errors of $\LSNet$ and $\SLSNet$, respectively. Long-time predictions are obtained iteratively considering 12 different initial conditions \eqref{eq:ICond}.}\label{fig:RB_Rand}
\end{figure}

In Figure \ref{fig:RB_Rand} we illustrate averaged error results obtained by averaging over all $10$ trained neural networks and, in addition, over the $12$ initial conditions \eqref{eq:ICond} in Figures \ref{fig:RB_Rand_LocS}--\ref{fig:RB_Rand_LocH} and \ref{fig:RB_Rand_SymS}--\ref{fig:RB_Rand_SymH}. Figures \ref{fig:RB_Rand_LocLoss}--\ref{fig:RB_Rand_LocH} show results of $\LSNet$, while Figures \ref{fig:RB_Rand_SymLoss}--\ref{fig:RB_Rand_SymH} illustrate results of $\SLSNet$. Comparing averaged MSE loss function \eqref{eq:loss} values for both neural networks, compare Figures \ref{fig:RB_Rand_LocLoss} and \ref{fig:RB_Rand_SymLoss}, we observe slightly smaller errors for the $\SLSNet$. Results of the averaged MSE accuracy \eqref{eq:acc} values are omitted since they do not vary too much from the loss values.    
   
We observe significant differences between both neural networks when the averaged absolute values of the kinetic energy \eqref{eq:KinEn} relative errors are compared in Figures \ref{fig:RB_Rand_LocH} and \ref{fig:RB_Rand_SymH}. With $\SLSNet$ we can reduce kinetic energy conservation relative errors by a significant magnitude. Unfortunately, independently of the $K$ or width value $m$ we observe linear growth in the kinetic energy even though all trained $\SLSNet$ have not produced any in-time exploding predictions over the whole prediction time interval $[0,1000]$. This is not true for $\LSNet$, see Figures \ref{fig:RB_Rand_LocS}--\ref{fig:RB_Rand_LocH}, where it is evident that some of the trained networks for some initial condition \eqref{eq:ICond} have produced unstable predictions. This may not necessarily be a surprise, since in Section \ref{sec:RB_Single} we already observed in long-time predictions, Figure \ref{fig:RB_Pred}, that occasionally some of the trained neural networks will fail to produce stable long-time predictions. This may be highly dependent on the given data and performance of the optimization algorithm, as well as considering that both constraints \eqref{eq:KinEn}--\eqref{eq:Iinv} are not satisfied by design.           

\begin{figure}[t]
\centering 
\subfigure[]{\label{fig:RB_RandSolt10}
\includegraphics[trim=1cm 0cm 0cm 0.5cm,clip=true,width=0.38\textwidth]{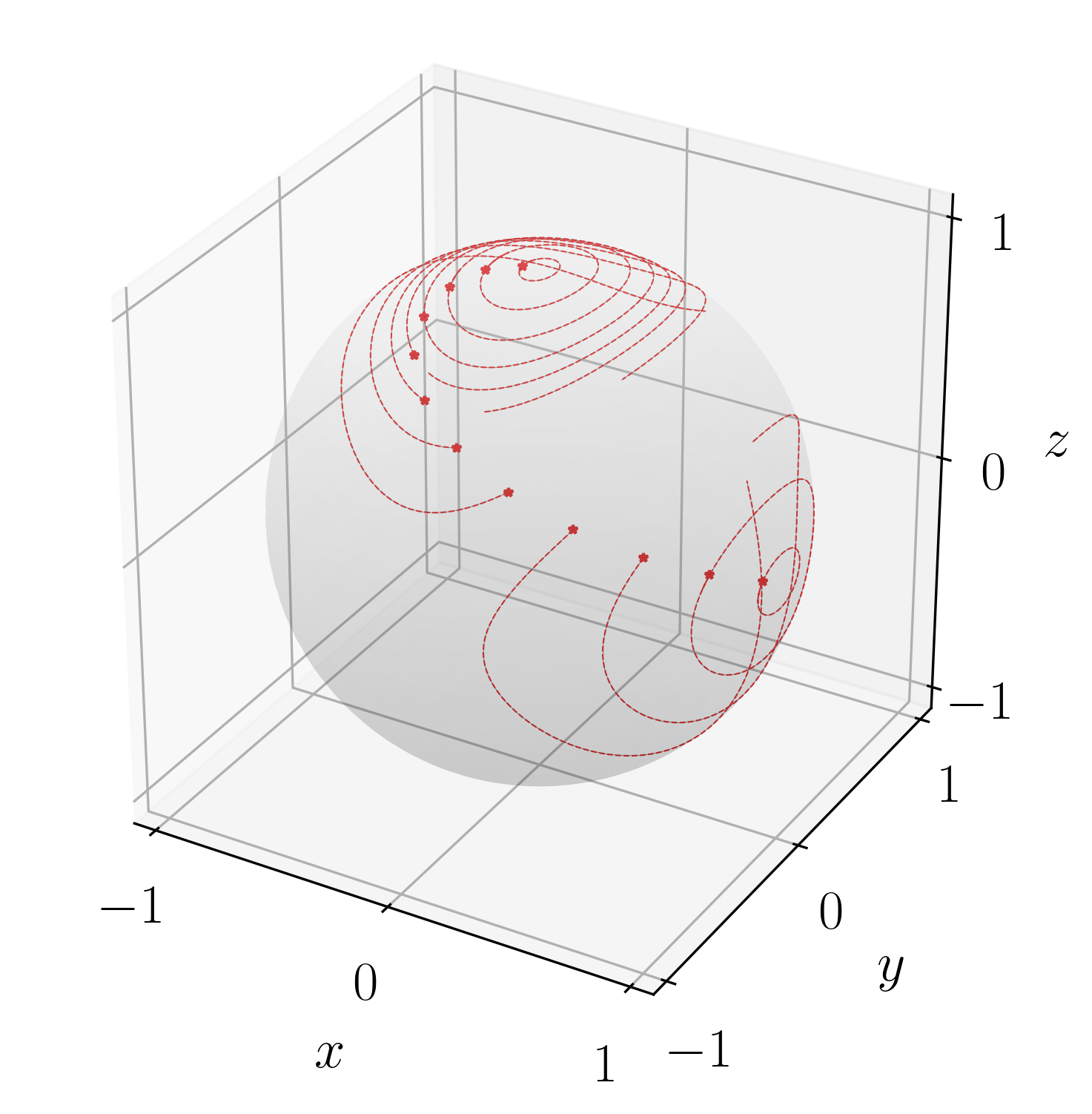}}\qquad
\subfigure[]{\label{fig:RB_RandSolt100}
\includegraphics[trim=1cm 0cm 0cm 0.5cm,clip=true,width=0.38\textwidth]{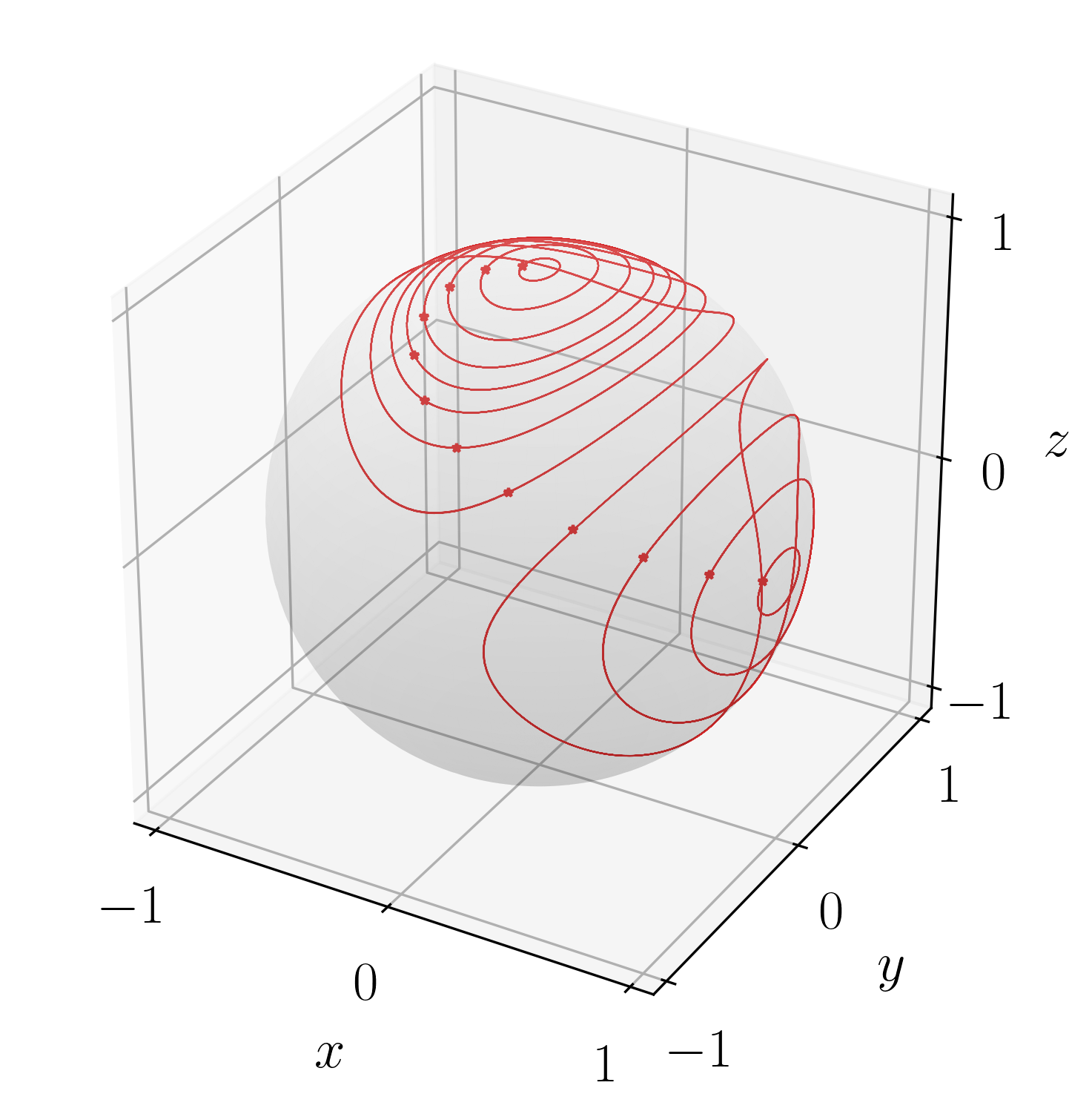}}
\subfigure[]{\label{fig:RB_RandSolt500}
\includegraphics[trim=1cm 0cm 0cm 0.5cm,clip=true,width=0.38\textwidth]{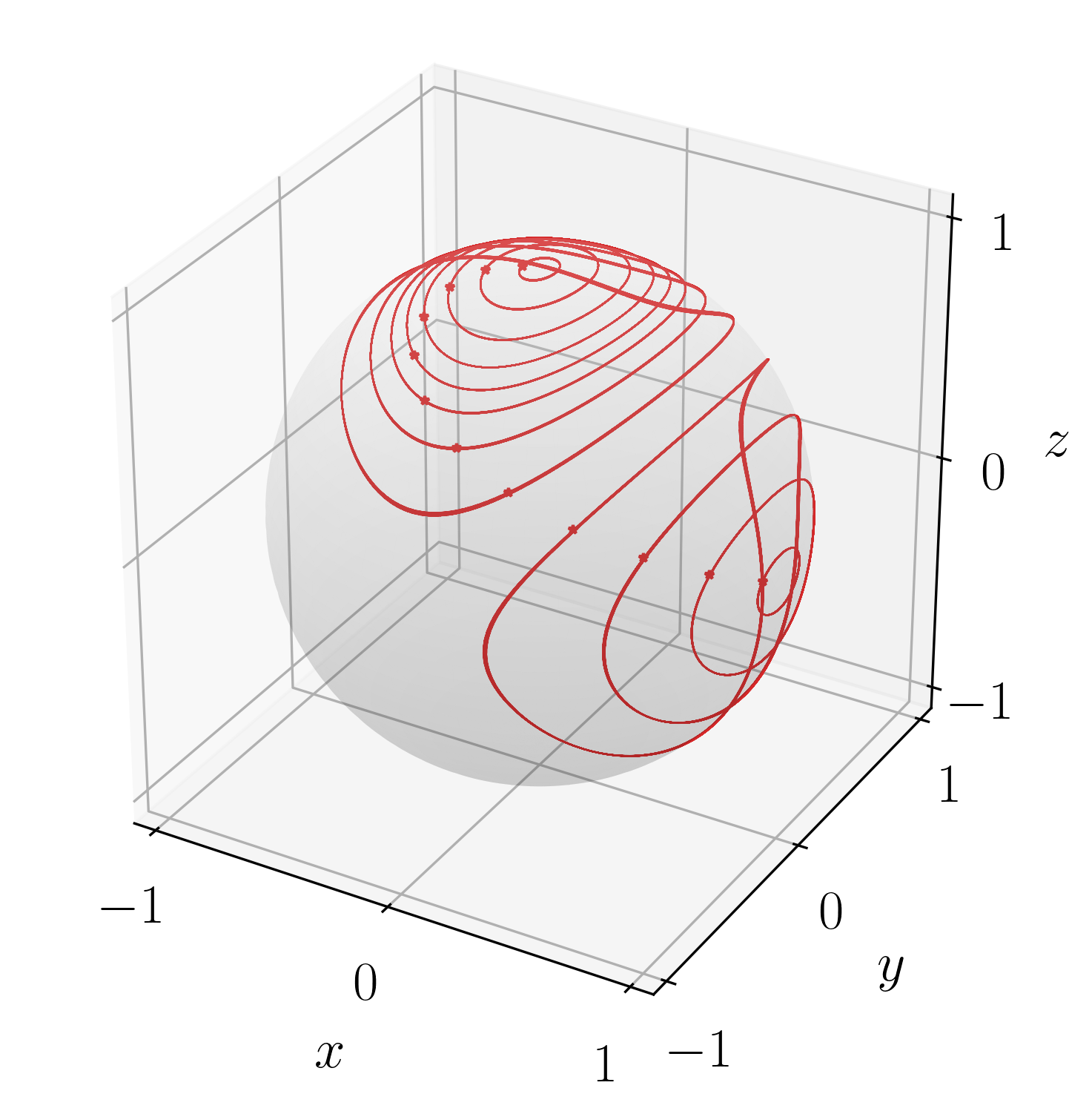}}\qquad 
\subfigure[]{\label{fig:RB_RandSolt1000}
\includegraphics[trim=1cm 0cm 0cm 0.5cm,clip=true,width=0.38\textwidth]{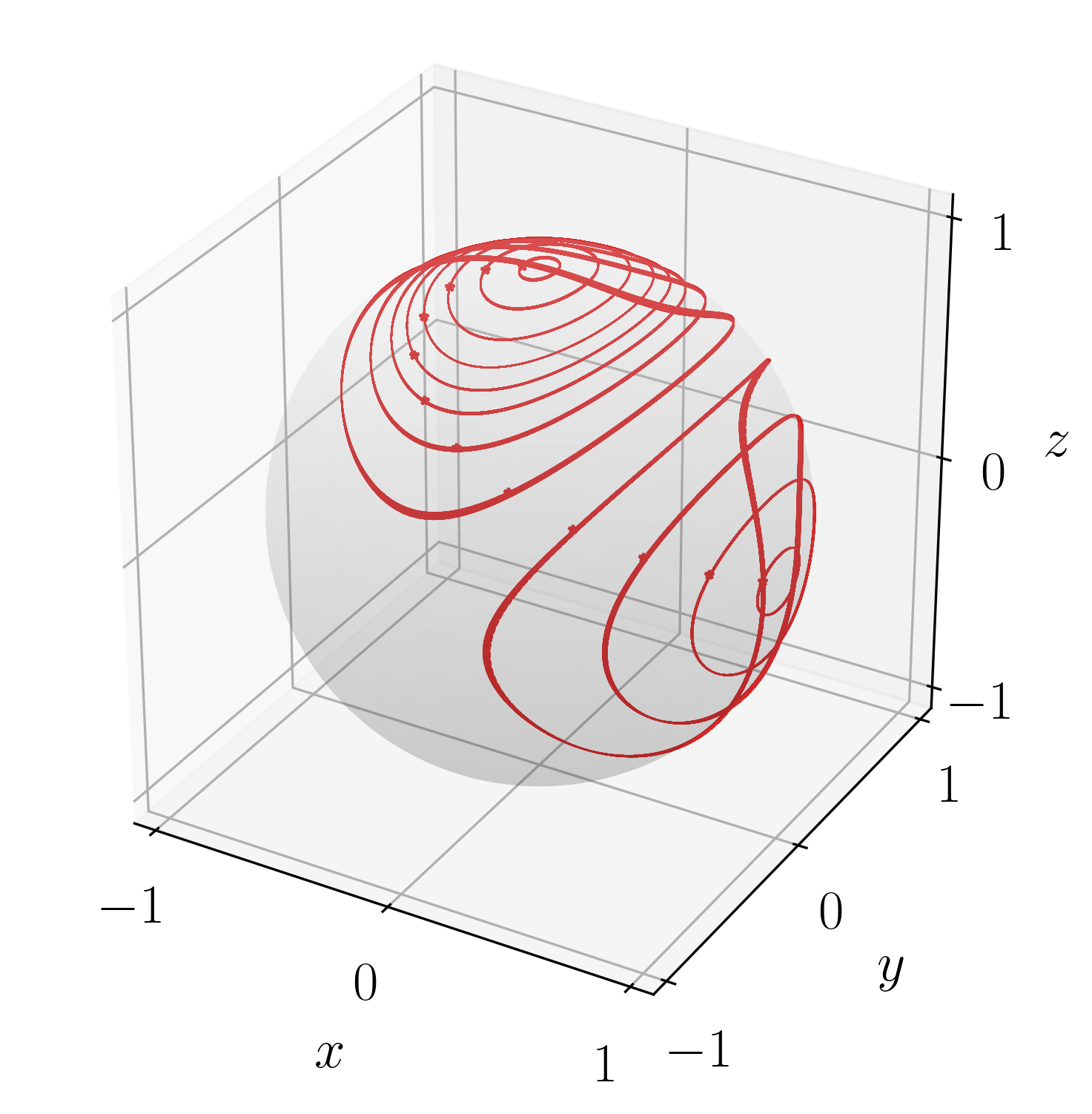}}
\caption{Long-time predicted $12$ solution trajectories of the rigid body problem \eqref{eq:RBody} for $t\in[0,T_{end}]$ by the symmetric phase volume-preserving neural network $\SLSNet$ \eqref{eq:SymLSNet} with $K=1$ and $m=32$, and time step $\tau=0.1$. Dots indicate the initial conditions \eqref{eq:ICond}. (a) $T_{end}=10$. (b) $T_{end}=100$. (c) $T_{end}=500$. (d) $T_{end}=1000$.}\label{fig:RB_RandSol}
\end{figure}

After closer inspection of the results in Figure \ref{fig:RB_Rand_SymH}, we find that $\SLSNet$ with $K=1$ and $m=32$ have the smallest averaged kinetic energy relative errors. Thus, in Figure \ref{fig:RB_RandSol} we demonstrate long-time predictions of one of the trained $\SLSNet$ with $K=1$ and $m=32$, where the predictions are performed considering all $12$ initial conditions \eqref{eq:ICond}. In Figure \ref{fig:RB_RandSol} all $12$ initial conditions are indicated by dots on the unit sphere. Predicted solution trajectories are shown for four different time intervals $t\in[0,T_{end}]$, i.e., when $T_{end}=10$, $100$, $500$, $1000$, see Figures \ref{fig:RB_RandSolt10}--\ref{fig:RB_RandSolt1000}, respectively. Notice a very good agreement between predicted and the exact solutions, depicted with the solid black line, for $t\in[0,100]$ and qualitatively good results over the whole time interval $[0,1000]$.

As time progresses some of the predicted trajectories gradually deviate from the analytic closed orbits, which explains the linear grows of errors in Figure \ref{fig:RB_Rand_SymH}. Despite that, in the example of Figure \ref{fig:RB_RandSol} the relative error values for both conserved quantities \eqref{eq:KinEn}--\eqref{eq:Iinv} over the whole time interval $[0,1000]$ are below $1.5\%$. Thus, predictions are only valid for relatively long-times. Since the use of a symmetric version of the locally-symplectic neural networks $\LSNet$ significantly improved the long-time predictions, it remains an open question if even more structure-preserving properties should be incorporated into the neural networks for them being able to learn the whole rigid body dynamics and produce stable long-time predictions for any initial condition given on the unit sphere. For example, preservation of $\rho$-reversibility of the vector field $f(y)$, i.e., $\rho f(y) = -f(\rho y)$ for all $y$, where $\rho$ is an invertible linear transformation in the phase space, which reverses the direction of time. For the rigid body problem \eqref{eq:RBody} $\rho(y)=-y$. Thus, more research in this direction is required.            

\subsubsection{Learning with different time steps and number of data samples}\label{sec:RB_TimeSpan}
In the previous three sections, we considered numerical examples of training neural networks $\LSNet$, $\SLSNet$, and $\VPNet$ with the data of time step $\tau=0.1$. In this section, we extend the above-demonstrated results by exploring the effects of different time step $\tau$ values and the number of training data samples $N$. Numerical results are illustrated for the symmetric locally-symplectic neural networks $\SLSNet$ with $K=1$ and $m=16$ for learning the single periodic trajectory of Section \ref{sec:RB_Single} and with $K=1$ and $m=32$ for learning the whole rigid body dynamics \eqref{eq:RBody}, see Section \ref{sec:RB_Whole}. Obtained results show that phase volume-preserving dynamics \eqref{eq:ODE} can be learned and well-predicted by $\SLSNet$ for different values of time steps and lengths of sampling time intervals.

\begin{figure}[t]
\centering 
\subfigure[]{\label{fig:RBTau_Loss}
\includegraphics[trim=0cm 0cm 0cm 0cm,clip=true,width=0.32\textwidth]{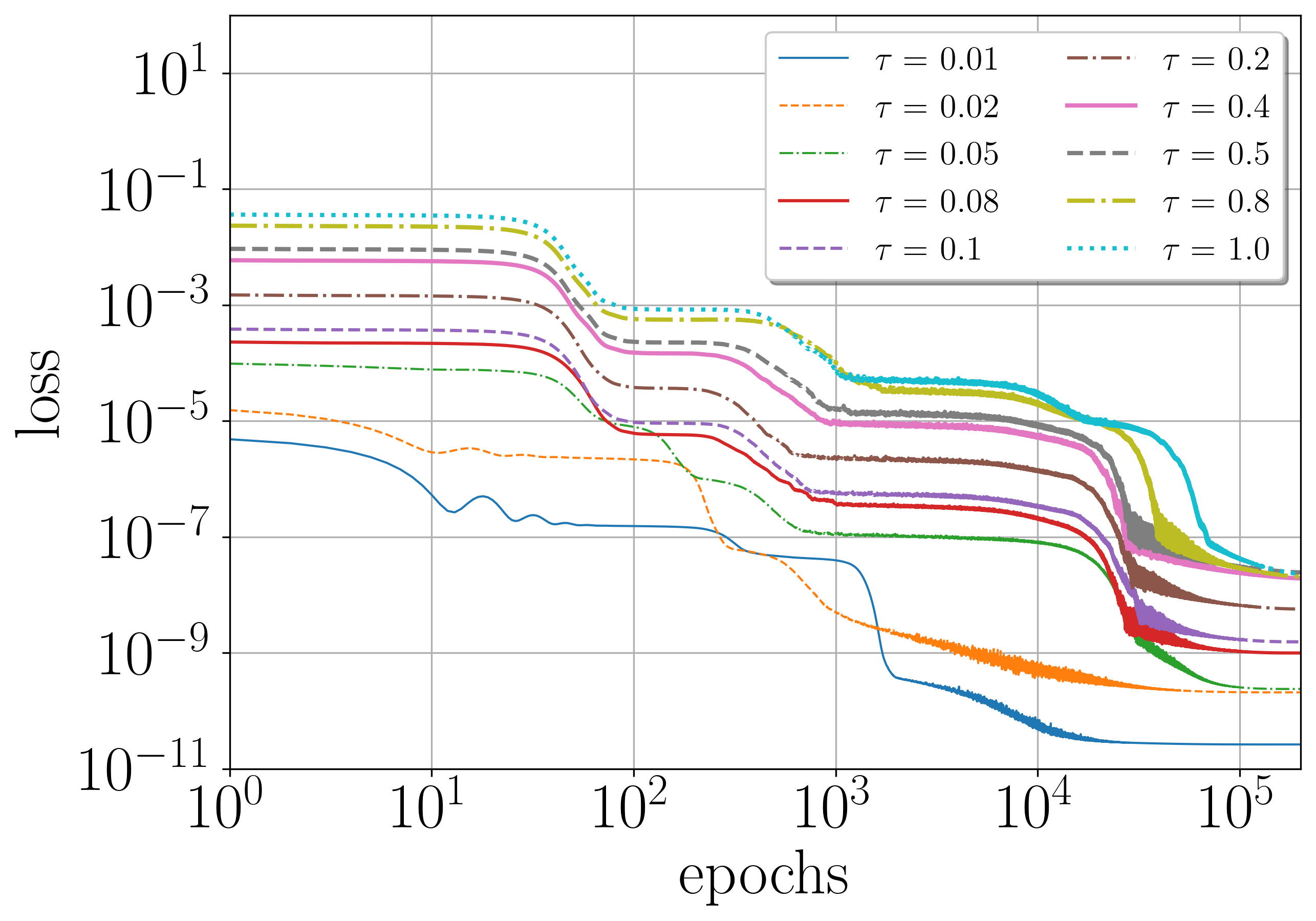}}
\subfigure[]{\label{fig:RBTau_RecS}
\includegraphics[trim=0cm 0cm 0cm 0cm,clip=true,width=0.32\textwidth]{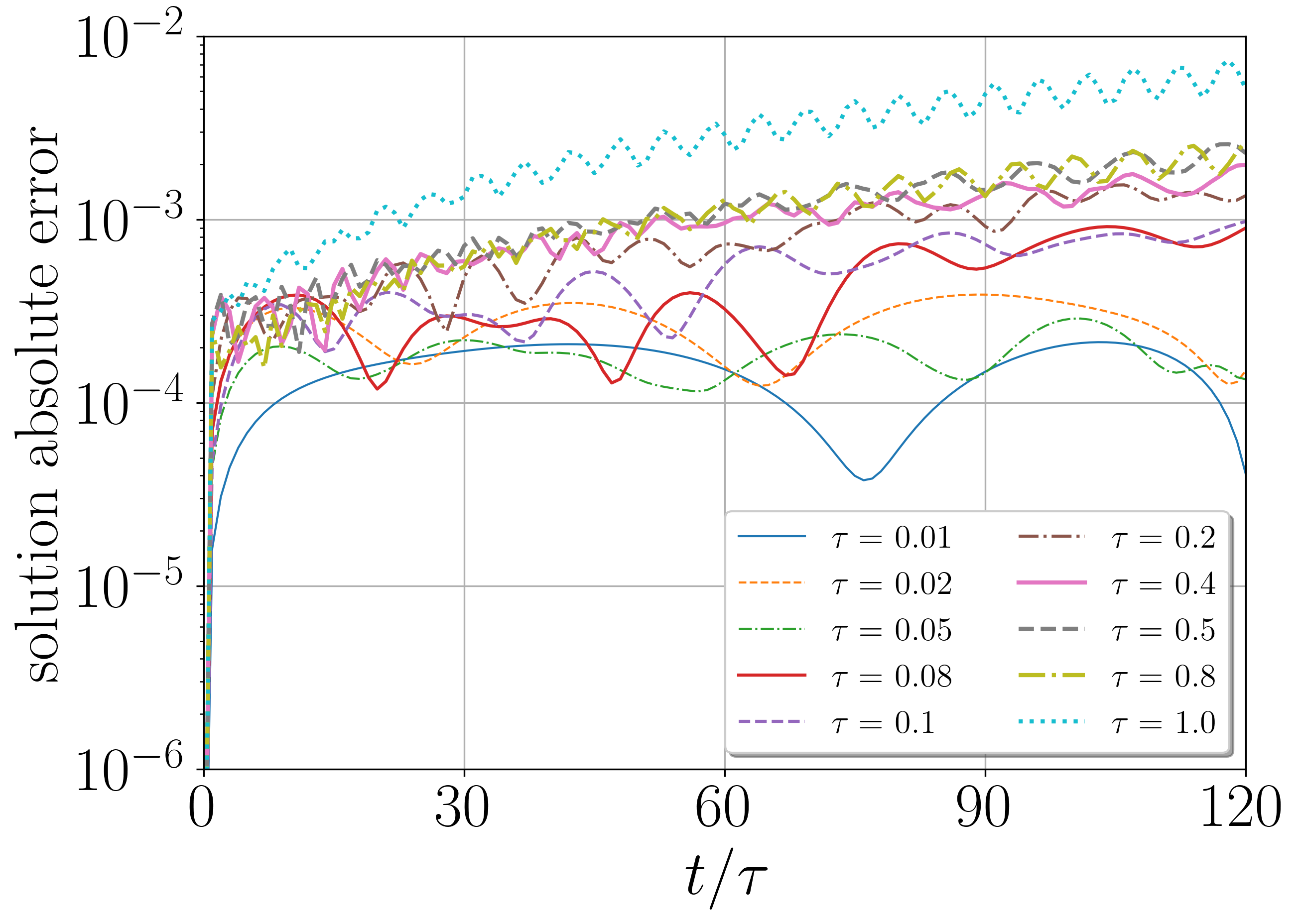}}
\subfigure[]{\label{fig:RBTau_RecH}
\includegraphics[trim=0cm 0cm 0cm 0cm,clip=true,width=0.32\textwidth]{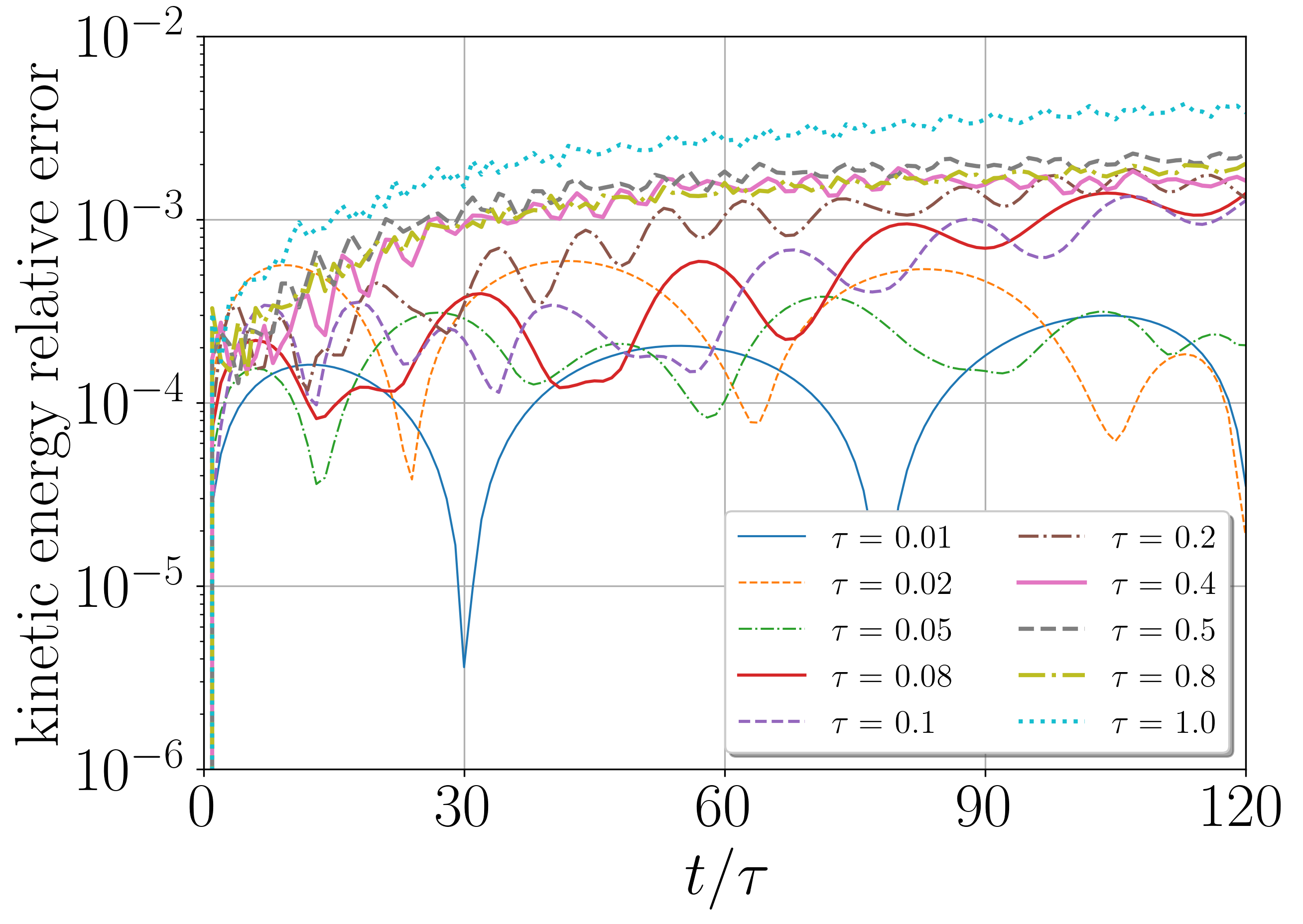}}
\subfigure[]{\label{fig:RBTau_Acc}
\includegraphics[trim=0cm 0cm 0cm 0cm,clip=true,width=0.32\textwidth]{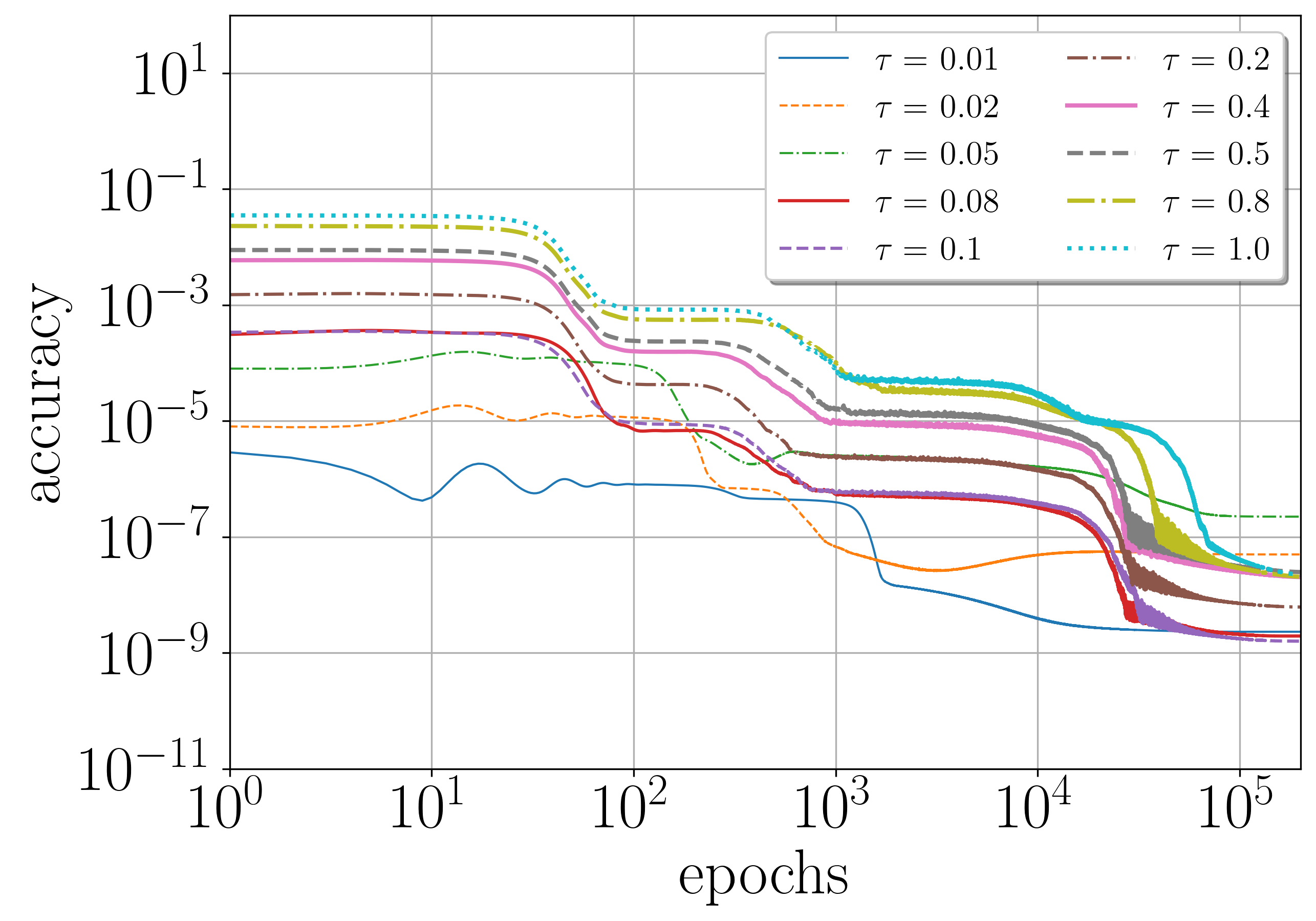}}
\subfigure[]{\label{fig:RBTau_PredS}
\includegraphics[trim=0cm 0cm 0cm 0cm,clip=true,width=0.32\textwidth]{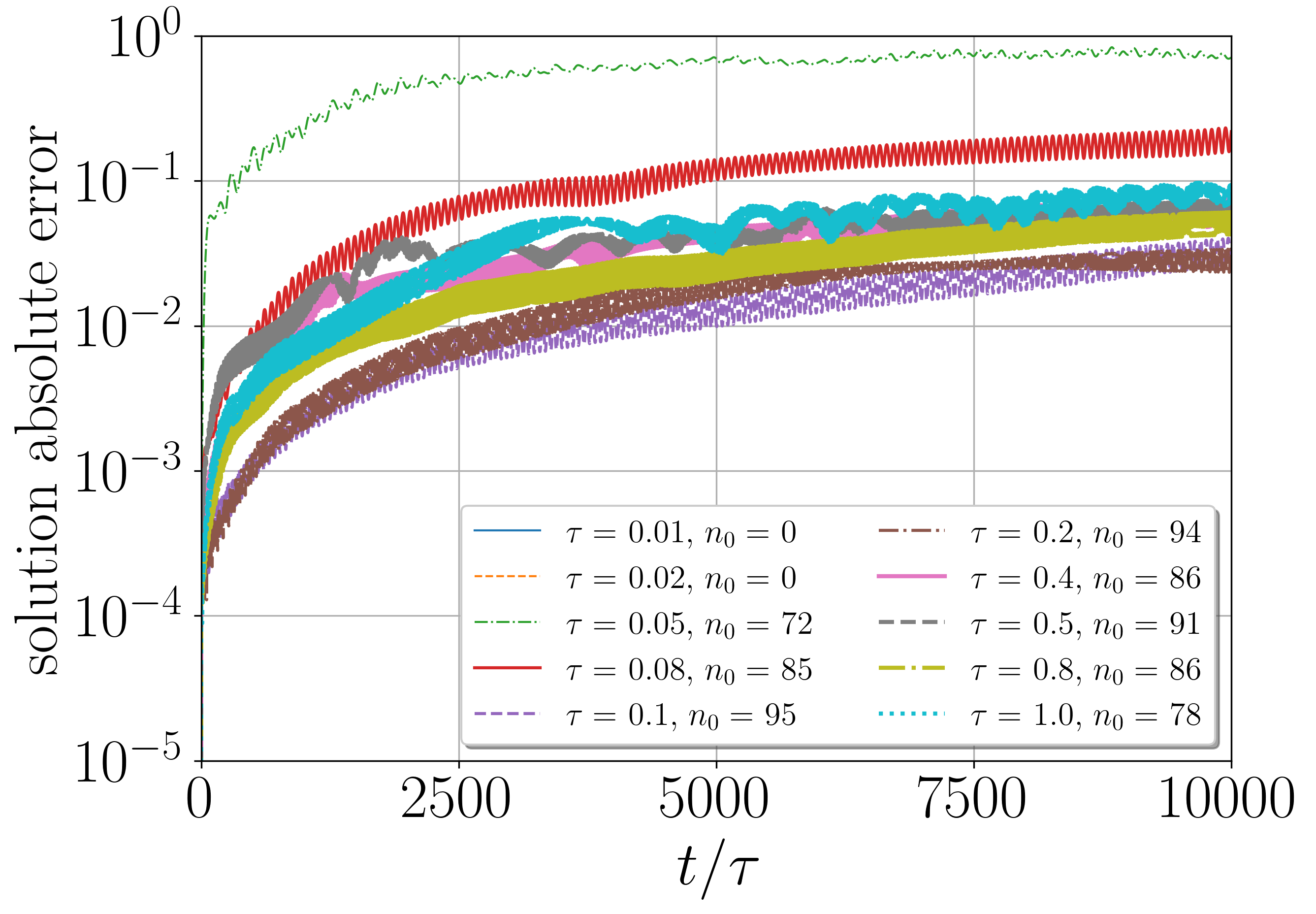}}
\subfigure[]{\label{fig:RBTau_PredH}
\includegraphics[trim=0cm 0cm 0cm 0cm,clip=true,width=0.32\textwidth]{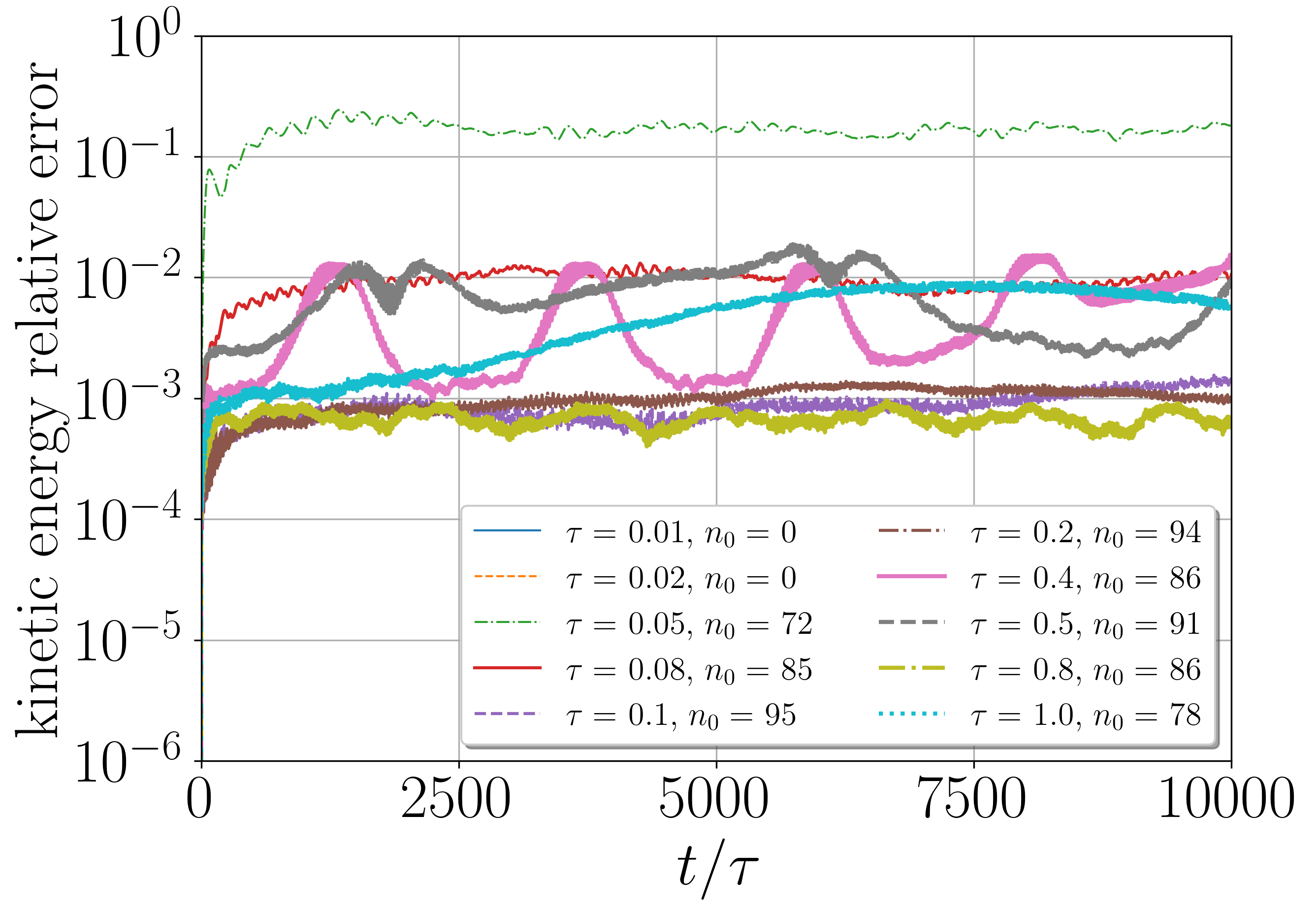}}
\caption{Numerical results of learning a single periodic trajectory of the rigid body dynamics \eqref{eq:RBody} with $\SLSNet$ ($K=1$ and $m=16$), $N=120$, and time step values $\tau=0.01$, $0.02$, $0.05$, $0.08$, $0.1$, $0.2$, $0.4$, $0.5$, $0.8$, $1$. Reconstructions of the solution are performed on the time interval $[0,T_{end}]$, where $T_{end}=N\tau$, with the initial condition $y_0$. Predictions are performed on the time interval $[0,\bar{T}_{end}]$, where $\bar{T}_{end}=10000\tau$, i.e., in a total of 10000 iterations by $\SLSNet$ of the initial condition $y(T_{end})$. (a) averaged MSE loss \eqref{eq:loss} function values. (b) averaged absolute errors of the reconstructed solutions. (c) averaged kinetic energy \eqref{eq:KinEn} relative errors of the reconstructed solutions. (d) averaged MSE accuracy \eqref{eq:acc} function values. (e) averaged absolute errors of the predicted solutions. (f) averaged kinetic energy \eqref{eq:KinEn} relative errors of the predicted solutions.}\label{fig:RBTau}
\end{figure}

\begin{figure}[t]
\centering 
\subfigure[]{\label{fig:RBTau_Rand_SymLoss}
\includegraphics[trim=0cm 0cm 0cm 0cm,clip=true,width=0.32\textwidth]{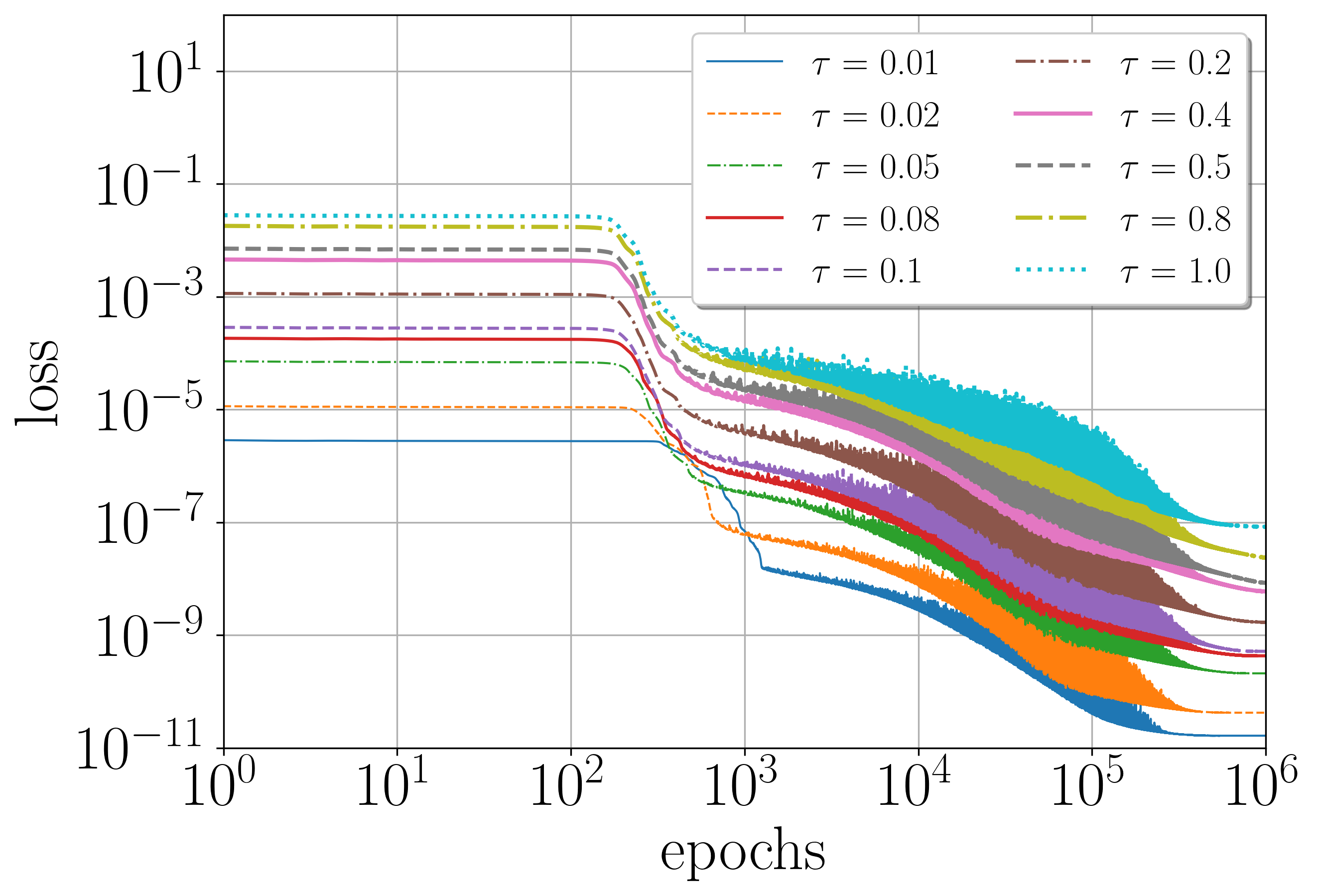}}
\subfigure[]{\label{fig:RBTau_Rand_SymS}
\includegraphics[trim=0cm 0cm 0cm 0cm,clip=true,width=0.32\textwidth]{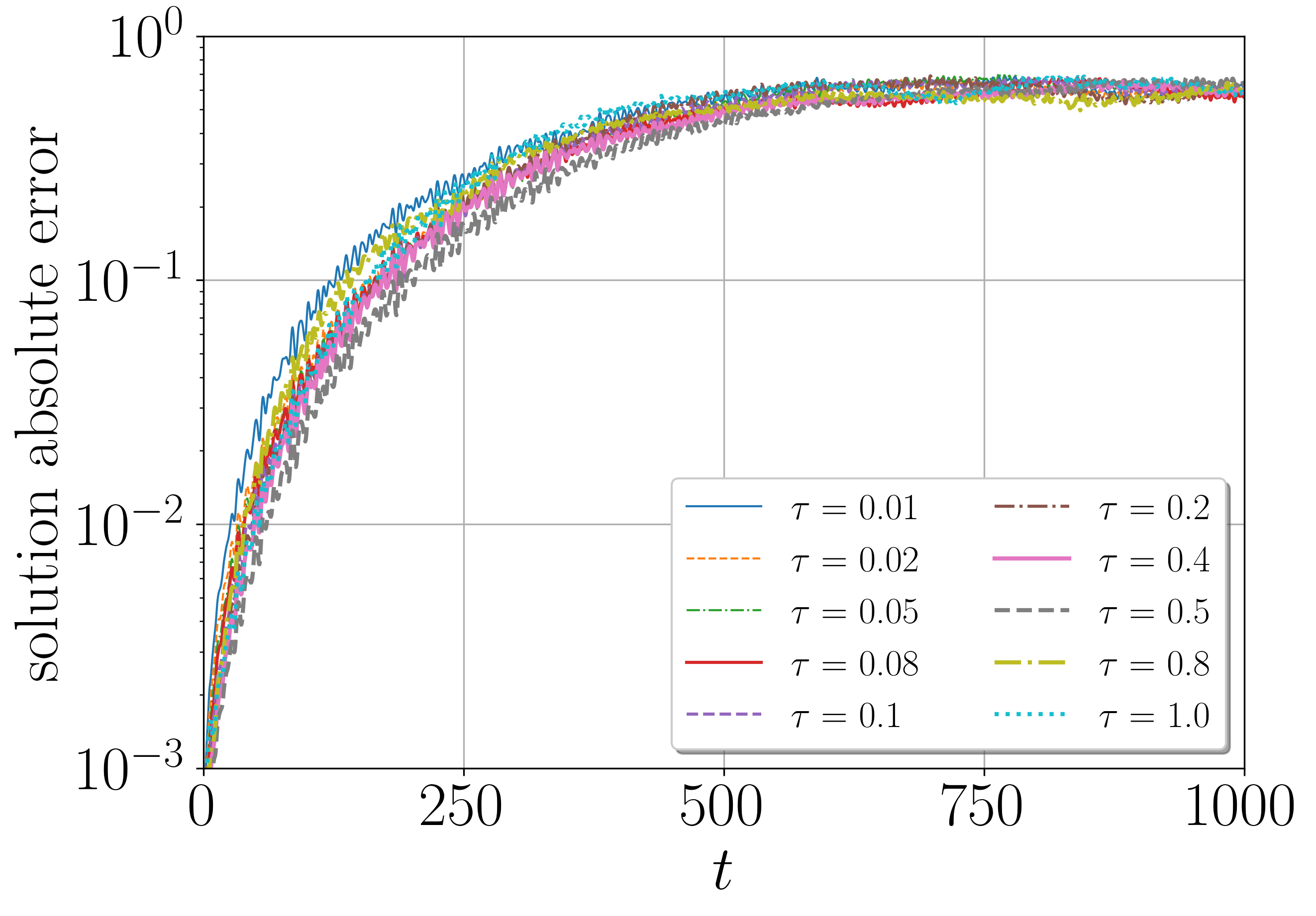}}
\subfigure[]{\label{fig:RBTau_Rand_SymH}
\includegraphics[trim=0cm 0cm 0cm 0cm,clip=true,width=0.32\textwidth]{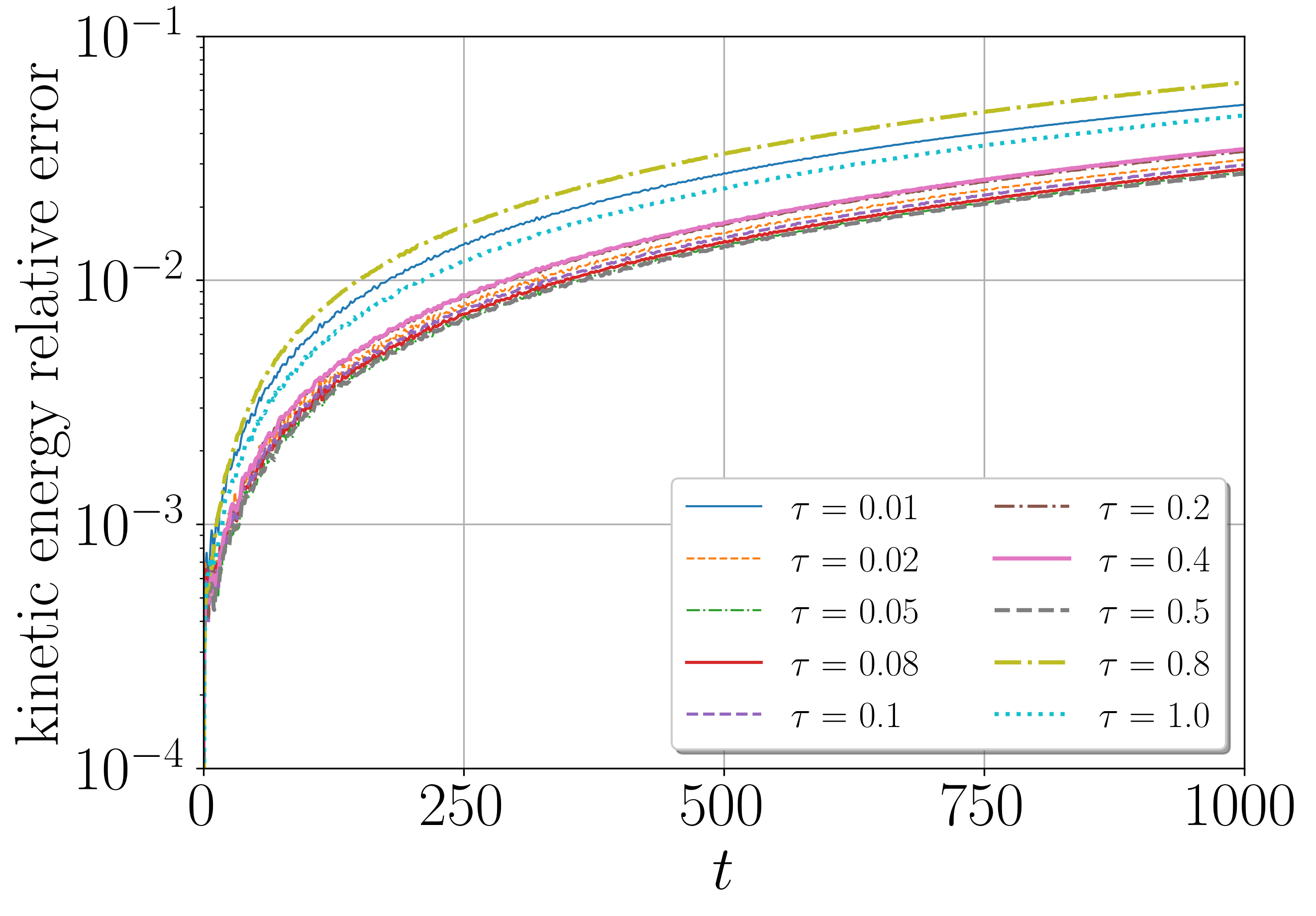}}
\caption{Numerical results of learning the whole rigid body dynamics \eqref{eq:RBody} with $\SLSNet$ ($K=1$ and $m=32$), $N=300$, and time step values $\tau=0.01$, $0.02$, $0.05$, $0.08$, $0.1$, $0.2$, $0.4$, $0.5$, $0.8$, $1$. Long-time predictions are obtained iteratively considering 12 different initial conditions \eqref{eq:ICond}. (a) averaged MSE loss \eqref{eq:loss} function values. (b) averaged absolute errors of the predicted solutions. (c) averaged kinetic energy \eqref{eq:KinEn} relative errors of the predicted solutions.}\label{fig:RBTau_Rand}
\end{figure}

For the first two examples, see Figures \ref{fig:RBTau} and \ref{fig:RBTau_Rand}, we consider 10 different time step values, i.e., $\tau=0.01$, $0.02$, $0.05$, $0.08$, $0.1$, $0.2$, $0.4$, $0.5$, $0.8$, $1$, with a fixed number of training data samples $N=120$ and $N=300$ for learning the single periodic trajectory and the whole dynamics, respectively. Accordingly, the validation data sets are formed of $M=40$ and $M=100$ data samples, as in Sections \ref{sec:RB_Single} and \ref{sec:RB_Whole}. In what follows, with $\tau$ value we identify each training data set and neural networks trained with these $\tau$-labeled data sets.

As in Section \ref{sec:RB_Single}, we trained 100 neural networks choosing different randomly generated initial weight values with each $\tau$-labeled training data set obtained from sampling the single trajectory, see Figures \ref{fig:RB_sol_A} and \ref{fig:RB_LongSol}, with the initial condition $y_0$ but using different time step values. Averaged numerical results are visualized in Figure \ref{fig:RBTau}. Considering a fixed number of training data samples $N=120$ while varying the time step $\tau$ we have sampled the trajectory from different time intervals, i.e., $t\in[0,T_{end}]$, where $T_{end}=N\tau=1.2$, $2.4$, $6$, $9.6$, $12$, $24$, $48$, $60$, $96$, $120$, respectively. Thus, the larger value of $\tau$, as longer the sampling (and reconstruction) time interval. Similarly, with the fixed number of neural network prediction iterations with larger $\tau$ value we have longer prediction time intervals $[0,\bar{T}_{end}]$, e.g., with $10000$ iterations, see Figures \ref{fig:RBTau_PredS}--\ref{fig:RBTau_PredH}, we have $\bar{T}_{end}=100$, $200$, $500$, $800$, $1000$, $2000$, $4000$, $5000$, $8000$, $10000$. 

\begin{figure}[t]
\centering 
\subfigure[]{\label{fig:RBTend_Loss}
\includegraphics[trim=0cm 0cm 0cm 0cm,clip=true,width=0.32\textwidth]{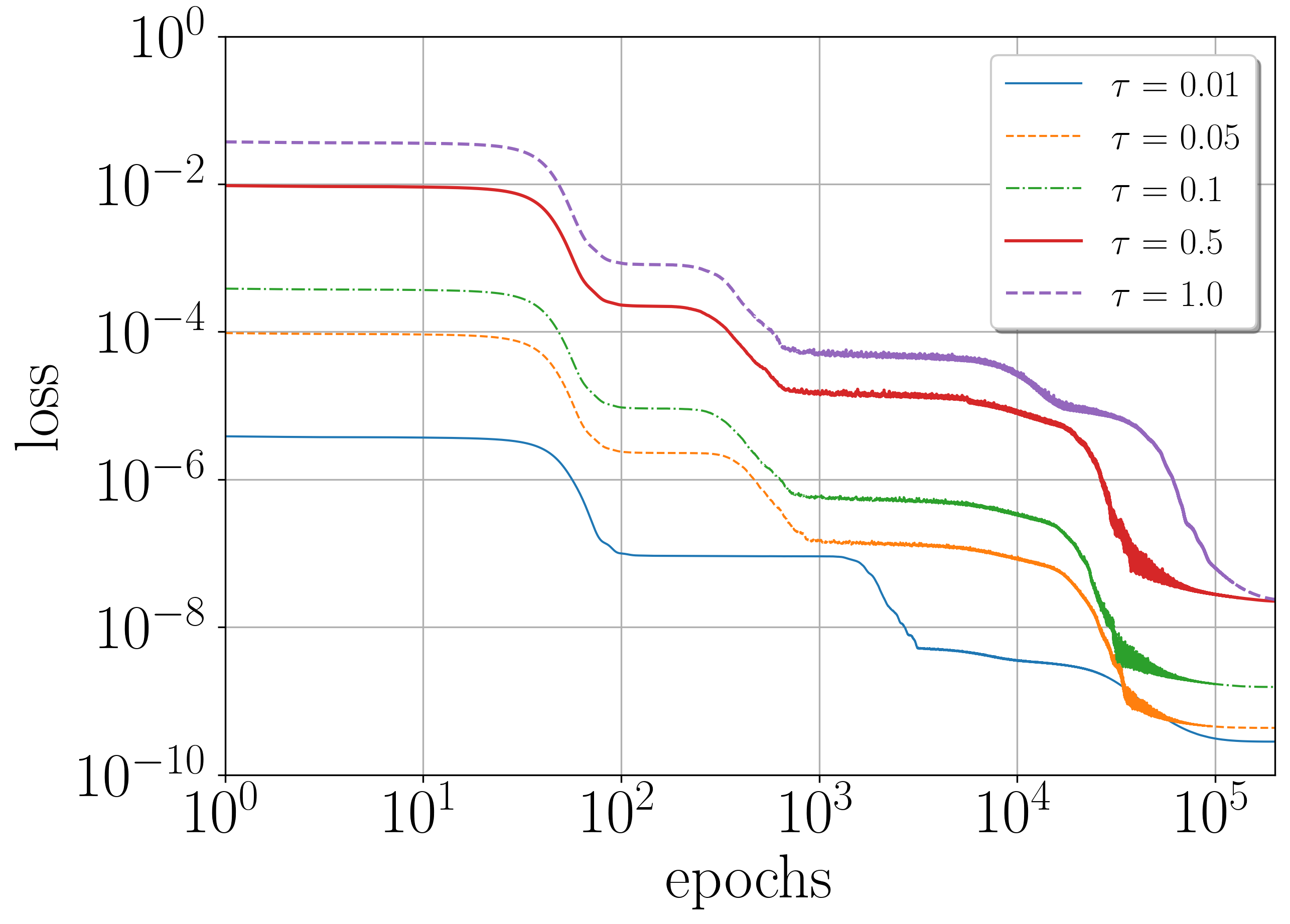}}
\subfigure[]{\label{fig:RBTend_RecS}
\includegraphics[trim=0cm 0cm 0cm 0cm,clip=true,width=0.32\textwidth]{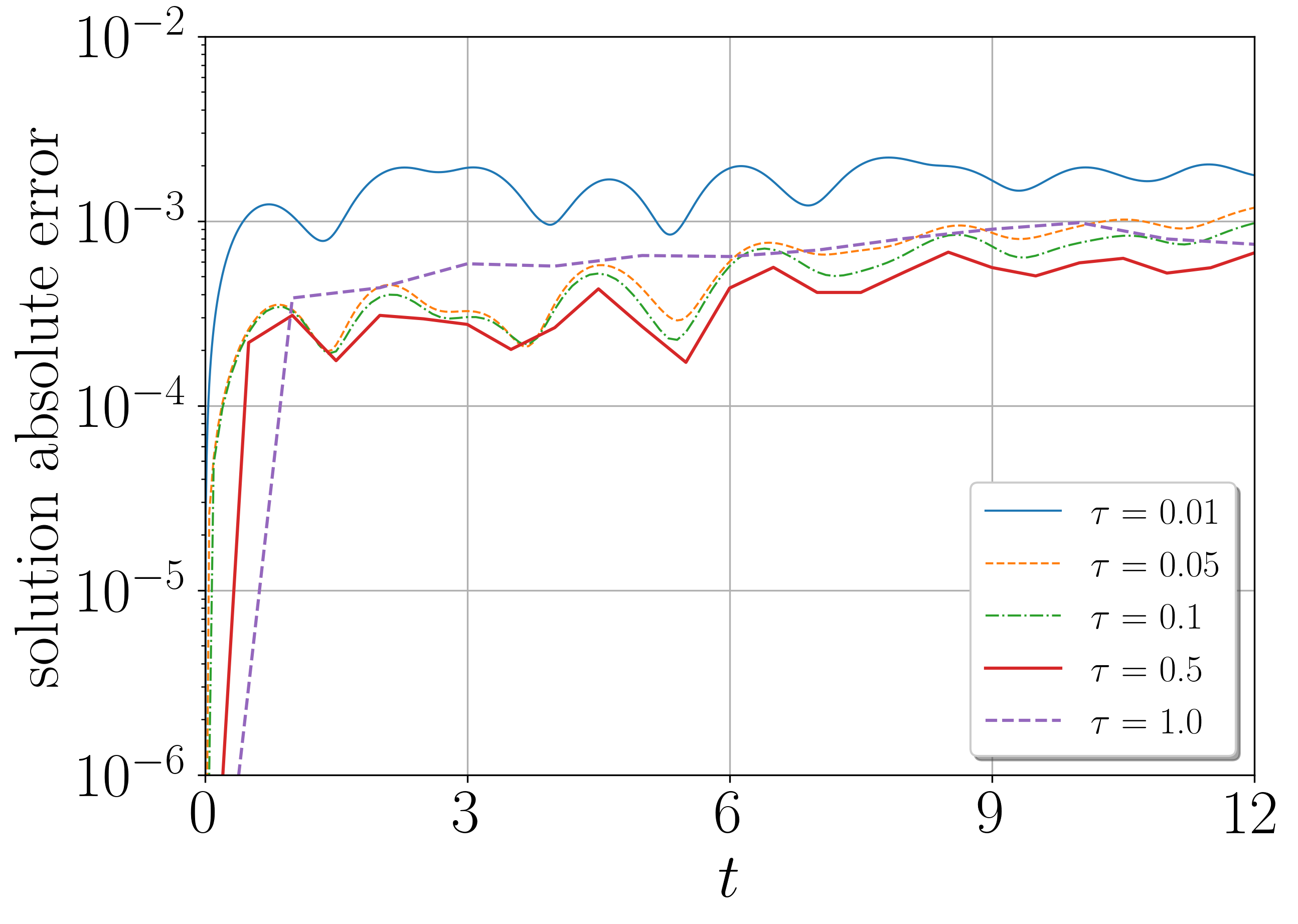}}
\subfigure[]{\label{fig:RBTend_RecH}
\includegraphics[trim=0cm 0cm 0cm 0cm,clip=true,width=0.32\textwidth]{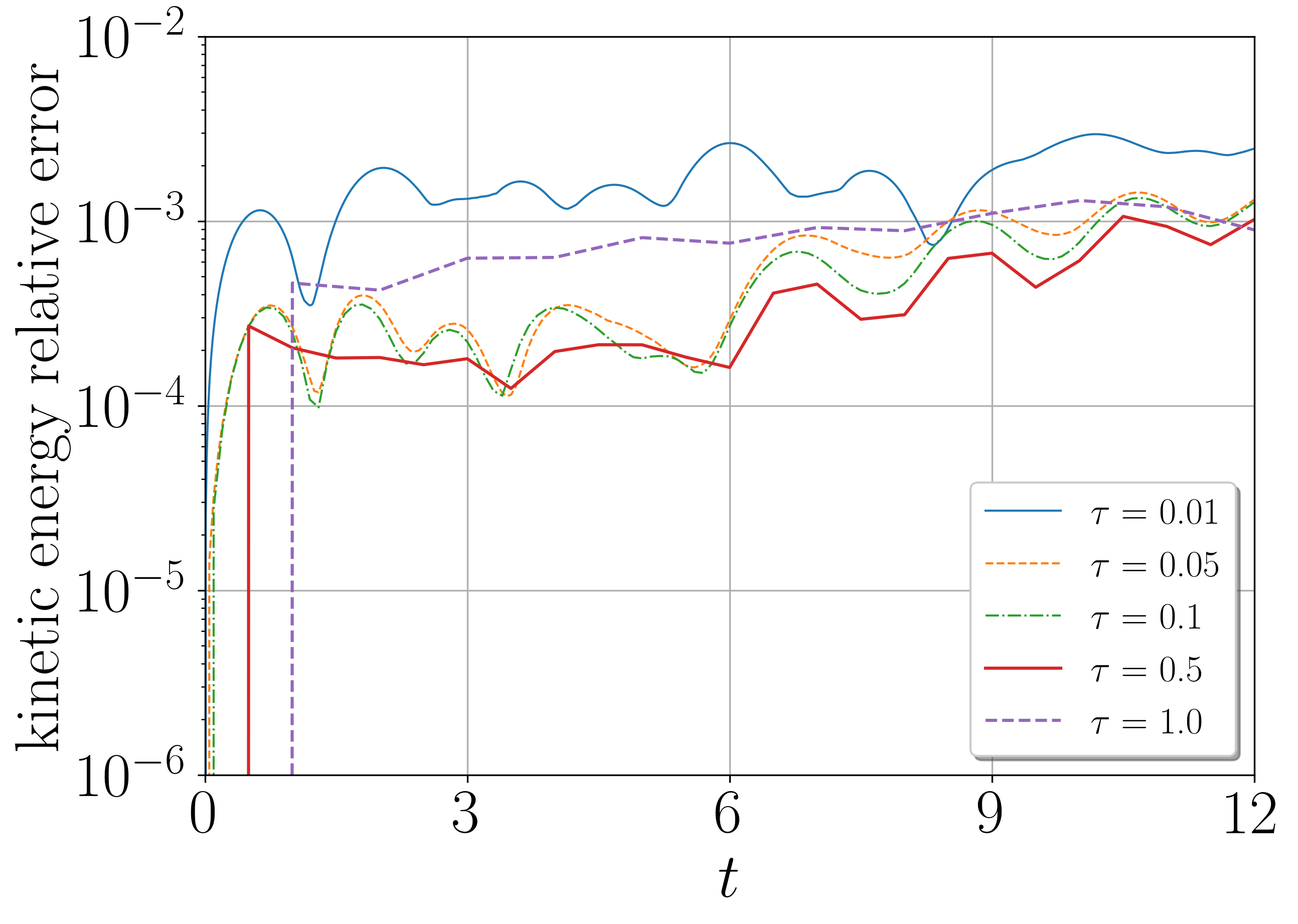}}
\subfigure[]{\label{fig:RBTend_Acc}
\includegraphics[trim=0cm 0cm 0cm 0cm,clip=true,width=0.32\textwidth]{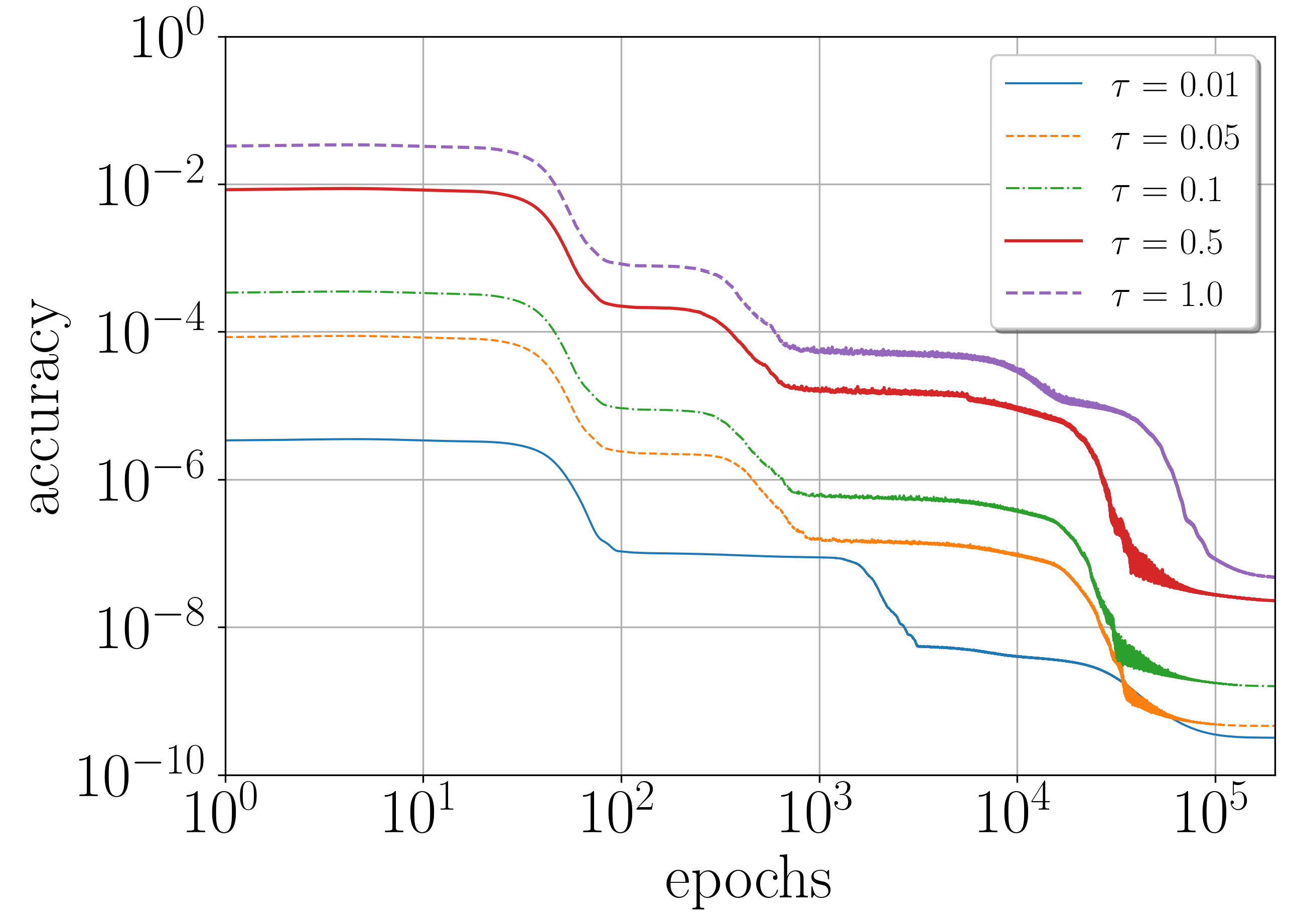}}
\subfigure[]{\label{fig:RBTend_PredS}
\includegraphics[trim=0cm 0cm 0cm 0cm,clip=true,width=0.32\textwidth]{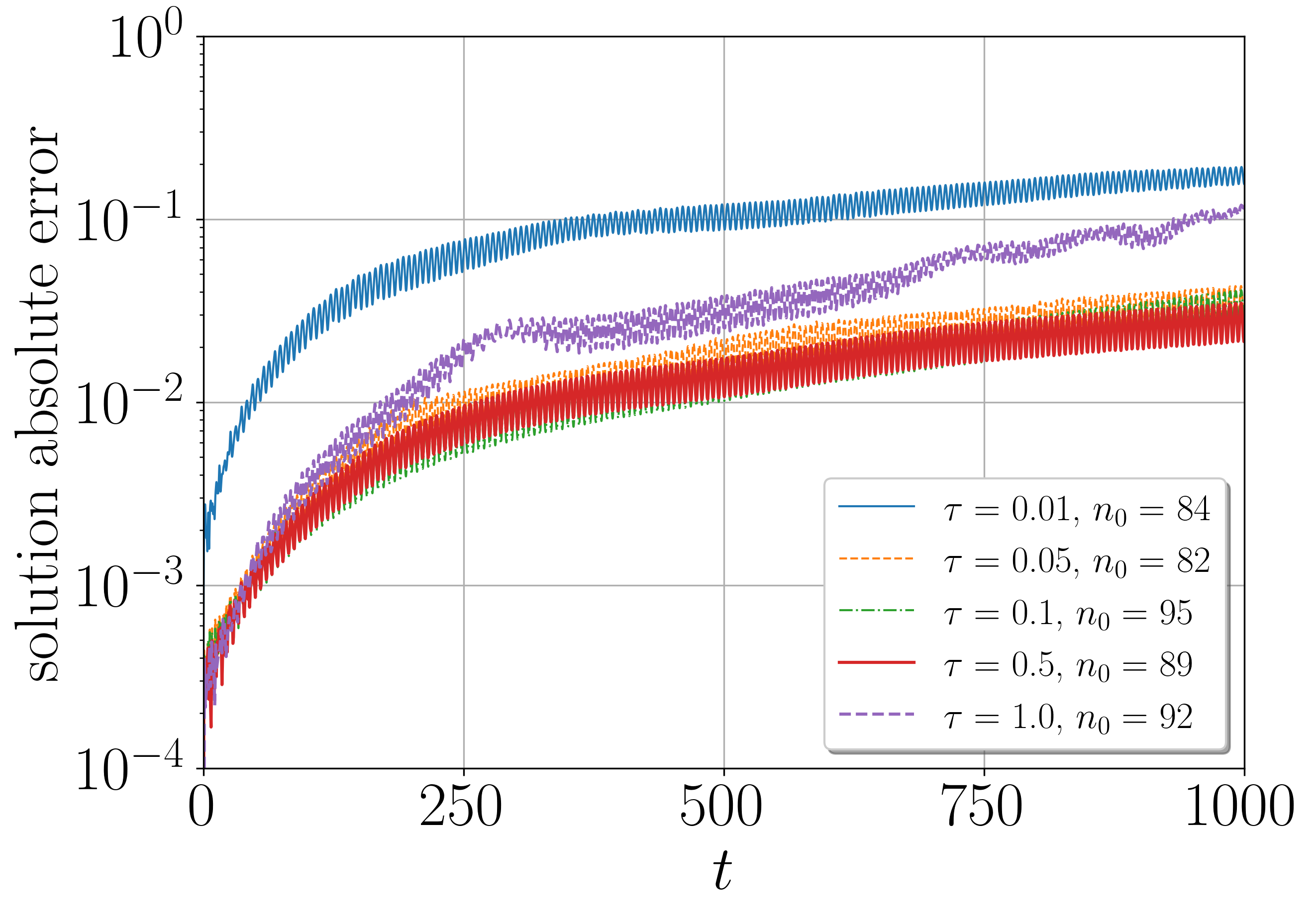}}
\subfigure[]{\label{fig:RBTend_PredH}
\includegraphics[trim=0cm 0cm 0cm 0cm,clip=true,width=0.32\textwidth]{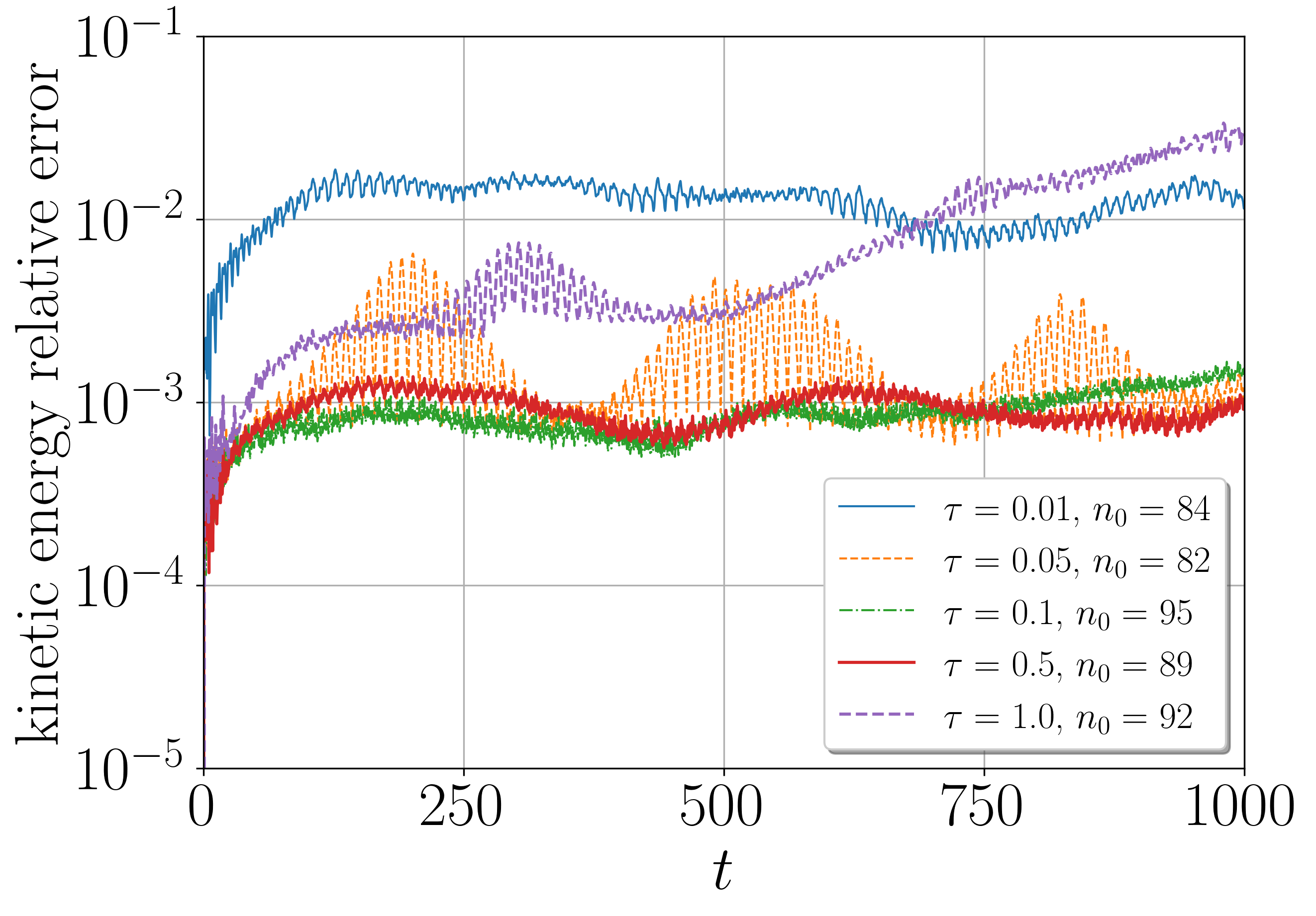}}
\caption{Numerical results of learning a single periodic trajectory of the rigid body dynamics \eqref{eq:RBody} with $\SLSNet$ ($K=1$ and $m=16$) and time step values $\tau=0.01$, $0.05$, $0.1$, $0.5$, $1$ collecting the training data on the time interval $[0,12]$. Reconstructions of the solution are performed with the initial condition $y_0$, while predictions with the initial condition $y_{12}$ on the time interval $[0,1000]$, with the different numbers of $\SLSNet$ iterations. (a) averaged MSE loss \eqref{eq:loss} function values. (b) averaged absolute errors of the reconstructed solutions. (c) averaged kinetic energy \eqref{eq:KinEn} relative errors of the reconstructed solutions. (d) averaged MSE accuracy \eqref{eq:acc} function values. (e) averaged absolute errors of the predicted solutions. (f) averaged kinetic energy \eqref{eq:KinEn} relative errors of the predicted solutions.}\label{fig:RBTend}
\end{figure}

Averaged MSE loss values \eqref{eq:loss} are illustrated in Figure \ref{fig:RBTau_Loss}, which shows that loss values decrease as the time step $\tau$ decreases. Thus, for larger time step $\tau$ values we may require training with more epochs. Investigating accuracy MSEs \eqref{eq:acc}, Figure \ref{fig:RBTau_Acc}, we observe a noticeable increase of errors for the neural networks trained on the data sets of time steps $\tau=0.01$, $0.02$, $0.05$ compared to other accuracy errors, which stay close to the associated loss errors. This suggests that the data sampling time intervals $[0,T_{end}]$ of the single trajectory for $\tau=0.01$, $0.02$, $0.05$ contain insufficient information on the dynamics and these neural networks will most likely generalize poorly, which is also evident in Figures \ref{fig:RBTau_PredS}--\ref{fig:RBTau_PredH}. Thus, with smaller time step values we require larger lengths of sampling time interval to include more information about the solution, see Figure \ref{fig:RBTend}, or more training data samples in learning the whole rigid body dynamics, see Figure \ref{fig:RBTau_Rand}.

Figures \ref{fig:RBTau_RecS}--\ref{fig:RBTau_RecH} demonstrate averaged solution and kinetic energy \eqref{eq:KinEn} reconstruction errors performing in total $N=120$ iterations by $\SLSNet$ of the initial condition $y_0$. Results of Figures \ref{fig:RBTau_RecS}--\ref{fig:RBTau_RecH} are consistent with the loss errors in Figure \ref{fig:RBTau_Loss}, i.e., we can observe smaller errors produced by the $\SLSNet$ trained on the data sets of smaller time step $\tau$ values. The same is not true for the long-time predictions performing 10000 neural network iterations of the initial condition given at time $t=T_{end}$, i.e., the first solution value in the validation data set of the single trajectory. In Figure \ref{fig:RBTau_PredS} we demonstrate averaged solution absolute errors, while in Figure \ref{fig:RBTau_PredH} we illustrate averaged kinetic energy relative errors. To both figure legends, we have added the number $n_0$ indicating the number of neural networks which produced stable long-time predictions. The averaging is only performed over these $n_0$ neural networks. We apply the same criteria for long-time stable predictions already stated in Section \ref{sec:RB_Single}. Since the neural networks trained with insufficient information of the dynamics, i.e., the cases with $\tau=0.01$, $0.02$, $0.05$, neural networks do not generalize well and in the cases with $\tau=0.01$ and $\tau=0.02$ there were no stable long-time predictions. The case with $\tau=0.05$ demonstrates that the information of the dynamics in the training data is limited and long-time prediction errors are significantly larger compared to other cases, which averaged kinetic energy relative errors are between $0.1$ and $1$ percent over 10000 iterations.        

When learning the whole rigid body dynamics, see Section \ref{sec:RB_Whole}, we also constructed 10 different training data sets using the same 10 different time step $\tau$ values above. We considered the same randomly generated $N=300$ (training) and $M=100$ (validation) data points illustrated in Figure \ref{fig:RB_RandIC} to compute the training and validation data sets with different time step values. In this example, we trained 10 different neural networks and performed predictions in time for 12 different initial conditions \eqref{eq:ICond} over the time interval $[0,1000]$. Thus, as smaller the time step $\tau$, more neural network iterations are required. In particular, in total $100000$, $50000$, $20000$, $12500$, $10000$, $5000$, $2500$, $2000$, $1250$, $1000$ iterations, respectively. 

Averaged loss, solution, and kinetic energy errors are illustrated in Figure \ref{fig:RBTau_Rand}. In Figure \ref{fig:RBTau_Rand_SymLoss} we observe a similar trend already seen in Figure \ref{fig:RBTau_Loss}, where the loss mean squared errors decrease with decreasing time step $\tau$ values. On the contrary, in this experiment, averaged accuracy MSEs \eqref{eq:acc} (not shown) have the same trend as for the loss values in Figure \ref{fig:RBTau_Rand_SymLoss}, indicating that neural networks will generalize well even with small values of $\tau$. We can see that in Figures \ref{fig:RBTau_Rand_SymS}--\ref{fig:RBTau_Rand_SymH}, where in Figure \ref{fig:RBTau_Rand_SymS} we illustrate averaged solution absolute errors as a function of time, while in Figure \ref{fig:RBTau_Rand_SymH} we demonstrate averaged kinetic energy relative errors. Compare Figures \ref{fig:RBTau_Rand_SymS}--\ref{fig:RBTau_Rand_SymH} to Figures \ref{fig:RB_Rand_SymS}--\ref{fig:RB_Rand_SymH}. Figure \ref{fig:RBTau_Rand_SymH} shows that we can obtain qualitatively good long-time predictions by the phase volume preserving neural networks $\SLSNet$ for a large range of time step $\tau$ values as long as there is a sufficient representation of dynamics in the training data set. 

In the third example of this section, we consider learning the single periodic trajectory of Section \ref{sec:RB_Single} by collecting training and validation data from the time intervals $[0,12]$ and $[12,16]$ with 5 different time step values $\tau=0.01$, $0.05$, $0.1$, $0.5$, $1$. Thus, the numbers of training and validation data samples vary. In particular, we have $N=1200$, $240$, $120$, $24$, $12$ and $M=400$, $80$, $40$, $8$, $4$, respectively. We already observed in Section \ref{sec:RB_Single} that collecting data from the time interval $[0,12]$ with the time step $\tau=0.1$ provided training data set with sufficient dynamics information for trained neural networks to produce good long-time predictions, e.g., see Figures \ref{fig:RB_LongSim} and \ref{fig:RB_Pred}.

Averaged errors of 100 trained neural networks $\SLSNet$ are visualized in Figure \ref{fig:RBTend}. MSE loss \eqref{eq:loss} and accuracy \eqref{eq:acc} values follow the same trend as in Figures \ref{fig:RBTau_Loss} and \ref{fig:RBTau_Rand_SymLoss}, i.e., we obtain smaller errors with smaller $\tau$ values. Notice that the accuracy errors for the cases $\tau=0.01$, $0.05$ in Figure \ref{fig:RBTend_Acc} are as small as the loss values in Figure \ref{fig:RBTend_Loss}, in contrast to what we observed in results of Figure \ref{fig:RBTau_Acc}. This can be easily explained by the fact that in this experiment the data sets labeled with $\tau=0.01$, $0.05$ contain more data points $N$ sampled on longer time interval $[0,12]$, which incorporates more dynamics information, e.g., see Figure \ref{fig:RB_sol_A}. 

Reconstruction solution and kinetic energy errors on the reconstruction time interval $[0,12]$ are shown in Figures \ref{fig:RBTend_RecS}--\ref{fig:RBTend_RecH}, respectively. Since the time steps are different, we have to iterate neural networks $\SLSNet$ of the initial condition $y_0$ different number of times. In this experiment, to reconstruct the solution on the reconstruction time interval we require exactly $N$ number of iterations, i.e., where $N$ is the number of data samples in each training data set. The smaller the value of $\tau$, as more iterations we have to perform. Thus, more errors may accumulate at each iteration, despite the loss values having smaller errors in Figure \ref{fig:RBTend_Loss}. On the contrary, with larger $\tau$ values we require fewer iterations by the neural networks but we have larger loss errors, which can be obtained smaller by increasing the number of epochs in training. Despite that, we are still able to obtain good long-time predictions, see Figures \ref{fig:RBTend_PredS}--\ref{fig:RBTend_PredH} and compare them to Figures \ref{fig:RB_PredSymS} and \ref{fig:RB_PredSymH}. 

Averaged solution and kinetic energy errors over the prediction time interval $[0,1000]$ are demonstrated in Figures \ref{fig:RBTend_PredS}--\ref{fig:RBTend_PredH}. Note the different number of iterations required for $\SLSNet$ of the initial condition $y_{12}$ with the different time step $\tau$ values, i.e., $100000$, $20000$, $10000$, $2000$, $1000$ iterations, respectively. The numerical results of Figure \ref{fig:RBTend} demonstrate that locally-symplectic neural networks $\SLSNet$ can learn phase volume-preserving dynamics not only with the different time step $\tau$ values but also with different number $N$ of dynamics data samples.  

\subsection{Learning quasi-periodic trajectories of volume-preserving dynamics}\label{sec:Charge}
For the final example of this numerical results section, we consider learning of a quasi-periodic motion of the charged particle in an electromagnetic field governed by the Lorentz force \cite{PoisNets,Zhu22}. The charged particle's position $y=(y_1,y_2,y_3)^T\in\R^3$ satisfies the following second-order ordinary differential equation:
\begin{equation}\label{eq:ChargeEq}
\mathrm{m} \dder{y}{t} = \mathrm{q} \left(E+\der{y}{t}\times B \right),
\end{equation}   
where $\mathrm{m}$ is the particle's mass, $\mathrm{q}$ is the electric charge, $E=-\nabla \varphi$ and $B=\nabla \times A$ are the electric and magnetic fields with the scalar electric potential $\varphi\in\R$ and the magnetic vector potential $A\in\R^3$, respectively, and $\nabla\times$ denotes the curl of a vector field. With an introduction of the momentum $p:=\mathrm{m}\der{y}{t}\in\R^3$ the equation \eqref{eq:ChargeEq} can be cast into the form \eqref{eq:HamJ} with the non-constant skew-symmetric matrix $J(y)$, i.e., the charged particle equations \eqref{eq:ChargeEq} are Poisson system \cite{Hairer,PoisNets} with the Hamiltonian
\begin{equation}\label{eq:ChargeH}
H(y,p) = \frac{1}{2\mathrm{m}}p^Tp + \mathrm{q}\varphi(y).
\end{equation}

Without loss of generality, we set $\mathrm{m}=1$ and $\mathrm{q}=1$, and consider the charged particle dynamics restricted to the motion on a plane with the potentials \cite{PoisNets,Zhu22}:
\[
\varphi(y) = \frac{1}{100\sqrt{y_1^2+y_2^2}}, \quad
A(y) = \frac{1}{3} \sqrt{y_1^2+y_2^2} \, (-y_2,y_1,0)^T. 
\]
Thus, the dimension-reduced charged particle motion on a plane is obtained by setting the initial conditions with $y_3(0)=0$ and $p_3(0)=0$. 

For learning the quasi-periodic motion of the dimension-reduced charged particle dynamics we consider the dimension-reduced initial conditions $\tilde{y}_0:=\tilde{y}(0)=(0.1,1)^T$ and $\tilde{p}_0:=\tilde{p}(0)=(1.1,0.5)^T$. We collect $N=200$ training data points with the time step $\tau=0.2$, i.e., from the time interval $[0,40]$, followed by $M=100$ solution values of the same trajectory to form the validation data set on the time interval $[40,60]$. Then neural networks are tested by reconstructing the solution on the time interval $[0,40]$ with the initial condition $(\tilde{y}_0,\tilde{p}_0)^T$, and performing predictions of the quasi-periodic trajectory for $t > 40$ with the dimension-reduced initial condition $(\tilde{y}_{40},\tilde{p}_{40})^T$, where $\tilde{y}_{40}:=\tilde{y}(40)$ and $\tilde{p}_{40}:=\tilde{p}(40)$.

\begin{figure}[t]
\centering 
\subfigure[]{\label{fig:ChargeLoss}
\includegraphics[trim=0cm 0cm 0cm 0cm,clip=true,width=0.32\textwidth]{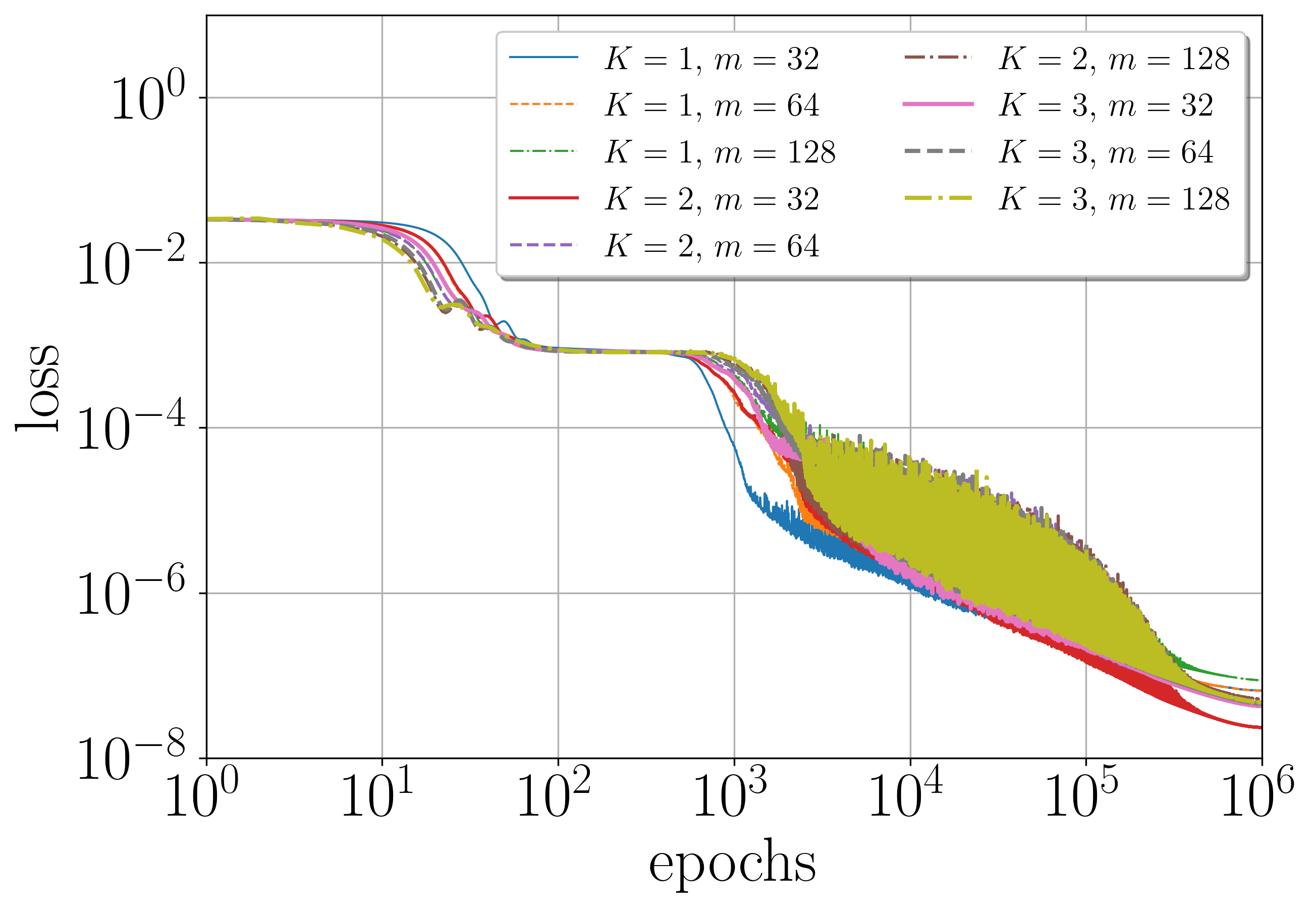}}
\subfigure[]{\label{fig:ChargeRecS}
\includegraphics[trim=0cm 0cm 0cm 0cm,clip=true,width=0.32\textwidth]{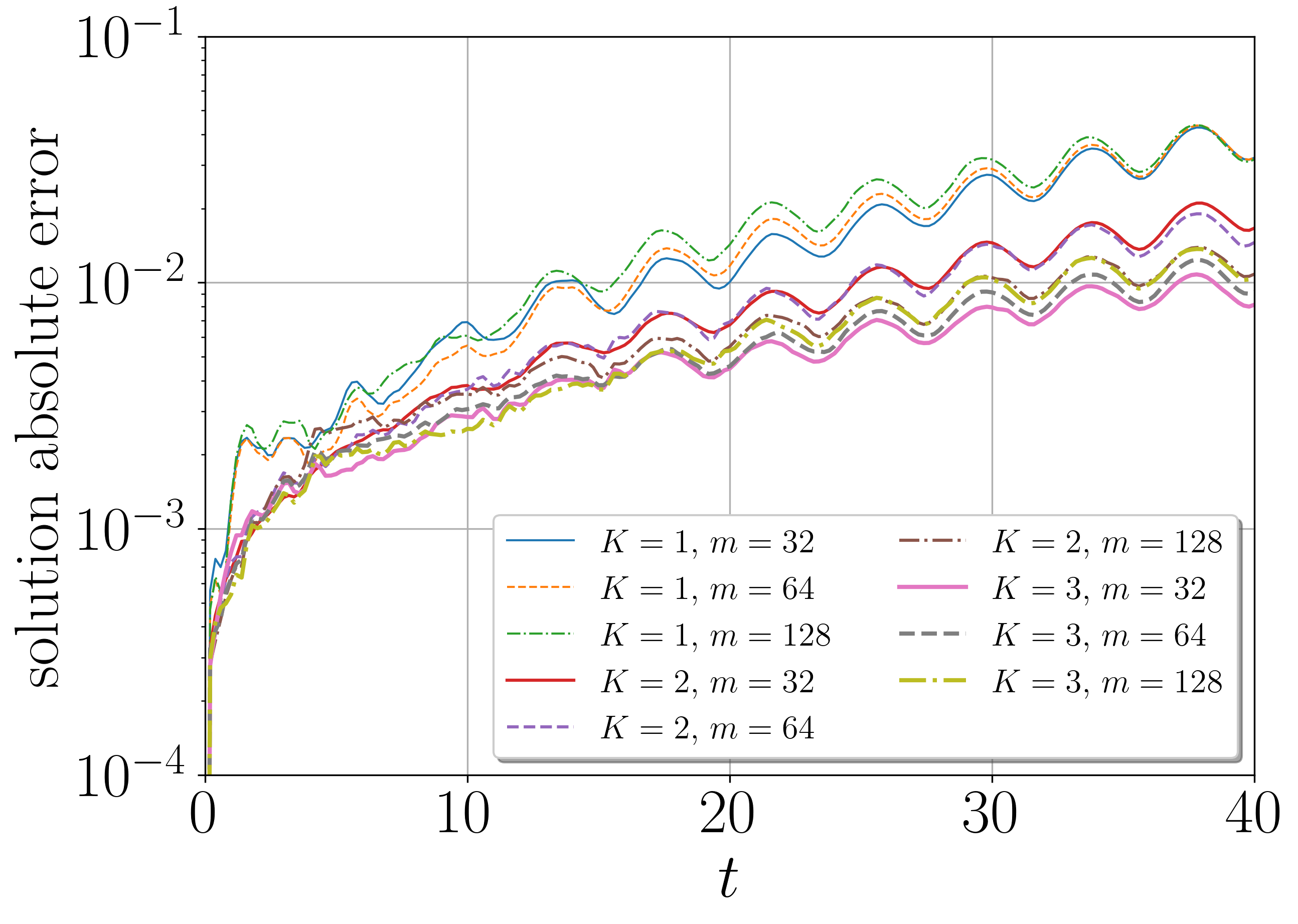}}
\subfigure[]{\label{fig:ChargeRecH}
\includegraphics[trim=0cm 0cm 0cm 0cm,clip=true,width=0.32\textwidth]{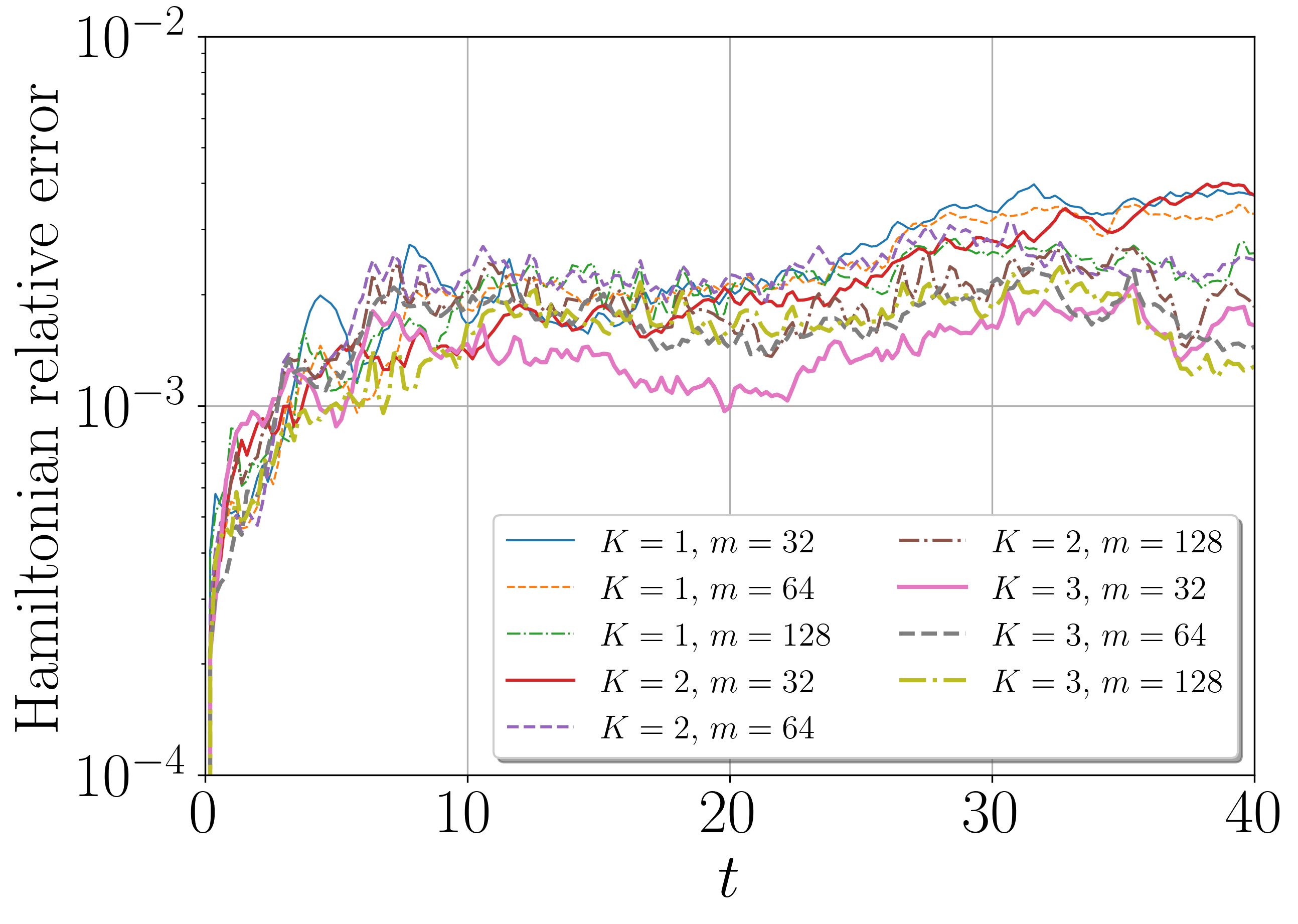}}
\subfigure[]{\label{fig:ChargeAcc}
\includegraphics[trim=0cm 0cm 0cm 0cm,clip=true,width=0.32\textwidth]{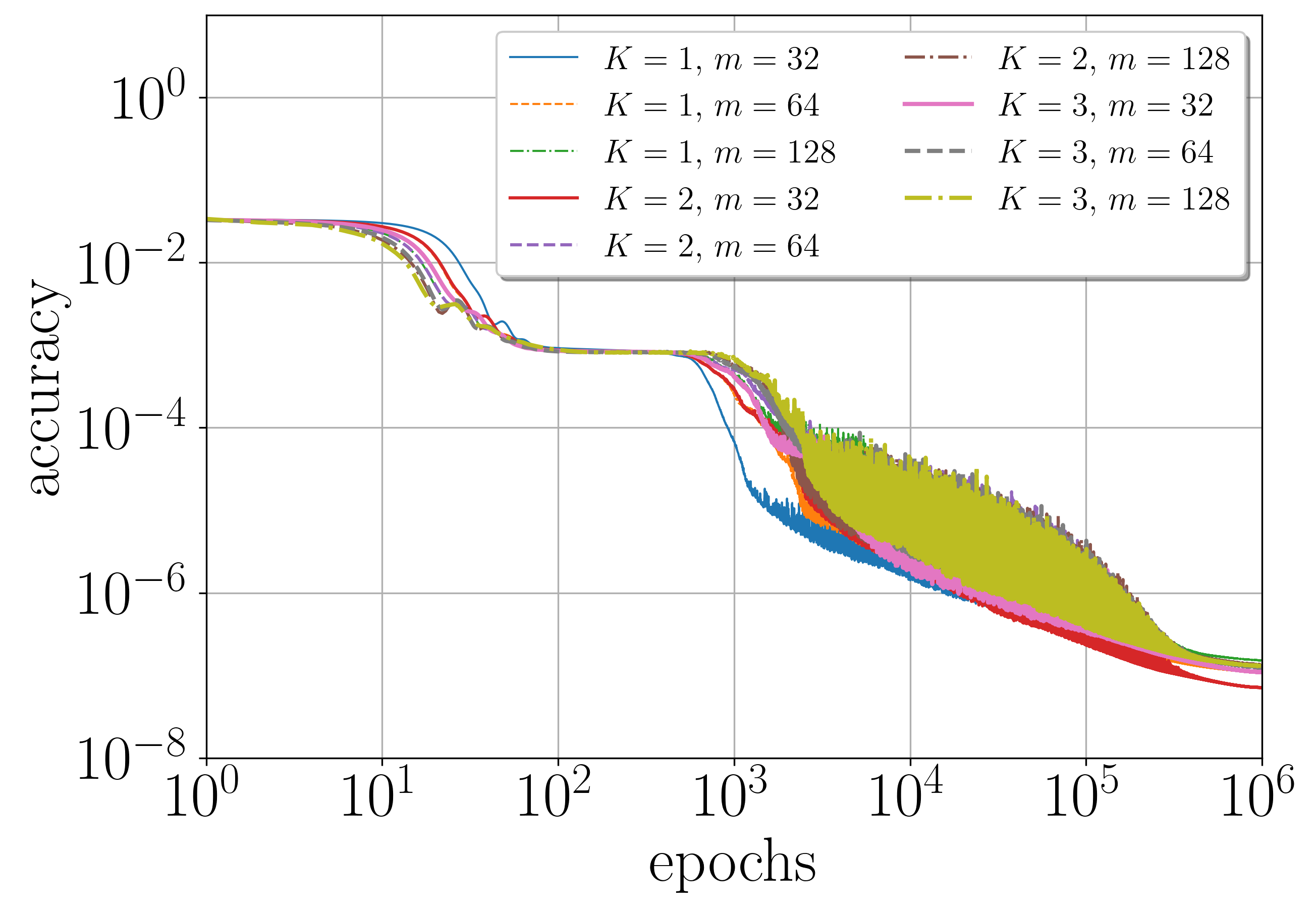}}
\subfigure[]{\label{fig:ChargePredS}
\includegraphics[trim=0cm 0cm 0cm 0cm,clip=true,width=0.32\textwidth]{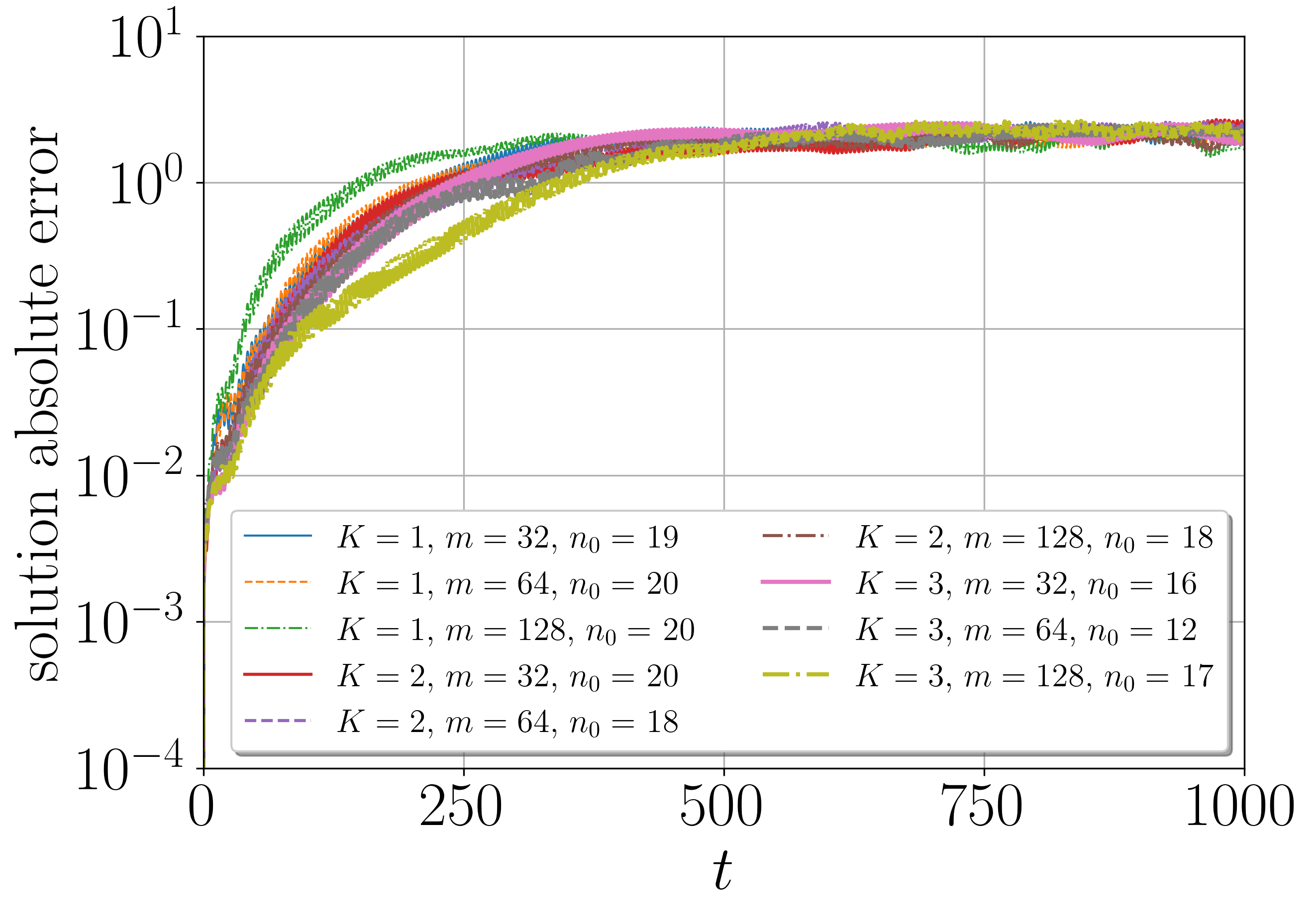}}
\subfigure[]{\label{fig:ChargePredH}
\includegraphics[trim=0cm 0cm 0cm 0cm,clip=true,width=0.32\textwidth]{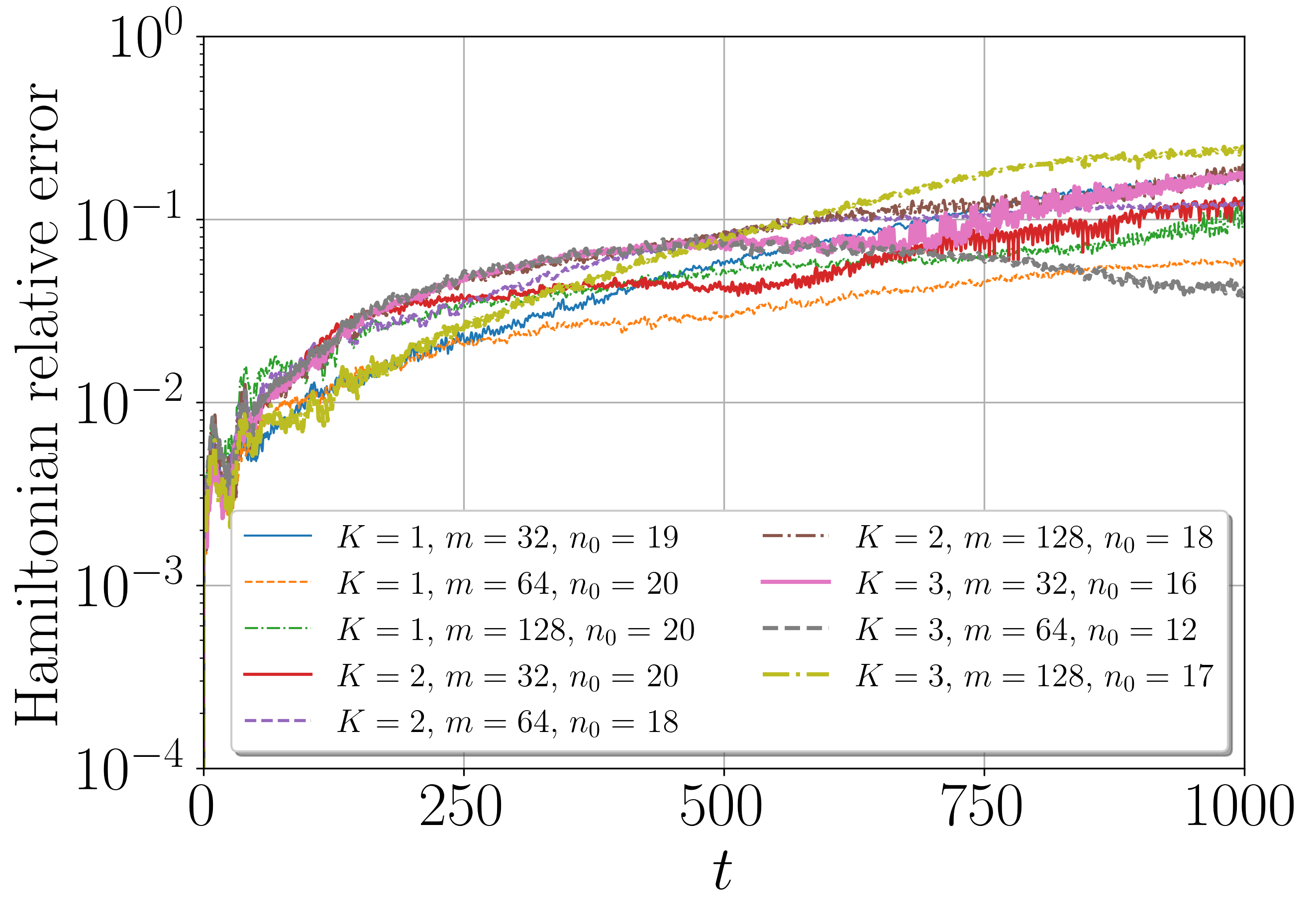}}
\caption{Numerical results of learning quasi-periodic motion of the charged particle in the electromagnetic field with $\SLSNet$. (a) averaged MSE loss \eqref{eq:loss} function values. (b) averaged absolute errors of the reconstructed solutions on the time interval $[0,40]$ with the initial condition $(\tilde{y}_{0},\tilde{p}_{0})^T$. (c) averaged Hamiltonian \eqref{eq:ChargeH} relative errors of the reconstructed solutions. (d) averaged MSE accuracy \eqref{eq:acc} function values. (e) averaged absolute errors of the predicted solutions on the long-time prediction time interval $[0,1000]$ with the initial condition $(\tilde{y}_{40},\tilde{p}_{40})^T$. (f) averaged Hamiltonian \eqref{eq:ChargeH} relative errors of the predicted solutions.}\label{fig:ChargeRes}
\end{figure}

\begin{figure}[t]
\centering 
\subfigure[]{\label{fig:ChargeRec}
\includegraphics[trim=0cm 0cm 0cm 0cm,clip=true,width=0.42\textwidth]{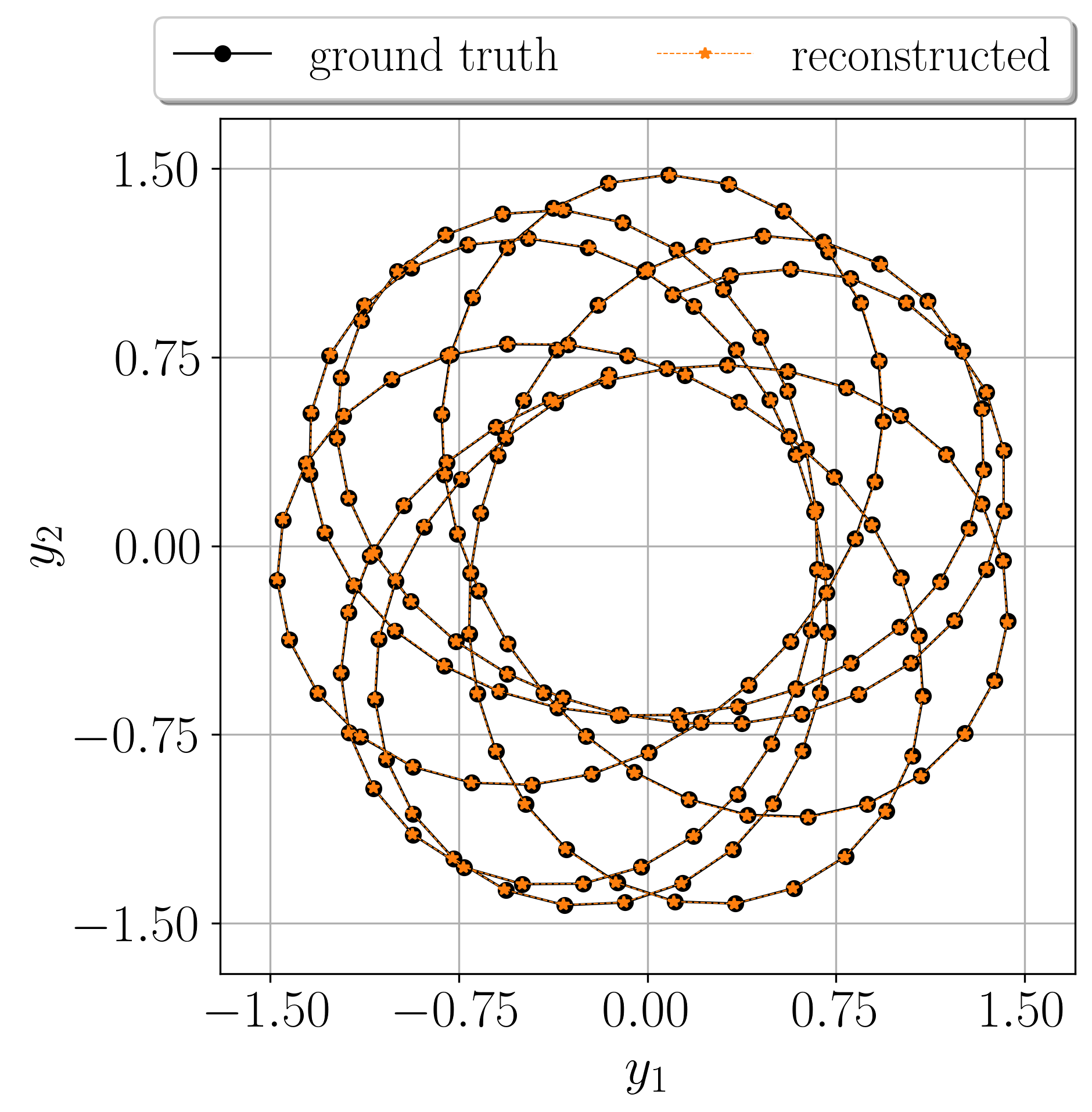}}\qquad
\subfigure[]{\label{fig:ChargeM100}
\includegraphics[trim=0cm 0cm 0cm 0cm,clip=true,width=0.42\textwidth]{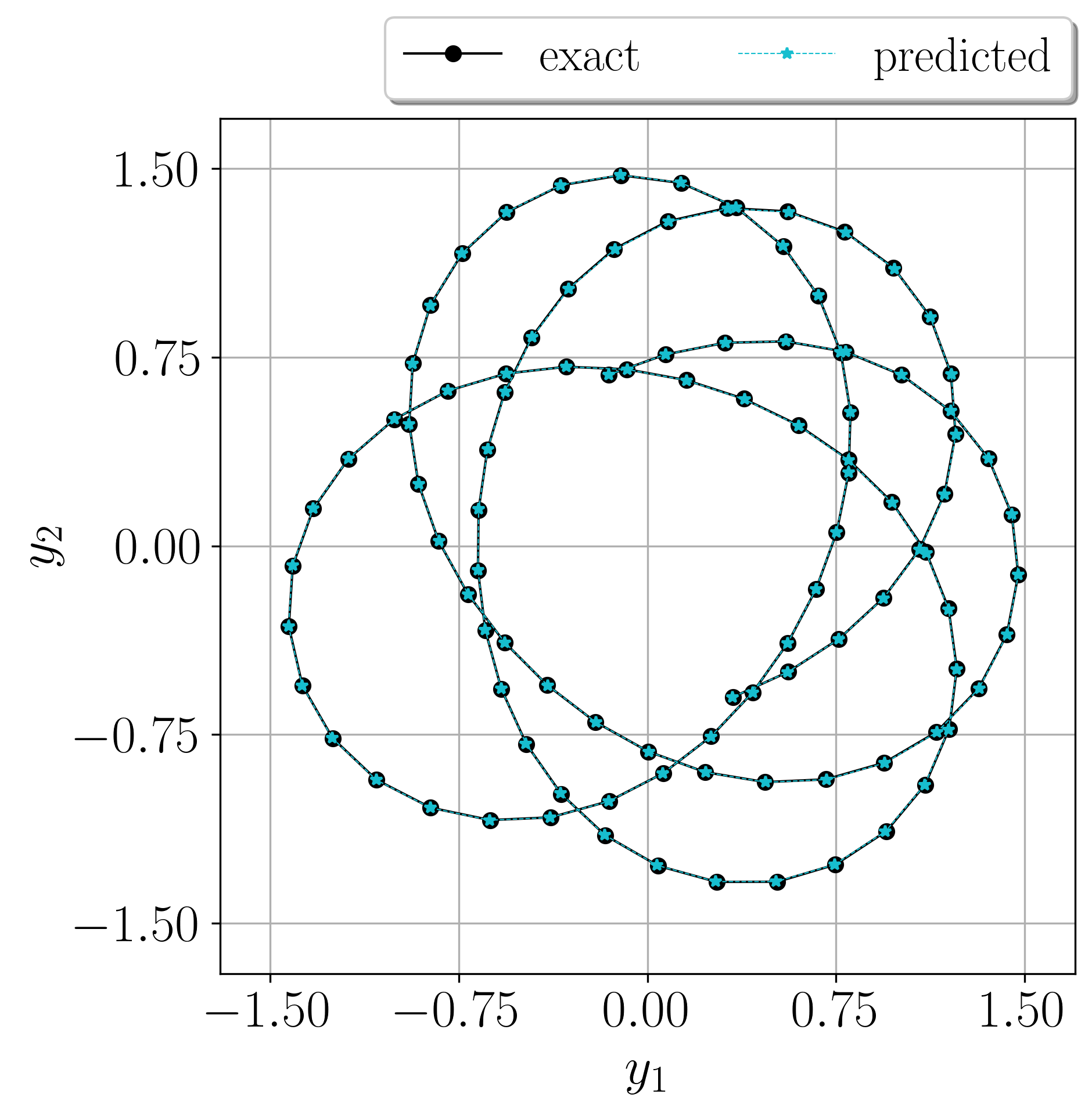}}
\subfigure[]{\label{fig:ChargeM500}
\includegraphics[trim=0cm 0cm 0cm 0cm,clip=true,width=0.42\textwidth]{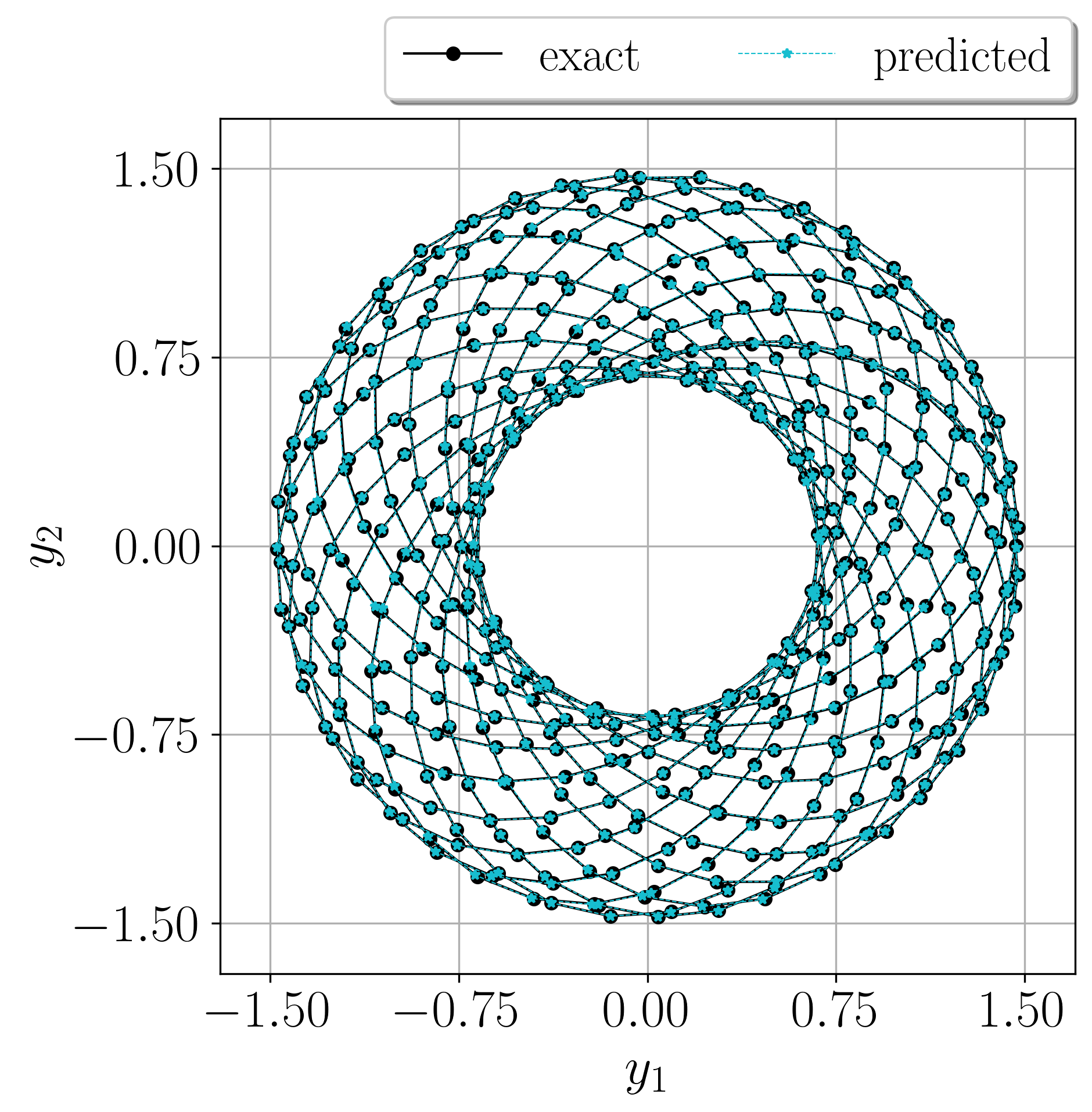}}\qquad
\subfigure[]{\label{fig:ChargeM1500}
\includegraphics[trim=0cm 0cm 0cm 0cm,clip=true,width=0.42\textwidth]{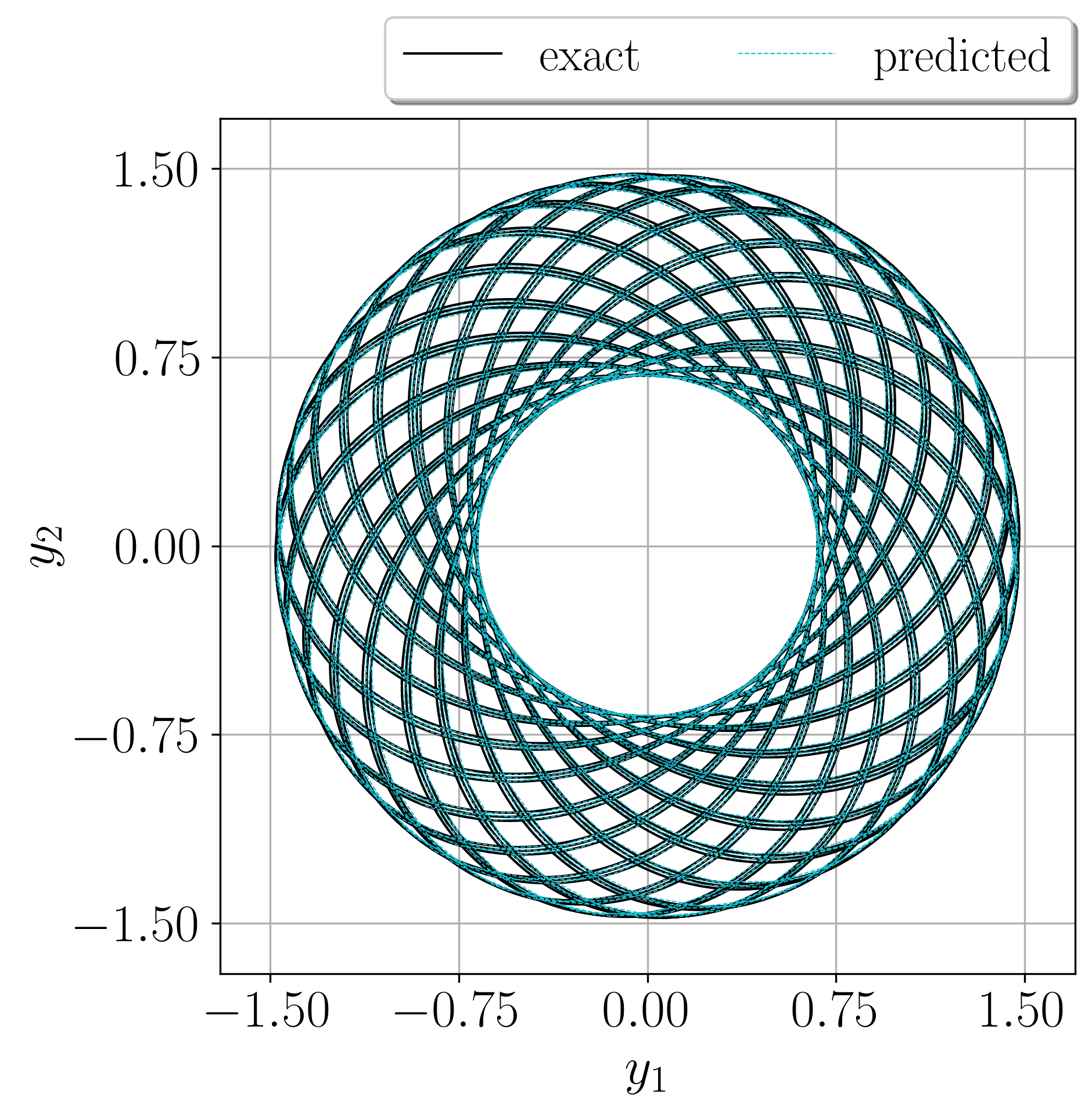}}
\caption{Reconstructed and predicted quasi-periodic trajectory of the motion of the charged particle by $\SLSNet$ with $K=2$ and $m=64$. Reconstructed solution on the time interval $[0,40]$ is computed with the initial condition $(\tilde{y}_0,\tilde{p}_0)^T$, while the predicted solution on the time interval $[0,T_{end}]$ is obtained with the initial condition $(\tilde{y}_{40},\tilde{p}_{40})^T$. (a) ground truth and reconstructed solution. (b) comparison between the exact and predicted solutions for $T_{end}=20$. (c) comparison between the exact and predicted solutions for $T_{end}=100$. (d) comparison between the exact and predicted solutions for $T_{end}=300$.}\label{fig:ChargeSol}
\end{figure}

To keep the presentation concise in Figures \ref{fig:ChargeRes}--\ref{fig:ChargeSol} we only visualize results obtained by the neural networks $\SLSNet$ while qualitatively similar results (not shown) were also obtained with $\LSNet$. In this example, symmetric locally-symplectic neural networks $\SLSNet$ were trained for $N_e= 10^6$ epochs using already stated batch Adam optimization algorithm with the exponential scheduling for the learning rate $\eta$. We considered $K=1$, $2$, $3$ network parameter values and three different network width values $m=32$, $64$, $128$. For each set of network parameter values, i.e., $K$ and $m$, in total we trained 20 neural networks with 20 different random initial weight values to produce and illustrate the averaged error results in Figure \ref{fig:ChargeRes}. Recall that in long-time predictions averaging is performed only over the number $n_0$ of the obtained stable solutions, see Figures \ref{fig:ChargePredS}--\ref{fig:ChargePredH}, i.e., predicted solution Hamiltonian \eqref{eq:ChargeH} relative errors are smaller than one.

In Figure \ref{fig:ChargeLoss} we show averaged MSE loss function \eqref{eq:loss} values, while in Figure \ref{fig:ChargeAcc} we demonstrate MSE accuracy function \eqref{eq:acc} values. Notice that the loss and accuracy values are ten and hundred times larger compared to the loss and accuracy values in learning the rigid body periodic solution, e.g., compare Figures \ref{fig:ChargeLoss} and \ref{fig:ChargeAcc} with Figures \ref{fig:RB_LossSym} and \ref{fig:RB_AccSym}. This may be attributed to the fact that the charged particle motion is quasi-periodic and this places additional challenges on the neural networks learning more general dynamics. Averaged absolute errors of the solution on the reconstruction time interval $[0,40]$ and on the prediction time interval $[0,1000]$ are illustrated in Figures \ref{fig:ChargeRecS} and \ref{fig:ChargePredS}, respectively. Linear growth of the averaged solution errors can be observed.

Averaged absolute values of the charged particle Hamiltonian \eqref{eq:ChargeH} relative errors are shown in Figures \ref{fig:ChargeRecH} and \ref{fig:ChargePredH}. On the reconstruction time interval, Figure \ref{fig:ChargeRecH}, the Hamiltonian \eqref{eq:ChargeH} errors are ten times larger compared to the rigid body kinetic energy \eqref{eq:KinEn} relative errors, compare the figure to Figure \ref{fig:RB_RecSymH}. On the prediction time interval $[0,1000]$ the averaged Hamiltonian conservation relative errors in Figure \ref{fig:ChargePredH} are around or below $10\%$ and exhibit gradual growth. Notice, as indicated by the numbers $n_0$ in the figure's legend, which is dependent on the length of the prediction time interval, that not all trained neural networks produced stable quasi-periodic solutions on the whole prediction time interval. Thus, indicating that in this numerical experiment only short-time predictions are physically reliable, which we visualize in Figures \ref{fig:ChargeM100}--\ref{fig:ChargeM1500}. 

In Figure \ref{fig:ChargeSol} we demonstrate performance of one trained neural network $\SLSNet$ with $K=2$ and $m=64$. In Figure \ref{fig:ChargeRec} we visualize reconstructed quasi-periodic trajectory of the motion of the charged particle on the time interval $[0,40]$. Observe excellent agreement between the ground truth solution and the neural network reconstructed solution from the initial condition $(\tilde{y}_0,\tilde{p}_0)^T$. Predicted quasi-periodic solution trajectory obtained iteratively by $\SLSNet$ from the initial condition $(\tilde{y}_{40},\tilde{p}_{40})^T$ is shown for three different time intervals $t\in[0,T_{end}]$, i.e., when $T_{end}=20$, $100$, $300$, see Figures \ref{fig:ChargeM100}--\ref{fig:ChargeM1500}, respectively.

In Figure \ref{fig:ChargeM100} notice very good agreement between the exact and predicted solutions on the validation time interval, which is the consequence of the small MSE accuracy function \eqref{eq:acc} values in Figure \ref{fig:ChargeAcc}. On longer prediction time interval $[0,100]$ we can already observe very small discrepancies between predicted and numerically computed exact solution of \eqref{eq:ChargeEq}, which can be attributed to the growth of the solution global error seen in Figure \ref{fig:ChargePredS}. Long-time predictions on the time interval $[0,300]$ are illustrated in Figure \ref{fig:ChargeM1500}. In Figure \ref{fig:ChargeM1500} we have removed markers from the visualization such that the qualitatively good predicted quasi-periodic motion of the charged particle can be appreciated.  

\begin{figure}[t]
\centering 
\subfigure[]{\label{fig:ChargeOrderLoss}
\includegraphics[trim=0cm 0cm 0cm 0cm,clip=true,width=0.32\textwidth]{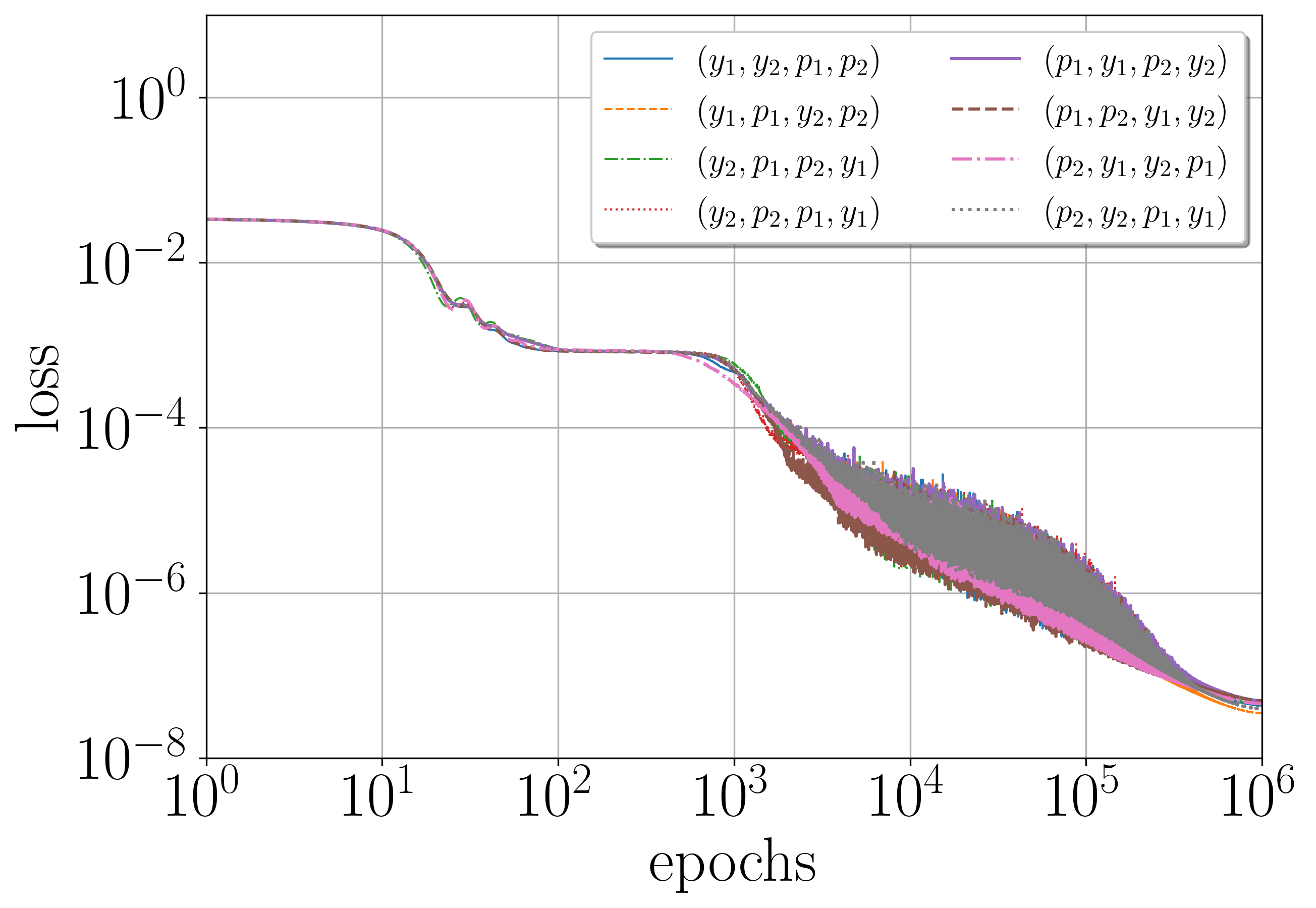}}
\subfigure[]{\label{fig:ChargeOrderRecS}
\includegraphics[trim=0cm 0cm 0cm 0cm,clip=true,width=0.32\textwidth]{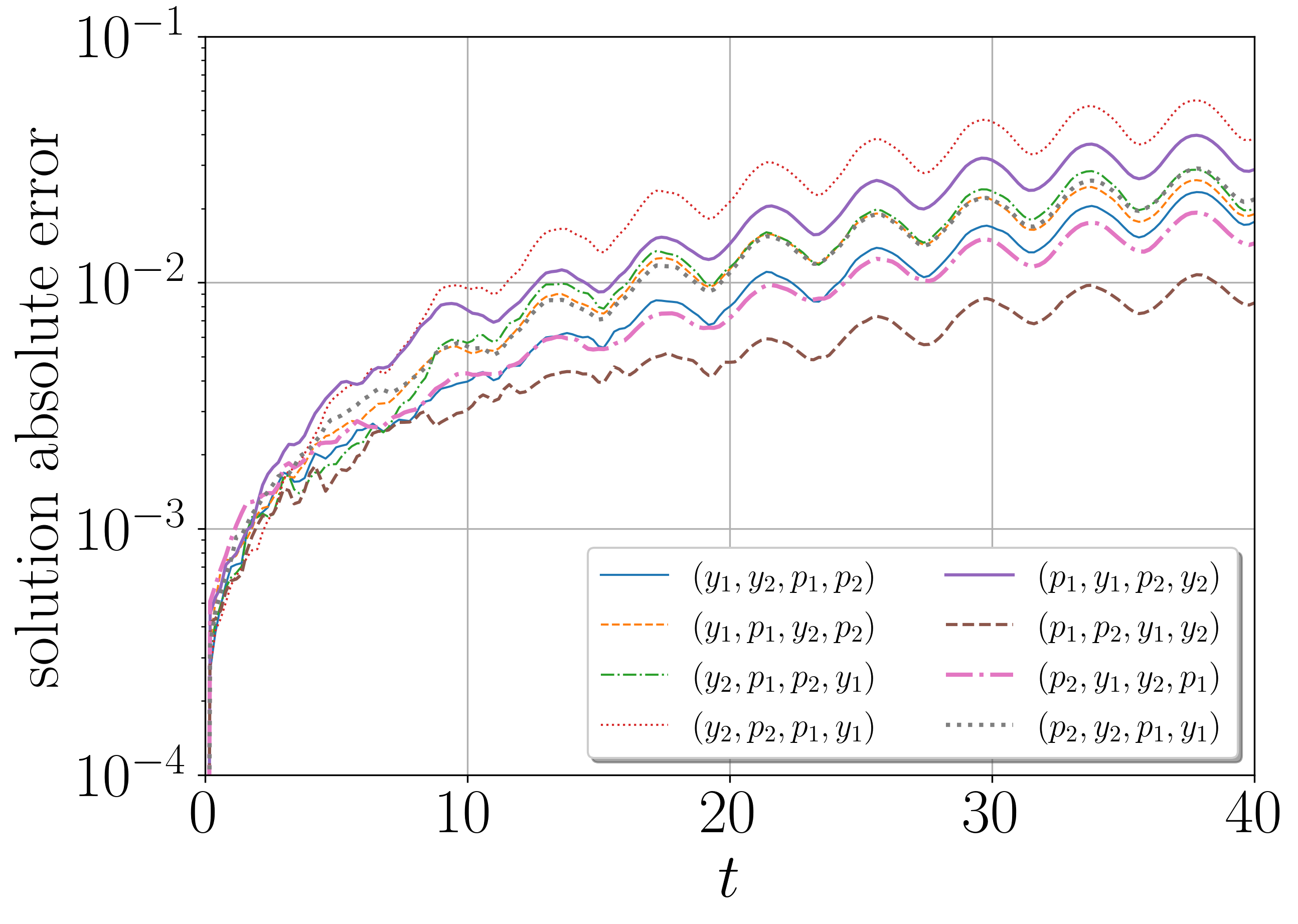}}
\subfigure[]{\label{fig:ChargeOrderRecH}
\includegraphics[trim=0cm 0cm 0cm 0cm,clip=true,width=0.32\textwidth]{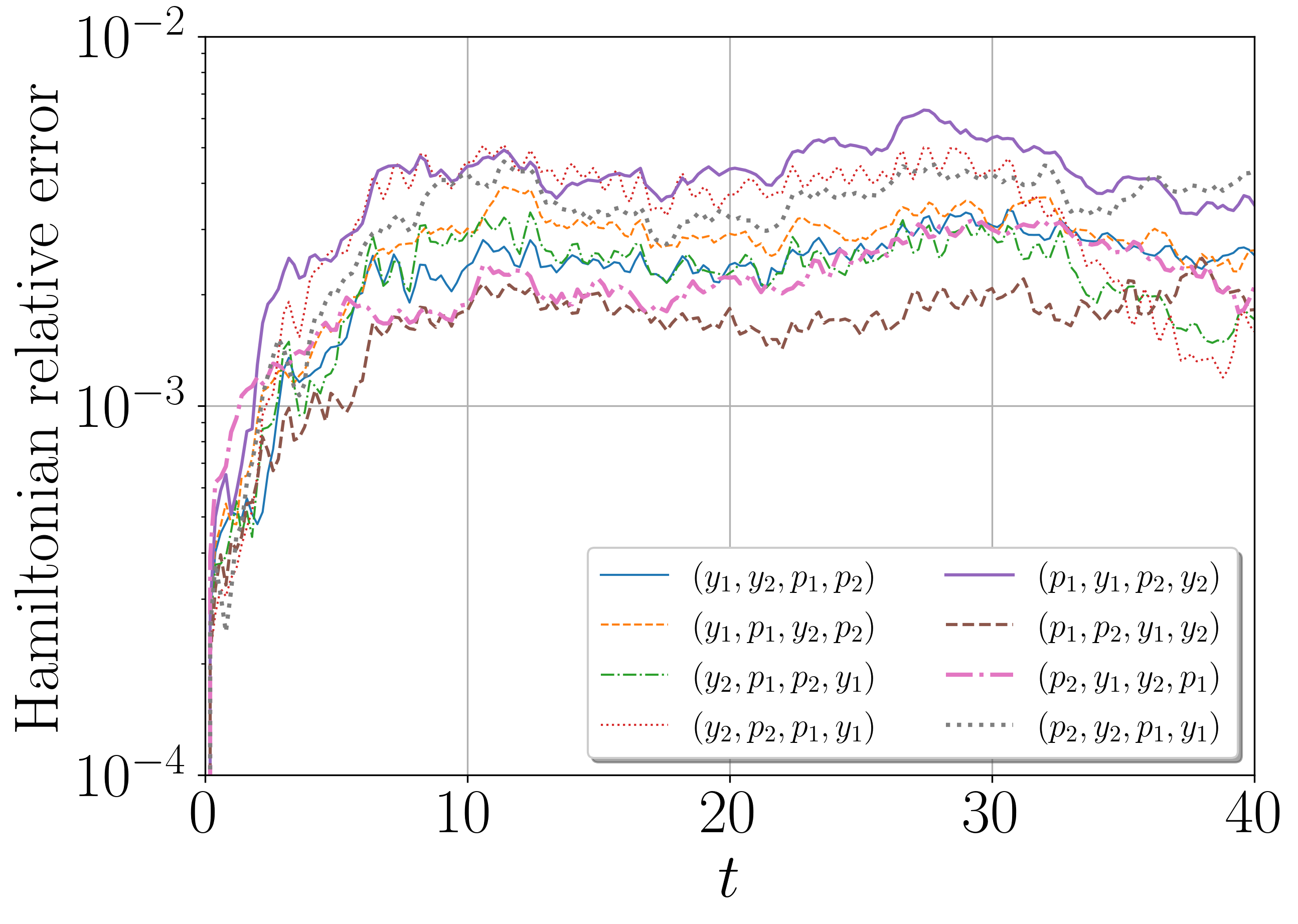}}
\subfigure[]{\label{fig:ChargeOrderAcc}
\includegraphics[trim=0cm 0cm 0cm 0cm,clip=true,width=0.32\textwidth]{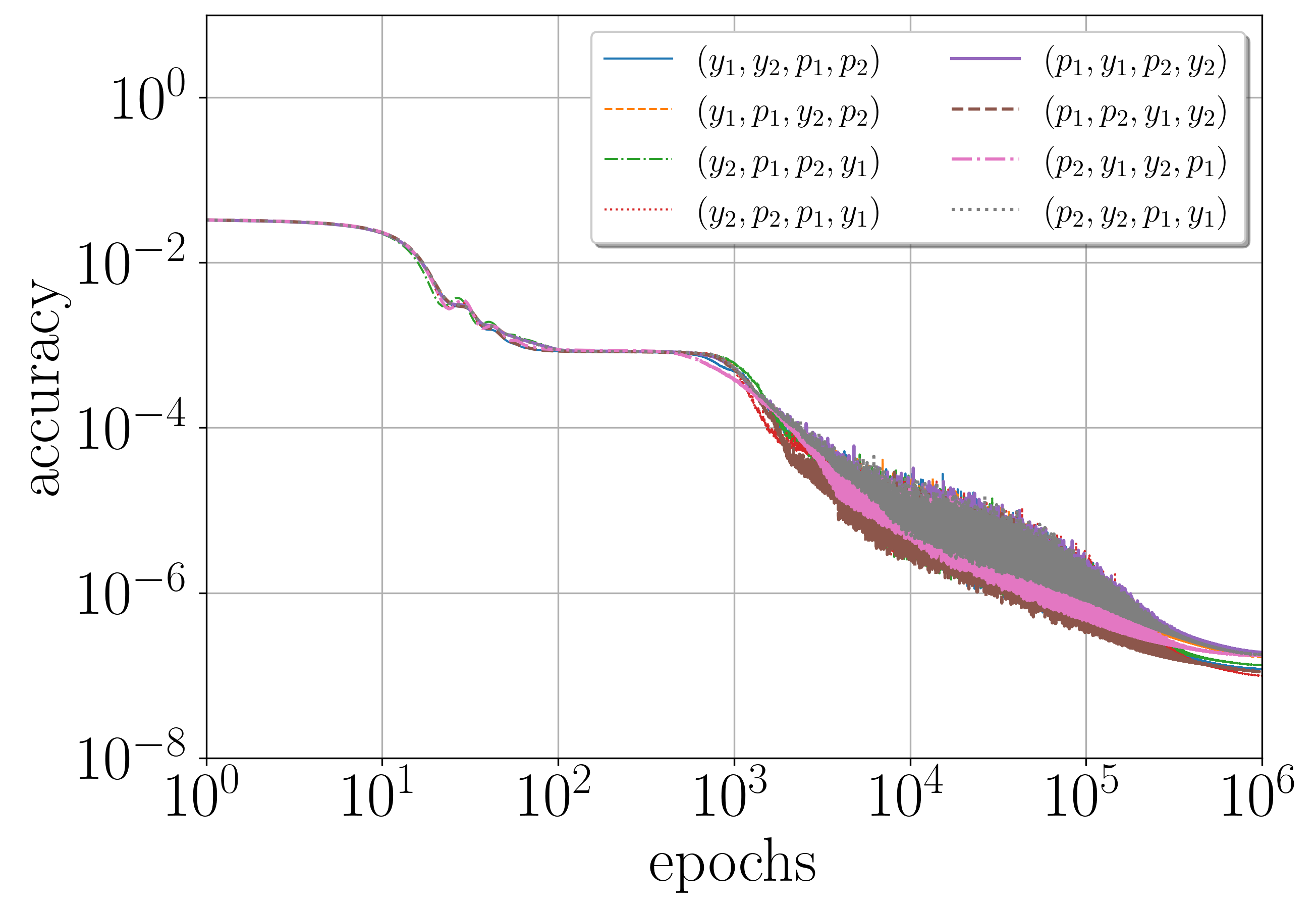}}
\subfigure[]{\label{fig:ChargeOrderPredS}
\includegraphics[trim=0cm 0cm 0cm 0cm,clip=true,width=0.32\textwidth]{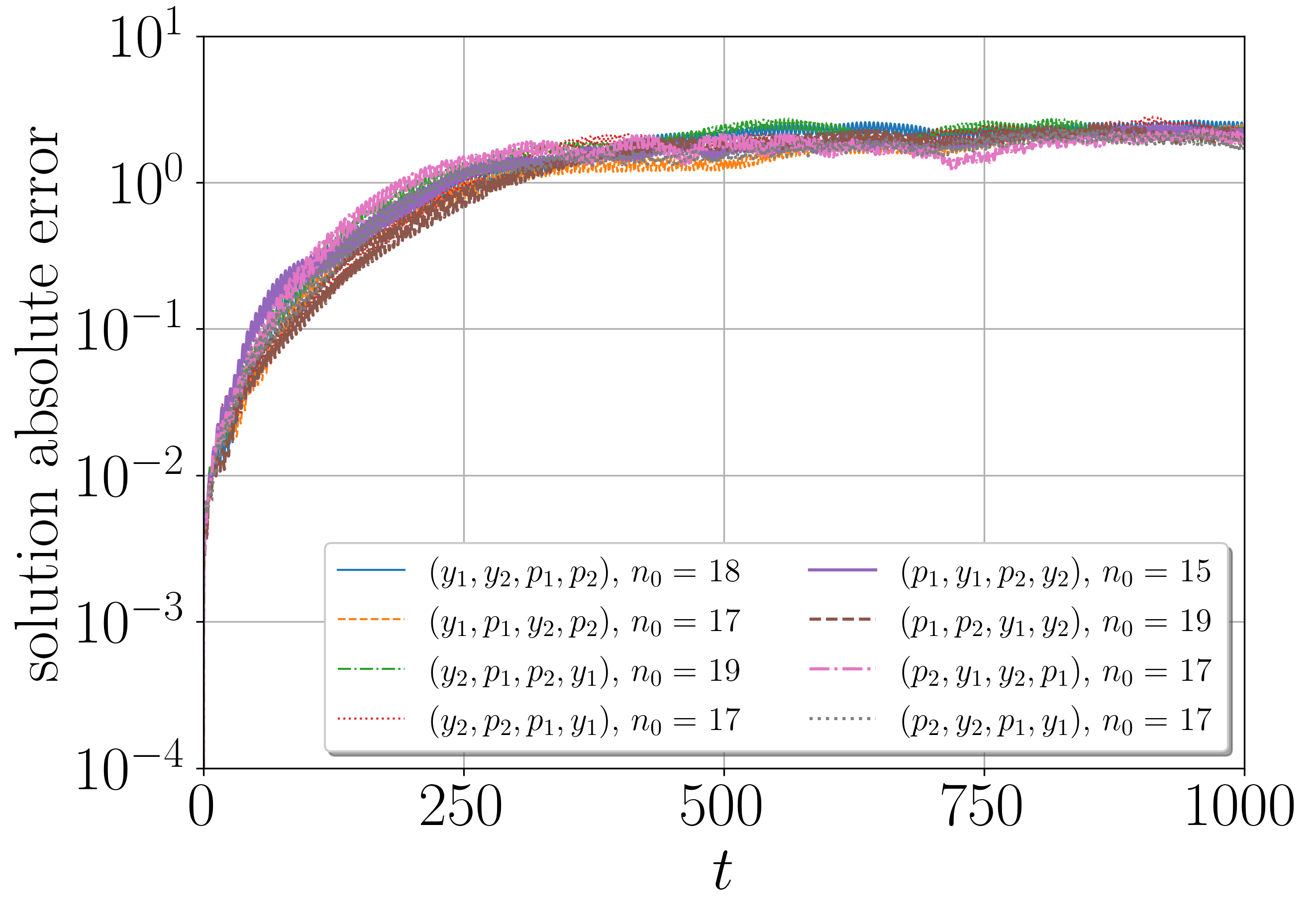}}
\subfigure[]{\label{fig:ChargeOrderPredH}
\includegraphics[trim=0cm 0cm 0cm 0cm,clip=true,width=0.32\textwidth]{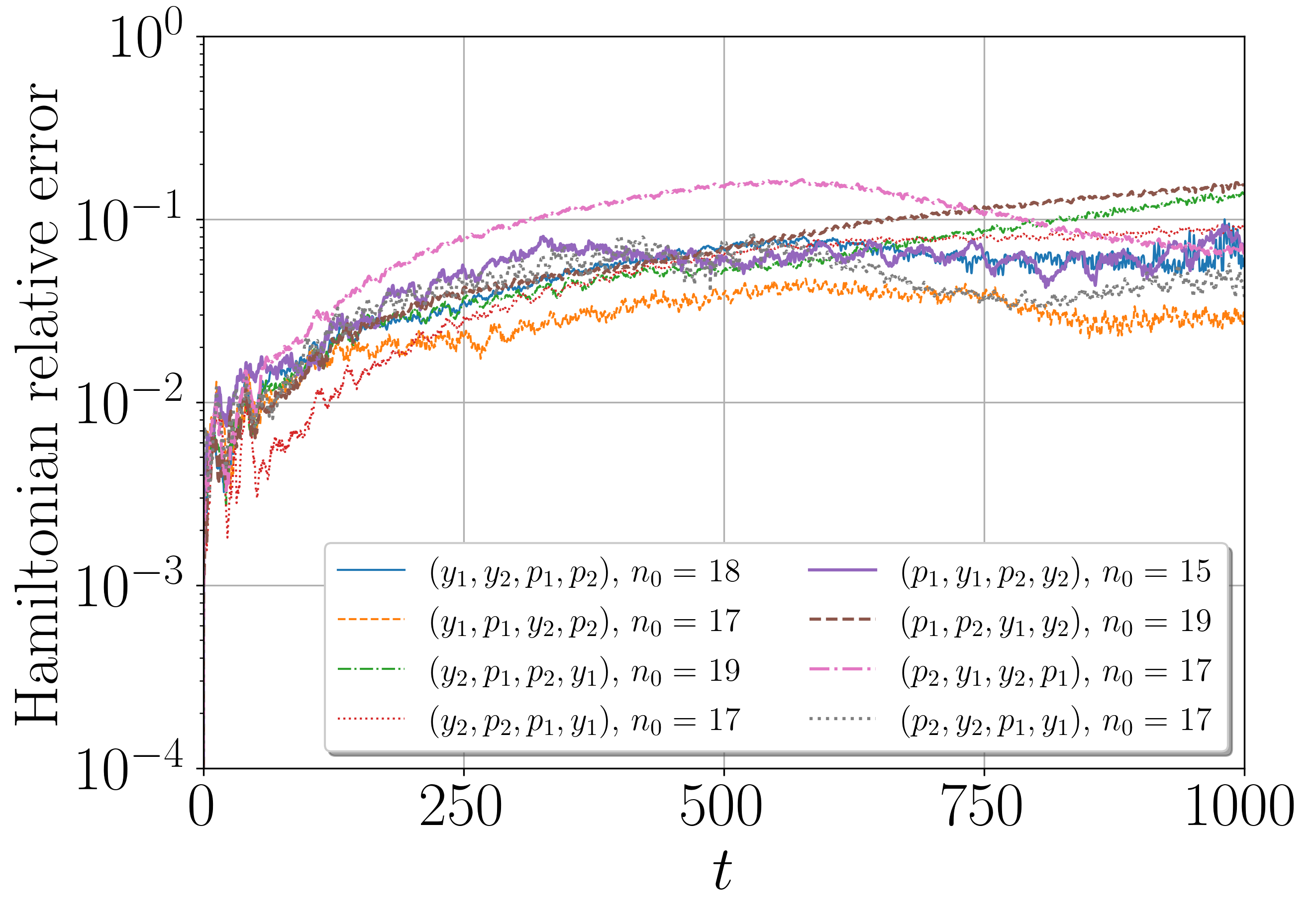}}
\caption{Numerical results of learning quasi-periodic motion of the charged particle in the electromagnetic field with $\SLSNet$ ($K=2$ and $m=64$) considering different order of variables indicated in legends. (a) averaged MSE loss \eqref{eq:loss} function values. (b) averaged absolute errors of the reconstructed solutions on the time interval $[0,40]$ with the initial condition $(\tilde{y}_{0},\tilde{p}_{0})^T$. (c) averaged Hamiltonian \eqref{eq:ChargeH} relative errors of the reconstructed solutions. (d) averaged MSE accuracy \eqref{eq:acc} function values. (e) averaged absolute errors of the predicted solutions on the long-time prediction time interval $[0,1000]$ with the initial condition $(\tilde{y}_{40},\tilde{p}_{40})^T$. (f) averaged Hamiltonian \eqref{eq:ChargeH} relative errors of the predicted solutions.}\label{fig:ChargeOrderRes}
\end{figure}

The flexibility of proposed locally-symplectic neural networks $\LSNet$ and $\SLSNet$ stems from the fact that even and odd dimensional phase volume-preserving dynamics \eqref{eq:ODE} can be learned and well-predicted, and the order of the variables is not essential. We validate this claim by considering the same training set-up used to produce numerical results of Figure \ref{fig:ChargeRes} but this time we vary the order of the vector $(y_1,y_2,p_1,p_2)^T$ components. In total, we consider 8 variations of the variables $y_1$, $y_2$, $p_1$, and $p_2$ as indicated by the legends in Figure \ref{fig:ChargeOrderRes}. For this experiment, we consider $\SLSNet$ with $K=2$ and $m=64$, and similarly to the experiments in Figure \ref{fig:ChargeRes} we train 20 different neural networks in each case with the same training and validation data sets. 

Averaged loss, accuracy, solution and the Hamiltonian \eqref{eq:ChargeH} errors are visualized in Figure \ref{fig:ChargeOrderRes}. Compare Figure \ref{fig:ChargeOrderRes} to Figure \ref{fig:ChargeRes}, and notice qualitative agreement between each subplot. The good agreement indicates that $\LSNet$ and $\SLSNet$ can learn dynamics independently of the system's dimension and variable order, and the order of variables does not affect the range of averaged reconstruction and prediction errors.  

\section{Discussion and conclusions} \label{sec:Conclusions}
In this work, we have proposed locally-symplectic volume-preserving neural networks $\LSNet$ for learning phase volume-preserving dynamics, which construction stems from the recently proposed symplectic neural networks $\SNet$ and the symplectic Euler numerical splitting methods. We have shown that the modules of the volume-preserving neural networks $\LSNet$ are locally-symplectic and their composition is phase volume-preserving. Considering that the locally-symplectic modules have efficiently computable inverse maps we have also extended $\LSNet$ to the symmetric locally-symplectic neural networks $\SLSNet$ such that the inverse of the neural network is equal to the feed-forward propagation of $\SLSNet$ with the negative time step, which is a general property of the flow of a dynamical system. We have numerically demonstrated that $\SLSNet$ outperform $\LSNet$ and from $\NICE$ adopted volume-preserving neural networks $\VPNet$, including the case of training with noisy data.     

Numerical validation of the proposed neural networks $\LSNet$ and $\SLSNet$ was performed considering linear and nonlinear phase volume-preserving dynamics. We have demonstrated learning of a dispersive wave solution of the linear semi-discretized advection equation, and, for learning nonlinear dynamics, we have considered two examples, i.e., the Euler equations of the periodic motion of a free rigid body and the charged particle quasi-periodic motion in an electromagnetic field governed by the Lorentz force. For the linear dynamics, learning objectives of linear volume-preserving neural networks, which follow from the construction, were discussed. While the performance of the linear $\LSNet$ and $\SLSNet$ for the linear dynamics is not optimal compared to the multivariate linear regression problem, locally-symplectic neural networks incorporate more information on the dynamics and allow training with non-constant time steps, if desired.

For the rigid body dynamics, three learning objectives were put forward, i.e., learning a single periodic trajectory, recovering periodic solutions in training with noisy data, and learning the whole dynamics from randomly sampled data. Neural networks $\LSNet$ and $\SLSNet$ performed exceptionally well in learning a single periodic trajectory since the neural networks were able to preserve both quadratic invariants of the dynamics up to high accuracy. Recall that the proposed neural networks do not preserve these invariants by design. We found that not all trained neural networks' produced predictions lead to good approximate kinetic energy conservation and, thus, to stable long-time predictions, but we have demonstrated quantitatively that the locally-symplectic neural networks on average produced qualitatively better long-time predictions compared to the volume-preserving neural networks $\VPNet$. In particular, $\SLSNet$ demonstrated the smallest solution errors as well as the smallest errors for both conserved quantities. We were able to draw the same conclusions considering learning of a single periodic trajectory with noisy data. We have found that volume-preserving neural networks are still able to produce good long-time predictions when a relative amount of noise has been added to the training data. In addition, we have observed that a large amount of noise in training data leads to reduced learning and fast-diverging predictions. When learning of the whole rigid body dynamics from irregularly sampled data was considered, $\SLSNet$ compared to $\LSNet$ produced significantly more accurate long-time predictions. The performance of $\SLSNet$ was also investigated by considering learning with training data sets sampled with different time step values, either with a fixed number of data points or on the fixed sampling time interval. In all cases, we found that $\SLSNet$ could produce qualitatively good long-time predictions for a large range of time step values as long as there is a sufficient dynamics representation in the training data set. 

We concluded numerical demonstrations with the second nonlinear problem, i.e., learning of the quasi-periodic motion of the charged particle in an electromagnetic field. We have found and demonstrated that quasi-periodic solutions are more difficult to learn for volume-preserving neural networks. Despite that, we have shown that $\SLSNet$ can produce qualitatively good short-time predictions. In addition, we have demonstrated that good results can be obtained independently of the order of variables, emphasizing, that $\LSNet$ and $\SLSNet$ can learn and well-predict phase volume-preserving dynamics being of even or odd dimension. After all performed numerical experiments, we can conclude that even greater challenges will be to obtain good predictions for the chaotic phase volume-preserving dynamics, and more research in this direction is required.   

For future work, we consider the following tasks. In this work, the proof of the universal approximation theorem was not provided. For sound theoretical justifications for using the proposed neural networks $\LSNet$ and $\SLSNet$ that would be very beneficial. Regarding learning linear systems we plan to investigate the trained matrices $B_1, B_2,\dots $, and how they are linked to the training data and the system's matrix $A$, and establish optimal $K$ and $m$ parameter values with respect to the system's dimension $n$, and when the non-constant time step $\tau$ values are considered. The choice of using the sigmoid function as the activation function was motivated by its use in \cite{SympNets}. Very good results, not shown, were also obtained with the Swish activation function and further investigation in this direction is required, either from a theoretical or numerical point of view. Current results demonstrate that the Adam optimization algorithm requires too many epochs to achieve desirable accuracy, which limits the full exploration of the proposed neural networks for learning high-dimensional dynamical systems. The use of a large number of epochs was also reported in \cite{SympNets,PoisNets} for learning Hamiltonian and Poisson dynamics. Thus, more research in faster converging optimization algorithms is still needed. We plan to explore how prior knowledge of $\rho$-reversibility property can be incorporated into the phase volume-preserving neural networks. Extensions of the proposed locally-symplectic neural networks $\LSNet$ and $\SLSNet$ for efficient learning of large-scale volume-preserving problems are highly desirable.

\section*{Acknowledgements}
J.~Baj\={a}rs acknowledges support from the lzp-2020/2-0267 grant funded by the Latvian Council of Science.

\bibliographystyle{unsrt}  
\bibliography{references}  

\begin{thebibliography}{10}

\bibitem{Hairer}
E.~Hairer, C.~Lubich, and G.~Wanner.
\newblock {\em Geometric numerical integration: structure-preserving algorithms
  for ordinary differential equations}.
\newblock Springer Science \& Business Media, 2006.

\bibitem{4ParaD}
T.~Hey, S.~Tansley, and K.M. Tolle.
\newblock {\em The Fourth Paradigm: Data-Intensive Scientific Discovery}.
\newblock Microsoft Research, Redmond, WA, 2009.

\bibitem{DDbook}
S.~Brunton and J.~Kutz.
\newblock {\em Data-Driven Science and Engineering: Machine Learning, Dynamical
  Systems, and Control}.
\newblock Cambridge: Cambridge University Press, 2019.

\bibitem{Montans19}
F.J. Montáns, F.~Chinesta, R.~Gómez-Bombarelli, and J.N. Kutz.
\newblock Data-driven modeling and learning in science and engineering.
\newblock {\em Comptes Rendus Mécanique}, 347:845--855, 2019.

\bibitem{Toth20}
P.~Toth, D.J. Rezende, A.~Jaegle, S.~Racanière, A.~Botev, and I.~Higgins.
\newblock Hamiltonian generative networks.
\newblock In {\em International Conference on Learning Representations}, 2020.

\bibitem{Bondesan19}
R.~Bondesan and A.~Lamacraft.
\newblock Learning symmetries of classical integrable systems.
\newblock In {\em ICML 2019 Workshop on Theoretical Physics for Deep Learning},
  6 2019.

\bibitem{Yang20}
S.~Yang, X.~He, and B.~Zhu.
\newblock Learning physical constraints with neural projections.
\newblock In {\em Proceedings of the 34th International Conference on Neural
  Information Processing Systems}, NIPS'20, Red Hook, NY, USA, 2020. Curran
  Associates Inc.

\bibitem{Sam19}
S.~Greydanus, M.~Dzamba, and J.~Yosinski.
\newblock Hamiltonian neural networks.
\newblock In H.~Wallach, H.~Larochelle, A.~Beygelzimer, F.~d\textquotesingle
  Alch\'{e}-Buc, E.~Fox, and R.~Garnett, editors, {\em Advances in Neural
  Information Processing Systems}, volume~32. Curran Associates, Inc., 2019.

\bibitem{SympNets}
P.~Jin, Z.~Zhang, A.~Zhu, Y.~Tang, and G.~Em Karniadakis.
\newblock Symp{N}ets: Intrinsic structure-preserving symplectic networks for
  identifying {H}amiltonian systems.
\newblock {\em Neural Networks}, 132:166--179, 2020.

\bibitem{PoisNets}
P.~Jin, Z.~Zhang, I.G. Kevrekidis, and G.~Em Karniadakis.
\newblock Learning {P}oisson systems and trajectories of autonomous systems via
  {P}oisson neural networks.
\newblock {\em IEEE Transactions on Neural Networks and Learning Systems},
  pages 1--13, 2022.

\bibitem{Zhong20}
Y.D. Zhong, B.~Dey, and A.~Chakraborty.
\newblock Symplectic {ODE}-{N}et: Learning {H}amiltonian dynamics with control.
\newblock In {\em International Conference on Learning Representations}, 2020.

\bibitem{Bertalan19}
T.~Bertalan, F.~Dietrich, I.~Mezi, and I.G. Kevrekidis.
\newblock On learning {H}amiltonian systems from data.
\newblock {\em Chaos}, 29:121107, 2019.

\bibitem{Xiong21}
S.~Xiong, Y.~Tong, X.~He, S.~Yang, C.~Yang, and B.~Zhu.
\newblock Nonseparable symplectic neural networks.
\newblock In {\em International Conference on Learning Representations}, 2021.

\bibitem{Weinan17}
E.~Weinan.
\newblock A proposal on machine learning via dynamical systems.
\newblock {\em Communications in Mathematics and Statistics}, 5:1--11, 2017.

\bibitem{Chen18}
T.Q. Chen, Y.~Rubanova, J.~Bettencourt, and D.K. Duvenaud.
\newblock Neural ordinary differential equations.
\newblock In {\em Advances in neural information processing systems}, pages
  6571--6583, 2018.

\bibitem{celledoni21}
E.~Celledoni, M.J. Ehrhardt, C.~Etmann, R.I. McLachlan, B.~Owren, C.-B.
  Schönlieb, and F.~Sherry.
\newblock Structure-preserving deep learning.
\newblock {\em European Journal of Applied Mathematics}, 32(5):888–936, 2021.

\bibitem{Haber17}
E.~Haber and L.~Ruthotto.
\newblock Stable architectures for deep neural networks.
\newblock {\em Inverse Problems}, 34(1):014004, 2017.

\bibitem{Chang18}
B.~Chang, L.~Meng, E.~Haber, L.~Ruthotto, D.~Begert, and E.~Holtham.
\newblock Reversible architectures for arbitrarily deep residual neural
  networks.
\newblock In {\em Proceedings of the Thirty-Second {AAAI} Conference on
  Artificial Intelligence, (AAAI-18), the 30th innovative Applications of
  Artificial Intelligence (IAAI-18), and the 8th {AAAI} Symposium on
  Educational Advances in Artificial Intelligence (EAAI-18), New Orleans,
  Louisiana, USA, February 2-7, 2018}, pages 2811--2818. {AAAI} Press, 2018.

\bibitem{Galimberti21}
C.L. Galimberti, L.~Furieri, L.~Xu, and G.~Ferrari-Trecate.
\newblock Hamiltonian deep neural networks guaranteeing non-vanishing gradients
  by design.
\newblock {\em arXiv preprint arXiv:2105.13205}, 2021.

\bibitem{MacDonald21}
G.~MacDonald, A.~Godbout, B.~Gillcash, and S.~Cairns.
\newblock Volume-preserving neural networks.
\newblock {\em arXiv preprint arXiv:1911.09576}, 2021.

\bibitem{Chen20}
Z.~Chen, J.~Zhang, M.~Arjovsky, and L.~Bottou.
\newblock Symplectic recurrent neural networks.
\newblock In {\em International Conference on Learning Representations}, 2020.

\bibitem{Zhu20}
A.~Zhu, P.~Jin, and Y.~Tang.
\newblock Deep {H}amiltonian networks based on symplectic integrators.
\newblock {\em Mathematica Numerica Sinica}, 42(3):370--384, 2020.

\bibitem{Tong20}
Y.~Tong, S.~Xiong, X.~He, G.~Pan, and B.~Zhu.
\newblock Symplectic neural networks in {T}aylor series form for {H}amiltonian
  systems.
\newblock {\em Journal of Computational Physics}, 437:110325, 2021.

\bibitem{Kadupitiya20}
J~C~S Kadupitiya, G.~C. Fox, and V.~Jadhao.
\newblock Solving {N}ewton's equations of motion with large timesteps using
  recurrent neural networks based operators.
\newblock {\em Machine Learning: Science and Technology}, 3(2):025002, 2022.

\bibitem{NICE}
L.~Dinh, D.~Krueger, and Y.~Bengio.
\newblock {NICE}: Non-linear independent components estimation.
\newblock {\em arXiv preprint arXiv:1410.8516}, 2014.

\bibitem{Feng95}
K.~Feng and Z.~Shang.
\newblock Volume-preserving algorithms for source-free dynamical systems.
\newblock {\em Numerische Mathematik}, 71:451--463, 1995.

\bibitem{Arnold}
V.I. Arnol'd.
\newblock {\em Mathematical Methods of Classical Mechanics}.
\newblock Springer-Verlag New York, 1989.

\bibitem{Zhu22}
A.~Zhu, P.~Jin, and Y.~Tang.
\newblock Approximation capabilities of measure-preserving neural networks.
\newblock {\em Neural networks: the official journal of the International
  Neural Network Society}, 147:72--80, 2022.

\bibitem{Xue14}
H.~Xue and A.~Zanna.
\newblock Generating functions and volume preserving mappings.
\newblock {\em Discrete \& Continuous Dynamical Systems}, 34(3):1229--1249,
  2014.

\bibitem{Adam}
D.P. Kingma and J.~Ba.
\newblock Adam: A method for stochastic optimization.
\newblock {\em arXiv preprint arXiv:1412.6980}, 2014.

\bibitem{Glorot10}
X.~Glorot and Y.~Bengio.
\newblock Understanding the difficulty of training deep feedforward neural
  networks.
\newblock In Yee~Whye Teh and Mike Titterington, editors, {\em Proceedings of
  the Thirteenth International Conference on Artificial Intelligence and
  Statistics}, volume~9 of {\em Proceedings of Machine Learning Research},
  pages 249--256, Chia Laguna Resort, Sardinia, Italy, 13-15 May 2010. PMLR.

\bibitem{Swish}
P.~Ramachandran, B.~Zoph, and Q.V. Le.
\newblock Swish: a self-gated activation function.
\newblock {\em arXiv preprint arXiv:1710.05941}, 2017.

\bibitem{LeVeque}
R.J. LeVeque.
\newblock {\em Finite Difference Methods for Ordinary and Partial Differential
  Equations: Steady-State and Time-Dependent Problems}.
\newblock SIAM, 2007.

\bibitem{Chen21}
R.-C. Chen and M.~Tao.
\newblock Data-driven prediction of general {H}amiltonian dynamics via learning
  exactly-symplectic maps.
\newblock {\em arXiv preprint arXiv:2103.05632}, 2021.

\end{thebibliography}

\end{document}